%% file: Tesis_main.tex
\documentclass[a4paper,12pt,twoside,openright]{book}
\usepackage{color}
\usepackage[utf8]{inputenc}
\usepackage[T1]{fontenc} 
\usepackage{lmodern}
\usepackage{aas_macros}

\usepackage[hang,flushmargin]{footmisc} 

\usepackage{bm}
\usepackage[skip=10pt,font=footnotesize,bf]{caption}
\usepackage{url}
\usepackage{amsmath,amssymb}
\usepackage[breaklinks,colorlinks,urlcolor=blue,citecolor=blue,linkcolor=red]{hyperref}
\usepackage{epsfig,graphicx,verbatim,xspace,multirow,mathtools}
\usepackage{booktabs}

\usepackage[british]{babel}
\usepackage{fancyhdr}
\usepackage{indentfirst}
\usepackage{epigraph}

\usepackage{CJKutf8}
\usepackage[export]{adjustbox}

\usepackage{comment}
\newcommand{\pablo}[1]{\textcolor{blue}{{\bf  #1}}}

\usepackage{array}

\usepackage[final]{pdfpages}

\def\eqref#1{{(\ref{#1})}}

\usepackage[hyperpageref]{backref}

\renewcommand*{\backref}[1]{}  
\renewcommand*{\backrefalt}[4]{
\ifcase #1 
No cited.
\or
{Cited on page} #2.
\else
{Cited on page} #2.
\fi}

\makeatletter
\def \cleardoublepage {\clearpage \if@twoside
\ifodd \c@page
\else
\null\thispagestyle{empty}\clearpage
\fi
\fi}
\makeatother

\renewcommand{\headrulewidth}{0.0pt}

\fancypagestyle{plain}{%
\fancyhead{} 
\fancyfoot[C]{\thepage}
\renewcommand{\headrulewidth}{0pt} 
}

\begin{document}
\sloppy	

\includepdf[pages=1,scale=1.3,pagecommand={}]{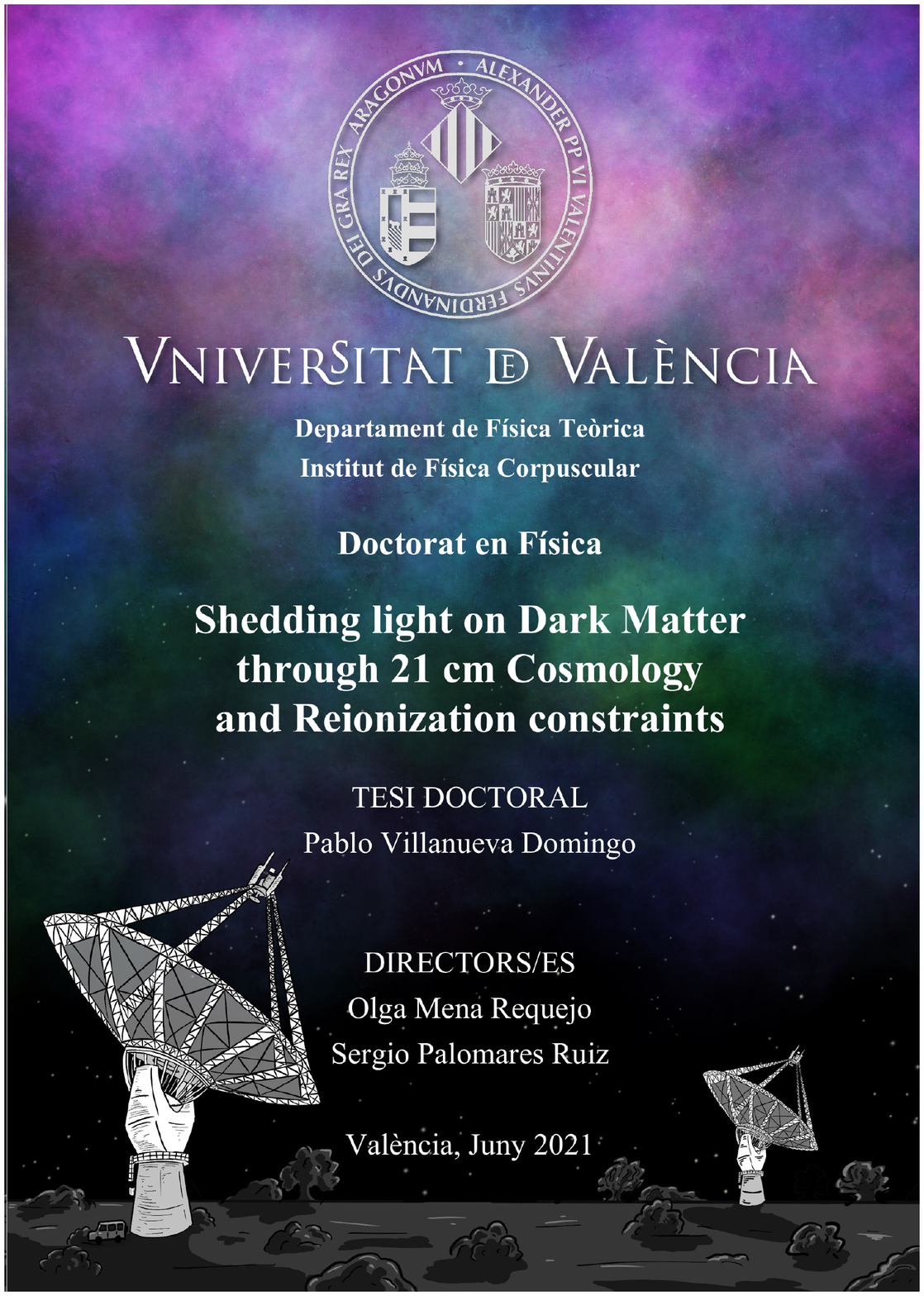}
\cleardoublepage

\include{Chapters/Others/PortadaInterior}

\include{Chapters/Others/Certifico}


\cleardoublepage
\vspace*{4cm} 
\hspace*{200pt} \\
\hspace*{250pt} \textit{A mis padres,} \\
\hspace*{250pt} \textit{a mis abuelos} \\
\hspace*{250pt} \textit{y a Inma.} \\
\cleardoublepage


\cleardoublepage
\vspace*{4cm} 
\hspace*{200pt} \\
\hspace*{200pt} \textit{…Y sentí vértigo y lloré,} \\
\hspace*{200pt} \textit{porque mis ojos habían visto} \\
\hspace*{200pt} \textit{ese objeto secreto y conjetural} \\
\hspace*{200pt} \textit{cuyo nombre usurpan los hombres,} \\
\hspace*{200pt} \textit{pero que ningún hombre ha mirado:} \\
\hspace*{200pt} \textit{el inconcebible universo.} \\
\hspace*{200pt} \\
\hspace*{200pt} ---\textsc{Jorge Luis Borges}, \textit{El Aleph} \\
\vspace*{4cm} 
\cleardoublepage

\include{Chapters/Others/Publications}

\input{Chapters/Others/Abbreviations}

\include{Chapters/Others/Preface}

\hypersetup{urlcolor=black}
\fancyfoot[C]{\thepage}
\include{Chapters/Others/Acknowledgements}
\hypersetup{urlcolor=blue}

\include{Chapters/Others/Contents}

\lhead[{\bfseries \thepage}]{ \rightmark}
\rhead[ Chapter \thechapter. \leftmark]{\bfseries \thepage}
\part{Introduction}\label{sec:theory}\thispagestyle{empty}
\label{partI}


\include{Chapters/Chapter_Intro/Chapter_Intro}


\include{Chapters/Chapter_DM/Chapter_DM}


\include{Chapters/Chapter_PBH/Chapter_PBH}


\include{Chapters/Chapter_21cm/Chapter_21cm}


\include{Chapters/Chapter_IGM/Chapter_IGM}



\cleardoublepage

\phantomsection

\addcontentsline{toc}{chapter}{{Bibliography}}

\renewcommand{\headrulewidth}{0.5pt}

\lhead[{\bfseries \thepage}]{Bibliography}
\rhead[{Bibliography}]{\bfseries \thepage}

\bibliographystyle{jhep}
\bibliography{biblio}


\setcounter{part}{1}
\part{Scientific Research}
\label{partII}\thispagestyle{empty}
\setcounter{chapter}{0}


\include{Chapters/Others/Summary}

\renewcommand{\headrulewidth}{0pt}

\vspace*{5cm}
The remaining of this part in the original edition includes the articles as they were published in the journals. Only the first pages of each publication are kept for this version. The complete texts can be found in the \href{https://roderic.uv.es/handle/10550/79956}{official thesis repository} of the University of València, as well as in Refs. \cite{Lopez-Honorez:2017csg, Villanueva-Domingo:2017lae, Villanueva-Domingo:2017ahx, Escudero:2018thh, Lopez-Honorez:2018ipk, Mena:2019nhm, 2021ApJ...907...44V}.

\include{Chapters/Papers/Papers}

\include{Chapters/Resumen_Tesis/Resumen_Tesis}

\cleardoublepage
\thispagestyle{empty}
\renewcommand{\headrulewidth}{0pt}
$\,$
\cleardoublepage

\end{document}

%% file: Chapters/Others/PortadaInterior.tex

\setlength{\unitlength}{1cm} 
\thispagestyle{empty}
\begin{center}

\begin{center}
\includegraphics[scale=0.08]{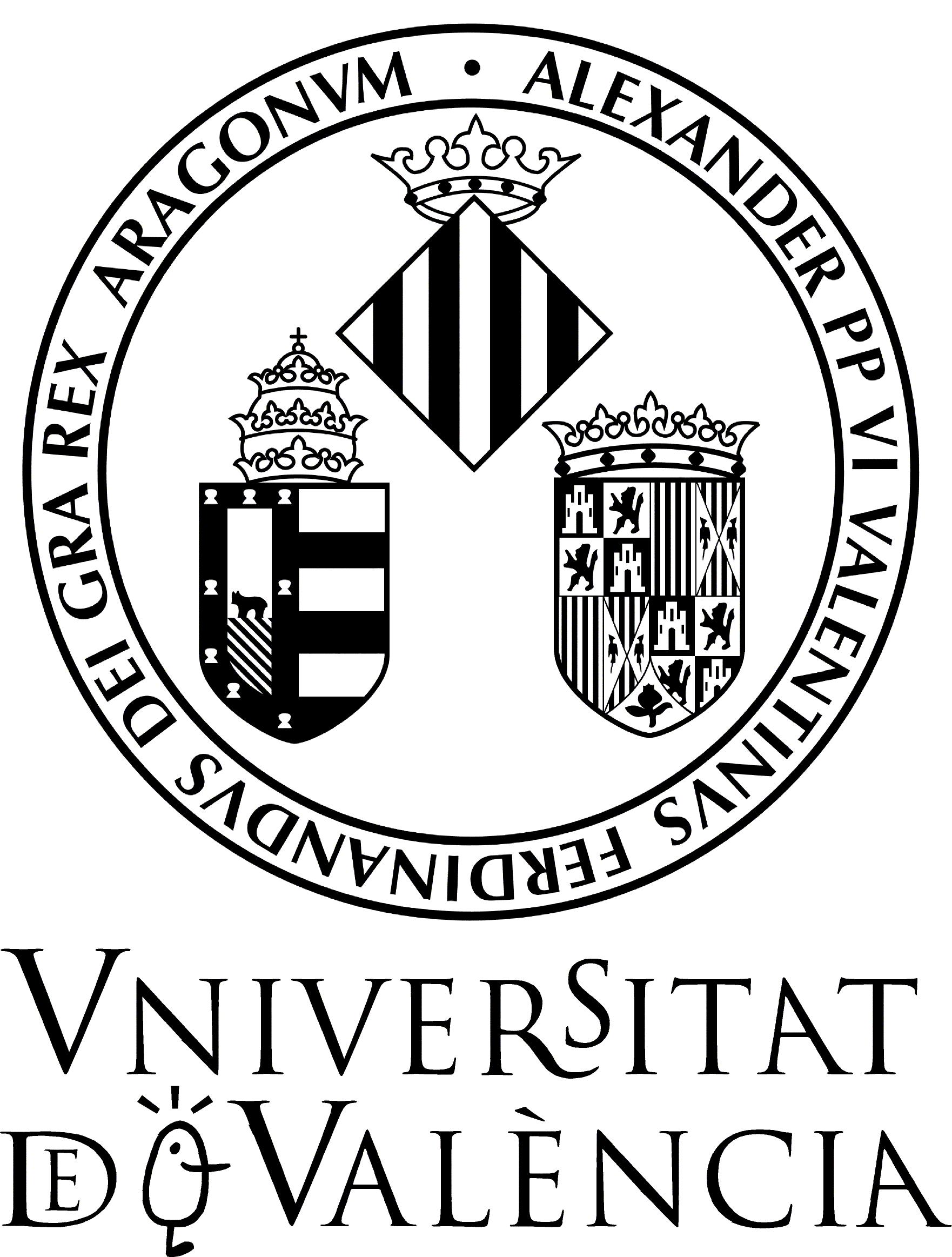}
\end{center}

\vspace{1cm}

{\Large \bf{Shedding light on Dark Matter}} \\[2ex]
{\Large \bf{through 21 cm Cosmology}} \\[2ex]
{\Large \bf{and Reionization constraints}} \\[2ex]

\vspace{0.5cm}

{\Large PhD Thesis}

\vspace{2cm}

\textbf{\Large{Pablo Villanueva Domingo}}\\ 

\vspace{0.5cm}

{IFIC - Universitat de Val\`{e}ncia - CSIC}\\
{Departament de Física Teòrica}\\
{Programa de Doctorat en Física}\\

\vspace{1cm}

\textbf{Under the supervision of}\\[3ex]

{\large \textbf{Olga Mena Requejo}} \\
{\textbf{and}} \\
{\large \textbf{Sergio Palomares Ruiz}} \\

\vspace{1.5cm}

{\large \bf{Val\`{e}ncia, June 2021}}

\end{center}

%% file: Chapters/Others/Certifico.tex
\begin{titlepage}
\cleardoublepage
\thispagestyle{empty}

\thispagestyle{empty}
\vspace*{3cm}

\noindent \textbf{Olga Mena Requejo},\\
investigadora científica del Consejo Superior de Investigaciones Cient\'ificas, y \vspace{0.2cm}

\noindent \textbf{Sergio Palomares Ruiz},\\
cient\'ifico titular del Consejo Superior de Investigaciones Cient\'ificas, \vspace{0.2cm}

\noindent \textbf{Certifican}:\\[2ex]
\noindent Que la presente memoria, \textbf{Shedding light on Dark Matter through 21 cm Cosmology and Reionization constraints}, ha sido realizada bajo su direcci{\'o}n en el Instituto de F{\'i}sica Corpuscular, centro mixto de la Universitat de València y del CSIC, por \textbf{Pablo Villanueva Domingo}, y constituye su Tesis para optar al grado de Doctor en F\'isica.\\[2ex]
\noindent Y para que as\'i conste, en cumplimiento de la legislaci\'on vigente, presenta en el Departamento de F\'isica Te\'orica de la Universidad de Valencia la referida Tesis Doctoral, y firman el presente certificado.\\[4ex]

\noindent València, a 29 de Abril de 2021,\\[12ex]

\noindent \hspace{2cm} Olga Mena Requejo \hspace{2cm} Sergio Palomares Ruiz

\newpage
\thispagestyle{empty}
$$ $$
\newpage
\thispagestyle{empty}

\end{titlepage}

%% file: Chapters/Others/Publications.tex
\frontmatter
\fancyfoot[C]{\thepage}
\chapter*{List of Publications}
\addcontentsline{toc}{chapter}{List of Publications}
\markboth{List of Publications}{List of Publications}

This PhD thesis is based on the following publications: \\

\begin{itemize}

\item Laura Lopez-Honorez, Olga Mena, Sergio Palomares-Ruiz and Pablo Villanueva-Domingo. \newline \textit{Warm dark matter and the ionization history of the Universe} \cite{Lopez-Honorez:2017csg}, \newline \href{https://journals.aps.org/prd/abstract/10.1103/PhysRevD.96.103539}{Phys. Rev., D96(10):103539, 2017.}

\item Pablo Villanueva-Domingo, Nickolay Y. Gnedin, and Olga Mena. \newline \textit{Warm Dark Matter and Cosmic Reionization} \cite{Villanueva-Domingo:2017lae}, \newline \href{https://iopscience.iop.org/article/10.3847/1538-4357/aa9ff5}{Astrophys. J., 852(2):139, 2018.}

\item Pablo Villanueva-Domingo, Stefano Gariazzo, Nickolay Y. Gnedin and Olga Mena. \newline \textit{Was there an early reionization component in our universe?} \cite{Villanueva-Domingo:2017ahx}, \newline  \href{https://iopscience.iop.org/article/10.1088/1475-7516/2018/04/024}{JCAP, 1804(04):024, 2018.}

\item Miguel Escudero, Laura Lopez-Honorez, Olga Mena, Sergio Palomares-Ruiz and Pablo Villanueva-Domingo. \newline \textit{A fresh look into the interacting dark matter scenario} \cite{Escudero:2018thh}, \newline \href{https://iopscience.iop.org/article/10.1088/1475-7516/2018/06/007}{JCAP, 1806(06):007, 2018.}

\item Laura Lopez-Honorez, Olga Mena and Pablo Villanueva-Domingo. \newline \textit{Dark Matter microphysics and 21 cm observations} \cite{Lopez-Honorez:2018ipk}, \newline \href{https://journals.aps.org/prd/abstract/10.1103/PhysRevD.99.023522}{Phys. Rev., D99(2):023522, 2019.}

\item Olga Mena, Sergio Palomares-Ruiz, Pablo Villanueva-Domingo, and Samuel J. Witte. \newline \textit{Constraining the primordial black hole abundance with 21-cm cosmology} \cite{Mena:2019nhm}, \newline \href{https://journals.aps.org/prd/abstract/10.1103/PhysRevD.100.043540}{Phys. Rev., D100(4):043540, 2019.}

\item Pablo Villanueva-Domingo and Francisco Villaescusa-Navarro. \newline \textit{Removing Astrophysics in 21 cm maps with Neural Networks} \cite{2021ApJ...907...44V}, \newline \href{https://iopscience.iop.org/article/10.3847/1538-4357/abd245}{The Astrophysical Journal, 907(1):44, 2021.}

\end{itemize}

Other works developed during the PhD but not included in the thesis:

\begin{itemize}

\item Samuel Witte, Pablo Villanueva-Domingo, Stefano Gariazzo, Olga Mena and Sergio Palomares-Ruiz. \newline \textit{EDGES result versus CMB and low-redshift constraints on ionization histories} \cite{Witte:2018itc}, \newline \href{https://journals.aps.org/prd/abstract/10.1103/PhysRevD.97.103533}{Phys. Rev., D97(10):103533, 2018.}

\item Pablo Villanueva-Domingo, Olga Mena and Jordi Miralda-Escudé. \newline \textit{ Maximum amplitude of the high-redshift 21-cm absorption feature} \cite{Villanueva_Domingo_2020},  \newline \href{https://journals.aps.org/prd/abstract/10.1103/PhysRevD.100.043540}{Phys. Rev. D101(8):083502, 2020.} 

\item Laura Lopez-Honorez, Olga Mena, Sergio Palomares-Ruiz, Pablo Villanueva-Domingo and Samuel J. Witte. \newline \textit{Variations in fundamental constants at the cosmic dawn} \cite{2020JCAP...06..026L}, \newline \href{https://iopscience.iop.org/article/10.1088/1475-7516/2020/06/026}{JCAP, 2006(06):026, 2020.} 

\item Pablo Villanueva-Domingo, Olga Mena and Sergio Palomares-Ruiz. \newline \textit{A brief review on primordial black holes as dark matter} \cite{Villanueva-Domingo:2021spv}, \newline
\href{https://www.frontiersin.org/articles/10.3389/fspas.2021.681084/full}{Front. Astron. Space Sci., 28 May 2021}

\item Pablo Villanueva-Domingo and Kiyotomo Ichiki. \newline \textit{21 cm Forest Constraints on Primordial Black Holes} \cite{Villanueva-Domingo:2021cgh}, \newline \href{https://arxiv.org/abs/2104.10695}{arXiv:2104.10695}

\end{itemize}

%% file: Chapters/Others/Abbreviations.tex
\chapter*{Abbreviations}
\markboth{Abbreviations}{Abbreviations}
\addcontentsline{toc}{chapter}{Abbreviations}  

\def\arraystretch{1.5}
\begin{tabular}{|c|c|}
\hline
Acronym & Meaning \\
\hline
$\gamma$DM & Photon Interacting Dark Matter \\
$\Lambda$CDM & $\Lambda$ - Cold Dark Matter model   \\
$\nu$DM & Neutrino Interacting Dark Matter \\
AGN & Active Galactic Nucleus \\
BAO & Baryonic Acoustic Oscillations \\
BBN & Big Bang Nucleosynthesis \\
BH & Black Hole \\
BSM & Beyond the Standard Model \\
CDM & Cold Dark Matter \\
CL & Confidence Level \\
CMB & Cosmic Microwave Background \\
DM & Dark Matter \\
EoR & Epoch of Reionization \\
FDM & Fuzzy Dark Matter \\
FLRW & Friedmann, Lemaître, Robertson, Walker \\
FZH & Furlanetto, Zaldarriaga, Hernquist \\
GP & Gunn, Peterson \\
H & Hydrogen \\
He & Helium \\
HeI & Neutral Helium \\
HeII & Single ionized Helium \\
\hline
\end{tabular}

\def\arraystretch{1.5}
\begin{tabular}{|c|c|}
\hline
Acronym & Meaning \\
\hline
HeIII & Double ionized Helium \\
HI & Neutral Hydrogen \\
HII & Ionized Hydrogen \\
IDM & Interacting Dark Matter \\
IGM & Intergalactic Medium \\
LSS & Large Scale Structure \\
LL & Lyman Limit \\
Ly$\alpha$ & Lyman $\alpha$ transition \\
MACHO & Massive Astrophysical Compact Halo Object \\
MHR & Miralda-Escudé, Haehnelt, Rees \\
MW & Milky Way \\
NFW & Navarro, Frenk, White \\
PBH & Primordial Black Hole \\
PS & Press, Schechter \\
QSO & Quasi-Stellar Object \\
SIDM & Self-Interacting Dark Matter \\
SM & Standard Model of particle physics \\
ST & Sheth, Tormen \\
WF & Wouthuysen, Field \\
WIMP & Weakly Interacting Massive Particle \\
\hline
\end{tabular}

%% file: Chapters/Others/Preface.tex
\chapter*{Preface}
\markboth{Preface}{Preface}
\addcontentsline{toc}{chapter}{Preface} 

During the last decades, our understanding of the universe has reached a remarkable level, being able to test cosmological predictions with astonishing precision. Observations of relic photons of the Cosmic Microwave Background (CMB), together with galaxy surveys, provide us with a deep comprehension of the geometry, components and chronology of the cosmos. Nonetheless, the nature of the Dark Matter (DM) still remains unknown. In this doctoral thesis, signatures of DM candidates which can leave an impact on the process of formation of structures and on the evolution of the Intergalactic Medium (IGM) are studied. The analysis of the state of ionization of the IGM, its impact on the CMB, and specially the 21 cm cosmological signal, can provide insightful information regarding the properties of the DM.

This thesis is organized in three parts. Part \ref{partI} is devoted to a broad introduction to the fundamentals describing the state of the art of the topics considered. The basics of the standard cosmological model and an overview of the cosmic timeline is presented in Chapter \ref{chap:Introduction}. In Chapter \ref{chap:DarkMatter}, the status and small-scale issues of the Cold Dark Matter (CDM) paradigm are discussed, together with the physics of two alternative non-standard DM scenarios: Warm Dark Matter (WDM) and Interacting Dark Matter (IDM). Chapter \ref{chap:PBHs} examines Primordial Black Holes (PBHs) as another DM candidate, focusing on their physical properties and observational effects. The details of the 21 cm cosmological signal and its current experimental bounds are reviewed in Chapter \ref{chap:21cm}. Finally, Chapter \ref{chap:IGM} is dedicated to the treatment of the IGM and the evolution of its ionization and thermal state. Part \ref{partII} includes the original scientific articles published during the development of the PhD, which constitute the main work of this thesis. Finally, Part \ref{partIII} contains a summary of the main results in Spanish.

%% file: Chapters/Others/Acknowledgements.tex
\chapter*{Acknowledgements}
\addcontentsline{toc}{chapter}{Acknowledgements}

Querría comenzar estos agradecimientos por mis directores, Olga y Sergio, sin cuya guía esta tesis no habría sido posible. Gracias, Olga, por estar siempre disponible cuando lo necesitaba. Por tu atención hacia mi bienestar, tanto en el trabajo como en lo personal. Por ponerte a trabajar codo con codo conmigo, especialmente al principio, y por aportarme las herramientas para ser capaz de trabajar autónomamente. Por tu trabajo incansable y tu pasión por lo que haces, que sin duda siempre me has contagiado. Por haber apostado siempre por mí. Gracias, Sergio, por tu rigor y meticulosidad en el trabajo, por tu atención al detalle. Por las extensas discusiones sobre física que hemos tenido, ahondando con profundidad en temas a veces pasados por alto, aportándome siempre una perspectiva diferente. Por tus minuciosas y concienzudas correcciones de los artículos y de la tesis, que siempre me animaban a dar lo máximo de mí. Por tus consejos y por tu entusiasmo por la física. Solo espero que se me haya quedado algo de estas cualidades. Ha sido un auténtico placer trabajar con los dos durante estos años. Esta tesis existe gracias a vuestro apoyo y vuestra guía.

During the development of this PhD, I have had the great pleasure of visiting several wonderful places. I would like to thank the people of Fermilab, and specially Nick Gnedin, for hosting me. Thank you, Nick, for your advice and expertise, for your orientation and for encouraging me to work autonomously. Estic molt agraït també a Jordi Miralda Escudé per les estades a la Universitat de Barcelona. Ha sigut un enorme plaer discutir de física amb tu i aprendre del teu coneixement. Gràcies també a la gent del ICCUB que vau fer la meua estada tan agradable, especialment a Raphael Sadoun, Albert Sanglas i Andreu Arinyo. I owe a great debt of gratitude to Masahiro Takada-san for the hospitality at the Kavli IPMU. And also for introducing me to Kiyotomo Ichiki-san, to who I am really grateful for inviting me to the Nagoya University, and starting a fruitful collaboration. \begin{CJK}{UTF8}{min}どうもありがとうございます\end{CJK}. Querría agradecer a Paco Villaescusa Navarro por su acogida en la Universidad de Princeton y en el CCA. Gracias por iniciarme en el fascinante mundo de las redes neuronales, y por nuestras absorbentes y esperadas reuniones, donde siempre aprendo algo nuevo. Estoy muy agradecido también a Laura Lopez Honorez por la invitación a la Universidad de Bruselas. Ha sido un placer poder trabajar contigo, aprender de ti y contagiarme de tu pasión por la física. And also thanks to the people of the ULB theory department, specially Matteo Lucca, Deanna Hopper and Rupert Coy, who made my stay really pleasing with a lot of coffee and chess games.

I would also like to thank my collaborators that I had the pleasure to work with: Stefano Gariazzo, Miguel Escudero and Sam Witte, together with the aforementioned Laura Lopez Honorez, Nick Gnedin, Jordi Miralda Escudé, Francisco Villaescusa Navarro and Kiyotomo Ichiki. Thank you all for your orientation, for the hard work and for sharing your experience and knowledge with me. I am really pleased to have had the opportunity of working with all of you.

Querría agradecer a la gente del grupo SOM por las fascinantes discusiones sobre física, tanto en los \textit{meeting groups} como en las comidas de los viernes, donde tanto he aprendido: Nuria Rius, Pilar Hernández, Andrea Donini, Carlos Peña, Jordi Salvadó, Sam Witte, Daniel G. Figueroa, Jacobo López y Verónica Sanz. Y por supuesto, al resto de estudiantes del grupo por las experiencias compartidas durante estos años: Miguel Escudero, Héctor Ramírez, Miguel Folgado, Fer Romero,  Andrea Caputo, Víctor Muñoz, Stefan Sandner y David Albandea.

Quiero agradecer a toda la gente que he tenido la oportunidad de conocer (o conocer mejor) durante los años de doctorado. Doy las gracias a Miguel Escudero y a Héctor Ramírez por vuestro apoyo en el doctorado y por las aventuras por Fermilab. Gràcies a Clara Murgui per ser una companya de despatx estupenda (encara que mai es volguera apuntar a berenar). Gracias a Brais y Lydia por haberme alegrado tantos momentos durante y después de la cuarentena. A Andrés \textit{Posete} por tu curiosidad y tus incisivas preguntas sobre física. Obrigado Eduardo da Silva pelas conversas em português. Gracias a Víctor Muñoz por las discusiones de física, las cervezas, y por iniciarme en la fascinante cultura chilena. A Stefan, mi guiri preferido (aunque ya más valenciano que alemán), el perfecto compañero de hamburguesas y meriendas. A Juan \textit{Lope} por nuestros locos proyectos artísticos (aunque a menudo no vayan a ninguna parte) y por avivar siempre mi lado creativo. A Miguel Folgado, por ser mi compañero de doctorado y de aventuras, por el inolvidable viaje a Japón y por los buenos momentos en el despacho y fuera de él, que sin duda continuaremos cuando lo abandonemos.

Quiero agradecer a todos los amigos que conocí en la carrera, y me han acompañado durante todos estos años, especialmente a Paco, Jaime, Bea, Dani, Laura, Álex, Kevin, Consuelo y Arturo. Gracias por vuestra amistad, ya fuese en el campus, en el camping o en los \textit{skypes}. Aunque hayamos tomado diferentes caminos en nuestras vidas, lo que ha unido la Física no lo ha separado el tiempo. A Dani y Cristian, por los conciertos y vuestra cálida acogida en la \textit{torre Eiffel} durante mis meses en Barcelona. A Jose y Álvaro por todas las risas y porque rodemos muchos cortos más. Y a Pablo, por ser un amigo con quien siempre he podido contar, reírme por algún motivo y evolucionar como persona.

No puedo olvidarme de mis amigos de toda la vida: David, Irene, Javi, Jesús, Mónica, Vicente, Miguel, Sofía, Sara, Sheila y Cris, y especialmente los imprescindibles Álvaro, Dani y Juan Carlos. Ya fuese en el \textit{cole}, en la plaza del Carmen, en el \textit{bajo} o en la \textit{kame}, siempre me habéis dado un lugar donde disfrutar, reírme, relajarme y desconectar de todo. Por haber crecido conmigo y haberme hecho crecer. Gracias por acompañarme y apoyarme durante tantos años.

Quiero agradecer a mi familia, tíos y primos, por todo el cariño y el apoyo que me habéis mostrado siempre. Gracias a mi yayo, que siempre lo tengo en la memoria. Agradezco a mi yaya por los veranos inolvidables en Bronchales, vividos y por venir. Al meu abuelo per les teues fascinants històries sobre la teua vida i per les teues preguntes sobre galàxies. A mi abuela por haber estimulado mi curiosidad por el mundo y la naturaleza desde que tengo uso de razón. Sin nuestras conversaciones sobre el universo, quizá esta tesis no existiría. Gracias a la gente de Novelda, y especialmente a Pili, a la Abueli y a Mamen, por hacerme sentir siempre como uno más de la familia (¡y por sus deliciosas comidas!). A mis padres por su apoyo incondicional y su cariño, por haber creído siempre en mí. Porque sin la educación que me habéis aportado, no sería lo que soy ahora. 

Y por supuesto, a mi compañera de vida y amiga, Inma. Gracias por haberte convertido en un pilar fundamental para mí, y por haberme premiado con compartir nuestra vida juntos. Por tu apoyo en los malos momentos y por hacerme reír tanto en los buenos. Ambos cerramos ahora un ciclo y empezamos otro, que sin duda será igual de bueno o mejor que el presente. Y finalmente, no puedo olvidarme de Amèlie y Odín, por su cariño y sus ronroneos, que han hecho de la escritura de esta tesis un proceso mucho más agradable.

Gracias a todos los que habéis hecho posible que haya llegado hasta aquí. Esta tesis también os pertenece.


\begin{figure}[h]
\adjustimage{width=.2\textwidth,right}{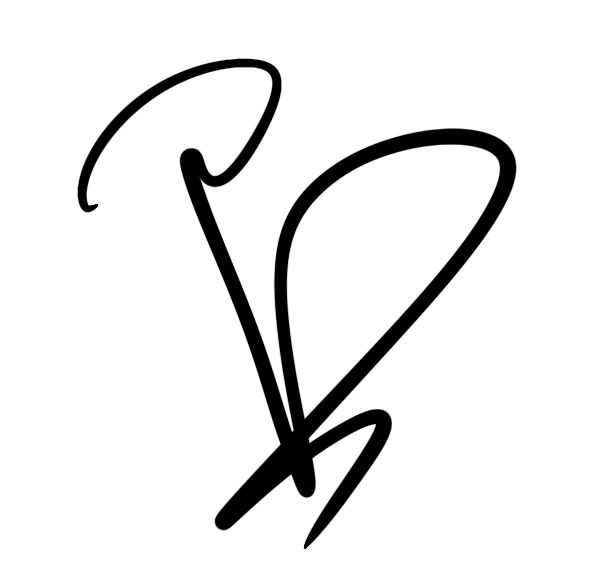}
\end{figure}




%% file: Chapters/Others/Contents.tex
\fancyfoot[C]{} 
\tableofcontents
\definecolor{linkcolour}{rgb}{0.85,0.15,0.15}
\hypersetup{linkcolor=linkcolour}

\mainmatter 
\renewcommand{\headrulewidth}{0.5pt} 

\addcontentsline{toc}{chapter}{Contents}

%% file: Chapters/Chapter_Intro/Chapter_Intro.tex
\end{comment}

\chapter{The $\Lambda$CDM Universe}

\label{chap:Introduction}

\section{Introduction}

An astounding progress in our understanding of the universe has been achieved since the birth of physical cosmology, with the 1917 Einstein's article, when the universe as a whole became the subject of study \cite{1917SPAW.......142E}. The correlation between distances of galaxies and their velocities found during the next decade by Lemaître \cite{1927ASSB...47...49L} and Hubble \cite{1929PNAS...15..168H} led to the acceptance that the space is expanding. After the first ideas about an initial singularity proposed by Lemaître \cite{1931Natur.127..706L}, it took several decades to develop the Big Bang theory. The model of a hot early universe filled by a plasma was first formulated by Alpher, Gamow and Herman \cite{PhysRev.73.803, 1948Natur.162..774A}. It implied the synthesis of light elements in the early universe, the Big Bang Nucleosynthesis (BBN), rather than from the posterior stellar nucleosynthesis. Furthermore, the hot cosmic plasma would have left a remnant radiation of relic photons from the early universe at a few Kelvin, the Cosmic Microwave Background (CMB). But it was not until the serendipitous discovery by Penzias and Wilson of an isotropic radiation at $2.73$ K \cite{1965ApJ...142.1149P} that  the scientific community took the Big Bang theory seriously.\footnote{Several upper bounds and even positive measurements of the CMB temperature were obtained since 1941, although their relevance was missed for decades due to the unawareness of the Big Bang theory by experimentalists, and the ignorance of these measurements by theorists, see, e.g., \cite{2009fbb..book.....P}.} The measurement of the CMB and the abundance of Helium and light elements became the unavoidable confirmations of the Big Bang model. Via the detection of the CMB anisotropies decades later, together with the galaxy surveys observing the Large Scale Structure (LSS), we have been able to accurately determine the amount of matter, and the intrinsic flat curvature of the observable cosmos.

On the other hand, rotational curves of galaxies and estimation of mass from the velocity dispersion through the virial theorem resulted in the awareness about a large amount of invisible matter in the universe, much more than the estimated one from stars \cite{2018RvMP...90d5002B}. This invisible component behaves as standard matter, in terms of gravitational interactions, but it does not radiate neither significantly interact with other known particles, and had to be of non-baryonic origin. Dark Matter (DM) soon became a fundamental ingredient of the models of structure formation and cosmic evolution. Nevertheless, the nature of the DM components still remains a conundrum. Light neutrinos were early suggested as possible candidates, but numerical simulations showed that cosmic structures would be formed in a different way than what is observed, ruling out this scenario. Several particle physics theories Beyond the Standard Model (BSM), such as Supersymmetry, predicted different candidates which may have been in thermal equilibrium with the thermal plasma, but decoupled at some moment, leaving a relic amount of weakly-interacting matter which would account for the DM. The generically known as Weakly Interacting Massive Particles (WIMPs) conform the standard DM scenario, although none of these candidates have been detected yet. Besides particles, macroscopic objects have been suggested to conform the DM, generally referred to as MACHOs. Despite its specific nature, mostly all observations seem to agree that the invisible matter is \textit{cold}, i.e., behaving like pressureless matter with negligible temperature. These evidences have given rise to the Cold Dark Matter (CDM) paradigm, which successfully explains most of the cosmic structures we are able to see.

At the end of the last century, observation of distant galaxies via type Ia supernovae showed the accelerated expansion of the universe. This fact pointed out the existence of the last unknown cosmic ingredient, coined as Dark Energy (DE) \cite{1998AJ....116.1009R, 1999ApJ...517..565P}. However, unlike DM, which in terms of gravitation, behaves as standard matter, the DE evolution is not consistent with any other kind of known matter. Its energy density keeps mostly constant, acting thus as a cosmological constant, as the $\Lambda$ term introduced by Einstein in his seminal article \cite{1917SPAW.......142E}. However, other possible and exotic equations of state cannot be ruled out yet. Later observations from the CMB anisotropies and LSS agreed and confirmed these results. The most precise joint analysis of the Planck collaboration results with LSS data shows that only 5\% of the energy content of the universe is in form of standard baryonic matter, while 26\% appears as DM and 69\% consists of DE \cite{Aghanim:2018eyx}. These results have promoted the $\Lambda$CDM model to the state of standard cosmic paradigm, which has successfully overcame many cosmological tests. The existence of such important contributions of DM and DE imply that most of the universe is, in some way, \textit{dark}, and its nature mostly unknown to us.

Thanks to the measurements of CMB anisotropies and the estimation of primordial nuclei abundances, we have a clear picture of the processes involved in the early universe. We mostly understand the composition of the cosmic plasma and its evolution prior to Recombination, happening at $z \sim 1100$, when electrons and protons got bound to form neutral atoms and the CMB radiation decoupled. Any new exotic physics affecting the primordial plasma may modify the CMB or BBN abundances, and thus could be potentially probed and constrained. On the other hand, galaxy surveys such as the Sloan Digital Sky Survey (SDSS) or space based observatories, like the Hubble Space Telescope (HST), provide us with unvaluable data from the galaxies and clusters surrounding us. An outstanding breakthrough was the discovery of Baryonic Acoustic Oscillations (BAOs), which is the enhancement on correlation of galaxies at distances corresponding to the sound horizon at the CMB decoupling, by SDSS \cite{2005ApJ...633..560E} and the 2dF Galaxy Redshift Survey \cite{2005MNRAS.362..505C}, which has been subsequently measured with remarkable precision up to redshift $z \sim 2$ \cite{Ata:2017dya, deMattia:2020fkb}. This phenomenon is consistent with our early universe picture, and yields relevant information regarding the clustering properties of galaxies. The observation of the Lyman $\alpha$ (Ly$\alpha$) forest, the distribution of small clouds of neutral gas producing absorption features in quasar spectra, gives us the best tool so far to probe small cosmological scales \cite{Dijkstra:2017lio, 2010gfe..book.....M}. All these observations can be contrasted with numerical simulations to infer how structure formation has proceeded, forming the current galaxies we see.

However, between the Recombination era and the more recent times reachable by galaxy surveys, there are some epochs in the history of the universe which still remain mainly unknown to us. It is the case of the so-called \textit{Dark Ages}, the period of time between the decoupling of the CMB from matter (around $3.8 \times 10^5$ years after the Big Bang) and the beginning of star formation. This epoch is considered as \emph{dark} due to the absence of observable sources of light in the universe. Despite that, its relevance in the understanding of the cosmic evolution is essential, since the onset of the growth of structure takes place in this epoch. At a given moment, approximately $10^8$ years after the Big Bang, some collapsed regions were so dense that Hydrogen could ignite, giving rise to the firsts stars. The so-called Cosmic Dawn, when the earliest galaxies started to shine, also remains very uncertain. The first generation of stars, known as Population III, could have been short-lived, with no metallicity, and much more massive than the stars surrounding us, but these are too far to be directly observable with current instruments.

Lastly, the primeval galaxies filled the Inter Galactic Medium (IGM) with energetic light, which heated and ionized the universe. The observations of quasar spectra suggest that at redshift $z \simeq 6$, when the universe was nearly 7\% its current age ($9 \times 10^8$ years after the Big Bang), most of the IGM became completely ionized, indicating the end of the Epoch of Reionization (EoR) \cite{2001AJ....122.2850B}. While the Ly$\alpha$ forest brings us information about the final stage of this event, how reionization started and proceeded is not completely understood yet. The nature of the sources capable of injecting so much energy into the medium is still unclear. Although galaxies could be capable of driving the process of reionization, a non-negligible contribution from quasars or from more exotic sources, such as decaying DM, cannot be ruled out. On the other hand, whereas the CMB is also modified by events after its decoupling, through the secondary anisotropies, it only provides indirect signs of these epochs. For instance, the increase in free electrons during reionization induces polarization of the CMB photons, together with a suppression in the anisotropy spectrum. Both effects are driven by the Thomson optical depth, which is an integrated quantity over all the ionization history, rather than probing specific redshifts. Therefore, measurements of the CMB do not provide us with enough information to have a precise and complete description of the EoR.

During the recent decades, a novel probe to explore the universe is gaining interest among researchers, based on the redshifted 21 cm line of Hydrogen. The hyperfine structure of the Hydrogen allows transitions with a rest frequency of 1420 MHz, or equivalently, a wavelength of 21 cm. This transition arises from the splitting of the ground state into two energy levels as a consequence of the coupling of the electron spin and the nucleus spin, in a similar way in which the coupling between the orbital angular momentum and the spin of the electron causes part of the fine structure. The interaction between both $1/2$ spins mix them in two possibles levels with definite spin $F$: the triplet state, with $F=1$ (parallel spins), and the singlet state, with $F=0$ (antiparallel spins). Therefore, the third component $m_F$ can take three values $m_F=-1,0,1$ in the triplet case, and only one value $m_F=0$ for the singlet case. Since its theoretical prediction by van de Hulst in 1945 \cite{1945NTvN...11..210V} and its discovery in 1951 by three different groups within a few weeks \cite{1951Natur.168..356E, 1951Natur.168..357M, 1951Natur.168..358P}, this atomic transition has been widely observed from astrophysical sources. 21 cm surveys provided the first maps of our galaxy and the most reliable rotation curves of galaxies, which yielded ineluctable evidence of the existence of DM during the 70's.

However, the measurements so far come from nearby galaxies or gas clouds. Measuring the neutral Hydrogen of the IGM at high redshift implies a much more ambitious goal. The lifetime of the excited state is about $10^{15}$ s, the transitions being somewhat unfrequent, and therefore the signal is very faint. However, the large amount of neutral Hydrogen (HI) atoms filling the IGM could compensate for this fact, resulting in a high enough number of transitions so that this signal could be observable. Moreover, most HI atoms are expected to be in its ground state, since the excited levels have lifetimes much shorter than the typical times needed for excitation. These facts may lead to a significant signal coming from hyperfine transitions at high redshift. Due to the expansion of the universe, photons emitted or absorbed at 21 cm at a given redshift $z$ reaching the Earth would decrease its energy by a factor $1+z$. Thus, unlike local 21 cm observations, one would seek the redshifted line, where the frequencies in the observed spectrum would indicate the epoch of absorption or emission of such photons. Notice that observations are not only restricted to galaxies, since neutral Hydrogen is spread over the medium around them, allowing to cartography the entire space. For these reasons, measurements of the redshifted 21 cm line could be an excellent way to map the IGM, allowing us to trace its three-dimensional history. The 21 cm signal is very sensitive to the temperature of the medium, and therefore it could give us information about the different stages of heating and cooling in the IGM evolution. Furthermore, since the signal would only be present if there is enough neutral Hydrogen in the universe, its detection can allow us to know the precise details of the reionization process as a function of the redshift. Therefore, 21 cm cosmology conforms a powerful and unique tool to study the Cosmic Dawn and the EoR, as well as to explore the nature of the sources of energetic radiation.

The interest in this research line has recently notably increased within the community, specially after the unexpected result of the Experiment to Detect the Global EoR Signature (EDGES) \cite{Bowman:2018yin}. The EDGES collaboration claimed the measurement of an absorption dip centered around 78 MHz which could be consistent with the detection of the 21 cm cosmological signal, but presenting an amplitude twice as that expected from the standard scenario. This fact may imply some sort of new physics able to further cool the medium, such as DM interacting with baryons, or an extra source of radio emission. Although this observation has not been confirmed by other experiments, and it has generated strong criticism regarding the systematics and foregrounds treatment, this measurement has triggered a lot of research exploring plausible non-standard models capable of explaining such a large amplitude. While the EDGES results come from the global averaged signal over all sky, other more ambitious designs based on interferometry would be able to observe the spatial inhomogeneities of the 21 cm radiation. There is plenty of work devoted to plan and build large radio interferometers to detect this signal. Some of them are already working and collecting data, such as the Murchison Widefield Array (MWA) \cite{Bowman_2013} or the LOw Frequency Array (LOFAR) \cite{van_Haarlem_2013}, which have placed stringent bounds on the maximum signal. However, the next generation of interferometric arrays such as the Hydrogen Epoch of Reionization Array (HERA) \cite{Beardsley:2014bea} and the Square Kilometer Array (SKA) \cite{Mellema:2012ht} may be able to reach positive detections in the future, allowing us the access to unvaluable cosmic information.

The Cosmic Dawn and reionization are not only interesting due to the involved astrophysical processes, but can also provide us with more fundamental information. Star formation is hosted within DM halos, and becomes stronger in more clustered regions. More massive halos allow an earlier fragmentation and collapse into stars than lighter ones. The DM distribution determines therefore how and when galaxies form. Hence, observing these epochs may yield us valuable information regarding the DM nature, such as different alternatives to the standard CDM scenario. While CDM provides an excellent fit to current large-scale observations, a few problems arise at galactic and sub-galactic scales, which can be summarized in the fact that the CDM model predicts more structures than observed. In order to account for these issues, several alternatives to CDM have been proposed presenting a suppression in the power at small scales which could reconcile theoretical models with measurements. Instead of presureless cold matter, one could consider particles with non-negligible dispersion velocities, named Warm Dark Matter (WDM). To successfully address the unsolved questions at small scales, WDM particles should have masses of the order of the keV, much lighter than the typical GeV range of masses for WIMPs. On the other hand, Interacting Dark Matter (IDM) particles via elastic scattering with light species, such as photons or neutrinos, would produce further collisional damping and induce oscillations in the power spectrum. As a consequence, small scale fluctuations would be partly washed out, reducing the number of low mass halos, in a similar way to WDM. Such DM models with suppression of fluctuations would delay structure formation processes, leading to later Cosmic Dawn and reionization epochs. Reionization data from quasar spectra and the redshifted 21 cm line are sensitive to the timing of the onset of ionizing radiation, so they represent important tools to probe WDM and IDM scenarios.

Besides the small scale crisis, some CDM candidates could also leave important signatures in the thermal evolution of the IGM. A fascinating possibility considers Black Holes (BHs) formed in the early universe, rather than from stellar collapse, to conform the DM. The interest on Primordial Black Holes (PBHs) has been revived after the first LIGO observation of gravitational waves from BH mergers \cite{PhysRevLett.116.061102}, which may be consistent with objects of primeval nature. The rich physics of PBHs triggers many observational effects to probe them, which allows us to constrain their masses and abundance. Besides the stringent bounds on the amount of PBHs, these still could account for part of the DM. PBHs with solar masses would present strong accretion mechanisms. It implies that the surrounding matter would infall onto them, releasing during the process large amounts of radiation into the medium. These high energetic photons could be absorbed by the IGM, ionizing and heating the gas, and hence modifying the thermal history. These effects may be seen as a suppression of the global average or of the fluctuations of the 21 cm cosmological signal. The 21 cm power spectrum appears to be an excellent and robust probe to constrain PBH scenarios.

In this PhD thesis, the thermal evolution of the IGM has been studied, specially via the 21 cm line. The understanding of the cosmic history has been employed as a tool to explore the constituents of the DM. Several DM models which may leave relevant signatures in the cosmic evolution have been considered. It is the case of WDM and IDM scenarios, which predict a suppression of fluctuations, or PBHs, which instead imply extra energy injections, contributing to the heating and ionization. Besides these model dependent approaches, the DM impact on thermal history and matter clustering can also be studied without assuming any specific DM scenario. The possibility of early reionization epochs which may be driven by exotic DM has been studied by using CMB data. On the other hand, novel deep learning methods have been employed to seek the link between 21 cm fields and the underlying 3D matter density maps, which would provide explicit information about the DM distribution. The works included here are a sample of the capacity of reionization data and the 21 cm signal to unveil the DM properties. With the forecoming radiointerferometers, 21 cm observations may shed light on the enigmatic nature of DM.

\section{Background cosmology}
\label{sec:background}

\begin{figure}
\centering
\includegraphics[scale=0.7]{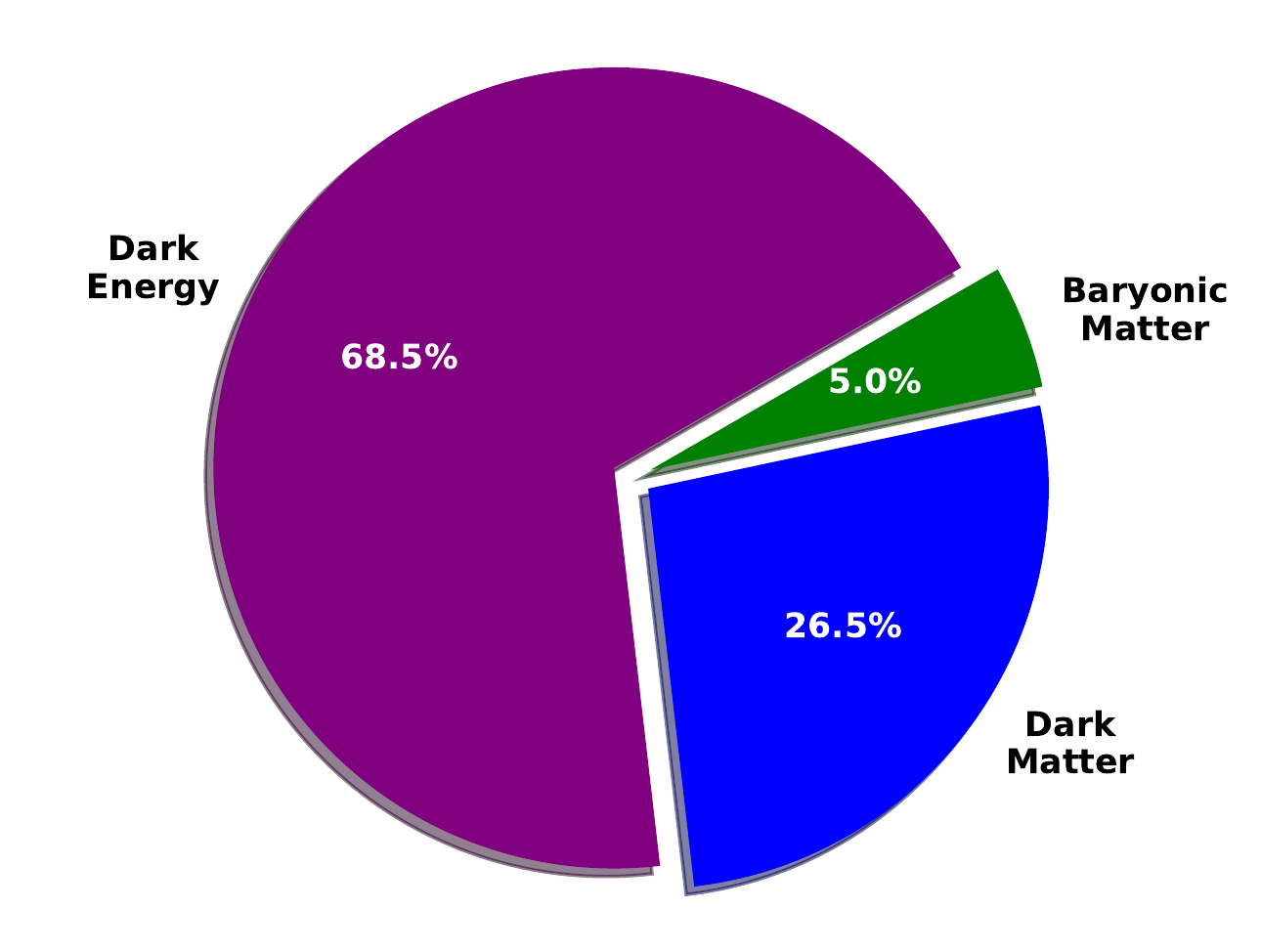}
\captionof{figure}{Contribution of each component to the current total energy density.}
\label{fig:cosmopie}
\end{figure}

It is worth it to start by reviewing the fundamentals about the background cosmology. In General Relativity, spacetimes are determined by their \textit{metric} tensors, $g_{\mu\nu}$, whose evolution is ruled by the Einstein Equations:
\begin{equation}
G_{\mu\nu} + \Lambda g_{\mu\nu} = 8\pi G_N T_{\mu\nu},
\label{eq:EinsteinEquations}
\end{equation}
where $G_N$ is the Newton gravitational constant, $G_{\mu\nu}$ is the Einstein tensor, which accounts for the curvature of the spacetime by a combination of derivatives of the metric, and $T_{\mu\nu}$ the energy-momentum tensor, which accounts for the matter content. The constant $\Lambda$ is known as the \textit{cosmological constant}, and it is a term allowed by the symmetries in the lagrangian, and thus shall be considered (unless the asymptotic limit to the flat Minkowski metric is required, in which case it must vanish). The geometry of spacetime, accounted for in the left-hand side of the above equation, is sourced and modified by the energy content, given by the right hand side. Equivalently, the evolution of the present matter is ruled by the curvature of spacetime.

One can apply these equations to a cosmological framework. As a first approximation, the universe appears to be statistically homogeneous and isotropic at the largest scales, which is known as the \textit{cosmological principle}. In General Relativity, the background spacetime can be characterized by the Friedman-Lemaître-Robertson-Walker (FLRW) metric, that describes an expanding spatially homogeneous and isotropic universe. It was independently developed by Friedman \cite{1922ZPhy...11..326E, 1924ZPhy...21..326F} and Lemaître \cite{1927ASSB...47...49L}, while in 1935, Robertson \cite{1935ApJ....82..284R} and Walker \cite{1937PLMS...42...90W} independently proved in a rigourous way that this metric is the unique spacetime spatially homogeneous and isotropic. The only degree of freedom present in the metric is the \textit{expansion factor} $a(t)$, which is only function of the cosmic time $t$. Spatial hypersurfaces have constant curvature, which can be positive, negative or zero, leading respectively to closed, open or flat spacetimes. Observational data is consistent with a flat universe  \cite{Aghanim:2018eyx}, as can be expected from inflation (see e.g. Ref. \cite{Baumann:2009ds}). Therefore, along this thesis, zero curvature is assumed. The metric, written in terms of the line element $ds^2$, reads
\begin{equation}
ds^2=g_{\mu \nu}dx^\mu dx^\nu = -dt^2 + a(t)^2 \left(dr^2 + r^2 d\Omega^2 \right),
\label{eq:FLRW}
\end{equation}
where $r$ is the radial comoving coordinate and $d\Omega$ is the differential solid angle. On the other hand, homogeneity and isotropy imply an energy momentum tensor of the form
\begin{equation}
T_{\mu\nu} = (\rho + P) U_\mu U_\nu + P g_{\mu \nu},
\label{eq:energytensor}
\end{equation}
where $U_\mu$ is the 4-velocity of the observers comoving with the fluid, and $\rho$ and $P$ the energy density and pressure measured in the frame of such observer. Substituting Eqs. \eqref{eq:FLRW} and \eqref{eq:energytensor} in the Einstein equations, Eq. \eqref{eq:EinsteinEquations}, it is straightforward to derive the evolution equations fulfilled by the expansion factor, the so-called Friedmann equations \cite{2003moco.book.....D}:
\begin{equation}
H^2\equiv\left(\frac{\dot{a}}{a}\right)^2=\frac{8\pi G_N}{3}\rho + \frac{\Lambda}{3}~,
\label{eq:Friedman1}
\end{equation}
\begin{equation}
\dot{H}+H^2 = \frac{\ddot{a}}{a}=-\frac{4\pi G_N}{3}(\rho + 3P) + \frac{\Lambda}{3}~,
\label{eq:Friedman2}
\end{equation}
where the dot corresponds to the derivative respect to the cosmic time, $\dot{f}=df/dt$.\footnote{These equations are specific cases of more general spacetimes: Eq. \eqref{eq:Friedman1} (sometimes referred to as \textit{the} Friedman equation) is the Hamiltonian constraint in the 1+3 decomposition of General Relativity, while Eq. \eqref{eq:Friedman2} is a simpler case of the Raychaudhuri equation, which drives the expansion of congruences of observers (see, e.g., Ref. \cite{2012reco.book.....E} for more details).} We have implicitly defined the \textit{expansion rate}, or Hubble parameter, as $H=\dot{a}/a$, which determines how fast the universe expands. To measure times and distances, it is customary to employ, rather than the cosmic time, the expansion factor $a$, or the \textit{redshift} $z$, related to $a$ and to the usual time  variable as $a=1/(1+z)$ and $dz=-cdt/(1+z)H(z)$, respectively. Either from the conservation of the energy-momentum tensor, or from Eqs. \eqref{eq:Friedman1} and \eqref{eq:Friedman2}, we can also derive the continuity equation, which states the conservation of the total energy
\begin{equation}
\dot{\rho}+3H(\rho+P)=0~.
\label{eq:cons_en}
\end{equation} 
The total energy density and pressure are given by a sum over the different species which contribute as a source for the expansion $\rho  = \sum_i\rho_i$, $P=\sum_iP_i$, with $\rho_i$ and $P_i$ the energy density and pressure of the species $i$. The equation of state $w_i=P_i/\rho_i$ determines the evolution of $\rho_i$, and it is characteristic for each species. In the following, we exclude interactions between different particles which may produce energy exchanges, and thus Eq. \eqref{eq:cons_en} applies to each species separately. In the case of \textit{radiation} (i.e., massless particles such as photons, or neutrinos in the early universe), the pressure is given by $P_{rad}=(1/3)\rho_{rad}$, and thus, Eq. \eqref{eq:cons_en} without interactions lead to a dependence with the scale factor as $\rho_{rad} \propto a^{-4}$. On the other hand, non-relativistic \textit{matter}, which encompasses both dark and baryonic matter, has negligible pressure, and hence, $\rho_m \propto a^{-3}$. The $\Lambda$ contribution can be regarded as some sort of \textit{dark energy}, with constant energy density, $\dot{\rho}_\Lambda =0$.

Eq. \eqref{eq:Friedman1} states that the expansion rate is given by the contribution of different species. It is customary to rewrite it in terms of the relative mass-energy densities with respect to the \textit{critical density} $\rho_c(z) = 3H(z)^2/8\pi G_N$. We thus define the \textit{density parameters} as $\Omega_i = \rho_{i,0}/\rho_{c,0}$, where the subscript $0$ indicates the current values. With this definition, the density parameters sum up to 1 in a flat universe, $\sum_i \Omega_i$, being this sum $<1$ ($>1$) in open (closed) universes. Taking into account radiation ($i=rad$), matter ($i=m$) and the contribution from the cosmological constant ($i=\Lambda$), on can write the expansion rate as the sum
\begin{equation}
H^2=H_0^2(\Omega_{rad}(1+z)^4 + \Omega_m(1+z)^3 + \Omega_{\Lambda})~.
\label{eq:hubble}
\end{equation}
The Hubble rate evaluated today, $H_0$, is customarily written as $H_0=100 \,h $ s$^{-1}$ km $Mpc^{-1}$, with $h$ the \textit{reduced Hubble constant}, which from CMB and BAO data takes the value of $h=0.693$ \cite{Pogosian:2020ded}.\footnote{In recent years, a controversy regarding the value of the Hubble constant has arisen, since estimates from early time probes (CMB, BAOs, BBN) present a tension with local
measurements from distances of galaxies, supernovae, etc. See e.g. Ref. \cite{DiValentino:2021izs} for a detailed review.} The $\Omega_i$ parameters are extracted from galaxy clustering measurements and CMB anisotropies observations, being the preferred values $\Omega_m=0.315$, $\Omega_{\Lambda}=0.685$ and $\Omega_{rad}=9.04 \times 10^{-5}$ \cite{Aghanim:2018eyx}. Since each of the components of the above equation has different time dependences, the cosmic timeline can be divided into eras accordingly to the dominant contribution. In the primordial universe, at high $z$, the $(1+z)^4$ factor of the radiation is the largest one, corresponding to the \textit{era of radiation domination}. This is followed by the \textit{era of matter domination}, which starts at the \textit{matter-radiation equality}, given by $1+z_{eq} = \Omega_m/\Omega_{rad} \sim 3500$. The cosmic expansion is ruled by non-relativistic matter until $1+z_\Lambda = (\Omega_\Lambda/\Omega_m)^{1/3} \simeq 1.3$, when the \textit{era of $\Lambda$ domination} begins. This energy era still holds until today, after 13.8 Gyr after the Big Bang. In this thesis, we mainly focus on phenomena which take place during the matter-dominated era. Therefore, in most of the cases of interest the radiation and $\Lambda$ contributions to Eq. \eqref{eq:hubble} can be safely neglected, writing $H^2 \simeq H_0^2 \Omega_m(1+z)^3$.

\begin{figure}
\centering
\includegraphics[scale=0.6]{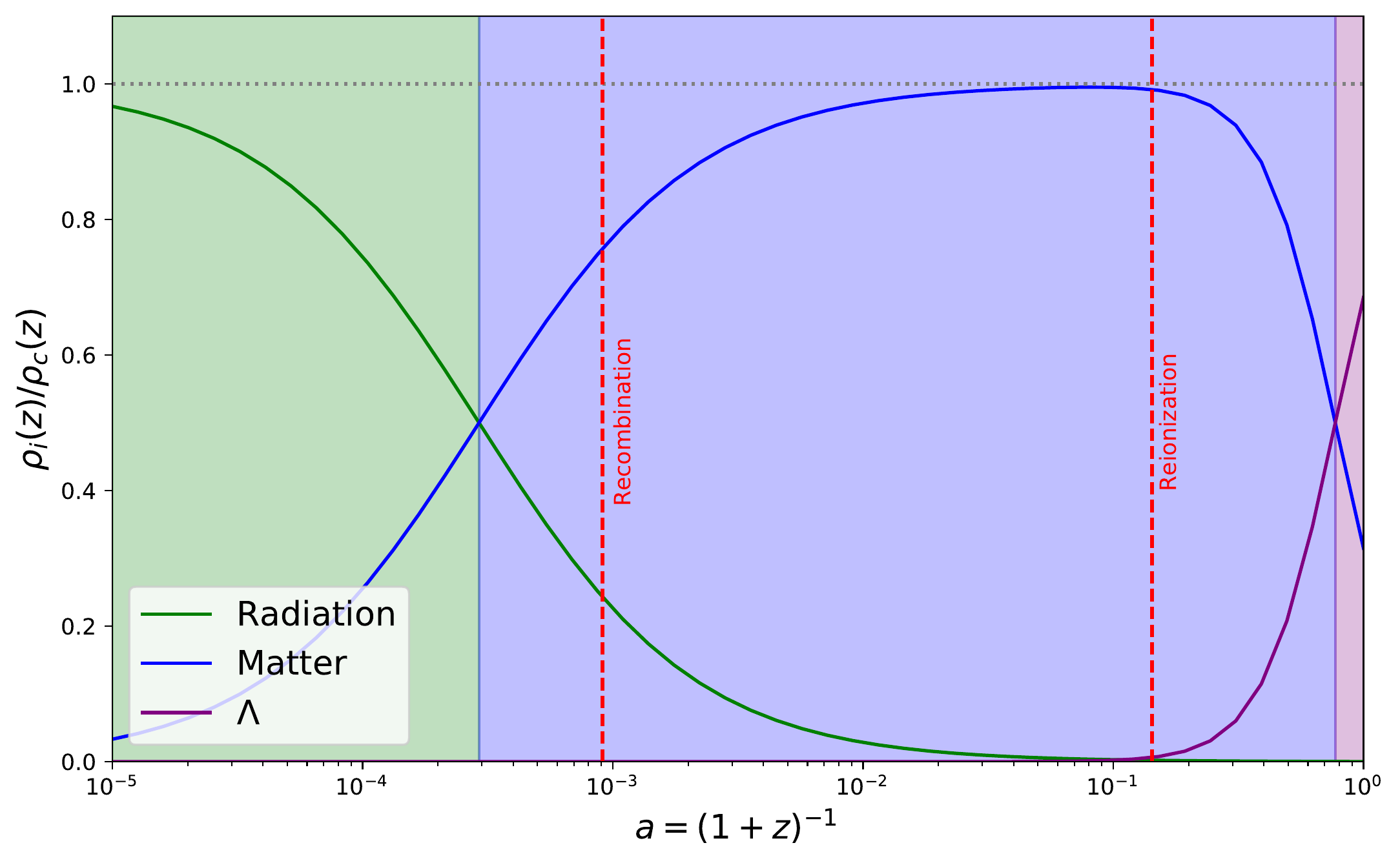}
\captionof{figure}{Evolution with the scale factor $a$ of the energy density of each component relative to the total density. Recombination and the end of the EoR are indicated with vertical red lines.}
\label{fig:density}
\end{figure}

The matter content can be splitted into its \textit{baryonic} and \textit{dark} components. We can write the baryon energy density as $\rho_b=\mu m_p n_b$, with $m_p$ the proton mass, $n_b$ the number density of baryons and $\mu$ the mean molecular weight, which accounts for the contribution of Helium, and takes the value $\mu=1.187$. Due to the conservation of the number of baryons, one can write $n_b=n_{b,0}(1+z)^3$, and thus the fraction to the critical density as $\Omega_b=\rho_{b,0}/ \rho_{c,0}$. From the joint analysis of CMB and BAOs, the value preferred by data is $\Omega_b=0.048$ \cite{Aghanim:2018eyx}. A diagram of the current energy density contribution is shown in Fig. \ref{fig:cosmopie}. On the other hand, Fig. \ref{fig:density} depicts the evolution of each energy component as a function of the scale factor, denoting the different energy domination eras.

Most of cosmological observations match astonishingly well with the $\Lambda$CDM model, which assumes a cosmological constant $\Lambda$ as the Dark energy component, and presureless Cold Dark Matter (CDM) as the dominant matter component. The standard cosmological scenario can be described by six independent parameters, namely the baryon $\Omega_bh^2$ and the CDM $\Omega_ch^2$ energy parameters (where $h$ is the dimensionless Hubble parameter, defined through $H_0= 100 h$ km s$^{-1}$Mpc$^{-1}$), the ratio between the sound horizon and the angular diameter distance at decoupling $\Theta_{s}$, the reionization optical depth $\tau_T$, and two parameters which determine the primordial power spectrum from inflation: the scalar spectral index $n_s$ and the amplitude of the primordial spectrum $A_{s}$, see Ref.~\cite{Aghanim:2018eyx} for their current best-fit values and uncertainties. 

\section{Overview of the cosmic history}

In this section we briefly review the most important milestones of the cosmic chronology, specifying the relevant epochs and the transitions among them. Simplifying the full picture, we can divide the cosmic timeline in the next phases (see, e.g., \cite{2005pfc..book.....M, Barkana:2000fd}):

\begin{itemize}
\item \textbf{Primordial Universe.} Until $\sim 1$ second after the Big Bang. This period encompassed many relevant phenomena, such as Inflation, Reheating, Baryogenesis, the Electroweak transition and the Hadronization, among others. Since we are interested in more recent times, we gloss over the details of this epoch. During the inflationary period, the energy content was dominated by some scalar field or other exotic species which led to a much faster expansion of the universe, by a factor of at least $\sim 10^{26}$, in a very short period of time, around $10^{-33}$ s. The fields which drove inflation decayed to the particle and radiation species in the so-called Reheating epoch, starting the radiation domination era. Thenceforth, the universe was composed by a highly homogeneous plasma, filled by a large number of species in thermal equilibrium. With the expansion and adiabatic cooling of the universe, many of these species decoupled from the cosmic plasma. The small fluctuations produced by Inflation settled the seeds where cosmic structures would grow from afterwards \cite{Baumann:2009ds}.

\item \textbf{Neutrino decoupling, $e^+e^-$ annihilation and Primordial Nucleosynthesis.} From $1$ second to $10^3$ minutes after the Big Bang. When the temperature of the cosmic plasma dropped to $\sim 1$ MeV, electroweak interactions were not strong enough to maintain neutrinos in equilibrium with the rest of species. Neutrinos decoupled and thereafter evolved independently \cite{Lesgourgues:2018ncw}. This process was closely followed by electron-positron annihilation, when photons were not energetic enough anymore to keep producing pairs through the reaction $\gamma + \gamma \rightarrow e^- + e^+$. All positrons annihilated, remaining a small fraction of electrons, and producing an extra heating of the cosmic plasma. Soon after, at $\sim 10$ seconds, the binding energy of Deuterium could not be longer overcome by radiation. This starts the production of deuterium, which enables the formation of light elements during the so-called Primordial, or Big Bang Nucleosynthesis (BBN). As a result, protons and neutrons became bounded in nucleons, mainly in a proportion of $\sim$75\% of Hydrogen and $\sim$25\% Helium nuclei, by mass fraction, as well as small traces of other light elements, such as Lithium and Berilium \cite{Iocco:2008va, Pitrou:2018cgg}.

\item \textbf{Matter-radiation equality.} Around $z\sim 3500$, $4.7 \times 10^4$ years after the Big Bang. Due to the different scaling of the densities with the scale factor, the matter energy content becomes as relevant as radiation at this time, becoming the dominant contribution thereafter. During the matter dominated era, small fluctuations grow much faster than during radiation domination, when they remained stalled.

\item \textbf{Recombination and CMB decoupling.} Around $z\sim 1100$, $3.8 \times 10^5$ years after the Big Bang. Two fundamental related processes happened at this time. The cosmic plasma continued cooling until the photons became not energetic enough to keep electrons and protons ionized. Thus, most of the electrons and protons combined to form neutral atoms which were no longer ionized. This is the first important phase transition in the ionization state of the IGM, known as Recombination.\footnote{Actually, they had never been \textit{combined} before, but this term persists for historical reasons.} On the other hand, the decreasing number of free electrons reduced the rate of the Thomson scattering with photons of the cosmic plasma, which kept the baryonic matter coupled to radiation. Therefore, photons decoupled from baryons, evolving separately without interacting among themselves. Photons released then conform what we observe now as the Cosmic Microwave Background (CMB) \cite{Hu:2002aa}.

\item \textbf{Dark Ages.} Since $z\sim 1100$ to $z\sim 30$, from $3.7 \times 10^5$ to $\sim 10^8$ years after the Big Bang. After the CMB decoupling, the universe became transparent to radiation and mostly neutral. Due to the absence of light sources, for which is known as the Dark Ages, this period is hardly observable, except maybe via the 21 cm signal from neutral hydrogen. During these times, the growth of structures becomes important, magnifying the inhomogeneities and the collapse of matter into DM halos.

\item \textbf{Cosmic Dawn.} Since $z\sim 30$ to $z\sim 10$, from $\sim 10^8$ to $\sim 5 \times 10^8$ years after the Big Bang. The first stars are formed within massive halos, yielding the most ancient galaxies and quasars. First stars are composed by nuclei formed during BBN, and thus present very low metalicity, probably conforming the so-called Population III. These stars would be more massive and short-lived than the succeeding stellar generations, the metal-poor Population II and metal-rich Population I stars. These first light sources emitted UV and $X$-ray radiation, heating, exciting and ionizing the surrounding medium, which should have left an imprint on the expected 21 cm signal from this period.

\item \textbf{Reionization.} Since $z\sim 10$ to $z\simeq 6$, from $\sim 5\times  10^8$ to $\sim 9 \times 10^8$ years after the Big Bang. The UV part of the spectrum emitted by the early galaxies ionized their local environment. These ionized (HII) bubbles grew until this epoch, when they started to overlap, leading eventually to a fully ionized IGM. This is the second great transition of the ionization state of the IGM after Recombination. This process leaves a suppression of the CMB power spectrum, erasing the anisotropies due to the increase of free electrons. Observation of absorption features in quasar spectra (the Ly$\alpha$ forest and the Gunn-Peterson effect) reveals valuable information regarding the end of the overlap period, at $z\simeq 6$, although the initial stages are still poorly known. Forthcoming 21 cm experiments could allow us to explore this epoch \cite{Ferrara:2014sda, Wise_2019}.

\item \textbf{$\Lambda$-Matter equality.} Around $z\sim 0.3$, $ \sim 10^{10}$ years after the Big Bang. The nearly constant energy density from the DE starts to dominate the energy content of the universe over matter, causing an accelerated expansion of the universe. Growth of fluctuations slowed down since then.

\end{itemize}

In Chapter \ref{chap:IGM}, the thermal and ionization history during the Dark Ages, Cosmic Dawn and the EoR is detaily studied, analyzing the impact on the 21 cm cosmological signal.

\begin{comment}

\bibliographystyle{../../jhep}
\bibliography{../../biblio}

%% file: Chapters/Chapter_DM/Chapter_DM.tex
\end{comment}

\chapter{Non-standard Dark Matter scenarios}


\label{chap:DarkMatter}

\hspace{1cm}

Although the unknown nature of Dark Matter (DM) has concerned us for decades, it is still one of the most important unsolved problems in modern physics. In order to explain structure formation at both large and small scales, several models of DM have been proposed, composed by different kinds of particles. In this chapter we review the steps which led to the requirement of DM, and the adoption of the Cold Dark Matter (CDM) paradigm. The growth of fluctuations and description of halos are summarized. Motivated by solving small scale problems present in the CDM scenario, several alternative models are reviewed, which lead to the suppression of fluctuations at small scales, discussing their impact on structure formation. We focus on two scenarios: Warm Dark Matter (WDM), such that DM particles have a non-negligible velocity dispersion, free-streaming at low scales; and Interacting Dark Matter (IDM), whose particles interact with photons or neutrinos, damping perturbations by collisional coupling.

\section{Why Dark Matter matters}

Before discussing specific DM models, we start by overviewing the historical progress of evidences of DM and the consolidation of the CDM paradigm. Some of its possible issues are outlined, motivating alternative DM candidates which are briefly summarized.

\subsection{A historical overview}
\label{sec:DMhistory}

The historical development of the ideas which led to the adoption of the DM as a constituent of the universe has been widely discussed in the literature (see, e.g., Refs. \cite{2018RvMP...90d5002B, 2010dmp..book.....S, 2017NatAs...1E..59D, 2012AnP...524..507F, PeeblesCDM}.) Although there were hints of the existence of non-visible matter as soon as in the early 20th century (see, e.g., Ref. \cite{2018RvMP...90d5002B}), the firsts evidences of the existence of such matter were found in the 30s. In 1933 and 1937, Fritz Zwicky made use of the virial theorem with dispersion velocities measured in the Coma cluster, finding the presence of mass that does not emit radiation, about $\sim$500 times more than the standard radiative one \cite{1933AcHPh...6..110Z, 1937ApJ....86..217Z}.\footnote{This ratio is, however, an overestimation of the actual value by a factor of $\sim 8$, due to a wrong estimate of the Hubble parameter at that time \cite{2018RvMP...90d5002B}.} A similar work was performed by Sinclair Smith in 1936 with data from the Virgo cluster, also finding $\sim$100 times more mass than expected \cite{1936ApJ....83...23S}. Horace Babcock, in his PhD thesis in 1939, presented the rotation curve of M31 (Andromeda) up to 20 kpc from its center, showing high values for the circular velocity (although he attributed it to a stronger absorption or dynamical effects in the outer parts of the galaxy) \cite{1939LicOB..19...41B}, and similar findings were drawn from the rotation curve of M33 by Mayall \& Aller in 1942 \cite{1942ApJ....95....5M}. In 1959, Kahn \& Woltjer considered the relative motion between the Milky Way and Andromeda, identifying much more mass than the observed one from stellar origins in order to explain how they are approaching each other \cite{1959ApJ...130..705K}. However, these first hints were not correctly interpreted by the scientific community during several decades.

It was not until the 1970s when strong evidences of the presence of invisible matter were found. Measurements of rotation curves of several galaxies in 21 cm and photometry suggested more mass than expected in the outer regions \cite{1970ApJ...159..379R, 1970ApJ...160..811F, 1972ApJ...176..315R, 1972ApJ...175..347W, 1973A&A....26..483R}. In 1973, Ostriker and Peebles performed early numerical N-body simulations, and noted that spiral rotating galaxies were unstable, unless a massive spherical halo were present \cite{1973ApJ...186..467O}. Shortly after, two influential papers brought together all the mass discrepancies, evidencing the need for invisible non-baryonic matter, which would be the dominating component, and concluding that the matter density was $\Omega_m \simeq 0.2$, contrarily to the widely assumed value of $\Omega_m = 1$ at that time \cite{1974ApJ...193L...1O, 1974Natur.250..309E}. A major breakthrough came in 1978 from the rotation curves of a set of galaxies, measured by Bosma in his PhD thesis with the 21 cm line \cite{1978PhDT.......195B}, and by Rubin, Thonnard, and Ford in optical observations \cite{1978ApJ...225L.107R}. Both groups found flat rotation curves well beyond the observed radii of galaxies, meaning that there was invisible mass exceeding the region occupied by stars and gas. At the end of the 70's, the existence of some sort of non-radiating Dark Matter seemed unavoidable \cite{1979ARA&A..17..135F}.

The question then was: which kind of particles compose such invisible matter? Neutrinos seemed to be the perfect candidate for composing such DM, since they had been already measured in experiments, they do not interact with radiation, and the first neutrino oscillation measurements by that time suggested their being massive. The possibility of neutrinos as constituents of the DM was firstly pointed out in 1972 by Cowsik and McClelland \cite{1972PhRvL..29..669C, 1973ApJ...180....7C}, and independently by Szalay and Marx in 1976 \cite{Szalay:1976ef}. Neutrino masses were found to be constrained from cosmological arguments. In 1966, in the first paper considering the role of neutrinos in cosmology, Gershtein and Zeldovich derived an upper bound on the sum of the neutrino masses comparing their energy density with the critical density of the universe around $\lesssim 400$ eV, improving by several orders of magnitude the upper bound in the muonic neutrino mass from earth-based experiments \cite{1966ZhPmR...4..174G}.\footnote{Cowsik and McClelland re-derived this bound 6 years later \cite{1972PhRvL..29..669C}, being thereafter known as the Cowsik-McClelland limit, despite presenting some mistakes in the computation \cite{Dolgov:2014xpa}. The Gershtein-Zeldovich bound with current data is $ 94 \;{\rm eV} \; \Omega_\nu h^2$ \cite{2002PhR...370..333D}.} On the other hand, from the Pauli exclusion principle and assuming neutrino DM as the main constituent of massive halos, Tremaine and Gunn derived a lower bound for the neutrino mass of about $m_\nu > 100$ eV \cite{1979PhRvL..42..407T} (although it depends on $\Omega_m$). This may be potentially inconsistent with the Gershtein-Zeldovich limit, constraining the range of neutrino masses if they constituted the DM. The announcement (later proven to be wrong) of the detection of an electron antineutrino mass around $\sim 30$ eV \cite{1980PhLB...94..266L} reinforced the possibility of neutrinos as the DM constituent \cite{PhysRevLett.45.1980}, specially in the Moscow's Zeldovich group, who further studied the impact of neutrino \textit{hot} DM (HDM)\footnote{The terminology distinguishing between Hot, Warm and Cold DM according to the velocity dispersion was proposed in the mid 80's \cite{1984ASIC..117...87B}.} on the growth of fluctuations \cite{1980PAZh....6..457D, 1981NYASA.375...32D}. HDM was found to present a large free-streaming scale, erasing perturbations below it, and thus providing a \textit{top-down} collapse, where big structures are formed before, and later fragmented to form smaller objects. However, increasingly better N-body numerical simulations during the early 80's contrasted with observations of the CfA, the first 3D galaxy survey \cite{1982ApJ...253..423D}, ruled out the possibility of neutrino DM, since HDM predicted much less small-scale structures than those observed in data \cite{1983ApJ...274L...1W}.

With light neutrinos not being a plausible candidate, different alternatives were required. Peebles was the first to study the impact on fluctuations of a \textit{cold} DM (CDM), i.e., with negligible free-streaming scale \cite{1982ApJ...263L...1P}. Contrary to HDM, in a CDM scenario, structure formation proceeds \textit{bottom-up}, presenting power at all scales, and thus forming small-size objects which later merge to form larger structures, in a hierarchical way. First simulations of structure formation within the CDM framework resembled the observed clustering properties of galaxies \cite{1985ApJ...292..371D}, promoting CDM to a promising candidate for the non-visible matter. Collapse of matter lead to the formation of DM halos, whose abundance was well described by analytical estimates of the halo mass function, such as the Press-Schechter formalism \cite{Press:1973iz}, or by the Sheth-Thormen prescription, which accounts for the ellipticity of halos \cite{Sheth:1999mn, Sheth:1999su}. N-body simulations showed that CDM halos have an universal density profile, well fitted by a double power law, now known as the Navarro-Frenk-White (NFW) profile after its authors \cite{1997ApJ...490..493N}. This profile, valid over a large range of halo masses, scales as $r^{-1}$ at small radii and as $r^{-3}$ at larger distances, and is completely characterized by its virial mass and radius, and the so-called concentration parameter.

Several particle physics models were able to predict a candidate behaving as this kind of cold, collisionless and non-radiating matter. The prototype of CDM particles are the so-called WIMPs (Weakly Interacting Massive Particles) (term coined in 1985 \cite{1985NuPhB.253..375S}). These are heavy-mass particles with mass $\gtrsim 1$ GeV in equilibrium with the thermal plasma in the early universe due to weak-like interactions, but decoupling at some moment, freezing out its abundance, which remained mostly constant until now. This mechanism, known as \textit{freeze-out}, allows obtaining the current observed DM density at current times.\footnote{The freeze-out of a heavy lepton was independently proposed in five papers published in 1977 during two months by the following groups: Hut \cite{1977PhLB...69...85H}; Lee and Weinberg \cite{1977PhRvL..39..165L} (Lee passing away shortly before the publication); Sato and Kobayashi
\cite{1977PThPh..58.1775S}; Dicus, Kolb, and Teplitz \cite{1977PhRvL..39..168D}; and Vysotskii, Dolgov, and
Zeldovich \cite{1977ZhPmR..26..200V}. However, none of them realized that its relic abundance may be the one needed to constitute the non-visible DM required from astronomical observations \cite{2018RvMP...90d5002B, PeeblesCDM}. It must be noted, nevertheless, that the freeze-out mechanism, as usually happened in cosmology, had already been studied by Zeldovich and the Moscow group a decade before \cite{1965AdA&A...3..241Z, 1965PhL....17..164Z} (see also Ref. \cite{Dolgov:2014xpa}).} The abundance depends mostly on the cross section of the interaction, which is required to be of the order of the weak interactions to produce the observed DM density, coincidence known as the WIMP miracle. Examples of such particles are heavy thermal remnants of annihilation appearing in Supersymmetry, such as neutralinos, the supersymmetric partners of the gauge bosons, which were first considered as DM particles in 1984 \cite{1984NuPhB.238..453E}.
Other popular candidates for CDM are scalar fields, such as axions \cite{Preskill:1982cy}, a hypothetical particle introduced through the so-called Peccei-Quinn mechanism to solve the strong CP problem in quantum chromodynamics \cite{PhysRevLett.38.1440}. These particles may be produced by non-thermal means, such as from the decay of topological defects or other parent particles. Other popular method is the so-called misalignment mechanism (or vacuum realignment), where the axion field is initially displaced from the vacuum and then relaxes to the potential minimum, behaving as non-relativistic matter \cite{2016PhR...643....1M}. A last group aspirant to constitute CDM, and perhaps the most obvious possibility, are MACHOs (Massive Astrophysical Compact Halo Objects) \cite{1993NYASA.688..390G}, already suggested during the 70's \cite{1975A&A....38....5M, 1978MNRAS.183..341W}. With this term, coined by Kim Griest as opposed to WIMPs \cite{2018RvMP...90d5002B}, a variety of objects are encompassed which would behave as non-relativistic and non-radiating matter, such as balls of Hydrogen and Helium not massive enough to initiate nuclear burning, like brown dwarfs with masses $\sim 0.01 M_\odot$ or Jupiter-like planets with masses $\sim 0.001 M_\odot$. Moreover, black hole remnants from massive stars, or Primordial Black Holes (PBHs) formed in the early universe are also included. Gravitational microlensing is one of the main tools to study them, and has strongly constrained their abundance. However, since MACHOs could only be present in the universe after the formation of first stellar and astrophysical objects, they are unable to successfully explain large scale matter fluctuations seen in the CMB and the number of baryons from BBN. An exception of that are PBHs, which represent a particularly exciting candidate, requiring a special treatment, and will be extensively discussed in Chapter \ref{chap:PBHs}.

On the other hand, between the hot and cold limiting cases, an intermediate \textit{warm} scenario was also plausible, with masses around $\sim $ keV which presented a non-negligible free-streaming scale, but still consistent with data and N-body simulations. The first proposals of such WDM particles were gravitinos of mass $\sim 1$ keV (the spin $3/2$ supersymmetric partner of the graviton) in 1982 \cite{PhysRevLett.48.223, PhysRevLett.48.1636, 1982Natur.299...37B}. Although standard neutrinos were ruled out as DM candidates, other similar species may account for that. It is the case of the right-handed \textit{sterile neutrinos}, non-interacting with Standard Model (SM) particles except by a small mixing with standard active neutrinos. Several mechanisms were suggested to produce them in the early universe from neutrino oscillations, such as the proposed by Dodelson and Widrow in 1993 through oscillations with active neutrinos out of resonance \cite{PhysRevLett.72.17}, or by Shi and Fuller in 1999 via resonant production \cite{1999PhRvL..82.2832S}. Those particles would have a mass $\gtrsim$ keV, and thus would be a good candidate for WDM (or even CDM). Simulations and data at that time were not accurate enough to discern between the warm and cold scenarios, but CDM started to become the preferred alternative in the community, until becoming the standard cosmological paradigm. However, as shall be reviewed in the following, during the 90's, several problems related to structure formation at small scales challenged the CDM success, revitalizing the WDM alternative.

\subsection{Small-scale crisis of the CDM paradigm}
 
\label{sec:smallscale}

The CDM model has shown a great success fitting the data from the large scale structure of the universe. However, there are some discrepancies between observations and N-body simulations at galactic and subgalactic scales, which are not very well explained within the CDM paradigm. Some of these problems arose during the 1990s, when the CDM model predicting hierarchical clustering started to become widely accepted, and N-body simulations improved their resolution to smaller scales. All of them are related to the fact that CDM scenarios predict more small scale fluctuations than those observed in data. Next, we shall review the most relevant issues. See, e.g., Refs. \cite{2017Galax...5...17D, 2017ARA&A..55..343B} for a comprehensive overview of the subject.

\begin{itemize}
\item \textbf{Missing satellite issue}

Due to the absence of a cutoff in its power spectrum, CDM models predict a lot of subhalos around massive galaxies. N-body simulations show DM self-bound clumps at all resolved scales, and many more low-mass halos than those present in observations, failing to reproduce the observed circular velocities \cite{1999ApJ...522...82K, 1999MNRAS.310.1147M}. Concretely, few dozens of dwarf spheroidal satellite galaxies of the Milky Way have been observed, in contrast to the $> 100$ satellites present in numerical simulations \cite{2012MNRAS.422.1203B, 2011MNRAS.415L..40B}. The observation of ultra-faint dwarf galaxies by galaxy surveys such as DES have alleviated the problem \cite{2015ApJ...807...50B, 2015ApJ...813..109D, Nadler:2020prv}. Many solutions to this issue have been proposed within the CDM scenario, most of them relying on the fact that not all the subhalos may be visible. Examples of them are based on a suppressed gas accretion in low-mass halos after the EoR \cite{2000ApJ...539..517B}, or considering supernovae feedback \cite{1986ApJ...303...39D}, facts which inhibit the formation of stars in small mass halos. Other proposals state that an empirical relation between stellar and halo masses can be used to correct the detection efficiency of galaxy surveys, providing the proper number of counts \cite{2018PhRvL.121u1302K}. 

\item \textbf{Cusp-core problem}

A robust prediction from the CDM model which is present in all N-body simulations is the \textit{cuspy} distribution of matter in the inner parts of halos, with density increasing abruptly at small distances from the center. More specifically, CDM density profiles usually rise as $\rho(r) \propto r^{-\gamma}$, with $\gamma$ between 0.8 and 1.4 over the central radii $r$ \cite{2010MNRAS.402...21N} ($\gamma \simeq 1$ in the widely used NFW profile \cite{1997ApJ...490..493N}). This appears to be in contradiction with the rotation curves of most of the observed dwarf galaxies, which suggest that they must have flatter central density profiles, i.e., with $\gamma \simeq 0$, coined as \textit{cores} \cite{1994Natur.370..629M, 1994ApJ...427L...1F,2015AJ....149..180O}. Hydrodynamic simulations show that it may be possible to settle the problem thanks to baryonic feedback from supernova explosions and stellar winds, which would erase the central cusps \cite{2008Sci...319..174M}. A flat core could also be obtained by considering stellar and gas dynamics, by kinematically heating up DM at the centers of galaxies \cite{Read:2018fxs}.

\clearpage

\item \textbf{``Too-big-to-fail'' problem}

While the number of low-mass satellites have already been shown to be problematic, the most massive satellite galaxies also present some issues. One naturally would asign the brightest Milky Way (MW) galaxy satellites to the most massive subhalos present in N-body simulations. However, the Aquarius and Via Lactea simulations of the MW showed a population of $\sim 10$ halos very massive and dense, by a factor of $\sim 5$, in such a way that they would be too massive not to host bright dwarf satellites of the MW, which would be more massive than the ones actually observed \cite{BoylanKolchin:2011dk, 2011MNRAS.415L..40B}. This could be understood by the fact that if those very massive subhalos host the brightest satellites, the deep potential wells would lead to circular velocities much larger than the observed dispersion velocity of the observed dwarf galaxies. While in low-mass halos, one can resort to baryonic effects to prevent star formation and then become non-visible, these too massive halos would be too big to fail producing stars and being visible (by baryonic feedback or any other known mechanism), and thus they should be observed. For this reason, this issue is known as the \textit{too-big-to-fail} problem. As in the cusp-core problem, this is related to the fact that CDM tends to produce too much mass in subhalos. Although this issue was originally identified in the MW, it has also been found in the Andromeda satellites \cite{2014MNRAS.440.3511T} and in field galaxies of the Local group, beyond the virial radius of its main galaxies \cite{2014MNRAS.439.1015K}. In order to solve this issue, as well as baryonic feedback, interactions between the MW and its satellites, such as disk shocking or tidal stripping, have been proposed, in order to reduce the central masses of the satellites (e.g., \cite{2012ApJ...761...71Z}). However, simulations able to properly capture these effects need to resolve very low masses and are very challenging numerically \cite{2017ARA&A..55..343B}.

\end{itemize}

As already stated, there are several ways to overcome the aforementioned discrepancies within the standard $\Lambda$CDM scenario. Baryonic physics, such as stellar winds or supernovae feedback, has been invoked to solve all the above problems, being plausible to account for all of them at once \cite{2014JCAP...12..051D}. Other solutions rely on the poorly known mass of the MW, interpreting thus the above issues as possible indicators of a lower mass for the MW than the one assumed \cite{2012MNRAS.424.2715W}. However, DM models different from CDM could also solve some or all of these problems, presenting an interesting and well motivated alternative. 

\subsection{Non-standard DM candidates}

Despite the current efforts on detecting WIMPs, axions, or other possible CDM constituent particles, they remain undiscovered in experiments \cite{Lin:2019uvt}. This fact, together with the observational discrepancies discussed above, motivate considering other DM models different from the standard cold paradigm. Some examples of these non-CDM candidates, and how they could account for the small-scale discrepancies, are briefly discussed in the following. These models are mostly characterized by their phenomenology in structure formation, rather than by specific particle physics theories. It is worth emphasizing that the term ``non-CDM'' is employed here to refer to DM scenarios which present different features at small scales, although behaving as CDM at large ones. With ``non-standard'', we also include candidates which can act as CDM with respect to structure formation, but differ from the archetypal WIMP scenario, as is the case of BHs formed in the early universe.

\begin{itemize}

\item \textbf{Warm Dark Matter (WDM)}

In a typical WDM scenario, DM particles with masses of $\sim$ keV would lead to a substantial velocity dispersion, driving these particles to free-stream and erase fluctuations at small scales. The missing satellite problem is naturally solved, since a cutoff in the power at small scales leads to an underabundance of small structures, compared to the CDM case. As is shown in simulations, WDM can predict the required quantity of subhalos around the most massive ones \cite{2012MNRAS.420.2318L}. Moreover, due to its dispersion velocities, WDM does naturally produce cores. However, to reproduce the observed cores, a WDM mass of $ \sim 0.1$ keV would be required, in a range already ruled out by Ly$\alpha$ forest analyses \cite{2012MNRAS.424.1105M}. Thus, non-ruled out particles with masses $\gtrsim 2$ keV would not be light enough to satisfy all the current galactic data. Finally, within a WDM scenario with $m_X \sim 2$ keV, the ``too-big-to-fail'' issue could be solved due to the relatively shallower profiles of the expected WDM dwarf galaxies compared to their CDM counterparts. However, thermally produced WDM particles with a higher mass particle may not be able to solve the problem satisfactorily \cite{2014MNRAS.441L...6S}.

\item \textbf{Interacting Dark Matter (IDM)}

On the other hand, collisions between DM particles and either photons or neutrinos may avoid the formation of substructures. This happens due to the collisional damping present in IDM scenarios, which, in a similar way to WDM, erase small-scale fluctuations. For this reason, IDM can also explain the low quantity of low mass halos, and thus reconcile expectations with MW satellite observations \cite{Boehm:2014vja}. Furthermore, as found in high resolution IDM simulations, the largest subhalos are less concentrated than those in the CDM scenario, presenting rotation curves which agree with observations for interaction cross sections of $\sigma \simeq 10^{-9} \sigma_T ({\rm GeV}/m_{\rm DM})$ \cite{Schewtschenko:2015rno}, and thus accounting for the ``too-big-to-fail'' problem. 

\item \textbf{Self-Interacting Dark Matter (SIDM)}

A widely discussed possibility considers Self-Interacting Dark Matter (SIDM), where, unlike standard collisionless CDM, DM particles present non-negligible interactions among themselves \cite{1992ApJ...398...43C, Tulin:2017ara}. These collisions would be possibly mediated by hidden gauge fields, and are a generic consequence of those models \cite{Loeb:2010gj}. Due to scattering, heat would be transferred from high to low velocity particles, enhancing the velocity dispersion of the central regions and reducing the cuspy densities of the halos. For that reason, SIDM was proposed to solve the cusp-core problem \cite{2000PhRvL..84.3760S}, which could be explained in this way, as shown in N-body simulations \cite{Rocha:2012jg, 2012MNRAS.423.3740V}. While the ``too-big-to-fail'' discrepancy may also be alleviated with SIDM \cite{2012MNRAS.423.3740V}, the amount of substructures predicted in simulations is almost identical to that in CDM, and thus the missing satellite problem would remain unsettled \cite{DOnghia:2002nap, 2012MNRAS.423.3740V}.

\item \textbf{Fuzzy Dark Matter (FDM)}

Another popular example considers DM composed by an ultra light scalar field, behaving as axion-like particles (although different from the QCD axion). A specially interesting case is the so-called Fuzzy Dark Matter (FDM), which is a limit of a scalar field DM with masses $\sim 10^{-22}$ eV without self-interactions, behaving as a classical scalar field at cosmological scales \cite{2000PhRvL..85.1158H, 2017PhRvD..95d3541H}. Its evolution is ruled by the Schrödinger equation in the expanding universe, which can be recasted in continuity and Euler-like fluid equations (the so-called Madelung equations), with an additional effective quantum potential term. This induces an effective Jeans scale (given by a macroscopic de Broglie wavelength), which further suppresses fluctuations at small scales, while FDM behaves as CDM at larger scales \cite{2018EPJWC.16806005L}. According to these effects, FDM has been proposed to account for the aforementioned small-scale problems \cite{2017PhRvD..95d3541H, Du:2016zcv}.

\item \textbf{Primordial Black Holes (PBHs)}

BHs formed in the early universe from the direct collapse of high density fluctuations conform an interesting candidate for DM, specially after the first measurement of gravitational waves from a merger of BHs by the LIGO collaboration \cite{PhysRevLett.116.061102}. Regarding its behavior in the formation of structures, these PBHs would mostly act as CDM, although solar mass PBHs may present an enhancement on the fluctuations at small scales due to their discrete distribution \cite{2003ApJ...594L..71A}. Besides that, it has been claimed that the missing satellite and too-big-to-fail problems may be also alleviated, since the presence of PBHs would imply a large population of ultra-faint dwarf galaxies, in order to be consistent with the LIGO merger rates \cite{Clesse:2016vqa, 2018PDU....22..137C}. Moreover, they would present unique features which may imply different observational effects, such as PBH evaporation or emission of energetic radiation due to accretion. Given the richness of the physics involved, and the variety of phenomenological effects in the evolution of structures and the IGM, Chapter \ref{chap:PBHs} is entirely dedicated to their study.

\end{itemize}

Driven by solving the above observational issues, these models become plausible candidates for DM. Additionally, some particle physics models may predict such particle candidates, also motivating their study. In this thesis, we focus on studying three of the aforementioned non-standard DM alternatives, WDM, IDM, and PBHs which can leave substantial imprints in the thermal evolution of the universe, the formation of first galaxies, the Reionization epoch and the 21 cm signal. The physical effects, constraints and impact on structure formation of WDM and IDM scenarios shall be discussed along this chapter, while PBHs are studied in the next one.

\section{Structure formation in a nutshell}

Before discussing in detail different non-CDM models, it may be useful to review the standard lore of formation of structures, summarizing the computation of linear fluctuations and the description of halos.

\subsection{Growth of linear perturbations}

\label{sec:lineargrowth}

In this section, we briefly review some key points regarding the growth of perturbations and the formation of structures, such as the Jeans instability and the evolution of the growth factor. The universe at the largest scales behaves as a homogeneous and isotropic fluid, as described in Sec. \ref{sec:background}. However, the real universe is highly inhomogeneous at smaller scales, and the study of the evolution of deviations from the background model is needed in order to understand how the galaxies and structures we see today have been formed. The common lore of formation of structures considers that the early universe was highly homogeneous, with small perturbations seeded from quantum fluctuations by Inflation \cite{2005pfc..book.....M}. These inhomogeneities grew along the evolution of the universe, eventually causing the formation of cosmological structures such as galaxies and clusters. Since these fluctuations were very small in the primordial universe, at early times they can be treated as linear perturbations of the Einstein Equations, which greatly simplifies their study. Although to properly treat the evolution of linear perturbations in the expanding universe, General Relativity is required, it is possible to obtain some quantitative correct results at scales inside the Hubble radius within a Newtonian framework, much easier to interpret.

Denoting with $\rho$ the matter density, and with $\bar{\rho}$ its mean background value, the \textit{density fluctuation}, \textit{density contrast} or \textit{overdensity}, $\delta$, is defined as
\begin{equation}
\delta({\bf x},t) = \frac{\rho({\bf x},t)-\bar{\rho}(t)}{\bar{\rho}(t)}.
\end{equation}
Continuity and Euler fluid equations for perturbations of the matter density $\delta$ and velocity ${\bf v} $ at first order in a expanding medium read \cite{1993ppc..book.....P, Kolb:1990vq, 2005pfc..book.....M,Baumann_lectures}
\begin{equation}
\dot{\delta} +\frac{1}{a}\nabla \cdot {\bf v} = 0, \; \; \; \; \dot{{\bf v}} + H{\bf v} = -\frac{1}{a \bar{\rho}}\nabla \delta P - \frac{1}{a}\nabla \Phi,
\label{eq:cont_euler}
\end{equation}
where $\delta P$ is the pressure perturbation and $\Phi$ the gravitational potential, which fulfills the Poisson equation,
\begin{equation}
\frac{1}{a^2}\nabla^2 \Phi = 4\pi G_N \bar{\rho} \delta \, .
\label{eq:poisson}
\end{equation}
The $a$ factors and the drag term $H {\bf v}$ account for the expanding background. The above equations can be combined, leading to an unique evolution equation for the matter overdensity $\delta$. It is more practical to work in Fourier space, where the gradient terms are replaced by $\vec{ \nabla} \rightarrow i {\bf k}$, having then algebraic rather than differential equations (in the spatial coordinates). Furthermore, in the linear regime, different Fourier modes evolve independently, which implies that a field at a scale ${\bf k}$ does not depend on other scales ${\bf k}'$. We define the Fourier transform of a field $\delta({\bf x})$ as
\begin{equation}
\tilde{\delta}({\bf k}) = \int d^3k \; \delta({\bf x}) e^{i{\bf k} \cdot {\bf x}} \, .
\end{equation}
Combining and taking the Fourier transform of Eqs. \eqref{eq:cont_euler} and \eqref{eq:poisson}, one obtains a single evolution equation for $\tilde{\delta}({\bf k})$ as \cite{Kolb:1990vq}
\begin{equation}
\ddot{ \tilde{\delta}} +2H\dot{\tilde{\delta}} +\left(\frac{c_s^2}{a^2}k^2 -4 \pi G_N \bar{\rho}\right) \tilde{\delta}=0~,
\label{eq:delta_evol}
\end{equation}
where we have used an adiabatic equation of state which relates the pressure $\delta P$ and the density through the speed of sound $c_s = \delta P/ \delta \rho$. Attempting exponential solutions of the form $\tilde{\delta} \sim \exp i \omega t$, we can recognize in this equation a damped oscillator, with a frequency given by $\omega^2=\frac{c_s^2}{a^2}k^2 -4 \pi G_N \bar{\rho}$, where the Hubble rate damps fluctuations. Two opposite terms determine this quantity: the first one is due to the thermal motion of the fluid through pressure, while the second one is a consequence of gravitational attraction. Therefore, there exist two possible behaviors: if $c_s^2k^2/a^2 >4 \pi G_N \bar{\rho}$, then $\omega^2>0$, and the perturbations oscillate, and are progressively dissipated due to the $H$ term. However, if $c_s^2k^2/a^2 < 4 \pi G_N \bar{\rho}$, the frequency acquires an imaginary value, so that the dominant solution grows exponentially. This fact indicates that the small fluctuations become unstable and grow. Matter tends to collapse by the gravitational force, increasing the density in some regions, breaking there the validity of the linear treatment. The separation between the oscillatory and collapsing regimes is governed by the (comoving) Jeans length $\lambda_J$, and its Fourier wavenumber $k_J = 2\pi/\lambda_J$, defined as
\begin{equation}
\lambda_J=\frac{2 \pi}{ k_J}=c_s \sqrt{\frac{\pi }{ G_N \bar{\rho}}}~.
\label{eq:lambdajeans}
\end{equation}
We can define the Jeans mass as the mass within a sphere of diameter $\lambda_J$ with a density equal to the background one: $M_J=(4\pi/3)\bar{\rho}(\lambda_J/2)^3$. These results can be easily interpreted: the regions of the universe with masses above the critical mass $M_J$ (or fluctuations with wavelengths larger than $\lambda_J$) eventually collapse. On the other hand, fluctuations smaller than the Jeans scale would be dissipated and washed out.

The above equations can apply to the standard CDM case, where pressure can be safely neglected. During matter domination, $H=2/(3t)$ and $4 \pi G_N \bar{\rho} = \frac32 H^2$. Thus, Eq. \eqref{eq:delta_evol} reads
\begin{equation}
\ddot{ \tilde{\delta}} +\frac{4}{3t}\dot{\tilde{\delta}}  - \frac{2}{3t^2} \tilde{\delta}=0~.
\end{equation}
It is possible to split the time and wavevector dependencies, defining the \textit{growth factor} $D(z)$ as
\begin{equation}
\tilde{\delta}({\bf k},t) = \frac{D(t)}{D(t_*)}\tilde{\delta}({\bf k},t_*),
\end{equation}
with $t_*$ any reference time. The growth factor is usually normalized to unity at redshift $0$, $D(z=0)=1$. Thus, it is straightforward to obtain the two solutions of the equation above, which presents a decaying mode as $ D(t) \propto t^{-1}$, and a growing solution as $D(t) \propto t^{2/3} \propto a$.

During radiation domination, one should take into account the effect of radiation fluctuations sourcing the potential term. While a correct evaluation of the growth of perturbations at these epochs requires General Relativity, it is still possible to estimate the evolution of $\delta$ arguing that the potential term in Eq. \eqref{eq:delta_evol} can be neglected, since $\ddot{\delta} \sim H\dot{\delta} \sim H^2 \delta $ and $\rho_m \ll \rho_\gamma \sim H^2$ \cite{Baumann_lectures}. Thus, since $H=1/(2t)$, one finds
\begin{equation}
\ddot{ \tilde{\delta}} +\frac{1}{t}\dot{\tilde{\delta}} \simeq 0~,
\end{equation}
whose solutions are a constant mode, and a logarithmic growing mode, $D(t) \propto \log(a)$. Contrary to the matter domination era, when matter fluctuations evolve linearly with the scale factor, during radiation domination they only slowly grow in a logarithmic way. The stagnation of growth previous to the matter-radiation equality is known as the \textit{Mészáros effect} \cite{1974A&A....37..225M}, and points out that structure formation only becomes relevant at later times, at the matter domination era.

Since linear perturbations evolve from Gaussian fluctuations, during the linear regime gaussianity still holds. It means that, as long as non-linearities are not important, since they are gaussian distributed, the density fluctuations are only characterized by the second statistical moment, given by the \textit{two-point correlation function}, $\xi(\mathbf{x},\mathbf{y}) = \langle \delta(\mathbf{x}) \delta(\mathbf{y}) \rangle$. It will be more useful to work with the Fourier transform of this quantity, known as the \textit{matter power spectrum} $P(k,z)$, which is given by
\begin{equation}
\langle  \widetilde{\delta} (\mathbf{k}, z)  \widetilde{ \delta}^* (\mathbf{k}^\prime, z) \rangle = (2\pi)^3 \delta_D (\mathbf{k} - \mathbf{k}^\prime) P(k,z) ~,
\label{eq:PS}
\end{equation}
where $\delta_D$ is the Dirac delta function and the brackets account for the average over different realizations. Statistical homogeneity enforces the appearance of the delta function, while statistical rotational invariance implies that the power spectrum depends only on the module of $\mathbf{k}$. Given the factorization of the growth function, the linear power spectrum can be written as $P(k,z)=D(z)^2P(k,0)$. It is also customary to work with the so-called dimensionless power spectrum
\begin{equation}
\Delta^2 (k,z) = \frac{k^3}{2 \pi^2} P(k,z).
\end{equation}
The initial power spectrum is determined by the primordial conditions of the universe, drawn by the inflation process. In the standard picture, inflation stretches the microscopic quantum fluctuations, which become perturbations of the fields at cosmological scales. A well-known prediction from inflationary models is the nearly scale-invariant spectrum, close to the Harrison-Zeldovich spectrum $P(k) \propto k^{n}$, with $n \simeq 1$ at large scales, suggested before independently by Harrison \cite{1970PhRvD...1.2726H} and Zeldovich \cite{1972MNRAS.160P...1Z} from other arguments. This scaling can be derived from the assumption that the gravitational potential field must be scale invariant, i.e., it presents the same power at all scales. This is a desirable property to ensure that it is finite at both small and large scales. Thus, $\Delta_\Phi^2 \sim constant$, where $\Delta_\Phi^2$ is the dimensionless power spectrum for the potential fluctuations. From the Poisson equation in Fourier space, Eq. \eqref{eq:poisson}, $k^2\Phi \propto \delta$, and thus $k^4 \Delta_\Phi^2 \propto \Delta^2$, implying that $P(k) \propto k$. The preferred value for the spectral index from CMB and LSS data of $n=0.9665 \pm 0.0038$ \cite{Aghanim:2018eyx} only slightly deviates from the Harrison-Zeldovich spectrum, as inflation predicts.

The study of linear perturbations in General Relativity shows that the Fourier modes of the gravitational potential whose wavelength is larger than the horizon at the epoch of matter-radiation equality (i.e., $k< a(z_{eq})H(z_{eq})$, with $ z_{eq}$ the matter-radiation equality redshift defined in Sec. \ref{sec:background}) were causally uncorrelated, and thus have remained mostly constant. As a consequence, their corresponding matter fluctuations keeps the same spectrum than the initial, roughly $\propto k$. On the other hand, those modes which had crossed the horizon before the matter-radiation equality (i.e., $k> a(t_{eq})H(t_{eq})$) become damped due to the stall of the growth of fluctuations during the radiation epoch. Moreover, baryon-photon coupling due to Thomson scattering induces some oscillations in the matter power spectrum, known as Baryon Acoustic Oscillations (BAOs). To properly predict the linear evolution and the exact shape of the power spectrum, numerical codes are required to solve the relativistic perturbation equations, such as the popular and public Boltzmann solvers CAMB (Code for Anisotropies in the Microwave Background) \cite{Lewis:1999bs} and CLASS (Cosmic Linear Anisotropy Solving System) \cite{Lesgourgues:2011re}. This description stands for standard CDM cosmologies, while other DM alternatives may present differences in the growth of the density contrast, leaving different signatures in the power spectrum. It is the case of the WDM and IDM models considered below, which present an additional cutoff or suppression at small scales, or PBHs, which may enhance the power at small scales due to shot noise, as shall be discussed in Sec. \ref{sec:shot}.

It is worth it to mention that the standard treatment of linear fluctuations considered here accounts for \textit{isentropic} (also called \textit{adiabatic}) perturbations, i.e., those where the entropy remains unperturbed. An additional kind of perturbations are \textit{isocurvature} modes, where fluctuations in entropy rather than in density fields are considered (see, e.g., \cite{2010gfe..book.....M}). The current understanding of the evolution of fluctuations explains the formation of cosmic structures from adiabatic modes. However, isocurvature modes may still be relevant in some scenarios, such as PBHs, as studied in the next chapter.

\subsection{Virial equilibrium}
\label{sec:virial}

When fluctuations go of order unity, $\delta \sim 1$, linear perturbation theory is no longer valid, and thus non-linear approaches are required in order to understand how cosmic structures and galaxies have formed from the small initial overdensities. Analytical approximate methods, such as the Zeldovich approximation \cite{1970A&A.....5...84Z}, can offer relatively accurate results, although N-body simulations are needed for precise calculations. A very simple analytical prescription, although still qualitatively useful, considers that roughly spherical overdense regions expand and eventually turnaround and collapse due to the gravitational force. This picture of spherical collapse, firstly outlined by Gunn \& Gott in 1972 \cite{1972ApJ...176....1G}, can be analytically solved, finding that the overdensity collapses to a point, reaching infinite density in a finite time. One can match the result from the extrapolation of linear theory evaluated at that collapse time, finding that, at redshift $z$, it corresponds to \cite{1993ppc..book.....P, 2010gfe..book.....M}
\begin{equation}
\delta_c(z) = \frac{1.686}{D(z)}
\label{eq:delta_c}
\end{equation}
for the Einstein-de Sitter universe. One can interpret this as assuming that all overdensities above that threshold at redshift $z$ would collapse. However, the collapse to a point would be an unrealistic situation provided a slight violation of spherical symmetry or pressure effects. Instead, matter collapses to a halo, reaching a self-gravitating virial equilibrium by violent relaxation \cite{Barkana:2000fd, 2010gfe..book.....M}. From spherical collapse and the virial theorem, one can estimate the overdensity that a virialized overdensity reaches at redshift $z$ relative to the critical density $\rho_c(z)$ defined in Sec. \ref{sec:background}, as \cite{Bryan:1997dn}
\begin{equation}
\Delta_{vir}(z) = 18\pi^2 + 82 \, (\Omega_{\textrm{m},z} - 1) - 39 \, (\Omega_{\textrm{m},z} - 1)^2 ~,  
\end{equation}
where we have defined
\begin{equation}
\Omega_{\textrm{m},z} = \frac{\Omega_{\rm m} \, (1+z)^3}{\Omega_{\rm m} \, (1+z)^3 + \Omega_\Lambda} ~.
\end{equation}
One can define the \textit{virial radius} $R_{\rm vir}$, related to the mass of the halo, $M$, by
\begin{equation}
M = \frac{4\pi}{3} \, \Delta_{vir}(z) \, \rho_c(z) \, R_{\rm vir}^3 ~.
\label{eq:Rvir}
\end{equation}
Then,
\begin{equation}
\begin{split}
R_{\rm vir}(M, z) & \simeq 0.784 \, h^{-1} \textrm{kpc} \,  \left(\frac{M}{10^8 M_\odot \, h^{-1}} \, \right)^{1/3} \\
& \times \left(\frac{\Omega_{\rm m}}{\Omega_{\textrm{m},z}} \, \frac{\Delta_{vir}(z)}{18\pi^2} \right)^{-1/3} \left(\frac{1+z}{10}\right)^{-1} \, .
\end{split}
\end{equation}
From the expression of the circular velocity, $V_c^2 = G_N M/R_{\rm vir}$, one can write the \textit{virial temperature} $T_{\rm vir}$ as
\cite{Barkana:2000fd}
\begin{equation}
\begin{split}
T_{\rm vir}(M, z) = \frac{\mu m_p V_c^2}{2 k_B} & \simeq 1.98 \times10^4 \, \textrm{K} \, \left(\frac{\mu}{0.6} \, \right) \left(\frac{M}{10^8 M_\odot \, h^{-1}} \, \right)^{2/3} \\
& \times \left(\frac{\Omega_{\rm m}}{\Omega_{\textrm{m},z}} \, \frac{\Delta_{vir}(z)}{18\pi^2} \right)^{1/3} \left(\frac{1+z}{10}\right)  ~,
\end{split}
\label{eq:Tvir}
\end{equation}
being $\mu  = 0.6$ the mean molecular weight of ionized IGM and $\mu  = 1.22$ for the neutral one. As shall be shown, these quantities are of great interest since they roughly describe the mass and length scales of DM halos.

\subsection{Halo Mass Function}
\label{sec:halos}

During the Dark Ages and later times, baryons and DM surrounding primordial seeds tend to collapse and form DM halos, as N-body simulations predict. These halos could eventually host star forming galaxies and quasars, responsible for altering the IGM. It is thus of great importance to understand the abundance and mass spectrum of existing halos. The \textit{halo mass function} accounts for the comoving number of halos per unit halo mass and volume, and can be written as~\cite{Press:1973iz}
\begin{equation}
  \frac{dn}{dM}=-\frac12 \, \frac{\rho_m}{M^2} \, f(\nu) \, \frac{d\ln \sigma^2}{d\ln M} ~,
\label{eq:dndM}
\end{equation}
where $\rho_m = \Omega_m \, \rho_c$ is the average matter density in the Universe at $z = 0$, $f (\nu)$ is the so called \textit{first-crossing distribution} and $\sigma^2 = \sigma^2(M,z)$ is the variance of the filtered density perturbations at a scale given by the mass $M$. The variance $\sigma^2$ is computed using linear perturbation theory, as discussed in Sec. \ref{sec:lineargrowth}. Its redshift dependence is given by the growth function, $D(z)$, normalized to 1 at $z = 0$, as $\sigma(M, z) = D(z) \sigma(M, z = 0) $. The variance at $z = 0$ and at scale $R$, $\sigma(M) \equiv \sigma(M, z = 0)$, reads
\begin{equation}
\sigma^2(M(R)) = \int \frac{d^3k}{(2\pi)^3} \, P(k) \, W^2(kR) ~, 
\label{eq:sig2}
\end{equation}
where $P(k)$ is the linear power spectrum at $z = 0$ and $W(kR)$ is the Fourier transform of a filter function. Several choices are possible, being usually taken as a top-hat (TH) function in real space (i.e., a Heaviside step function), whose Fourier transform reads
\begin{equation}
W_{\rm TH}(kR) = \frac{3}{kR} \, \left(\sin(kR)- kR\cos(kR)\right) ~.
\label{eq:TH}
\end{equation}
The relation of the mass $M$ with the scale $R$ depends on the choice of filter function. In the top-hat case, it is given by $M = \frac{4\pi}{3} \, \rho_m \, R^3$. It is worth it to note that depending on the type of filter, the mass might not be uniquely defined in terms of the scale.

On the other hand, the first-crossing distribution, $f (\nu)$, depends on the specific details on the collapse model, and is a function of $\nu = \delta_c^2(z)/\sigma^2(M)$, with $\delta_c(z)$ the linearly extrapolated overdensity for spherical collapse from Eq. \eqref{eq:delta_c}. From the extended Press-Schechter, or excursion set formalism \cite{Bond:1990iw, ZENTNER_2007, Maggiore_2010}, it is possible to derive the function $f (\nu)$. The easiest scenario, the Press-Schechter (PS) prescription, assumes spherical collapse and the condition that only overdensities above the critical density $\delta_c(z)$ are large enough to collapse to a halo. This threshold can be understood as a constant barrier, since it does not depend on the mass $M$. For gaussian distributed fluctuations, the former assumptions imply a first crossing distribution of the form \cite{Press:1973iz}
\begin{eqnarray}
f(\nu) =\sqrt{\frac{2 \, \nu}{\pi}} \, e^{- \nu/2} ~, 
\label{eq:PShmf}
\end{eqnarray}
Although there is general agreement of this result with the halo abundances from N-body simulations, it underpredicts the number of more massive halos, while overpredicts the low mass ones. The problem becomes even more pronounced at higher redshifts \cite{2005Natur.435..629S}. However, further improvements on the prediction of the halo mass function can be achieved by considering instead a more realistic ellipsoidal collapse model. In the excursion set-formalism, it implies a moving barrier, where the density threshold depends on the mass $M$. Although it cannot be computed exactly by analytical means, the Sheth-Tormen (ST) halo mass function is based on this assumption and provides a fit for the first-crossing distribution \cite{Sheth:1999mn, Sheth:1999su, Sheth:2001dp}
\begin{eqnarray}
f(\nu) = A \, \sqrt{\frac{2 \, q \, \nu}{\pi}} \left(1 + (q \, \nu)^{-p}\right) \, e^{-q \, \nu/2} ~, 
\label{eq:SThmf}
\end{eqnarray}
with $p = 0.3$, $q = 0.707$ and $A =(1 + 2^{-p}\Gamma(1/2-p)/\sqrt{\pi} )^{-1} = 0.3222$. The above computations are only strictly valid for a hierarchical CDM universe. Instead, standard spherical (or elliptical) collapse does not longer apply to DM scenarios with structure suppression by free-streaming or collisional damping, effects which need to be taken into account. The proper modifications to the formalism for WDM and IDM are discussed in the following sections.

\begin{figure}
\begin{center}
\includegraphics[scale=0.55]{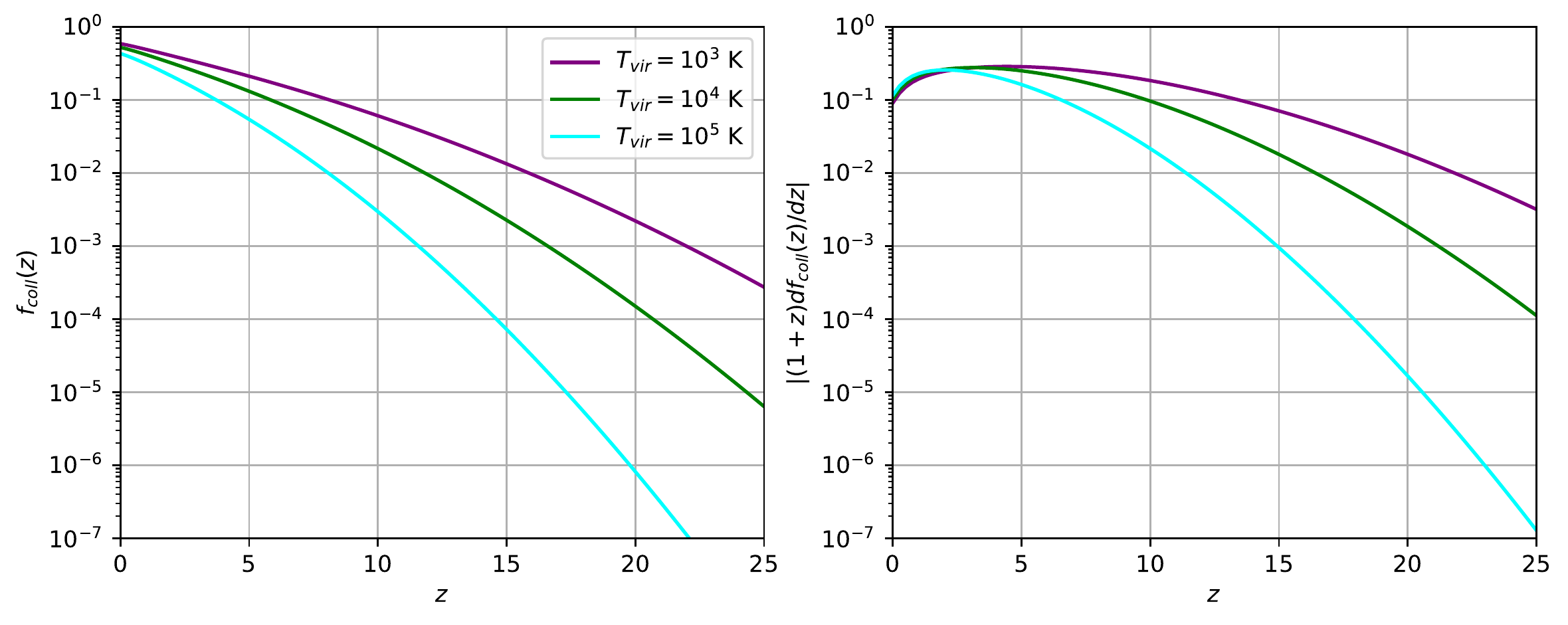}
\caption{\textit{Left:} evolution of the fraction of mass collapsed in halos above a mass corresponding to a virial temperature $T_{\rm vir}$ via Eq. \eqref{eq:Tvir}, assuming a ST prescription. \textit{Right:} derivative of the same quantity than in the left panel times $1+z$.}
\label{fig:fcoll}
\end{center}
\end{figure}

As shall be seen in Chapter \ref{chap:IGM}, stellar formation and radiation fields responsible for heating and ionizing the IGM can be written in terms of the amount of halos capable of hosting star formation. Thus, a quantity of interest is the fraction of mass collapsed in halos above a mass $M$, which can be written in terms of the halo mass function, Eq. \eqref{eq:dndM}, as
\begin{equation}
f_{coll}(z, M) = \frac{1}{\rho_m}
 \int_{M}^{\infty} dM' \, M' \, \frac{d n(M')}{dM'}~.
 \label{eq:fcoll}
\end{equation}
The relevant threshold mass for the IGM evolution is related to the minimum mass required to host star formation in a halo, as shall be discussed in Chapter \ref{chap:IGM}. Using the PS prescription, Eq. \eqref{eq:PShmf}, the collapsed fraction can be computed analytically, obtaining
\begin{equation}
f_{coll, PS}(z, M)= erfc \left(\frac{\delta_c(z)}{\sqrt{2}\sigma(M)}\right),
\label{eq:fcollPS}
\end{equation}
where $erfc$ is the complementary error function, whereas for the ST function, Eq. \eqref{eq:SThmf}, it must be computed numerically. Figure \ref{fig:fcoll} shows the evolution of the fraction of mass collapsed and its redshift derivative for different threshold masses given by their virial temperatures.

\section{Warm Dark Matter}
\label{sec:WDM}

As already discussed, a well motivated DM scenario considers particles with masses of the order of $\sim 1$ keV, coined as WDM. The distinction with CDM cosmologies comes basically from its behavior at the decoupling period. In the WDM case, particles decoupled while they were still relativistic, while in the CDM case, particles had become already non-relativistic at its decoupling time. As a consequence of that, WDM particles still have some velocity dispersion, being negligible in the CDM case. Whereas at large scales, both models are indistinguishable, at small scales, due to the random thermal motion, WDM shows a suppression of power, above some cutoff scale, which washes out fluctuations. This causes a delay in structure formation and avoids a large population of small-size collapsed objects, or low-mass halos. 

There are several possible candidates from particle physics theories which may behave as WDM. Supersymmetric theories predict the existence of a fermionic partner of the graviton, the gravitino, with spin $3/2$ and mass $\sim 1$ keV, which was the earliest proposal of a WDM particle as a thermal relic \cite{PhysRevLett.48.1636, 1982Natur.299...37B}. On the other hand, right-handed sterile neutrinos are present in many Beyond the Standard Model (BSM) scenarios, and can behave as WDM candidates \cite{Abazajian:2012ys, Adhikari:2016bei,Abazajian:2017tcc}. In this decade, there was a renewed interest in sterile neutrino WDM when two independent groups reported a X-ray excess at an energy of 3.55 keV in the Chandra and XMM-Newton data from galaxy clusters \cite{2014ApJ...789...13B, 2014PhRvL.113y1301B}. Since this signal is not known as an atomic line in the spectra of galaxies or clusters, it may come from a different source. It has been speculated to be originated from the decay of a sterile neutrino of mass $m_s = 7.1$ keV to an active neutrino, through the decay $\nu_s \rightarrow \nu + \gamma$. But the existence of this signal as well as its interpretation are still under discussion: further observations of galactic clusters with the Chandra ACIS \cite{2016A&A...592A.112H} and with the XMM-Newton satellite \cite{Bhargava:2020fxr} have found no evidence of such an excess. In addition, the signal could be due to an incomplete subtraction of atomic lines~\cite{2015MNRAS.450.2143J}. Although nowadays there is not a definite conclusion about the nature of this signal, it is expected that upcoming X-ray surveys will clarify its status.

In the following, kinetic properties of WDM, possible candidates, the free-streaming effect and the impact on the formation of cosmological structures are discussed.

\subsection{Kinetic properties}

Next, we review the general equilibrium properties of the WDM distribution. We assume that WDM is composed of particles of mass $m_{X}$ which were thermally coupled to the plasma in the early universe, but became decoupled through a freeze-out mechanism while they were still relativistic \cite{Kolb:1990vq, 2001ApJ...558..482B, Bode:2000gq}. This is contrary to the standard WIMP scenario, where DM particles decoupled when they were already non-relativistic. Considering this kind of particles as fermions, the initial distribution in phase space is given by the Fermi-Dirac function\footnote{Although the following computations stand for fermionic DM, the final results can also apply to bosons, multiplying the number density by a factor $4/3$ (since the fermionic number density is $3/4$ times the bosonic one when particles are relativistic \cite{2003moco.book.....D}).}
\begin{equation}
f_X(p)=\frac{1}{e^{pc/(k_B T_X)}+1}~,
\label{eq:FD}
\end{equation}
with $p$ the momentum and $T_X$ the WDM temperature. While interactions were strong enough to maintain the equilibrium with the thermal plasma, WDM particles became thermally distributed, with the temperature $T_X$ equal to the one of the plasma. After decoupling, the functional form of Eq. \eqref{eq:FD} becomes frozen, remaining self-similar, with the temperature redshifting as $\propto (1+z)$. Integrating the distribution function Eq. \eqref{eq:FD} over momentum one gets the number density of WDM particles, which reads
\begin{equation}
n_X=g_X \frac{3\zeta(3)}{4\pi^2}T_X^3~,
\label{eq:numbwdm}
\end{equation}
where $\zeta$ is the Riemann zeta function and $g_X$ is the internal number of degrees of freedom, chosen as 2 for spin 1/2 particles. On the other hand, the energy density when WDM particles are non-relativistic (as currently they should behave) simply reads $\rho_X = m_X n_X$. Thus, we can relate their number density today $n_{X,0}$ with their energy density, or equivalently, with the fraction relative to the critical density, $\Omega_{X}=m_{X}n_{X,0}/\rho_c$. In addition, from Eq. \eqref{eq:FD} the root-mean-square velocity of the particles $v_{rms}$ can be computed as
\begin{equation}
v_{rms}=\frac{\sqrt{\langle p^2\rangle}}{m_X}=\frac{1}{m_X}\left( \frac{\int g_X d^3p \, p^2 f_X}{\int g_X d^3p f_X}\right)^{1/2}=3.597 \; \frac{k_B T_X}{m_X}~.
\end{equation}
Since $T_X \propto (n_X/g_X)^{1/3}$, we finally can write the above result as
\begin{equation}
v_{rms}=0.0437(1+z) \left( \frac{\Omega_Xh^2}{0.15} \right)^{1/3} \left( \frac{m_X}{1 \textrm{keV}} \right)^{-4/3} \textrm{km s}^{-1}~,
\label{eq:rms}
\end{equation}
which exhibits the scaling of the velocity dispersion with the mass and the abundance, explicitly showing that less massive particles present a larger velocity dispersion.

We can relate $T_X$ with the neutrino temperature considering the entropy conservation before and after the thermal decoupling, as it is usually done with neutrino decoupling \cite{2003moco.book.....D}. The entropy goes as $s \sim g_* T^3$, with $g_*$ the \textit{effective thermal number of degrees of freedom}. If thermally coupled, bosons contribute just as $g$ to $g_*$, while  fermions as $7/8 \, g$, due to the proper phase space integral. Before DM decoupling, it takes a value $g_*(t_{dec})$. After DM decoupling, but before electron-positron annihilation, the degrees of freedom are given by electrons (2), positrons (2), neutrinos (3), antineutrinos (3) and photons (2), summing thus to $g_* = 43/4=10.75$. From the conservation of comoving entropy, one then finds
\begin{equation}
\frac{T_X}{T_\nu}= \left(\frac{43/4}{g_*(t_{dec})}\right)^{1/3}.
\end{equation}
To relate $T_X$ to the CMB temperature, one has to take into account that the thermal plasma is heated up by the electron-positron annihilation after neutrino decoupling, while the comoving neutrino temperature remains unchanged, leading to a relation $T_\nu = T_\gamma (4/11)^{1/3}$ \cite{2003moco.book.....D, Kolb:1990vq}.
 
It is worth it to add a cautionary note regarding thermal WDM. As happens with standard neutrinos, one can constrain the mass and number of WDM particles thermally decoupled in the early universe via the Gershtein-Zeldovich limit, enforcing that they cannot overclose the universe, i.e., $\Omega_{X} \lesssim 1$. The fraction of WDM can be written in terms of the mass and the temperature as
\begin{equation}
\Omega_{X} = \frac{m_X n_X}{\rho_{c,0}} \simeq \frac{m_X}{94.10 \, {\rm eV} h^2} \left(\frac{T_X}{T_\nu}\right)^3.
\end{equation}
In order to have a thermal WDM candidate with mass of the order of $\sim 1$ keV accounting for all the DM ($\Omega_{\rm WDM} \simeq 0.25$), one needs $\left(T_X/T_\nu\right)^3 \sim 0.01$, implying $g_*(t_{dec}) \sim 10^3$.\footnote{In the original gravitino literature, this bound is relaxed to $g_*(t_{dec}) \sim 200$, but assuming $\Omega_{\rm WDM}=1$, which we now know it must be lower \cite{PhysRevLett.48.223}.} Such a big number is much larger than the expected effective number of degrees of freedom before the electroweak transition in the SM, which is roughly $\sim 100$ (see, e.g., Ref. \cite{Kolb:1990vq}). Thus, in order to be consistent with the Gershtein-Zeldovich limit, WDM particles should have decoupled in the very early universe, when some kind of new physics would present many degrees of freedom at high energies. Hence, thermal WDM, although in principle achievable, requires extra exotic physics. Despite that, it is the benchmark WDM scenario, at which most of constraints are placed. As discussed in the next section, other candidates with a thermal-like spectrum are usually considered, which may avoid the above restriction.

An additional warning must be stressed. Light particles with a thermal distribution behaving like radiation during the BBN era would enhance the radiation energy density, which has a great impact in the nucleon abundances resulting from BBN. It is quantified through the \textit{effective number of neutrinos} $N_{eff}$, which takes the value of $N_{eff}=3.044$ in the standard scenario, coming from the three neutrino species plus quantum electrodynamics corrections \cite{Iocco:2008va, Lesgourgues:2018ncw} (see Ref. \cite{Bennett:2020zkv} for a recent and accurate computation). The joint analysis of BBN and CMB data from Planck agrees with that previous quantity, with an uncertainty of around the 10\%, being thus well constrained \cite{Aghanim:2018eyx}. Additional radiation species during BBN would contribute to $N_{eff}$ with a factor $(T_X/T_\nu)^4 \propto g_*(t_{dec})^{-4/3}$, which must take a low value in order to be consistent with BBN and CMB results. This implies an additional reason why $g_*(t_{dec})$ must be large enough, independently on the total abundance and mass. Note however that thermal relics with $\Omega_X \sim 0.25$ and $m_X \sim 1$ keV able to be compatible with the Gershtein-Zeldovich limit would also overcome this restriction.

\subsection{Free-streaming}

\begin{figure}
\begin{center}
\includegraphics[scale=0.75]{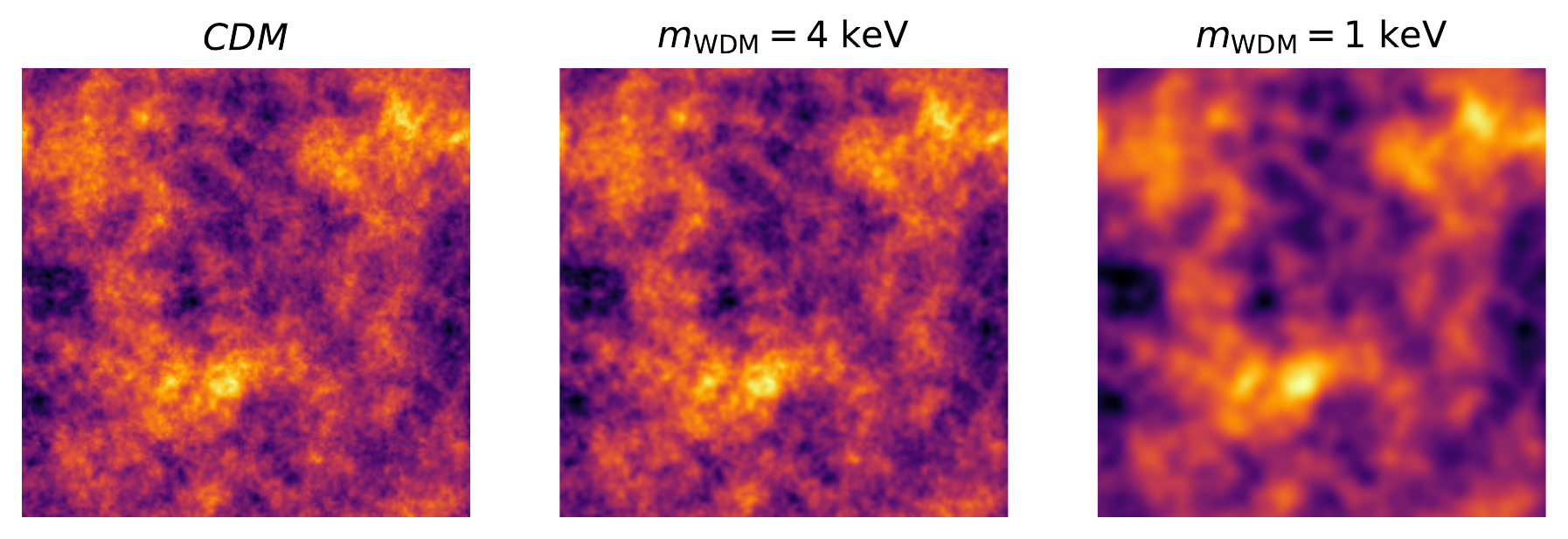}
\caption{Random gaussian fields in a window of $10$ Mpc$/h$ generated by the matter power spectrum at redshift $0$, for CDM (left), and WDM, with $m_{\rm WDM}=4$ keV (middle) and $m_{\rm WDM}=1$ keV (right), employing the power spectrum with the transfer function from Eq. \eqref{eq:twdm}. Note that lower particle masses increase the suppression scale, which erases smaller structures, and only larger fluctuations remain.}
\label{fig:wdmgaussian}
\end{center}
\end{figure}

If WDM particles are supposed to be thermally produced in the early universe (as in the WIMP scenario), they are assumed to be still relativistic at the epoch of their decoupling from the thermal plasma, but expected to become non-relativistic around the time of matter radiation equality, $t_{\rm eq}$, when the matter perturbations start to grow substantially, as discussed in Sec. \ref{sec:lineargrowth}. The free-streaming scale can be estimated as the distance a particle can travel until the matter-radiation equality $t_{\rm eq}$ \cite{Kolb:1990vq, Schneider:2013ria}. While the particle is relativistic, its velocity is close to the speed of light $v \sim c$. However, after a time $t_{\rm nr}$, when the temperature drops below $T_X(t_{\rm nr}) \sim m_Xc^2/3$, the particle becomes non-relativistic, and hereafter its velocity decays as $v \propto a^{-1}$. Thus, the comoving distance travelled up to $t_{\rm eq}$ can be written as \cite{Kolb:1990vq}
\begin{equation}
\lambda_{fs} = \int_{0}^{t_{\rm eq}} dt' \frac{v(t')}{a(t')} \simeq \int_{0}^{t_{\rm nr}} dt' \frac{c}{a(t')} + \int_{t_{\rm nr}}^{t_{\rm eq}} dt' \frac{v(t')}{a(t')}.
\end{equation}
Since this process takes place in the radiation domination era, the scale factor evolves as $a(t)=a_{\rm nr}(t/t_{\rm nr})^{1/2}$, and the free-streaming length reads
\begin{equation}
\lambda_{fs} \simeq r_{\rm H}(t_{\rm nr})\left( 1 + \frac12 \log \left(  \frac{t_{\rm eq}}{t_{\rm nr}} \right) \right),
\end{equation}
where $r_{\rm H}(t_{\rm nr}) = c\int_{0}^{t_{\rm nr}} dt'/a(t') = 2 t_{\rm nr}/a(t_{\rm nr})$ is the comoving horizon at the time $t_{\rm nr}$. From the condition $T_X(t_{\rm nr}) \sim m_Xc^2/3$, since the WDM and photon temperatures evolve with the same dependence, $T_X \propto T_\gamma \propto a^{-1}$, one finds that $a_{\rm nr} \propto t_{\rm nr}^{1/2} \propto m_X^{-1} (T_X/T_\gamma)$, and thus, the $\lambda_{fs} \propto m_X^{-1} (T_X/T_\gamma)$. Given that, from Eq. \eqref{eq:numbwdm}, $n_X \propto g_X T_X^3$, and $\Omega_X = m_X n_{X,0}/\rho_{c,0}$, the scaling of the free-streaming scale goes as \cite{Schneider:2011yu}
\begin{equation}
\lambda_{fs} \simeq 0.4 \left( \frac{m_X}{\rm keV} \right)^{-4/3} \left( \frac{\Omega_X h^2}{0.135} \right)^{1/3} h^{-1} {\rm Mpc}.
\label{eq:lambda_fs}
\end{equation}
%
One can apply this estimate to different free-streaming species. For instance, in the case of a massive neutrino species of $m_\nu \sim $ eV, since the neutrino temperature is related to the CMB one by $T_\nu = T_\gamma (4/11)^{1/3}$ (as discussed in the previous section), one has $\Omega_\nu h^2 \simeq m_\nu/(91 {\rm eV})$, and one obtains $\lambda_{fs, \nu} \simeq 10^3 \; {\rm Mpc} \; (1 {\rm eV}/m_\nu)$. If these particles formed most of the DM, which enters in the category of HDM, this estimate predicts that structures with sizes below that scale would not be formed, in contradiction with observations. However, for particles with masses around $\sim$ keV, one finds a free-streaming length of the order of $\lambda_{fs, {\rm WDM}} \sim$ 10-100 kpc, which corresponds to subgalactic sizes. This range is precisely the one required to explain the small-scale problems discussed in Sec. \ref{sec:smallscale}. To further illustrate this effect, Fig. \ref{fig:wdmgaussian} shows realizations of random gaussian fields, as the cosmological density field behaves in the linear regime, generated with the power spectrum corresponding to WDM (commented in the next section) with $m_{\rm X}=4$ keV and WDM with $m_{\rm X}=1$ keV, together with the CDM case, within a box of 10 Mpc/$h$. The small-sized fluctuations are washed out in WDM scenarios, with a greater effect the lower the mass.

The estimation of the free-streaming scale can also be carried out from the Jeans instability analysis discussed in Sec. \ref{sec:lineargrowth}. While in CDM models, pressure can be safely neglected, leading to small Jeans lengths, thermal motion of WDM models provides a larger scale above which perturbations grow. Substituting the speed of sound in Eq. \eqref{eq:lambdajeans} by the velocity dispersion from Eq. \eqref{eq:rms}, one finds a scaling of the Jeans length of $\lambda_J \propto T_X/m_X \propto m_X^{-4/3}\Omega_X^{1/3}$, which is the same as the one obtained above.

The mass of WDM particles and the free-streaming scale are not univocally related, since the momentum distribution and the number density depend on the particle production mechanism. So far, throughout all the discussion, thermal relics have been assumed to constitute the WDM. However, besides thermal production through a freeze-out decoupling process from the thermal plasma, as in the WIMP scenario, WDM candidates may be produced by other mechanisms. For instance, sterile neutrinos produced at non-resonant oscillations from active neutrinos have also been proposed, by the so-called Dodelson-Widrow mechanism \cite{PhysRevLett.72.17}. The distribution function for particles produced in this way has the same functional form as a thermal distribution, but suppressed by a factor which depends on the masses of the active and sterile neutrinos. Although the original Dodelson-Widrow model as 100\% of DM has been completely ruled out from Local Group galaxy counts plus X-ray observations \cite{Horiuchi:2013noa}, it is still considered as the archetypal sterile neutrino DM scenario. On the other hand, resonantly produced sterile neutrinos have also been proposed through the so-called Shi-Fuller mechanism, which would have a much cooler distribution function (i.e., skewed to low energies) than a thermally produced particle with the same mass \cite{1999PhRvL..82.2832S}. Although thermal production is usually assumed for most of the bounds, it is sometimes possible to relate the constraints for non-thermal WDM to those for thermal WDM \cite{Colombi:1995ze}. This happens, for instance, with non-resonantly-produced sterile neutrinos (as the Dodelson-Widrow mechanism), where bounds on the sterile neutrino mass $m_s$ can be derived from constraints on the equivalent thermal relic mass $m_X$ by the relation \cite{2016JCAP...08..012B}
\begin{equation}
m_s = 3.90 \; {\rm keV} \; \left( \frac{m_X}{\rm keV} \right)^{1.294} \; \left( \frac{0.25 \times 0.7^2}{\Omega_X h^2} \right)^{1/3} .
\label{eq:sterilemass}
\end{equation}
Constraints on resonantly produced sterile neutrinos (e.g., due to the Shi-Fuller mechanism) cannot be translated from thermal relic bounds and need to be studied separately. This is because there is not a single function relating the masses of both scenarios, since the relation also depends on the mixing angle \cite{2016JCAP...04..059S, 2017JCAP...12..013B}. Generally, for the same free-streaming scale, the Shi-Fuller mechanism provides lower masses than non-resonantly produced neutrinos \cite{Abazajian:2021zui}. 

The most stringent observational constraints on the WDM particle mass are obtained from observations of the Ly$\alpha$ forest, based on the comparison of high redshift quasar spectra with hydrodynamical simulations. From QSOs spectra observed with the Keck High Resolution Echelle Spectrometer (HIRES) and the Magellan Inamori Kyocera Echelle (MIKE) spectrographs, together with SDSS data from the DR1 and DR2 data releases, this procedure has led to a limit on the mass of a WDM thermal relic of $m_X > 3.3$ keV (at the 2$\sigma$ level)~\cite{2013PhRvD..88d3502V}. Making use of the SDSS-III/BOSS data, previous lower bounds where improved to $m_X > 4.09$ keV for a thermal relic, and to $m_s > 24.4 $ keV for a non-resonantly produced sterile neutrino, both at the 95 \% CL (although slightly weakened when Planck 2016 CMB data is also considered) \cite{2016JCAP...08..012B}. A recent analysis employing eBOSS and XQ-100 Ly$\alpha$ data found the most stringent bounds on the thermal relic mass of $m_X > 5.3$ keV, or a limit on the non-resonantly-produced sterile neutrino mass of $m_s > 34$ keV (both at 95 \% CL) \cite{2020JCAP...04..038P}. The drawback of these methods is that they rely on model-dependent numerical simulations, where assumptions regarding the thermal history are taken, such as a relation between the temperature and the density, which may not be completely accurate. Besides these Ly$\alpha$ forest constraints, the number of satellite galaxies of the Milky Way has also been used to explore the allowed range of masses. Since WDM predicts fewer satellite galaxies than the standard scenario, this allows rejecting masses below $m_X = 2.02$ keV (95 \% CL) for a thermal relic, regardless of the specific assumptions about galaxy formation processes. This limit improves up to $3.99$ keV when modeling Reionization, which suppresses dwarf galaxies formation \cite{2020arXiv201108865N}. A recent DES analysis, with new observed satellite galaxies, reject masses below $6.5$ keV \cite{Nadler:2020prv}. Anyway, MW satellite constraints strongly depend on the MW mass, whose value still presents some uncertainties \cite{Wang:2015ala, 2020MNRAS.494.4291C, 2020MNRAS.494.5178F}. It is worth it to keep in mind that the aforementioned bounds stand for all DM composed by WDM, but in mixed scenarios with CDM plus WDM in some amount, the bounds would weaken. In Part \ref{partII} of this thesis, WDM scenarios are studied with simulations of the thermal evolution of the universe, exploring how Reionization data and the 21 cm signal can constrain such models \cite{Lopez-Honorez:2017csg, Villanueva-Domingo:2017lae}.

\subsection{Impact on structure formation}
\label{sec:warm-dark-matter}

\subsubsection{Linear power spectrum}

\begin{figure}
\begin{center}
\includegraphics[scale=0.4]{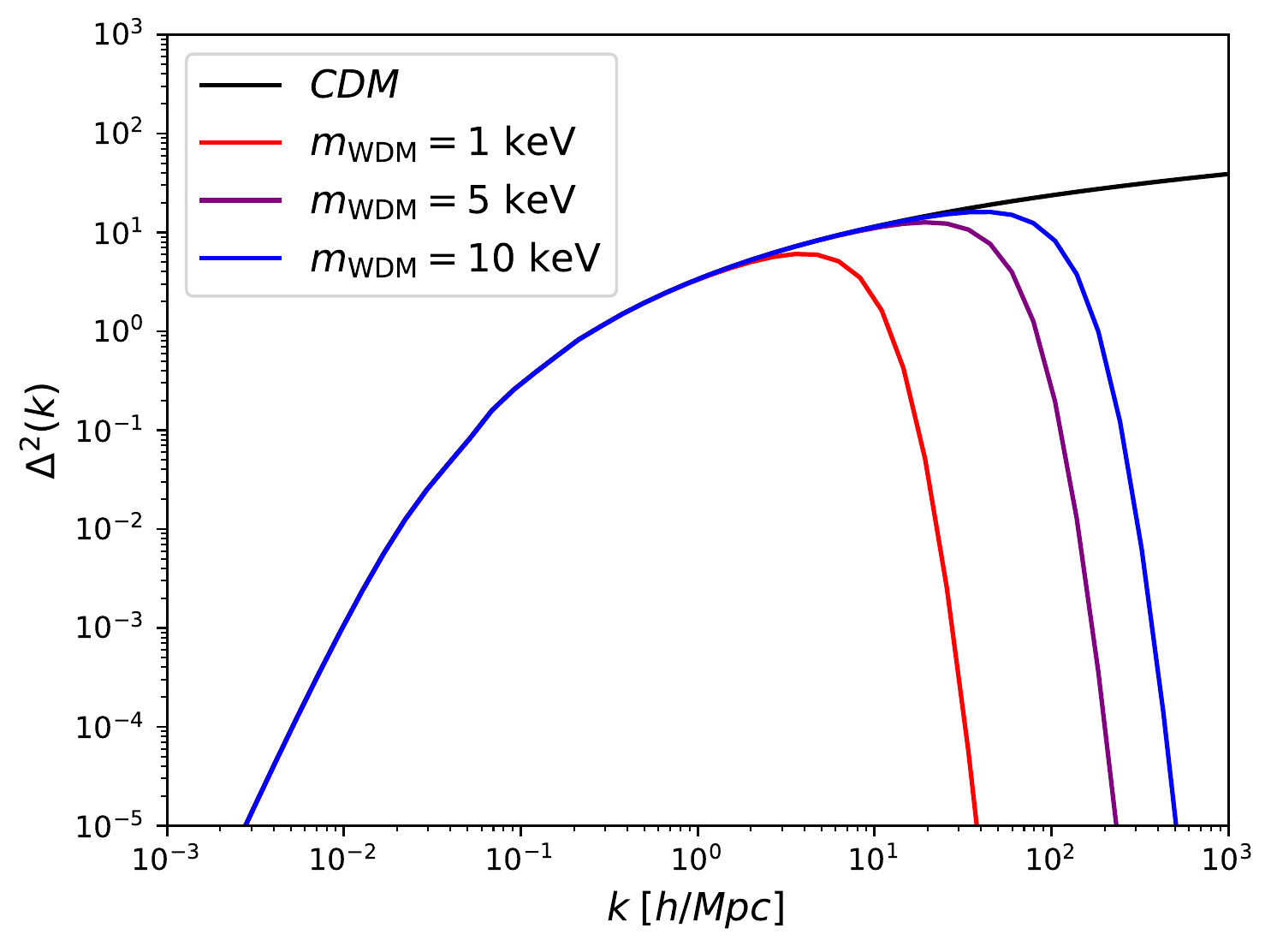}
\includegraphics[scale=0.4]{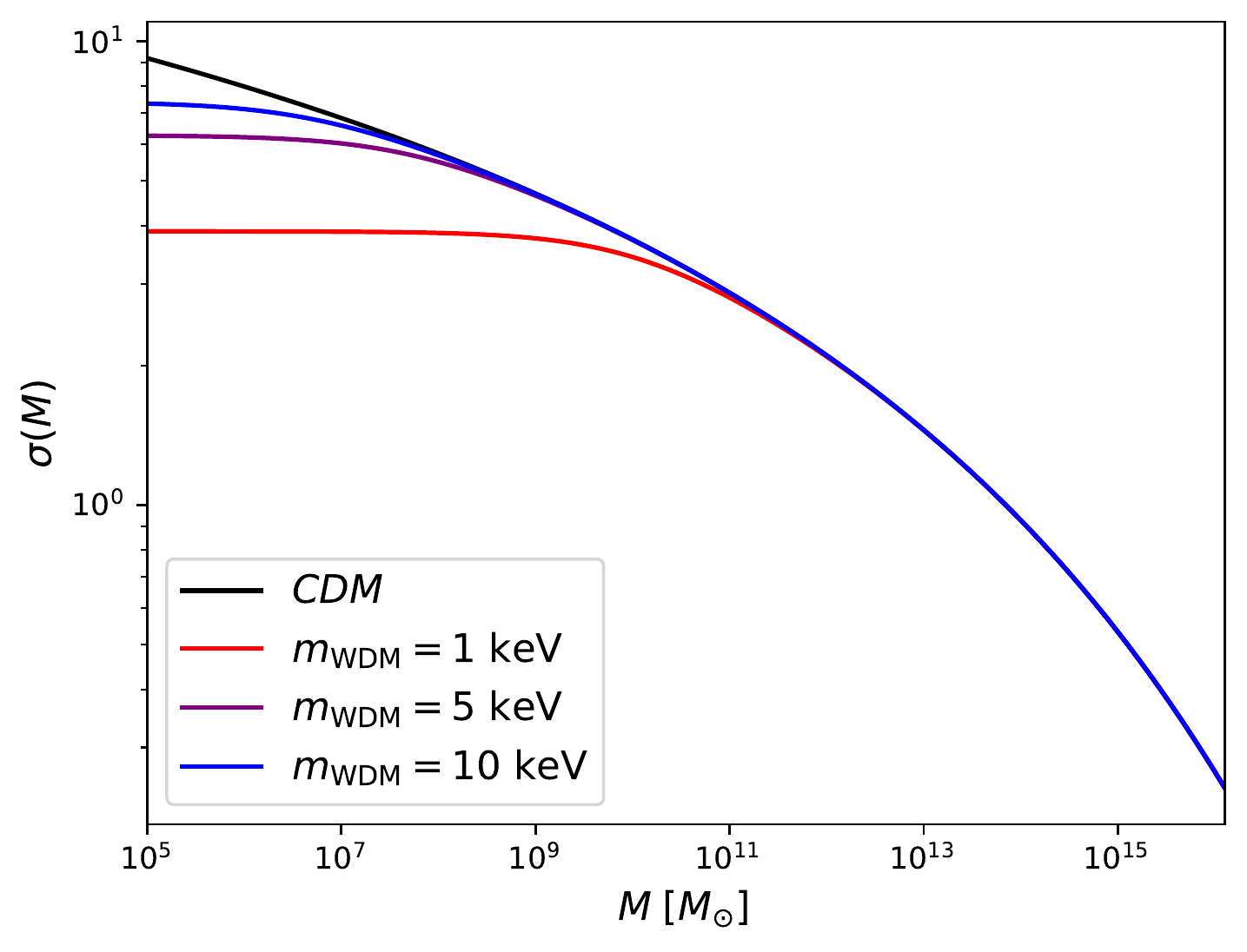}
\includegraphics[scale=0.4]{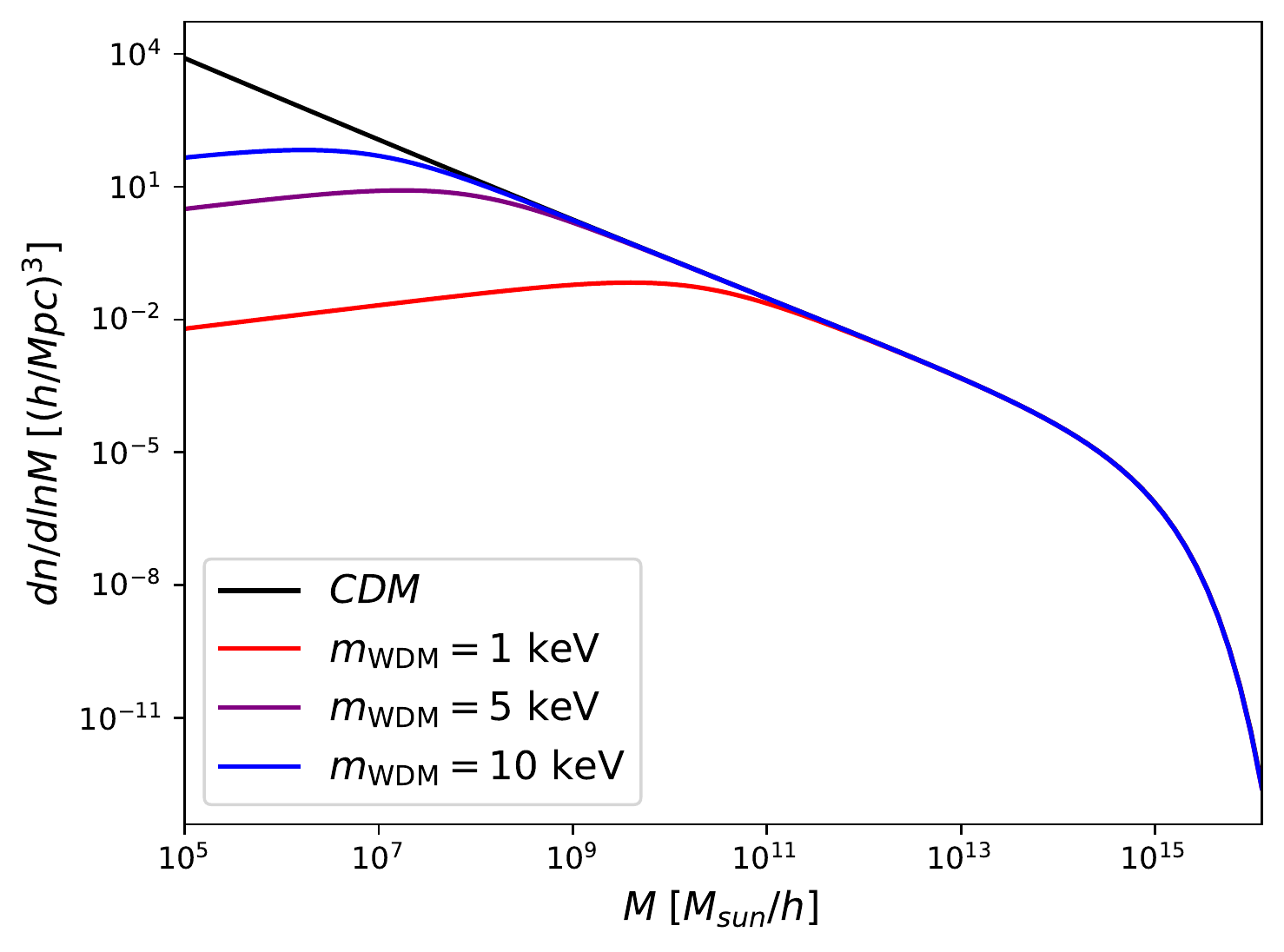}
\caption{Dimensionless power spectrum (top left), variance $\sigma(M)$ (top right) and halo mass function (bottom) for CDM and three WDM models, evaluated at $z=0$. Note the suppression at small scales, corresponding to high $k$ in the power spectrum, and at low masses $M$ in the variance and halo mass function.}
\label{fig:wdm_hmf}
\end{center}
\end{figure}

The above simplified sketch allows one to understand the scaling of the free-streaming length with the mass $m_X$, and roughly estimate the wavelength of modes which would be affected. However, it lacks two relevant points. On the one hand, it does not take into account the logarithmic growth of perturbations during the radiation dominated era. On the other hand, the free streaming does not instantaneously switch off after $t_{\rm eq}$. A precise evaluation of the free-streaming damping requires solving numerically the evolution of linear perturbations \cite{Bode:2000gq}. One can write the WDM linear power spectrum in terms of a suppression relative to the CDM case characterized by the transfer function $\mathcal{T}^2_{ X}(k)$,
\begin{equation}
\label{eq:pwdm}
P_{ X}(k) =   \mathcal{T}^2_{ X}(k) \, P_{\rm CDM}(k).
\end{equation}
Numerical computations from Boltzmann solver codes showed that the WDM transfer function can be fitted  to the function \cite{Viel:2005qj}
\begin{equation}
  \mathcal{T}_{ X}(k) = (1+ (\alpha k)^{2\nu})^{-5/\nu} ~, 
  \label{eq:twdm}
\end{equation}
with $\nu=1.12$, and the breaking scale
\begin{equation}
  \alpha= 0.049 \left(\frac{{\rm keV}}{m_X}\right)^{1.11}\left(\frac{\Omega_X}{0.25}\right)^{0.11}\left(\frac{h}{0.7}\right)^{1.22} \, {\rm Mpc}/h ~.
  \label{eq:alpha_wdm}
\end{equation}
The parameter $\alpha$ determines the scale where the suppression due to free streaming appears, and thus, it must be directly related to $\lambda_{fs}$. Note that our previous naive estimation of $\lambda_{fs}$, Eq. \eqref{eq:lambda_fs}, provides a decent estimation to the accurate computation of Eq. \eqref{eq:alpha_wdm}, with a similar scaling of the parameters $\Omega_X$ and $m_X$. It must be stressed that this fitting formula corresponds to a thermal relic, although as it has been shown, it can be translated to some non-thermal cases via Eq. \eqref{eq:sterilemass} \cite{Viel:2005qj}. The dimensionless power spectrum for several WDM cases is plotted in left panel of Fig. \ref{fig:wdm_hmf}, clearly showing the damping in fluctuations above the free-streaming scale. The CDM, contrarily, keeps growing at small $k$, a definitory characteristic of its hierarchical behavior.

On the other hand, it is customary to define the half-mode length $\lambda_{\rm hm} = 2\pi/k_{\rm hm}$, as the scale for which $\mathcal{T}_{X}/\mathcal{T}_{\rm CDM} = 1/2$. From the transfer function, Eq. \eqref{eq:twdm}, this quantity reads
\begin{equation}
\lambda_{\rm hm} =  2\, \pi \, \alpha_X \left(2^{\nu/5} - 1 \right)^{-1/(2 \, \nu)} ~,
\end{equation}
while its corresponding half-mode mass $M_{\rm hm}$ is thus defined by
\begin{equation}
M_{\rm hm}  = \frac{4 \, \pi}{3} \, \rho_m \, \left( \frac{\lambda_{\rm hm}}{2}\right)^3 = \frac{32 \, \pi^4}{3} \, \rho_m \, \alpha_X^3 \left(2^{\nu/5} - 1 \right)^{-3/(2 \, \nu)} .
\label{eq:Mhm}
\end{equation}
Together with $\alpha$, these scales are usually employed to reflect the lengths at which suppression effects become relevant.

\subsubsection{Halo mass function}

In order to obtain the halo mass function, one could naively employ the Press-Schechter formalism outlined in Sec. \ref{sec:halos}, making use the PS or ST halo mass functions with the WDM power spectrum. However, numerical simulations show that when doing that, N-body results are not correctly recovered, but always present a further suppression at low masses. To understand why this procedure fails, one has to recall that in the excursion set formalism, there is a mapping between linear fluctuations and non-linear ones, which can be computed from simple pressureless spherical or ellipsoidal collapse. This approach has a remarkable success for CDM, where structure formation progresses hierarchically. However, in WDM, the problem is more complicated, since pressure effects cannot be longer neglected at small scales, preventing hierarchical collapse, and the computations of simplified spherical or elliptical collapse become invalid. In order to properly treat the problem, one should re-evaluate the excursion set approach, modifying it according with the WDM nature, which involves considering a moving barrier \cite{2013MNRAS.428.1774B}. An easier approach consists on replacing the top-hat window function by a sharp-$k$ function (a Heaviside step function in Fourier space).\footnote{Besides this WDM issue, there are good reasons for considering the sharp-$k$ filter. Actually, in set excursion formalism, exact analytic results for the first crossing PS distribution can only be obtained with the sharp-$k$ filter, see, e.g., Refs. \cite{Bond:1990iw, Maggiore_2010}.} While in the CDM case the chosen filter does not significantly matter, with this choice in the WDM scenario, unlike with other window functions such as the top-hat filter, the halo mass function naturally tends to zero at low masses, as expected from the nature of structure formation with the free-streaming effect \cite{Schneider:2013ria}. This procedure has a remarkable success at low redshifts, with the advantage of being simple and well motivated. However, sharp-$k$ filters present the drawback that the mass is not univocally defined, besides the scaling $M \propto R^3$ with the filtering radius $R$. It requires to introduce a free parameter $c$ so that $M=4\pi/3 c \bar{\rho}R^3$, which has to be calibrated with simulations, finding the best match to be $c \simeq 2.5$ \cite{Schneider:2014rda}.

Nevertheless, to phenomenologically match the N-body simulations, the easiest way is by modifying the ST halo mass function with the appropriate factor which gives the best fit. The authors of Ref. \cite{Schneider:2011yu} provided a functional form for that which reads
\begin{equation}
\frac{dn^{\rm WDM}}{dM} = \left(1 + \frac{M_{\rm hm}}{b \, M}\right)^{a}  \frac{dn^{\rm ST, \, WDM}}{dM} ~.
\label{eq:hmf_wdm}
\end{equation}
The WDM halo mass function is thus given by the ST prescription $\frac{dn^{\rm ST, \, WDM}}{dM}$ (employing Eq. \eqref{eq:SThmf}), with the linear matter power spectrum corresponding to the WDM case, Eq. \eqref{eq:pwdm}, multiplied by a mass-dependent term. This correction factor further suppresses the abundance at masses below $M_{\rm hm}$, and is  determined by two parameters, $a$ and $b$. Although the initial formulation worked with $b=1$, it was later shown that the fit improved with $b = 0.5$, whereas $a = -0.6$ \cite{Schewtschenko:2014fca, Moline:2016fdo}.

As a final comment, it is worth emphasizing that modeling WDM with numerical simulations presents additional challenges with respect to CDM, such as the appearence of artificial numerical fragmentation. The existence of shot noise due to the initial random
thermal velocities typical in WDM can seed unphysical overdensities at small
scales. As the simulation evolves, these artificial fluctuations lead to the formation of spurious structures, producing an overabundance
of low-mass halos, which is just the opposite of what is
expected in a WDM scenario. This effect has been extensively discussed in the literature, and different ways to get rid off these fake halos have been proposed \cite{Wang:2007he,Schneider:2013ria, Angulo:2013sza, Hobbs:2015dda, Banerjee:2016zaa}.

\section{Interacting Dark Matter}

\label{sec:IDM}

Among the fundamental requirements of any DM candidate, it is the absence of interactions with radiation. However, it may be the case that such interactions actually exist but are so weak that are unobservable at large scales, although they could become important at smaller subgalactic scales. This exciting possibility is explored in the so-called Interacting Dark Matter (IDM) models, term which usually encompasses interactions with light particles, such as photons or neutrinos. In analogy with baryon-photon interactions via Thomson scattering, one could assume that DM particles can scatter elastically with radiation. This would induce similar effects in the growth of fluctuations of DM to those observed in the baryon sector, such as an oscillatory spectrum given by sound waves driven by radiation pressure.
Furthermore, small scales would be suppressed by \textit{collisional damping}, the analogous of Silk damping due to Thomson scattering
\cite{1968ApJ...151..459S}. Thus, as in the WDM case, it would result in the suppression of small-size cosmological structures, potentially having an impact on structure formation. As we shall discuss, IDM scenarios can be closely related to WDM models at the linear level, given that cosmological effects are similar. However, this sort of degeneracy between models is broken when non-linear effects are taken into account, such as in the study of DM halos.

Photon-DM interacting scenarios have been extensively studied in the literature, either studying its impact on the CMB, large scale structure and growth of linear perturbations \cite{Boehm:2000gq, Boehm:2001hm, Boehm:2004th, Wilkinson:2013kia, 2018JCAP...10..009S, Stadler:2019qno} or performing N-body simulations and comparing with the observed Milky Way satellites \cite{2014MNRAS.445L..31B, Schewtschenko:2015rno, 2015MNRAS.449.3587S, Moline:2016fdo, Schewtschenko:2016fhl}. On the other hand, the signatures of neutrino-DM interactions on the CMB and large scale structure have also been widely examined \cite{2006PhRvD..74d3517M, Wilkinson:2014ksa, Escudero:2015yka, Schewtschenko:2016fhl, 2019JCAP...08..014S, Stadler:2019qno, Mosbech:2020ahp}. Henceforth in this thesis, we focus on photon-DM interactions, although we shall comment on simmilarities with the neutrino-DM counterpart. Throughout this section, we physically motivate these scenarios, discuss the collisional damping effects, and the signatures that IDM may leave in structure formation.

\subsection{Particle physics motivation}

An IDM scenario can arise from different fundamental or effective particle physics models. Consider, as a simple example, the case of a \textit{hidden photon}, i.e., a vectorial gauge boson for a $U(1)$ gauge symmetry, represented by $B_\mu$. Such a boson may arise, for instance, from Supersymmetric models or string theories \cite{Dienes:1996zr, Mirizzi:2009iz}. These particles would not interact with the SM particles, being for that reason \textit{hidden}, but instead they could interact with DM. However, this field may also be mixed with the standard photon $A_\mu$. Defining the field strengths as $F_{\mu\nu}=\partial_\mu A_\nu - \partial_\nu A_\mu$ and $B_{\mu\nu}=\partial_\mu B_\nu - \partial_\nu B_\mu$, one can write the lagrangian of the photon and hidden photon terms as \cite{Holdom:1985ag, Mirizzi:2009iz}
\begin{equation}
\mathcal{L} = -\frac{1}{4}F_{\mu\nu}F^{\mu\nu} -\frac{1}{4}B_{\mu\nu}B^{\mu\nu} + \frac{g}{2}B_{\mu\nu}F^{\mu\nu} + \mathcal{J}^\mu_{\rm em}A_\mu + \mathcal{J}^\mu_{\rm DM}B_\mu.
\end{equation}
The first two terms correspond to the kinetic lagrangians of the gauge fields, while the third term accounts for a possible mixing between the standard and the hidden photon ruled by a mixing parameter $g$, which is allowed by the symmetries. The last terms provide the interaction between the SM and DM particles through the currents $\mathcal{J}^\mu_{\rm em}$ and $\mathcal{J}^\mu_{\rm DM}$, with $A_\mu$ and $B_\mu$ respectively. The mixing coupling is sometimes interpreted as a mixing angle, as $g=\sin(\chi)$ \cite{Mirizzi:2009iz}. It is possible, however, to get rid of the mixing term by a convenient redefinition of the fields, or equivalently, changing the base $A, B$ to diagonalize the kinetic part, as
\begin{equation}
A'_\mu = \sqrt{1-g^2}A_\mu \; ; \; \; \; B'_\mu = B_\mu -gA_\mu \; .
\end{equation}
Under the above change of variables, one finds that the mixing term is removed and the kinetic lagrangian becomes diagonalized:
\begin{equation}
\begin{split}
\mathcal{L} = & -\frac{1}{4}F'_{\mu\nu}F'^{\mu\nu} -\frac{1}{4}B'_{\mu\nu}B'^{\mu\nu} \\
&+ \frac{1}{\sqrt{1-g^2}} \mathcal{J}^\mu_{\rm em}A'_\mu + \mathcal{J}^\mu_{\rm DM}\left(B'_\mu + \frac{g}{\sqrt{1-g^2}} A'_\mu\right),
\end{split}
\end{equation}
with $F'_{\mu\nu}$ and $B'_{\mu\nu}$ the field strengths for the new fields $A'_\mu$ and $B'_\mu$. Note, however, that now the DM current gets coupled to the new photon field $A'_\mu$. In this way, an effective interaction between DM particles and photons arises, determined by the coupling $g$. Since these processes have not been measured, this constant must be very small. Actually, predictions from Supersymmetry and string theory range between $10^{-16}$ and $10^{-2}$ \cite{Dienes:1996zr}. The effective electric charge of the DM field would be a small non-integer quantity given by $g$, and the DM particles would then become \textit{millicharged}. Note that, if the DM are composed of fermionic fields, the $\gamma$DM coupling would have the same form that the interaction term in Quantum Electrodynamics, the same processes being allowed. More concretely, elastic scattering of DM particles with photons would be present, with an energy-independent cross section at low energies, in analogy with the Thomson scattering \cite{Peskin:1995ev}. Following that argument, in millicharged DM models, the $\gamma$DM cross section can be expressed as
\begin{align}
\sigma_{\gamma {\rm DM}} = \epsilon^2 \sigma_T \left(\frac{m_e}{m_{\rm DM}}\right)^2 ~, 
\end{align}
where $\sigma_T=6.65\times 10^{-25}$ cm$^2$ is the Thompson cross section, $\epsilon = |q|/e$ with $q$ the DM electric charge ($q = e g/\sqrt{1-g^2}$ in the example above), $e$ the electron charge and $m_e$ its mass. The mass factor in the expression above comes from the fact that the Thomson cross section scales as $\sigma_T \propto m_e^{-2}$. In models such as the outlined above, however, baryon-DM scattering would also be possible, which strongly constrains these scenarios using CMB and Ly$\alpha$ forest \cite{McDermott:2010pa, Dvorkin:2013cea}. The upper bound obtained on the millicharge from the aforementioned analyses can be summarized as
\begin{align}
\epsilon < 1.8\times 10^{-6}\left(\frac{m_{\rm DM}}{\rm GeV}\right)^{1/2} ~,
\label{eq:millichargedbound}
\end{align}
which is valid for $m_{\rm DM} \gtrsim$ MeV. Other bounds can be found from solar data \cite{Vinyoles:2015khy} or from accelerators \cite{Agrawal:2021dbo}. Since we mostly focus on the impact of collisional damping in structure formation, the IDM scenario assumed in this thesis only considers photon-DM scattering, without baryon-DM scattering, which would have additional effects on the thermal history of the universe. Therefore, the above limit may not be applicable to our case if non-gravitational baryon-DM interactions could be neglected. \footnote{Nevertheless, depending on the specific millicharged model, there may be still allowed regions in the parameter space where bounds could be relaxed, see, e.g., Fig. 1 of Ref. \cite{Vinyoles:2015khy}.} Non-millicharged IDM scenarios are still possible, however. It is the case, for instance, of DM composed by axion-like particles, whose lagrangian interaction term reads $ \kappa \, a \, F_{\mu\nu}F^{\rho \sigma}\epsilon^{\mu\nu\rho\sigma}$, with $a$ the axion-like scalar field, $\kappa$ the coupling constant and $\epsilon^{\mu\nu\rho\sigma}$ the Levi-Civita symbol. These scenarios present $\gamma$DM scattering with interaction amplitudes scaling as $\sim \kappa^4$, while axion-baryon scattering can only happen at higher order, $\sim \kappa^4 e^4$ (at one loop), in the case that axion-baryon interaction terms are absent or negligible.\footnote{Note, however, that such $\gamma$DM cross section would scale with the energy as $\propto E^2$, and thus our formalism for Thomson-like scattering would not be valid, thus it must be regarded just as an illustrative example.} On the other hand, $\nu$DM interactions would not be related to millicharged particles, and in that case the bound from Eq. \eqref{eq:millichargedbound} does not apply. In any case, the IDM scenario treated in this thesis should be regarded as a phenomenological scenario rather than a specific particle physics model, in order to explore the consequences of collisional damping from DM interactions with light or massless particles, which may leave an imprint on the evolution of the universe and the growth of structures.

\subsection{Collisional damping}

In this section, the damping scale due to $\gamma$DM interactions is derived from physical arguments.\footnote{Note that in the literature, damping effects by DM self-interactions are also sometimes considered \cite{Boehm:2000gq, Boehm:2004th}, although these are not covered in this thesis, restricting ourselves only to photon-induced damping.} Consider a DM particle immersed in the photon fluid, where both species interact. The path traveled after several successive collisions can be regarded as a random walk (see, e.g., Refs. \cite{Hu:2008hd, Kolb:1990vq}), as firstly proposed by Silk in his seminal paper where collisional damping of the CMB due to baryon-photon scattering was described \cite{1968ApJ...151..459S}. A photon travels a distance $\vec{r}_i$ before the $i$-th scattering. The net displacement $\vec{R}$ after $N$ free paths reads $\vec{R} = \sum_i^N \vec{r}_i$. While its average vanishes (since it is a vector composed by random walks without a preferred direction), its square leads to a non-zero value. Writing the variance of each step as $\lambda_{\rm mfp}^2 = \langle |\vec{r}_i|^2 \rangle$, thus, $\langle |\vec{R}|^2 \rangle = N \lambda_{\rm mfp}^2$ (where products between different steps average to zero since they are not correlated) \cite{1986rpa..book.....R}. The (physical) collisional damping scale $\lambda_{\rm cd,phys}$ would be given by the total traveled distance until the decoupling time $t_{\rm dec}$, $\lambda_{\rm cd,phys}^2 \sim N \lambda_{\rm mfp}^2$. For a interaction rate of $\Gamma = n_{\rm DM} \sigma_{\gamma \rm DM}$, the mean-free path of a photon reads $\lambda_{\rm mfp} = \Gamma^{-1}\;$.\footnote{The photon mean free path is always much larger than the DM one, by a factor $n_\gamma/n_{\rm DM}$. For this reason we are estimating the photon damping scale rather than the DM one. As long $\gamma$DM interactions are strong, DM particles remain coupled to photons and radiation fluctuations drag matter with them. Thus, the photon damping length is the relevant quantity \cite{Kolb:1990vq}.} On the other hand, the average number of collisions during a time interval $\Delta t$ is written as $N \sim c\Delta t/\lambda_{\rm mfp}$. Thus, the total distance traveled gives the comoving collisional damping scale $\lambda_{\rm cd}$ as \cite{Kolb:1990vq,Boehm:2000gq,2002dmap.conf..333B}
\begin{equation}
\lambda_{\rm cd}^2 \simeq \int_0^{t_{\rm dec}}  \; \frac{dt}{\lambda_{\rm mfp}} \; \frac{\lambda_{\rm mfp}^2}{a^2} = \int_0^{t_{\rm dec}}  \; \frac{dt}{n_{\rm DM} \sigma_{\gamma \rm DM}a^2},
\end{equation}
where the $a^2$ factor makes it a squared comoving distance.
The integral can be easily performed, taking into account that $\gamma$DM scatterings occur during the radiation-domination era, where $a \propto t^{1/2}$. The decoupling time $t_{\rm dec}$ is defined by the relation $H(t_{\rm dec}) \sim n_{\rm DM} \sigma_{\rm \gamma DM}$, from which one finds $a(t_{\rm dec}) \propto \Omega_{\rm DM}\sigma_{\gamma DM}/m_{\rm DM}$. Plugging in the numbers, we find a characteristic length of
\begin{equation}
\lambda_{\rm cd} \sim 0.01  \, {\rm Mpc}/h \, \left[10^8 \, \left(\frac{\sigma_{\gamma \rm{DM}}}{\sigma_T}\right) \, \left(\frac{{\rm GeV}}{m_{\rm DM}}\right)\right].
\label{eq:colldampscale}
\end{equation}
This is a rough but still decent estimation of the correct scale, providing the same order of magnitude for the relevant cross sections, as shall be seen in Sec. \ref{sec:linIDM}. Note that the above scale grows with the combination $\sigma_{\gamma \rm{DM}}/m_{\rm DM}$, rather than with the cross section alone. This is because it depends upon the interaction rate, which goes as $n_{\rm DM}\sigma_{\rm \gamma DM}\propto \sigma_{\rm \gamma DM}/m_{\rm DM}$. Therefore, this is the relevant quantity which rules the damping effects, while in the WDM case, it is the mass of the particle. Stronger scattering (i.e., larger cross sections) imply a longer damping scale and thus deeper impact on structure formation, washing out fluctuations below it. Note that cross sections of the order $\sigma_{\gamma {\rm DM}}/ \sigma_T \times \left(m_{\rm DM}/\textrm{GeV}\right)^{-1} \sim 10^{-8}$ would lead to damping lengths of the order of subgalactic sizes, required to account for the CDM issues discussed in Sec. \ref{sec:smallscale}. It must be stressed that the computation of Eq. \eqref{eq:colldampscale} is too simplistic since it gloss over several additional effects which may become relevant to determine the damping scale, such as heat conduction or shear effects \cite{Boehm:2004th}. Finally, Eq. \eqref{eq:colldampscale} may essentially work for other light or massless particles interacting with DM in thermal equilibrium with the thermal plasma in the early universe, such as neutrinos.

Based on this effect, CMB data from the Planck satellite have allowed to obtain upper bounds on the allowed elastic $\gamma {\rm DM}$ cross sections of
$\sigma_{\gamma {\rm DM}}/ \sigma_T \times \left(m_{\rm DM}/\textrm{GeV}\right)^{-1} \lesssim 10^{-6}$ \cite{Wilkinson:2013kia, 2018JCAP...10..009S}. However, more stringent constraints are provided from the study of satellite galaxies of the MW, which lie around 
$\sigma_{\gamma {\rm DM}}/ \sigma_T \times \left(m_{\rm DM}/\textrm{GeV}\right)^{-1} \lesssim 10^{-9}$ (value that depends on the MW mass) \cite{Boehm:2014vja}. On the other hand, elastic neutrino-DM scattering has been also constrained, resulting in the stringent constraints from Ly$\alpha$ forest $\sigma_{\nu {\rm DM}}/ \sigma_T \times \left(m_{\rm DM}/\textrm{GeV}\right)^{-1} < 1.5 \times 10^{-9}$ \cite{Wilkinson:2014ksa}.  In Part \ref{partII} of this thesis, we show how Reionization data, MW satellite galaxies and the 21 cm EDGES signal are able to provide stronger constrains on the elastic $\gamma$DM cross section \cite{Escudero:2018thh,Lopez-Honorez:2018ipk}.

\subsection{Imprints on structure formation}

\subsubsection{Linear power spectrum}

\label{sec:linIDM}

\begin{figure}[t]
\centering
\includegraphics[width=0.85\textwidth]{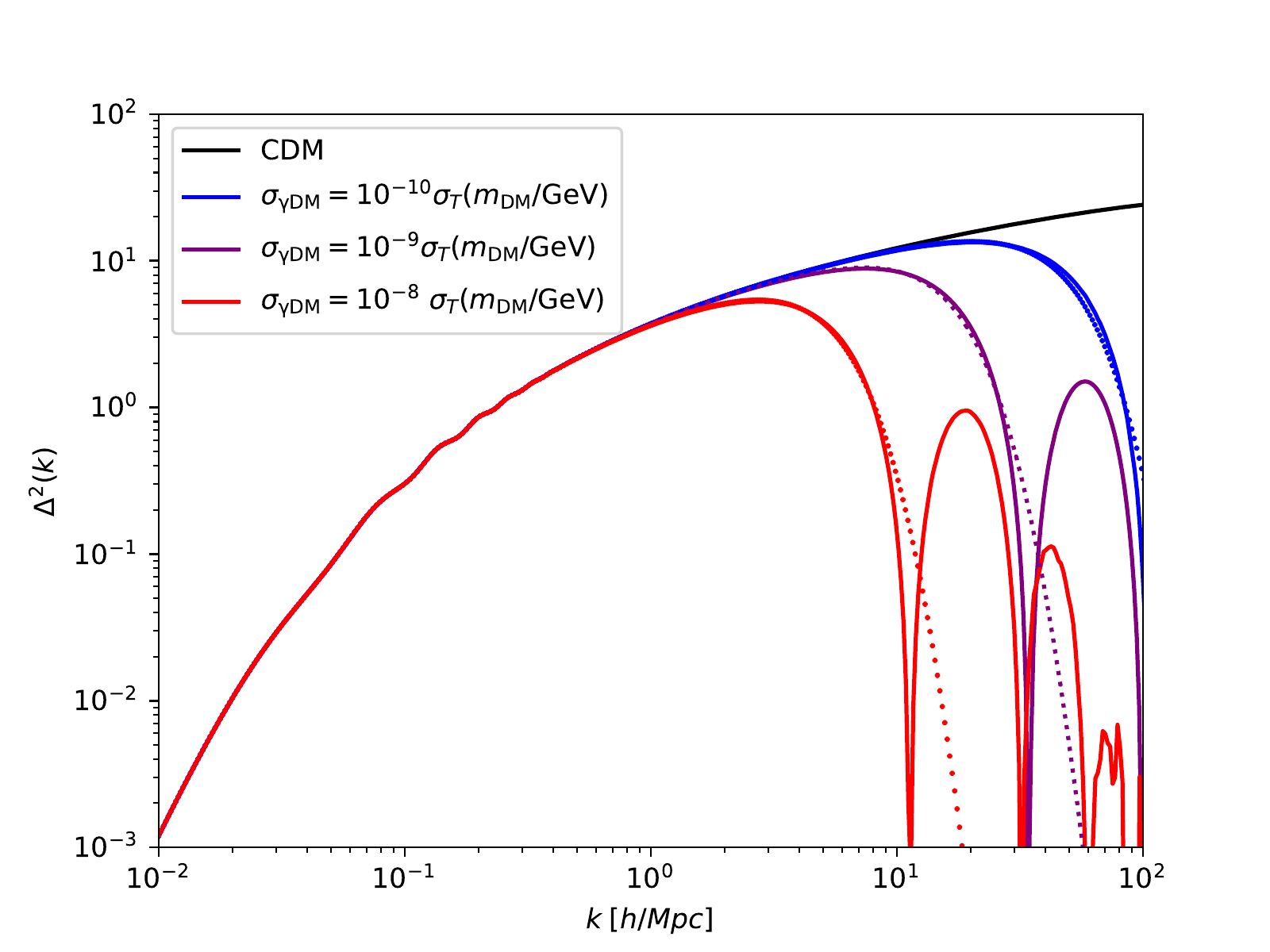} 
\caption{Dimensionless power spectrum for three $\gamma$DM scenarios calculated with a modification of the Boltzmann solver CLASS \cite{Lesgourgues:2011re} (solid curves), with the corresponding approximation using the fitting formula in Eqs. \eqref{eq:twdm} and \eqref{eq:alphIDM} (dotted curves). The transfer function approach successfully matches the first damping, and agrees with the exact computation up to the oscillatory effects appear. The equivalent WDM masses, following the correspondence from Eq. \eqref{eq:W-Icorr-lin}, are $0.7, 1.8, 4.7$ keV for $\sigma_{\gamma {\rm DM}}/ \sigma_T \times \left(m_{\rm DM}/\textrm{GeV}\right)^{-1} = 10^{-8}, 10^{-9}, 10^{-10}$ respectively.}
\label{fig:ps_IDM}
\end{figure}

While the above estimate qualitatively explains the effects of collisional damping, a precise evaluation of the impact on fluctuations requires the computation of the kinetic equations including the scattering term. More specifically, interactions between photons and DM particles induce a coupling of their respective velocity fluctuations. As is derived from the integration of the Boltzmann equations, the Euler equation from Eq. \eqref{eq:cont_euler} for DM needs to be modified, including a term accounting for scattering, as \cite{Boehm:2001hm}
\begin{equation}
 \dot{{\bf v}} + H{\bf v} = - \frac{1}{a}\nabla \Phi +S^{-1}\,\sigma_{\gamma {\rm DM}}\,n_{\rm DM}\left({\bf v}_\gamma - {\bf v} \right),
\end{equation}
where $S=(\rho_{\rm DM}+p_{\rm DM})/(\rho_{ \gamma}+p_{ \gamma})=(3/4)\rho_{\rm DM}/\rho_{ \gamma}$ accounts for the ratio between momenta. While interactions are strong, the scattering term couples fluctuations of DM and photons, and the latter drag DM perturbations. On the other hand, photons present a large radiation pressure, which counteracts gravitational collapse. The compensation of both effects leads to sound waves, which results in an oscillatory behavior of the spectrum at small scales. A similar effect happens in the CDM model in the baryon-photon sector. Due to Thomson scattering, baryonic fluctuations get coupled to radiation, causing wiggles on the matter spectrum. This effect is the cause of the well-known Baryon Acoustic Oscillations (BAOs), which have been measured with large scale structure observations \cite{2005ApJ...633..560E, 2005MNRAS.362..505C, deMattia:2020fkb}. For the same reason, in $\gamma$DM scenarios, DM fluctuations also present an oscillatory pattern in the power spectrum, due to the coupling of DM particles to photons, at scales which entered the horizon when these interactions were still strong. After $\gamma$DM decoupling, the DM spectrum approaches the standard CDM case.

To properly predict the linear evolution of fluctuations, a numerical code is required, such as CLASS \cite{Lesgourgues:2011re}, which solves the Boltzmann equations hierarchy for every species in linear perturbation theory. To study IDM scenarios, the code must be properly modified, including the interaction terms between photons and DM in the hierarchy of kinetic equations, as described in Refs. \cite{Boehm:2001hm, Wilkinson:2013kia, 2018JCAP...10..009S}. The results of the modification of CLASS for three IDM models are shown in Fig. \ref{fig:ps_IDM} in solid lines. As expected, oscillations appear below the collisional damping scale, matching the CDM case at larger scales.

Since the main effect of $\gamma$DM scattering is the suppression of small scale fluctuations, an approximation of this damping can be emulated by employing the transfer function fit from Eq. \eqref{eq:twdm}, properly modified. It has been shown that this fitting formula works for IDM models, choosing $\nu = 1.2$ and parameterizing the suppression scale in terms of the $\gamma$DM scattering cross section as \cite{Boehm:2001hm}
\begin{equation}
\alpha_{\rm{IDM}}= 0.073 \, \left[10^8 \, \left(\frac{\sigma_{\gamma \rm{DM}}}{\sigma_T}\right) \, \left(\frac{{\rm GeV}}{m_{\rm DM}}\right)\right]^{0.48} \left(\frac{\Omega_{\rm{IDM}}} {0.4}\right)^{0.15}\left(\frac{h}{0.65}\right)^{1.3} \, {\rm Mpc}/h ~.
\label{eq:alphIDM}
\end{equation}
Note that this length agrees with the order-of-magnitude estimation of Eq. \eqref{eq:colldampscale} at the relevant cross sections $\sigma_{\gamma {\rm DM}}/ \sigma_T \times \left(m_{\rm DM}/\textrm{GeV}\right)^{-1} \sim 10^{-8}$ (although the scaling of Eq. \eqref{eq:colldampscale} with the cross section is slightly different). The power spectrum resulting from using this approach is depicted with dotted lines in Fig \ref{fig:ps_IDM}. Notice that it accurately matches the exact IDM power spectrum from CLASS until the damped oscillatory effects dominate at small scales. However, the transfer function approximation can provide an accurate enough description where the damping scale appears, when IDM and CDM differences show up. Furthermore, as can be seen in Fig. \ref{fig:ps_IDM}, the second and subsequent maxima are suppressed by more than one order of magnitude, and thus their effects in structure formation are expected to be subdominant.

Given that both IDM and WDM scenarios can be described by the same fitting formula, one can consider a correspondence between the two models which lead to an equivalent damping in the power spectrum. By equating Eqs. \eqref{eq:alpha_wdm} and \eqref{eq:alphIDM}, one finds that the suppression present in a WDM model with $m_\text{WDM}$ matches the IDM scenario with cross section $\sigma_{\gamma \rm{DM}}$ and mass $m_{\rm DM}$ via
\begin{equation}
\left(\frac{\sigma_{\gamma \rm{DM}}}{\sigma_T}\right) \, \left(\frac{{\rm GeV}}{m_{\rm DM}}\right) \simeq 4.1 \times 10^{-9} \, \left( \frac{\text{keV}}{m_\text{WDM}} \right)^{2.4} ~.
\label{eq:W-Icorr-lin}
\end{equation}
Thus, IDM and WDM scenarios related by the correspondence above give rise to similar suppression of the linear power spectrum. For instance, IDM scenarios spanning the range $\sigma_{\gamma {\rm DM}} \in [10^{-11} - 10^{-8}] \, \sigma_T \times \left(m_{\rm DM}/\textrm{GeV}\right)$ roughly correspond to  WDM scenarios with $m_{\rm DM} \in [1-12]$~keV. For the models employed in Fig. \ref{fig:ps_IDM}, cross sections of $\sigma_{\gamma {\rm DM}}/ \sigma_T \times \left(m_{\rm DM}/\textrm{GeV}\right)^{-1} = 10^{-8}, 10^{-9}, 10^{-10}$ correspond to WDM masses of $0.7, 1.8, 4.7$ keV respectively. The relation above allows mapping IDM to WDM scenarios, so constraints based on the linear power spectrum can be translated from one scenario to the other. However, as will be seen next, the equivalence between both scenarios is broken at the level of the halo mass function.

Although we focus on photon-DM interactions, the effects of neutrino-DM elastic scattering are found to be similar, and the induced small-scale suppression in the power spectrum can also be parameterized with the same transfer function, correcting the breaking scale as $\alpha_{\rm{\nu DM}}\simeq 0.8 \times \alpha_{\rm{\gamma DM}}$~\cite{Schewtschenko:2014fca}.

\subsubsection{Halo mass function}

\begin{figure}
\begin{center}
\includegraphics[scale=0.7]{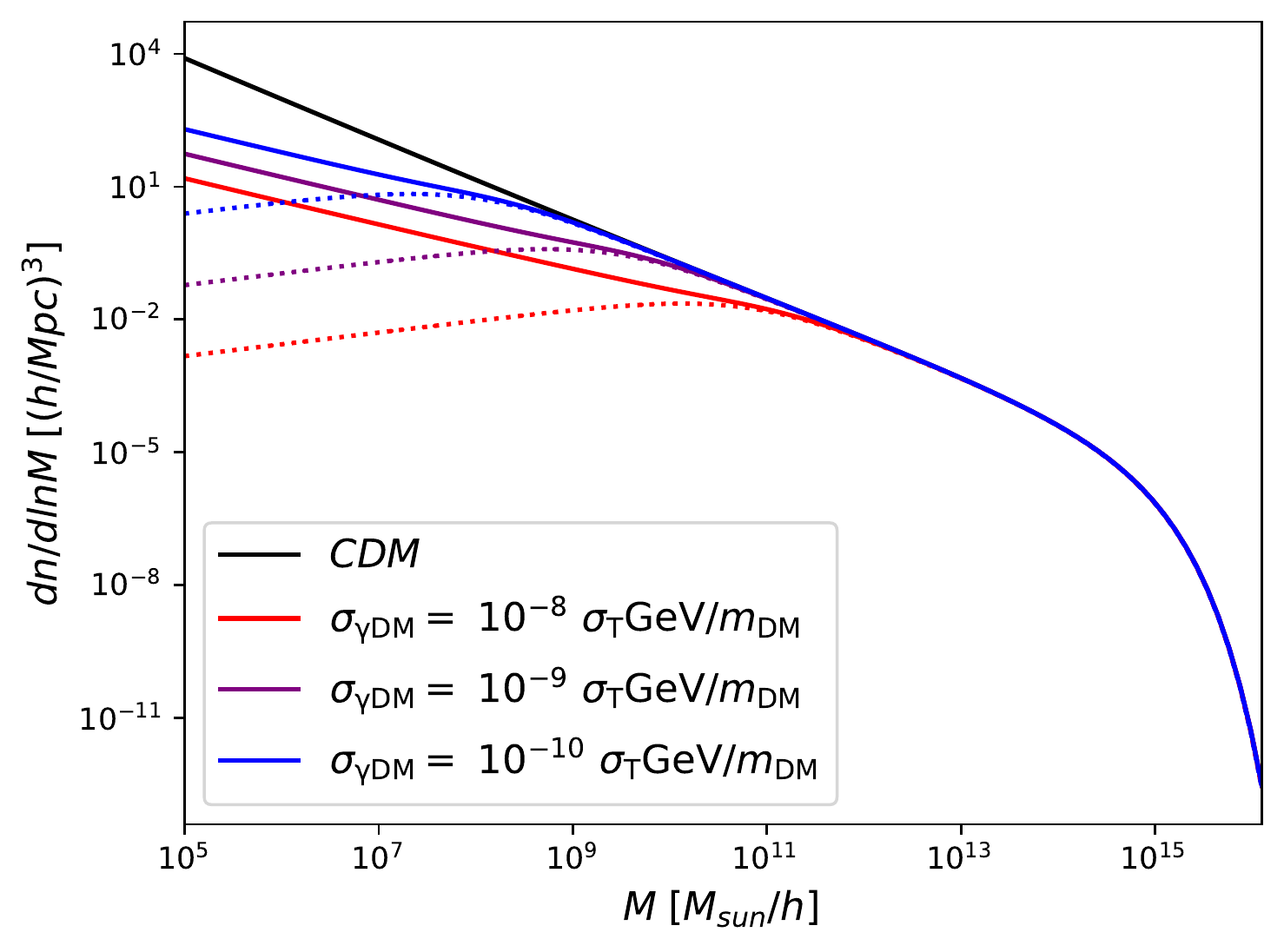}
\caption{Halo mass function for CDM and three IDM models as a function of the mass, evaluated at $z=0$. Dotted lines correspond to the WDM halo mass function with the same suppression scale.}
\label{fig:idm_hmf}
\end{center}
\end{figure}

To describe the halo mass function in an IDM universe, a similar procedure to the employed with WDM (Eq. \eqref{eq:hmf_wdm}) can be followed, although including some further modifications. Although the first suppression of the linear power spectrum in the IDM scenario looks very similar to the WDM case, and therefore can be parameterized by similar means, a key distinction between these models is that IDM presents oscillations at even smaller scales, enhancing and suppressing fluctuations at growing wavenumbers. This effect induces a different behavior in the non-linear regime, affecting to the abundance of halos. As a result of these oscillations, for the same half-mode mass, low-mass halos are more abundant for IDM than for WDM \cite{Schewtschenko:2014fca}. Thus, to match IDM N-body simulations, an additional mass-dependent correction must be added to the halo mass function \cite{Moline:2016fdo},
\begin{equation}
\frac{dn^{\rm IDM}}{dM} =\left(1 + \frac{M_{\rm hm}}{b \, M}\right)^{a} \left(1+\frac{M_{\rm hm}}{g \, M}\right)^{c} \, \frac{dn^{\rm ST,\, CDM}}{dM} ~,
\label{eq:hmf_idm}
\end{equation}
where the adopted best-fit parameters are $a = -1$, $b = 0.33$, $g = 1$, $c = 0.6$. The halo mass function $dn^{\rm ST, \, CDM}/dM$ corresponds to the standard ST first-crossing distribution considering the CDM linear power spectrum, instead of the IDM power spectrum. Note that this is different from the WDM case, where the WDM linear power spectrum explicitly appears in the halo mass function (integrated in the variance). In the IDM case, instead, the damping effects are only accounted for in the mass-dependent suppression factors of the above fit. This approach successfully matches the results from numerical simulations at $z = 0$ for cross sections around $ \sigma_{\gamma {\rm DM}} \sim \times 10^{-9} \, \sigma_T \, (m_{\rm DM}/{\rm GeV})$ \cite{Schewtschenko:2014fca, Schewtschenko:2016fhl}. The halo mass function for three IDM models is shown in Fig. \ref{fig:idm_hmf} (continous colored lines), compared to the CDM (black line) and to the corresponding WDM cases, with the same half-mode mass (dotted lines). Notice that, while both scenarios present a suppression of low-mass halos, it is more pronounced in the WDM cases, being milder for IDM. This is a consequence of the IDM oscillatory nature at small scales, which enhances low-mass halos with respect to WDM when non-linearities become important.

It has been shown that the halo mass function in $\nu$DM scenarios can be also well described by the procedure above \cite{Schewtschenko:2014fca}. This is because $\nu-$DM and $\gamma$DM give rise to similar physics, and thus the damping effect results to be closely equivalent. More concretely, numerical simulations at $z=0$ showed approximately the same halo mass function for the cases $\sigma_{\gamma {\rm DM}} = 2.9 \times 10^{-9}\, \sigma_T \, (m_{\rm DM}/{\rm GeV})$ and $\sigma_{\nu {\rm DM}} = 2.0 \times 10^{-9}\,\sigma_T \, (m_{\rm DM}/{\rm GeV})$, for a fixed half-mode mass \cite{Schewtschenko:2014fca}. This correspondence allows establishing a map between both scenarios, and estimating bounds on neutrino-DM cross sections.

\begin{comment}

\bibliographystyle{../../jhep}
\bibliography{../../biblio}

%% file: Chapters/Chapter_PBH/Chapter_PBH.tex
\end{comment}

\chapter{Primordial Black Holes}


\label{chap:PBHs}

Primordial Black Holes (PBHs) conform a natural candidate for being at least one of the components of the DM. In this chapter, we shall review the basics on their formation, abundance and other properties. Some of their effects are discussed, such as the accretion of the surrounding matter or the modification of the matter fluctuations at small scales, which could leave an impact in the evolution of the universe and the formation of structures. Lastly, the most relevant constraints on their masses and abundances are discussed. For further details, see, e.g., Refs. \cite{Sasaki:2018dmp, Green:2020jor, 2020ARNPS..7050520C, 2010PhRvD..81j4019C}.

\section{Foundations on PBHs}

\subsection{Motivation for the existence of PBHs}

The hypothesis of the formation of PBHs in the early universe was first suggested in 1967 by Zeldovich and Novikov, who argued that PBHs would accrete radiation with catastrophic consequences and this would conflict with observational data \cite{1967SvA....10..602Z}. In 1971, ignoring that previous work, Hawking proposed the standard mechanism of formation of BHs in the radiation era, by direct collapse of a highly overdense region \cite{1971MNRAS.152...75H}. Detailed computations of its formation mechanism by Carr and Hawking in 1974 \cite{1974MNRAS.168..399C} showed that the initial mass of those collapsed objects would not substantially grow by accretion, contrary to the first estimation by Zeldovich and Novikov. Later, Carr in 1975 \cite{1975ApJ...201....1C} computed the abundance, mass spectrum and conditions of formation of such objects. These facts made PBHs possible candidates to account for at least part of the DM \cite{1975Natur.253..251C, 1975A&A....38....5M}, which at those years, as discussed in Sec. \ref{sec:DMhistory}, started to be outlined as one fundamental problem in cosmology.

DM (partly) composed by PBHs constitutes an exciting possibility, and have attracted a great attention in the scientific community. One of the reasons is the enormous number of probes which can constrain the parameter space of PBHs, as it will be detailed in Sec. \ref{sec:constraintsPBH}. The variety of phenomenological effects produced by PBHs allows placing stringent bounds on their abundance, usually indicated by the energy fraction of DM as PBHs, defined as $f_{\rm PBH}=\Omega_{\rm PBH}/\Omega_{\rm DM}$, with $\Omega_{\rm PBH}$ and $\Omega_{\rm DM}$ the energy parameters respect to the critical density. Moreover, since PBHs are usually expected to be formed before nucleosynthesis, BBN constraints on the baryon abundance do not apply to them, and thus can be regarded as non-baryonic DM \cite{2020ARNPS..7050520C}. PBH DM could be considered as a MACHO candidate, and thus constraints applying over the MACHO abundance also affect PBHs, as shall be seen in Sec. \ref{sec:constraintsPBH}. Moreover, PBHs present unique signatures which would differentiate them from other MACHO species, such as brown dwarfs or Jupiter-like planets. Moreover, due to their primordial nature, PBHs can be accommodated with the LSS data from the CMB and the BBN bounds on the abundance of baryons, contrary to standard astrophysical MACHOs, which are unable to explain such phenomena.

Shortly after the first detection of gravitational waves from a merger of $\sim 30 \, M_\odot$ BHs by LIGO \cite{PhysRevLett.116.061102}, the question whether these could be of primordial nature was raised \cite{2016PhRvL.116t1301B}. It has been claimed that analyses of the posterior data from gravitational wave detectors LIGO and Virgo show that the detected mergers are compatible with the hypothesis of their components being of primordial nature, although there is no strong preference over stellar BHs \cite{2018PDU....22..137C, 2020ARNPS..7050520C, 2020arXiv201013811G}.

Standard astrophysical BHs are formed from the collapse of massive stars. Stellar evolution models and general relativity predict that those BHs can only be formed with masses above some threshold around $\sim 3 M_\odot$, the so-called Tolman–Oppenheimer–Volkoff limit \cite{1939PhRv...55..364T, 1939PhRv...55..374O}. This threshold arises from the condition that the degeneracy pressure of neutrons in a neutron star cannot balance its gravitational force, collapsing to a BH. In contrast, since PBHs are not formed from stars but from the direct collapse of fluctuations, in principle they could be produced with any mass. Thus, a positive measurement of a BH with a mass lower than $\sim 3 M_\odot$ would be a confirmation of the existence of primordial non-stellar BHs \cite{2018PDU....22..137C}.

It is generally claimed that DM formed by PBHs has the advantage that it does not require a new kind of particle, such as WIMPs, axions or sterile neutrinos. Actually, their formation is already present in standard cosmologies, although extremely unlikely. However, their production usually requires some specific inflationary scenarios or physics Beyond the Standard Model (BSM) in order to obtain a large enough abundance. The typically considered formation mechanism of PBHs arises from the direct collapse of primordial fluctuations, whose power is enhanced at small scales as a consequence of non-slow roll inflationary dynamics, as we shall further comment on below. There are, however, other scenarios which naturally predict a population of PBHs as a result of phase transitions in the early universe. A spontaneously broken symmetry produces a phase transitions, which may lead to the formation of different topological defects, depending on the specific broken symmetry and vacuums state. Eventually, those topological defects could collapse and form PBHs, as is the case of domain walls \cite{2016JCAP...02..064G, 2017JCAP...04..050D} or cosmic string loops \cite{Hawking:1987bn, PhysRevD.43.1106}, as well as colliding vacuum bubbles \cite{PhysRevD.26.2681, 2016JCAP...02..064G, 2017JCAP...12..044D}. Hence, the existence of PBHs would provide valuable hints from the still unknown physics of the very early universe.

Depending on their mass, BHs can be classified in three different groups. Stellar BHs, those which are formed after the collapse of a star, with masses from 5 to tens of solar masses. On the other hand, it is well known the existence of Super Massive Black Holes (SMBH) in most of the nuclei of galaxies, with masses ranging from $10^5$ to $10^{10} \, M_\odot$, and already existing at redshifts $z > 6$. For instance, in the case of the Milky Way, its central BH has a mass of about $\sim 4 \times 10^6 \, M_\odot$ \cite{2019A&A...625L..10G}. A third group lies between the former, and thus called Intermediate-Mass Black Holes (IMBH), whose masses range from $\sim 10^2$ to $10^5 \, M_\odot$. Contrarily to the other BH classes, there are fewer observational evidences of such objects. Nonetheless, recent LIGO and Virgo observations have found a merger event of BHs with masses $\sim 60$ and $\sim 80 M_\odot$, producing a remnant BH of $\sim 150 \, M_\odot$, in the so far unobserved range of masses of IMBHs \cite{PhysRevLett.125.101102}.

One of the main motivations to consider PBHs is that they could constitute the seeds for the aforementioned SMBH \cite{2018MNRAS.478.3756C}. Such massive objects cannot be produced by the collapse of the stellar interior after a supernova explosion, whose remnants BH usually have masses ranging from 5 to tens solar masses. Moreover, although BHs can grow by accreting surrounding matter, standard accretion mechanisms can hardly explain such a large increase in the mass from stellar BHs. Although their seeds could have formed at early times ($z \gtrsim 20$), the accretion rates would be suppressed in the relatively shallow potential wells of their host halos, as well as decreased by radiative feedback effects (see, e.g., Ref. \cite{Volonteri_2010} for more details). However, the existence of massive enough PBHs may act as seeds for the SMBHs, from which they could have grown by accretion.

It has also been argued that PBHs could naturally explain some of the small scale problems outlined in Sec. \ref{sec:smallscale}. PBH DM with merger rates compatible with the LIGO measurements would lead to the formation of a large population of ultra-faint dwarf galaxies without star formation, which could naturally explain the missing satellite and too-big-to-fail problems \cite{Clesse:2016vqa, 2018PDU....22..137C}.

\subsection{Formation and conditions of collapse}

The mass of a BH collapsed in the early universe depends on its formation time. A BH can be characterized by an extremely dense amount of matter in a very compact region, lying within the known as Schwarzschild radius, $R_S = 2GM_{\rm PBH}/c^2 \sim 3 M_{\rm PBH}/M_\odot$ km. Thus, the mean density inside that region can be estimated to be $\rho_S = M_{\rm PBH}/(4\pi R_S^3/3) \sim 10^{18} (M/M_\odot)^{-2}$ g cm$^{-3}$. On the other hand, the mean density of the universe in the radiation era goes as $\rho_c \sim 10^6 (t/{\rm s})^{-2}$ g cm$^{-3}$. In order to have PBH formation, densities at least of the order of the mean BH density, $\rho_c \sim \rho_S$, are needed. Therefore, the mass of the resulting PBHs should be of the order of the horizon mass at that time, i.e., the mass within a region of the size of the Hubble horizon, $M_{\rm PBH} \sim M_{\rm H}$ \cite{1975ApJ...201....1C}, which is defined as
\begin{equation}
M_{\rm H} = \frac{4}{3}\bar{\rho}\left(\frac{c}{H}\right)^3 = \frac{c^3}{2G H} \sim 10^{15} \; {\rm g}  \; \left( \frac{t}{10^{-23} {\rm s}} \right).
\label{eq:MH}
\end{equation}
Thus, PBHs with masses of $\sim M_\odot \simeq 2 \times 10^{33}$ g would have been formed at around the QCD phase transition, at $t \sim 10^{-6}$ s, while those born at the BBN epoch, $t\sim 1$ s, would be as massive as $10^4 M_\odot$. Since the PBH mass is roughly given by the horizon one, it means that fluctuations entering the horizon can collapse into PBHs. A detailed calculation shows that $M_{\rm PBH} = \gamma M_{\rm H}$, where the proportionality factor $\gamma$ depends on the details of gravitational collapse, and gets values lower than 1. Early estimates showed that it can approximated to $\gamma \simeq c_s^{3} \simeq 0.2$, with $c_s = \frac13$ the sound speed in the radiation epoch \cite{1975ApJ...201....1C, 2014arXiv1403.1198G}.

\begin{figure}[t]
\begin{center}
\includegraphics[scale=0.23]{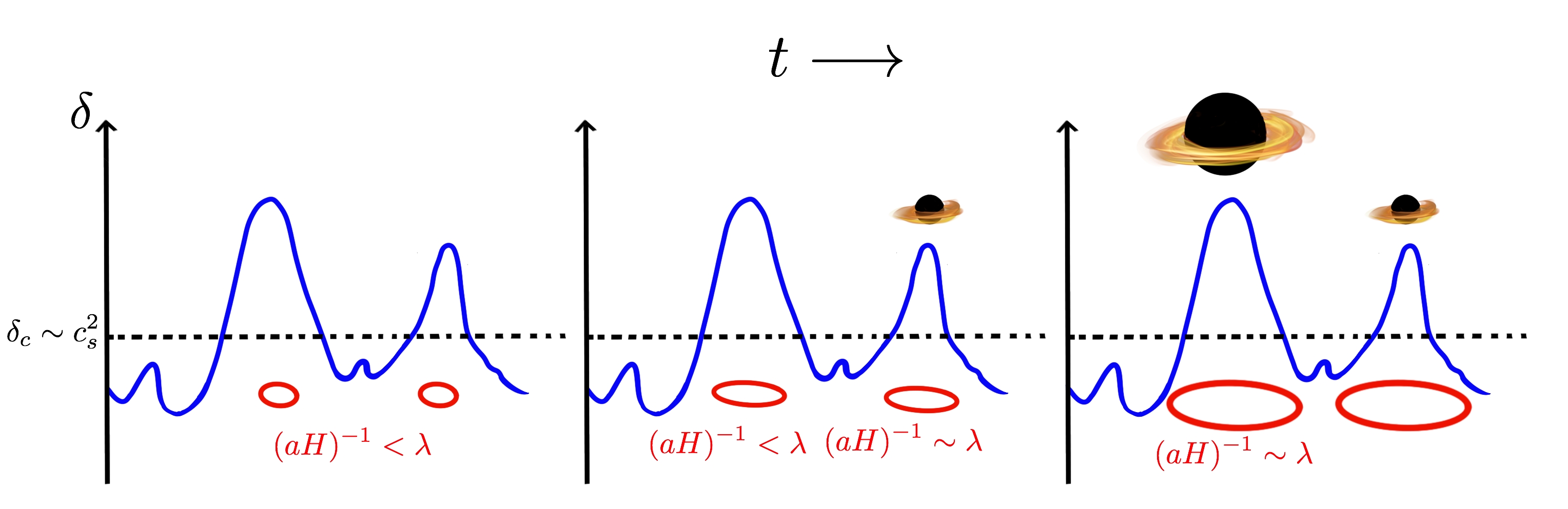}
\captionof{figure}{Sketch of the formation of PBHs from overdensities for three different successive moments. When fluctuations larger than a critical threshold $\delta_c \sim c_s^2$ enter the horizon, i.e., their comoving wavelength $\lambda=2\pi/k$ (which characterizes the size of the perturbation) is of the order of the Hubble horizon $(aH)^{-1}$, the overdense region collapses and a PBH is produced. As can be seen in Eqs. \eqref{eq:MH} and \eqref{eq:Mk}, longer modes (large $\lambda$, low $k$) enter the horizon later and lead to more massive PBHs.}
\label{fig:pbhformation}
\end{center}
\end{figure}

We can gain insight into the process and criterion of formation of PBHs by considering a simple picture of collapse in General Relativity \cite{Sasaki:2018dmp}. To simplify the problem, we can assume a spherical overdense region embedded in the expanding FLRW background. Hence, we can adopt a metric of the form
\begin{equation}
ds^2 = -dt^2 + a(t)^2\left(\frac{dr^2}{1-K(r)r^2} + r^2d\Omega^2\right),
\end{equation}
with $K(r)$ a positive curvature smoothly varying with the radial coordinate. This is the form of a locally closed universe, with a curvature which approaches asymptotically to $K(r) \rightarrow 0$ at large distances. In that limit, the flat FLRW limit is recovered. Since this overdense region should be much larger than the Hubble horizon, a leading order gradient expansion can be applied. Thus, neglecting derivatives of $K(r)$ in the 3-curvature of the hypersurface with constant time, one can obtain a generalized Friedman equation from the time-time component of the Einstein equations (or Hamiltonian constraint),
\begin{equation}
H^2 + \frac{K(r)}{a^2} = \frac{8\pi G}{3}\rho(t,r)
\end{equation}
with $H= \dot{a}/a$. Therefore, the overdensity can be written in terms of the curvature as
\begin{equation}
\delta = \frac{\rho - \bar{\rho}}{\bar{\rho}} = \frac{K(r)}{a^2H^2}
\end{equation}
From perturbation analysis, we know that the Jeans scale $k_J\simeq aH/c_s$, with $c_s$ the sound speed, sets the critical length for collapse, as seen in Sec. \ref{sec:lineargrowth}. Only perturbations at scales larger than the Jeans length $2\pi/k_J$ (\textit{i.e.}, with $k < k_J$) may be able to overcome pressure gradients and collapse. Thus, when the collapse leads to non-linear regime, $\delta \sim 1$, 
\begin{equation}
\delta \sim 1 \Rightarrow  \frac{K}{a^2H^2} =  \frac{K}{c_s^2 k_J^2} \sim 1,
\end{equation}
and hence we should identify $K \sim c_s^2 k_J^2$ at the collapse regime.\footnote{To gain insight into this fact, one can compare it to the situation in perturbation theory, where instead of the curvature $K$, one has the laplacian of the gravitational potential, which in Fourier space takes the form $k^2 \phi$. This is related to the density fluctuation through the Poisson equation, $k^2 \phi = -4\pi G\bar{\rho} a^2 \delta = -(c_sk_J)^2 \delta$. Thus, when $\delta \simeq 1$, $k^2 \phi \simeq (c_sk_J)^2$. In relativistic perturbation theory, the Newtonian potential is related to the curvature perturbation \cite{2003moco.book.....D}.} On the other hand, as stated before, the collapse is expected to happen when perturbations enters the horizon. For a fluctuation of wavenumber $k$, it occurs at a time $t_k$, implicitly defined from $k = H(t_k)a(t_k)$. Thus, we can write the overdensity when the perturbation enters the horizon as
\begin{equation}
\delta(t_k) \sim \frac{K}{a(t_k)^2H(t_k)^2} \sim c_s^2\frac{k_J^2}{k^2}.
\end{equation}
The condition of collapse $k < k_J$ enforces that $\delta(t_k) > c_s^2$. This implies that only overdensities above a certain threshold can lead to form a PBH. This minimum required overdensity is given by the speed of sound of the medium, which in the radiation dominated era, it is $c_s^2=1/3$. Although this is a very rough estimate, more detailed computations only slightly change the value of the threshold, which lies between $\delta_c \sim 0.3$ and $0.5$ \cite{2004PhRvD..70d1502G, 2019PhRvD.100l3524M}. Figure \ref{fig:pbhformation} shows a sketch of the process of PBH formation.

Since PBHs are formed when the fluctuation crosses the horizon, the mass of the PBH is mostly determined by the time of formation $t_f$ given by Eq. \eqref{eq:MH}, and equivalently, can be related to the wavelength of the perturbation. When the mode of wavenumber $k$ crosses the horizon, the condition $a(t_f) \, H(t_{f}) =k$ holds. Since the mass of the PBH is proportional to the horizon mass at the moment of formation, $M_{\rm PBH} \propto \gamma H^{-1}$. In the radiation dominated era, $H \propto a^{-2}$, and thus, $H_f \propto k^{2}$, obtaining the relation between the mass and the wavenumber as \cite{Sasaki:2018dmp},
\begin{equation}
M_{\rm PBH} \simeq 30\;  M_\odot \; \left( \frac{\gamma}{0.2} \right) \left( \frac{2.9 \times 10^5 {\rm Mpc}}{k} \right)^2.
\label{eq:Mk}
\end{equation}
From the equation above, one can see that probing a given scale $k$ could constrain PBH population of its corresponding mass. Furthermore, an enhancement in the power spectrum around that scale would result in a larger number of PBHs of such masses than in the standard scenario. This is the key for some inflation models with a potential which induces larger fluctuations at a range of very small scales. Different inflationary models may lead to completely different PBH scenarios. For instance, inflection points in a plateau of a single inflaton potential can lead to broad peaks in the power spectrum, which imply extended PBH mass functions \cite{2017PDU....18...47G, 2017JPhCS.840a2032G}. Other models, such as hybrid inflation, where two scalar fields are coupled, can also lead to broader peaks \cite{2015PhRvD..92b3524C}. Furthermore, models such as the chaotic new inflation may produce two inflationary periods, giving raise to relatively narrow peaks \cite{PhysRevD.58.083510, 2008JCAP...06..024S}. We comment more on the PBH mass function in Sec. \ref{sec:pbh_massfunc}. See, e.g., Ref. \cite{Sasaki:2018dmp} for more information regarding inflation models.

In this thesis, we only consider PBHs formed during the radiation era. Those produced  before inflation ends would have been diluted due to their negligible density during the inflationary accelerated expansion. PBHs formed at the matter era, or in an early matter domination era previous to the radiation era, have also been considered in literature, and may have different imprints, since the conditions of collapse are less restrictive, allowing the formation from smaller inhomogeneities (see, e.g., \cite{Green:2020jor}).

\subsection{Abundance and mass function of PBHs}
\label{sec:pbh_massfunc}

It is possible to estimate the initial abundance of PBHs at the moment of formation, taking into account all overdensities above the threshold of collapse $\delta_{\rm c, PBH} \simeq 1/3$. Assuming a Gaussian probability distribution $P(\delta)$ for the overdensities with variance $\sigma^2(M)$ at mass scale $M$ (as defined in Eq. \eqref{eq:sig2}), the initial abundance, defined as $\beta(M_{\rm H}) = \rho_{\rm PBH}(t_f)/\rho_{\rm tot}(t_f)$, reads \cite{Sasaki:2018dmp, Green:2020jor}
\begin{align}
\beta(M_{\rm PBH}) &= 2 \gamma \int_{\delta_{\rm c, PBH}}^\infty P(\delta) d\delta = \gamma \; {\rm erfc} \left(\frac{\delta_{\rm c, PBH}}{\sqrt{2}\sigma(M_{\rm PBH})} \right) \\
& \simeq \gamma \sqrt{\frac{2}{\pi}} \frac{\sigma(M_{\rm PBH})}{\delta_{\rm c, PBH}} \exp \left( -\frac{\delta_{\rm c, PBH}^2}{2\sigma^2(M_{\rm PBH})} \right),
\end{align}
where in the last equality, $\sigma(M_{\rm H}) \ll \delta_c$ has been assumed. The factor of $2$ in the first equality of the above equation accounts for a fudge factor, as done in the original Press-Schechter halo mass function prescription \cite{Press:1973iz}. For the standard cosmological scenario with an initial scale invariant power spectrum, at CMB scales, the amplitude of the fluctuations is around $\sigma(M_{\rm PBH}) \sim 10^{-5}$, leading to $\beta \sim 10^{-5} \exp(-10^{10})$, which is completely negligible \cite{2014arXiv1403.1198G}. Therefore, in order to have a relevant population of PBHs, larger values of the initial power spectrum are needed. On the other hand, the assumption of a Gaussian distribution may not be consistent with enhanced fluctuations and the presence of PBHs, and deviations from that could be unavoidable \cite{2018JCAP...03..016F, 2019JCAP...07..048D} (except for specific inflation models presenting an inflection point \cite{Atal:2018neu}). Non-gaussianities could have a great impact on the initial fraction and lead to a larger population, as well as leaving further detectable signatures in gravitational waves \cite{Cai:2018dig}.

Nonetheless, although the initial fraction $\beta$ is a very small quantity, since matter and radiation scale differently with redshift, the PBH contribution can become relevant at current times. To relate that value to the current density parameters of PBHs and radiation, $\Omega_{\rm PBH}$ and $\Omega_\gamma$, respectively, one has to take into account that, since PBHs behave as standard matter (dust), they scale as $\rho_{\rm PBH} \propto (1+z)^3$, while radiation goes as $\rho_\gamma \propto (1+z)^4$. Thus, relating the formation redshift with its corresponding PBH mass, we get \cite{2010PhRvD..81j4019C},
\begin{equation}
\Omega_{\rm PBH}(M) = \beta(M) (1+z_f) \Omega_\gamma \simeq \gamma^{1/2} \left( \frac{\beta(M)}{1.15 \times 10^{-8}} \right) \left( \frac{M}{M_\odot} \right)^{-1/2}.
\label{eq:abundanceomega}
\end{equation}
Hence, with initial fractions as low as $\beta \sim 10^{-8}$ of solar mass BHs, the fraction of energy in PBHs would be order unity.

In principle, depending on the specific mechanism of formation, a population of PBHs with different masses could be generated. It is thus relevant to consider the mass distribution function. As commented above, the specific shape of the enhancement in the fluctuations determines the mass distribution function. Sharp peaks in the power spectrum imply approximately monochromatic PBH mass functions, while broad peaks lead to extended mass distribution functions. For simplicity, it is usually assumed in literature, as done along this thesis, that all PBHs share the same value for their masses, having therefore what is called a \textit{monochromatic} distribution. In this case, the distribution is only determined by the fraction $f_{\rm PBH}$ and its mass $M_{\rm PBH}$, easing the computations. However, this premise may not be valid, and actually many inflationary models able to produce PBHs predict, instead, \textit{extended} mass functions, which can span over a large range of masses. Besides the central mass and total fraction of PBHs as DM, these distributions require more parameters, such as the variance, which determines the broadness. Some common parameterizations, such as lognormal or power law functions, have been employed to effectively model generic extended mass functions, translating the usual constraints on monochromatic PBHs to these broader distributions. Besides having more freedom due to a large parameter set to fit, the constraints on extended mass functions can also be even more stringent than in the single mass case \cite{2017PhRvD..96b3514C, 2018JCAP...01..004B}. Nonetheless, even with constraints forbidding $f_{\rm PBH} \sim 1$ in the monochromatic case, there are choices of the mass function which allow having all DM formed by PBHs \cite{Lehmann_2018}.
 
To gain insight about the relation between the fluctuations and the mass function, consider a very sharp feature around a scale $k_*$ in the matter power spectrum, modeled as a delta function. We then can write that contribution as $\Delta^2(k) =\Delta_*^2\, k \, \delta_D(k-k_*)$, with $\Delta_*^2$ the amplitude. From Eq. \eqref{eq:sig2}, it is straightforward to obtain the variance as $\sigma^2(M)=\Delta_*^2 \tilde{W}^2(k_*R)$, with $M \propto R^3$, and $\tilde{W}^2(k_*R)$ the square of the Fourier transform of the window function. Regardless of the specific choice of this window function, the variance, and therefore the abundance $\beta$, present their maximum value when $R \sim k_*^{-1}$ or lower. Following the Press-Schechter prescription, one can write the mass function $df_{\rm PBH}(M)/d{\rm ln}M$ relating it to the abundance from Eq. \eqref{eq:abundanceomega}, as \cite{Sasaki:2018dmp}
\begin{equation}
  \frac{d f_{\rm PBH}(M)}{d{\rm ln}M}d{\rm ln}M \simeq \frac{\delta_{\rm c, PBH}^2}{\sigma^2(M)} \,f_{\rm PBH}(M) \, \left| \frac{d\ln \sigma^2}{d\ln M} \right|d{\rm ln}M~,
\end{equation}
where $f_{\rm PBH}(M) = \Omega_{\rm PBH}(M)/\Omega_{\rm DM} \propto \beta(M)$. The derivative $\frac{d\ln \sigma^2}{d\ln M}$ depends on the window function but peaks at $R \sim k_*^{-1}$ (actually leading to a delta function $\delta_D(R-k_*^{-1})$ if the sharp-$k$ filter is chosen), and therefore on the mass function too.

\subsection{BH evaporation}

In 1974, Hawking realized that, due to quantum effects in curved spacetimes, BHs may emit particles at their event horizon \cite{1974Natur.248...30H} (although Zeldovich had already proposed that rotating BHs would radiate due to the uncertainty principle \cite{1988bhtb.book.....H}). The emitted radiation would have a thermal black body spectrum, with a temperature given by \cite{1974Natur.248...30H,2010PhRvD..81j4019C}
\begin{equation}
T_{\rm BH} = \frac{\hbar c^3}{8\pi k_B GM} \sim 10^{-7} \; {\rm K} \; \frac{M_\odot}{M},
\label{eq:HawkingT}
\end{equation}
which is known as \textit{Hawking temperature}. Since a BH is emitting particles, it would slowly lose mass until completely evaporate. The rate of energy loss can be estimated as the product of the area of emission $A$ times the energy density \cite{2003PhTea..41..299L}, as $dM/dt = -A \; \sigma_{\rm SB} T_{\rm BH}^4$, with $\sigma_{\rm SB}=\pi^2 k_B^4/(60 \hbar^3 c^4 )$ the Stefan-Boltzmann constant. Given that the area is that corresponding to the event horizon surface, i.e., at the Schwarzschild radius $R_S = 2 G M/c^2$, it reads $A= 4\pi R_S^2 = 16\pi G^2M^2/c^4$. Making use of the Hawking temperature from Eq. \eqref{eq:HawkingT}, one finds that the rate of mass loss is given by
\begin{equation}
\frac{dM}{dt} = - \frac{1}{15360 \pi} \frac{\hbar c^4}{ G^2 M^2},
\end{equation}
which can be easily integrated to obtain the lifetime of a PBH of initial mass $M$, which reads
\begin{equation}
\tau(M) =5120 \pi \frac{G^2 M^3}{\hbar c^4} \sim 10^{64} \; {\rm yr} \; \left( \frac{M}{M_\odot} \right)^3.
\end{equation}
The above result imply that PBHs with smaller masses, which have higher Hawking temperature, evaporate earlier. Those with masses of $\sim 10^{15}$ g or below would have already evaporated by now, having lifetimes shorter than the age of the universe, and therefore cannot be part of the current DM. Detailed computations have to take into account the different emission rates for each particle species, as was firstly done for massless particles by Page in 1976 \cite{1976PhRvD..14.3260P}. Currently, there are specific codes devoted to the prediction of the emitted spectra, such as {\tt BlackHawk} \cite{Arbey:2019mbc}.

\subsection{Clustering and spin}
\label{sec:spin}

It has been demonstrated that, if the fluctuations are originally gaussian distributed and around a relatively narrow peak, PBHs are not expected to be originated in clusters \cite{2018PhRvL.121h1304A, 2018PhRvD..98l3533D}. It means that, initially, PBHs are Poisson (randomly) distributed on small scales, rather than presenting spatial correlations. However, either primordial non-gaussianities or a broad peak in the power spectrum could lead to a significant initial clustering \cite{2019PTEP.2019j3E02S, 2018JCAP...10..043B} (although it has also been argued that broad spectra would not produce appreciable clustering, see Ref. \cite{2019JCAP...11..001M}). Anyway, PBHs could become bounded as the universe evolves. A proper determination of their clustering at later times is of great importance, for instance, in order to estimate their merger rates \cite{2018PhRvD..98l3533D}.

Since PBHs would be formed from the collapse of high density peaks relatively spherically symmetric, their torques and angular momentum are expected to be small \cite{Chiba:2017rvs, Mirbabayi:2019uph}. It is usually quantified with the dimensionless spin parameter, $\mathcal{S} = S/(GM_{\rm PBH}^2)$, where $S$ is the spin. Estimations of $\mathcal{S}$ for PBHs show that it is a small quantity, equal or lower than $0.01$ \cite{2019JCAP...05..018D}. Contrary to this, astrophysical BHs are expected to have substantially larger spin parameters since the angular momentum must be conserved during the collapse of its star of origin, which are often rotating. Hence, the spin can serve as a good proxy to distinguish the nature of a population of BHs. The measurement of low spin parameters could suppose a hint for the detection of BHs of primordial formation. The latest Bayesian analysis of LIGO/Virgo mergers suggest that low values for the spin parameter are strongly preferred by data, regardless of the priors considered \cite{2020arXiv201013811G}.

\section{Signatures of PBHs}
\label{sec:signatures}

In this section, we review the most relevant effects predicted from PBHs which could leave a strong signature in the evolution of the IGM and formation of structures. These are the accretion process, which would lead to an injection of energy into the medium, and the shot-noise contribution to the power spectrum, responsible of enhancing small scale fluctuations.

\subsection{Accretion onto PBHs}
\label{sec:acc}

Accretion of matter onto BHs is one of the most relevant effects which can strongly impact the surrounding medium, and thus leave significant observable signatures. The process of accretion was first studied by Hoyle and Lyttleton as a ballistic limit, neglecting the hydrodynamical, thermodynamical and pressure effects \cite{1939PCPS...35..405H, 1940PCPS...36..325H, 1940PCPS...36..424H}, and later by Bondi and Hoyle \cite{1944MNRAS.104..273B}. In these pioneer works, the accretion rate was found to depend on the relative velocity of the central point mass and the medium. The first to properly address the problem of spherical accretion, including the pressure effects, was Bondi in 1952 \cite{1952MNRAS.112..195B}. In the context of PBHs accreting matter in the early universe, it was shown that the expansion rate and Compton interactions played an important role \cite{2007ApJ...662...53R}. The accretion process in PBHs would result in the injection of energy into the surrounding medium, which could affect its properties. Early analyses of CMB data, sensitive to changes in the IGM properties, provided stringent constraints on the PBH abundance \cite{2008ApJ...680..829R}. However, posterior detailed computations showed that bounds are actually much weaker \cite{Ali-Haimoud:2016mbv, 2016arXiv161207264H, Poulin:2017bwe} (see Sec. \ref{sec:constraintsPBH} for more information regarding the observational constraints).

\subsubsection{Spherical accretion}

The physics of accretion is highly complex, but one can attempt an approximate simplified approach considering the non-relativistic limit and spherical symmetry, following the seminal work by Bondi \cite{1952MNRAS.112..195B, Ali-Haimoud:2016mbv}. In this framework, the BH is treated as a point mass surrounded by matter, embedded in a medium which tends to constant density and pressure, $\rho_\infty$ and $P_\infty$, far enough from the BH. At the relevant scales, one can address the problem within Newtonian fluid mechanics. The equations governing the evolution of the density $\rho$ and velocity $ {\bf v}$ of the matter surrounding the PBH are the continuity and Euler equations, respectively,
\begin{equation}
\frac{\partial \rho}{\partial t} + \nabla \cdot \left( \rho {\bf v} \right)=0,
\end{equation}
\begin{equation}
\frac{\partial {\bf v}}{\partial t} + {\bf v} \cdot \nabla  {\bf v}= - \frac{1}{\rho} \, \nabla P  -\frac{G M_{\rm PBH}}{r^2} {\bf\hat{r} },
\end{equation}
with $P$ the pressure, and the second term of the latter equation is the Newtonian force of gravity sourced by the central PBH, at a distance $r$ and pointing at the radial unity vector $\bf\hat{r}$. The pressure can be related to the gas temperature $T_K$ as $P = \rho \,  T/m_p$ (neglecting the helium contribution for the sake of simplicity), while its gradient can be related to the speed of sound $c_s$ as $\nabla P = c_s^2 \nabla \rho$. The simplest case to treat is considering spherical accretion, where the symmetry assumption greatly simplifies the picture, which allows extracting actual qualitative conclusions. We also assume stationary solutions, where the time derivatives can be neglected. The continuity equation can now be integrated, obtaining
\begin{equation}
4\pi r^2 \, \rho \, |v|  = \dot{M}_{\rm PBH} = {\rm constant} ~,
\label{eq:fluid1}
\end{equation}
while the Euler equation reads
\begin{equation}
v \, \frac{dv}{dr} = - \frac{1}{\rho} \, \frac{dP}{dr} -\frac{G M_{\rm PBH}}{r^2} ~, 
\label{eq:fluid2}
\end{equation}
Assuming a polytropic equation of state, $P=P_\infty(\rho/\rho_\infty)^{\gamma}$, it is straightforward to integrate Eq. \eqref{eq:fluid2}, obtaining  \cite{1952AJ.....57R..31W}
\begin{equation}
\frac12 v^2 -\frac{GM}{r} + \frac{c_{s,\infty}^2}{\gamma -1} \left[ \left(\frac{\rho}{\rho_\infty}\right)^{\gamma-1}-1 \right] =0,
\label{eq:bernoui}
\end{equation}
where the speed of sound at infinity is $c_{s,\infty}^2=\gamma P_\infty/\rho_\infty$. By scrutinizing the above equations, one finds the relevant scales to be the so-called Bondi velocity, Bondi radius, and Bondi time, defined by
\begin{equation}
\label{eq:rBndi}
v_{\rm B} = c_{s,\infty}, \, 	\; \;
r_{\rm B} = \frac{G M_{\rm PBH}}{v_{\rm B}^2}, \, \; \; t_B = \frac{r_B}{v_B}= \frac{G M_{\rm PBH}}{v_{\rm B}^3}.
\end{equation}
These scales have a direct interpretation: at distances around $ \sim r_B$, the accretion process starts to become important. Thus, at $\sim r_B$, the velocity reaches $v \sim c_{s, \infty}$, when the density is still close to the boundary value. It is customary to define the dimensionless accretion rate $\lambda$, as
\begin{equation}
\lambda = \frac{\dot{M}_{\rm PBH}}{4\pi  r_B^2 \rho_\infty v_B},
\end{equation}
which in the limit commented above, should take a value $\sim 1$. Its actual value depends on the equation of state and other non-gravitational forces such as pressure and viscosity, but in these cases of interest, it is always of order unity.

The asymptotic values at $r \rightarrow \infty$ are $\rho \rightarrow \rho_\infty$, and by Eq. \eqref{eq:fluid1}, the velocity at large radius decays as $|v| \simeq  \dot{M}_{\rm PBH}/(4\pi r^2 \rho_\infty) = \lambda v_B (r_B/r)^2$. The radial profiles vary depending on the equation of state considered. For instance, when the temperature evolves adiabatically, one has $T \propto \rho^{2/3}$ (see, e.g., Sec. \ref{sec:heating}), and thus $\gamma=5/3$. Examining Eqs. \eqref{eq:fluid1} and \eqref{eq:bernoui}, it can be found that the maximum accretion rate allowed is given by $\lambda_{ad}=1/4(3/5)^{3/2}\simeq 0.12$ \cite{Ali-Haimoud:2016mbv}. At low radii, the radial profiles evolve as $\rho \simeq \lambda \rho_\infty /\sqrt{2} (r_B/r)^{3/2}$, $v \simeq v_B \sqrt{2r_B/r}$ and $T \simeq T_\infty \lambda^{2/3}/2^{1/3} (r_B/r)$, where the temperature at the boundaries is related to the speed of sound as $c_{s,\infty}^2=\gamma T_{\infty}/m_p$. Note that the velocity does not depend on the accretion rate $\lambda$.

The other case of interest at high redshift is when the gas temperature is coupled to the CMB one, $T_{\gamma}$, by efficient Compton scattering. In this case, the gas can be regarded as isothermal (it has the same temperature, regardless of the density), $T=T_{\gamma}$ and then $\gamma=1$. In analogy with the previous case, if a solution exists, the accretion rate must be lower than a maximum value given by $\lambda_{iso}=e^{3/2}/3\simeq 1.12$. From Eqs. \eqref{eq:fluid1} and \eqref{eq:bernoui}, can be easily shown that the velocity and density radial profiles are the same as in the adiabatic case,\footnote{Actually, this is a coincidence, other equations of state would lead to different radial profiles.} while the temperature also decays as $r^{-1}$, but with a different prefactor \cite{Ali-Haimoud:2016mbv}.

In the physical relevant cases at high redshifts, the gas temperature would be between the CMB and the adiabatic evolution cases, and thus one would expect to have accretion rates between $\lambda_{iso}$ and $\lambda_{ad}$. As a reference value, in the adiabatic cooling regime, $T_K \propto (1+z)^2$ (see Sec. \ref{sec:heating}), which leads to a Bondi radius of $r_{\rm B} \sim 3\times 10^{-5} \, (M_{\rm PBH}/M_\odot) (30/(1+z))^2$~kpc. It is worth noting that this distance is much larger than the scales where general relativistic effects are relevant, which can be estimated as the Schwarzschild radius of the BH, defined as $r_S = 2GM_{\rm PBH}/c^2 \simeq 3 \, (M_{\rm PBH}/M_\odot)$ km, confirming the validity of the Newtonian approach.

As can be seen, the spherical symmetric case can be treated by analytic means given the equation of state, but its simplicity limits its range of validity. Firstly, it does not take into account the velocity of the BH relative to the medium, $v_{rel}$. The seminal Bondi's paper \cite{1952MNRAS.112..195B} already suggests a way to include that effect and match with the Hoyle-Lyttleton and Bondi-Hoyle results, replacing the Bondi velocity by an effective quadratic sum of velocities, $c_{s,\infty}^2 \rightarrow c_{s,\infty}^2 + v_{rel}^2$. One can thus write the accretion rate as
\begin{equation}
\dot{M}_{\rm PBH} = 4\pi \lambda \rho_\infty  \frac{(GM_{\rm PBH})^2}{(c_{s,\infty}^2 + v_{rel}^2)^{3/2}}.
\label{eq:mdotvlin}
\end{equation}
In the early universe, $v_{rel}$ can be estimated as the baryon-DM relative velocity computed in linear theory. Its root-mean-square value is approximately constant before Recombination, dropping linearly with $1+z$ at later times. Hence, it can be written as \cite{Ali-Haimoud:2016mbv}
\begin{equation}
\sqrt{\langle v_L^2 \rangle} \simeq
\begin{cases}
      30 \; {\rm km/s} & , \hspace{5mm} 1+z \gtrsim 10^3 \\
      30 \; {\rm km/s} \left( \frac{1+z}{10^3} \right) & , \hspace{5mm} 1+z \lesssim 10^3,
    \end{cases}
\end{equation}
where the amplitude $30 \; {\rm km/s}$ corresponds to the variance at the epoch of the CMB decoupling. On the other hand, real BHs spin, and thus form an accreting disk, being the spherical symmetric case not applicable. Even though PBH spins are expected to be small, as discussed in Sec. \ref{sec:spin}, the accreted matter may possess an angular momentum high enough to prevent the free-fall straight onto the BH and rotate around it. During the falling, angular momentum must be conserved, which leads to the formation of an accretion disk. More specifically, an accretion disk would form if the angular momentum is large enough to keep matter orbiting at Keplerian orbits at distances much larger than the innermost stable orbits, which are roughly given by the Schwarzschild radius \cite{Agol:2001hb}. Applying this criterion, it has been argued that PBHs formed in the early universe would form an accretion disk if the condition $f_{\rm PBH} M_{\rm PBH}/M_\odot \ll ((1+z)/730)^3$ is fulfilled \cite{Poulin:2017bwe}, which is satisfied for $M_{\rm PBH} \gtrsim M_\odot$ and a large enough abundance at the epoch of the CMB decoupling or at later times. However, some of the results outlined above are still valid as long as that the dimensionless accretion rate is properly modified. Accounting for viscosity effects and the matter outflows through jets, the typical values of order unity for $\lambda$ in spherical accretion are suppressed by roughly two orders of magnitude \cite{Poulin:2017bwe}.

\subsubsection{Accretion luminosity}

The relevance of the accretion mechanism resides in the fact that part of the energy carried by the inflowing matter is released later to the surrounding medium. The matter falling onto the BH is greatly accelerated, 
producing radiative emission of high energy photons by processes such as bremsstrahlung. The luminosity of the accreting BH is proportional to its accretion rate, and can be written as \cite{Poulin:2017bwe}
\begin{equation}
L_{\rm acc} = \epsilon(\dot{M}_{\rm PBH}) \, \dot{M}_{\rm PBH} ~,
\label{eq:lacc}
\end{equation}
where $\epsilon(\dot{M}_{\rm PBH}) $ denotes the radiative efficiency of the accretion process. This is in general a complicated function of the accretion rate $\dot{M}_{\rm PBH}$, depending on the details and geometry of the accretion mechanism. Given the luminosity, the rate of energy injected into the medium per unit volume is given by \cite{Ali-Haimoud:2016mbv, Poulin:2017bwe}
\begin{equation}
\left(\frac{dE}{dV\,dt}\right)_{\rm inj} = L_{\rm acc} \, n_{\rm PBH} = L_{\rm acc} \, \frac{f_{\rm PB
H} \, \rho_{\rm DM}}{M_{\rm PBH}} ~,
\label{eq:enginj}
\end{equation}
As a benchmark value, it is usually considered the \textit{Eddington limit}, which is the resulting luminosity from balancing the gravitational force by pressure radiation (e.g., \cite{2010gfe..book.....M}),
\begin{equation}
L_{\rm Edd} = \frac{4\pi G m_p c}{\sigma_T} = 1.26 \times 10^{38} \left( \frac{M_{\rm PBH}}{M_\odot} \right) \; {\rm erg/s}.
\label{eq:lEd}
\end{equation}
The energetic efficiency of the accretion process strongly depends upon the geometry, viscosity and other hydrodynamical considerations. If the accretion disk is optically thin, most of the energy released through viscous dissipation is radiated away, and the luminosities obtained can be close to $L_{\rm Edd}$. This was the basis of the thin disk model proposed by Shakura and Sunyaev in 1972 \cite{Shakura:1972te}, which is able to explain the extreme brightness of many far AGNs. However, nearby BHs appear to radiate in a much less efficient way \cite{2002luml.conf..405N}. Other mechanisms have been proposed to explain such behavior, such as the Advective-Dominated-Accretion-Flow (ADAF), firstly proposed independently by Refs. \cite{Ichimaru:1977uf, Rees:1982pe} and rediscovered years later by Refs. \cite{Narayan:1994is, 1994ApJ...428L..13N, 1995ApJ...438L..37A} (see, e.g., Ref. \cite{Yuan:2014gma} for a review). In this scenario, the dynamics is ruled by advective currents, forming a hot thick disk or torus. Most of the emitted energy is deposited in the same accretion disk, heating it up. Thus, only a small portion of energy is injected into the surrounding medium, being the radiative process inefficient. In the ADAF scenario, the efficiency function can be fitted by a broken power-law formula of the form
\begin{equation}
\epsilon = \epsilon_0 \, \left(\frac{100 \, \dot{M}_{\rm PBH}}{\dot{M}_{\rm Edd}} \right)^a,
\label{eq:efficiency}
\end{equation}
where $\dot{M}_{\rm Edd}= 10 \, L_{\rm Edd}$, and the slopes and amplitudes $\epsilon_0$ and $a$ depend on the mass range and the specific modeling of the viscosity effects \cite{Xie:2012rs}. Note that some previous studies in the literature assumed the case $a=1$ \cite{2008ApJ...680..829R, Ali-Haimoud:2016mbv}, while for such low accretion rates, it has been shown that $a \lesssim 0.6$ \cite{Xie:2012rs}. At a given range of masses, the final luminosity would be thus proportional to some power of the mass accretion rate. From Eqs. \eqref{eq:lacc} and \eqref{eq:efficiency}, one finds $L_{\rm acc} \propto \dot{M}_{\rm PBH}^{a+1}$, which, from Eq. \eqref{eq:mdotvlin}, is also proportional to the inverse velocity factor $\left(c_{s,\infty}^2 + v_{\rm rel}^2\right)^{-3(a+1)/2}$. Since the DM-baryon relative velocity components are random variables gaussian distributed (and thus its modulus follows a Maxwellian distribution), one must average the total luminosity. This allows to introduce an effective velocity $v_{eff}$, such that $L_{\rm acc} \propto v_{eff}^{-3 \, (1+a)/2}$, defined as \cite{Mena:2019nhm, 2008ApJ...680..829R}
\begin{align} 
	\label{eq:veffparam}
	v_{eff} & \equiv \left< \frac{1}{\left(c_{s,\infty}^2 + v_{ rel}^2\right)^{3 \, (1+a)/2}}\right>^{-\frac{1}{3 \, (1+a)}}  \\
	&= \left[ \int \frac{d^3v_{ rel}}{(2\pi\langle v_L^2 \rangle)^{3/2}} \frac{\exp \left(-\frac{v_{ rel}^2}{2\langle v_L^2 \rangle} \right)}{\left(c_{s,\infty}^2 + v_{ rel}^2\right)^{3 \, (1+a)/2}} \right]^{-\frac{1}{3 \, (1+a)}} \\
	&= c_{s,\infty}  \,  \left[ \left(\frac{3}{2}\right)^{\frac{3}{2}} \, U\left(\frac{3}{2}, 1 - \frac{3 \, a}{2}, \frac{3}{2} \mathcal{M}^{-2}\right) \, \mathcal{M}^{-3} \right]^{- \frac{1}{3 \, (1+a)}},
\end{align}
with $\mathcal{M} \equiv \frac{\sqrt{\left<v_L^2\right>}}{c_{s,\infty}}$ and $U (x ; y ; z)$ the confluent hypergeometric function of second kind, also known as Tricomi's function. Two limiting cases can be obtained up to first order, depending on the ratio between velocities:
\begin{equation}
v_{\rm eff} \simeq
\begin{cases}
 c_{s, \infty} \, {\cal{M}}^{1/(1+a)} \, \left[3 \, \sqrt{\frac{3}{2 \, \pi}} \, B\left(\frac{3 \, a}{2} , \frac{3}{2}\right) \right]^{-\frac{1}{3 \, (1+a)}}  & , \hspace{5mm} {\cal{M}} \gg 1 \\
 c_{s, \infty} & , \hspace{5mm} {\cal{M}}  \ll  1  ~,
\end{cases} 
\end{equation}
with $B(x, y)$ the beta function. The limit ${\cal{M}} \ll 1$ holds when the gas is very hot and thus the relative velocity with respect to the environment is negligible, which is fulfilled at early times $z > 10^4$, but also after the Cosmic Dawn, when the IGM is heated up by X-ray radiation from X-ray binaries and other sources (as shall be studied in Sec. \ref{sec:heating}). The opposite limit, ${\cal{M}} \gg 1$, depends upon the specific slope $a$. In the usual oversimplified scenario considered in the literature, $\epsilon \propto \dot{M}_{\rm PBH}$, and thus $a=1$, which leads to $v_{\rm eff} \sim \sqrt{c_{s, \infty}\left<v_L^2\right>^{1/2}}$ \cite{2008ApJ...680..829R, Ali-Haimoud:2016mbv}.

Finally, the energy emitted in the accretion processes is deposited through different channels into the medium. Rather than the rate of energy injected, the relevant quantity is thus the energy deposition rate for each channel, which reads
\begin{equation}
\label{eq:eninj}
\left(\frac{dE_c}{dV dt}\right)_{{\rm dep}} = f_c(z) \, \left(\frac{dE}{dV dt}\right)_{\rm inj} ~,
\end{equation}
where the subscript $c$ denotes the channel in which energy is deposited, namely: ionization of neutral Hydrogen or Helium, heating of the medium, or atomic excitations (where the ${\rm Ly}\alpha$ transitions are the most relevant). The energy deposition factors $f_c(z)$ quantify the fraction of energy which goes to the different channels, and are obtained from
\begin{equation}\label{eq:energydeposit}
f_c(z) = \frac{H(z)\int \frac{d\ln(1+z^\prime)}{H(z^\prime)} \int d\omega \, T_c(z,z^\prime,\omega) \, L_{\rm acc}(z^\prime, \omega)}{\int d\omega \, L_{\rm acc}(z, \omega)} ~,
\end{equation}
where the $T_c(z,z^\prime,\omega)$ are the transfer functions computed in Ref.~\cite{Slatyer:2015kla}. The above formula integrates over the energy spectrum of the emitted luminosity, which in ADAF models can be simply parameterized as \cite{Yuan:2014gma}
\begin{equation}\label{eq:L_spec}
L_{\rm acc}(\omega) \propto \Theta(\omega - \omega_{\rm min}) \, \omega^{\beta} \, \exp(-\omega / \omega_s) ~,
\end{equation}
where $\omega_{\rm min} \equiv (10 \, M_\odot / M_{\rm PBH})^{1/2}$ eV,  $\omega_s = 200$~keV (or more generally, $\omega_s \sim \mathcal{O}(m_e)$), and $\beta$ ranging in the interval $\beta \in [-1.3, -0.7]$, with the fiducial value $\beta = -1$.

\subsection{Shot noise}
\label{sec:shot}

As commented above, a large enough population of PBHs would require an enhancement of the fluctuations at very small scales, produced by some inflationary mechanism, which is highly model dependent. However, the presence of PBHs can also modify the matter power spectrum by other reasons at larger scales. By the point-source nature of PBHs, one may expect a shot noise contribution to the power spectrum constant in wavenumber. This is a general feature of a discrete distribution of points with negligible spatial extension, as PBHs can be approximated. In order to understand how it arises, consider the number density of a discrete distribution of $N$ points, given by
\begin{equation}
n({\bf x}) = \sum_i^{N} \delta_D({\bf x}-{\bf x}_i)~.
\end{equation}
Writing the mean density in a volume $V$ as $\bar{n}=N/V$ and defining the fluctuation as $\delta({\bf x}) = (n({\bf x})- \bar{n})/\bar{n}$, its Fourier transform reads
\begin{equation}
\tilde{\delta}({\bf k}) = \int d^3x \; \delta({\bf x}) \; e^{i{\bf k}\cdot {\bf x}} = \frac{1}{\bar{n}}\sum_i^{N} e^{i{\bf k}\cdot {\bf x}_i} - (2\pi)^3\delta_D({\bf k}).
\end{equation}
Thus, for ${\bf k} \neq {\bf 0} $, and writing the ensemble average $\langle ... \rangle$ as a spatial average (as implicitly done in Sec. \ref{sec:lineargrowth}), one has
\begin{align}
\langle  \tilde{\delta}({\bf k}) \tilde{\delta}({\bf k'}) \rangle &=
 \int \frac{d^3 x_i}{V} \frac{1}{\bar{n}^2}\sum_{i}^{N} e^{i ({\bf k} + {\bf k'})\cdot{\bf x}_i } + \left( i \neq j~term  \right) \\
 &= (2\pi)^3 \delta_D^{(3)}({\bf k} + {\bf k'})\frac{1}{\bar{n}} + \left( i \neq j~term  \right) \, .
\end{align}
While the second term gives the \textit{proper} power spectrum for non-zero spatial separation, from the first one, a contribution arises for ${\bf x}_i={\bf x}_j$, i.e., at zero spatial separation. Performing the integral of the first term, one gets the delta function of the second line which ensures translational invariance. Comparing to Eq. \eqref{eq:PS}, the shot noise power spectrum therefore reads
\begin{equation}
P_{\rm SN}(k) = \frac{1}{\bar{n}},
\label{eq:ps_pois}
\end{equation}
which is constant over wavenumbers (and thus a white noise), and inversely proportional to the number density of discrete sources. This power spectrum can be related to its real-space counterpart, the correlation function, which is a delta function in position space centered at zero distances.

On the other hand, it has been argued that these PBHs fluctuations give rise to isocurvature modes \cite{2003ApJ...594L..71A, 2006PhRvD..73h3504C,Inman:2019wvr}. In general, the entropy fluctuation is defined as
\begin{equation}
S = \frac{\bar{\rho}_{\rm DM}\delta_{\rm DM}}{\bar{\rho}_{\rm DM} + \bar{P}_{\rm DM}} - \frac{\bar{\rho}_{rad}\delta_{rad}}{\bar{\rho}_{rad} + \bar{P}_{rad}} = \delta_{\rm DM} - \frac{3}{4}\delta_{rad}\, .
\end{equation}
Isentropic (also called adiabatic) perturbations are those which $S=0$ initially, and can be regarded as fluctuations in the density field, leaving unperturbed the equation of state of the multi-component fluid. Standard CDM composed by entirely particles behave as isentropic modes, and in the case of no PBH, then $\delta_{\rm DM} = \delta_{\rm pDM}= \frac{3}{4}\delta_{rad}$, where $\rm pDM$ denote particle DM. Contrarily, isocurvature perturbations are those which arise from perturbations in the equation of state, rather than in the metric, producing thus entropy, instead to the isentropic ones. Consider now a mix of particle DM and PBHs. Since the total DM density is given by its sum, $\rho_{\rm DM}=\rho_{\rm pDM} + \rho_{\rm PBH}$, the joint DM perturbation reads $\delta_{\rm DM} = (1-f_{\rm PBH})\delta_{\rm pDM} + f_{\rm PBH}\delta_{\rm PBH}$, with the particle and PBH fluctuations uncorrelated. While the particle DM would still be adiabatic, and thus $\delta_{\rm pDM}=3\delta_{rad}/4$, the PBH perturbations may have a mix of adiabatic and isocurvature modes. Computing the entropy fluctuation, one finds $S=\delta_{\rm DM} - \frac{3}{4}\delta_{rad} = f_{\rm PBH}(\delta_{\rm PBH} - \delta_{\rm pDM})$. Given that $\delta_{\rm pDM}$ and $\delta_{\rm PBH}$ are uncorrelated, $S$ will be in general non-zero, and therefore, an isocurvature mode arises. Since by assuming gaussian fluctuations, primordial clustering can be neglected, as discussed in Sec. \ref{sec:spin}, the isocurvature part $\delta_{\rm PBH} - \delta_{\rm pDM}$ would correspond to an initial power spectrum given by Eq. \eqref{eq:ps_pois}.

For general isocurvature perturbations, the transfer function, which relates the evolved power spectrum to its primordial form, can be approximated as \cite{Peacock:1999ye}
\begin{equation}
T_{\rm iso} \simeq
\begin{cases}
  \frac32  \, (1+z_{\rm eq}), \; & {\rm for} \; k>k_{\rm eq} \\
  0, &{\rm otherwise}
\end{cases}
\end{equation}
where $z_{\rm eq}$ is the redshift of matter-radiation equality, and $k_{\rm eq} = H(z_{\rm eq})/(1+z_{\rm eq})$. Therefore, the contribution of isocurvature modes only affects scales smaller than those corresponding to the matter-radiation equality era. Given that PBHs contribute to the matter overdensity as $f_{\textrm{PBH}} \delta$, and including the growth factor defined in Sec. \ref{sec:lineargrowth}, the final contribution from the shot-noise to the matter power spectrum at redshift $z$ can be written as~\cite{2003ApJ...594L..71A}
\begin{align}
\label{eq:Pnoise}
\Delta P(k,z) &= T^2_{\rm iso} \, \frac{f_{\textrm{PBH}}^2}{n_{\textrm{PBH}}} \, D^2(z)
= \frac{9 \, (1+z_{\rm{eq}})^2 f_{\textrm{PBH}} \, M_{\rm PBH}}{4 \, \Omega_{\rm DM} \, \rho_c} D^2(z) \\
&\simeq 2.5 \times 10^{-2} f_{\textrm{PBH}} \left(\frac{M_{\rm PBH}}{30 \, M_\odot}\right) D^2(z)\ \textrm{Mpc}^3~,\quad k>k_{\rm{eq}} ~,
\end{align}
where standard values for $z_{\rm eq}$ and $\Omega_{\rm DM}$ have been employed (see Sec. \ref{sec:background}). It is noteworthy that the above power spectrum only depends on the joint product of the PBH fraction and the mass, instead of more complicated dependences as happens with accretion effects (see Sec. \ref{sec:acc}). Thus, the shot-noise contribution is degenerate on these two parameters.

\begin{figure}
\begin{center}
\includegraphics[scale=0.43]{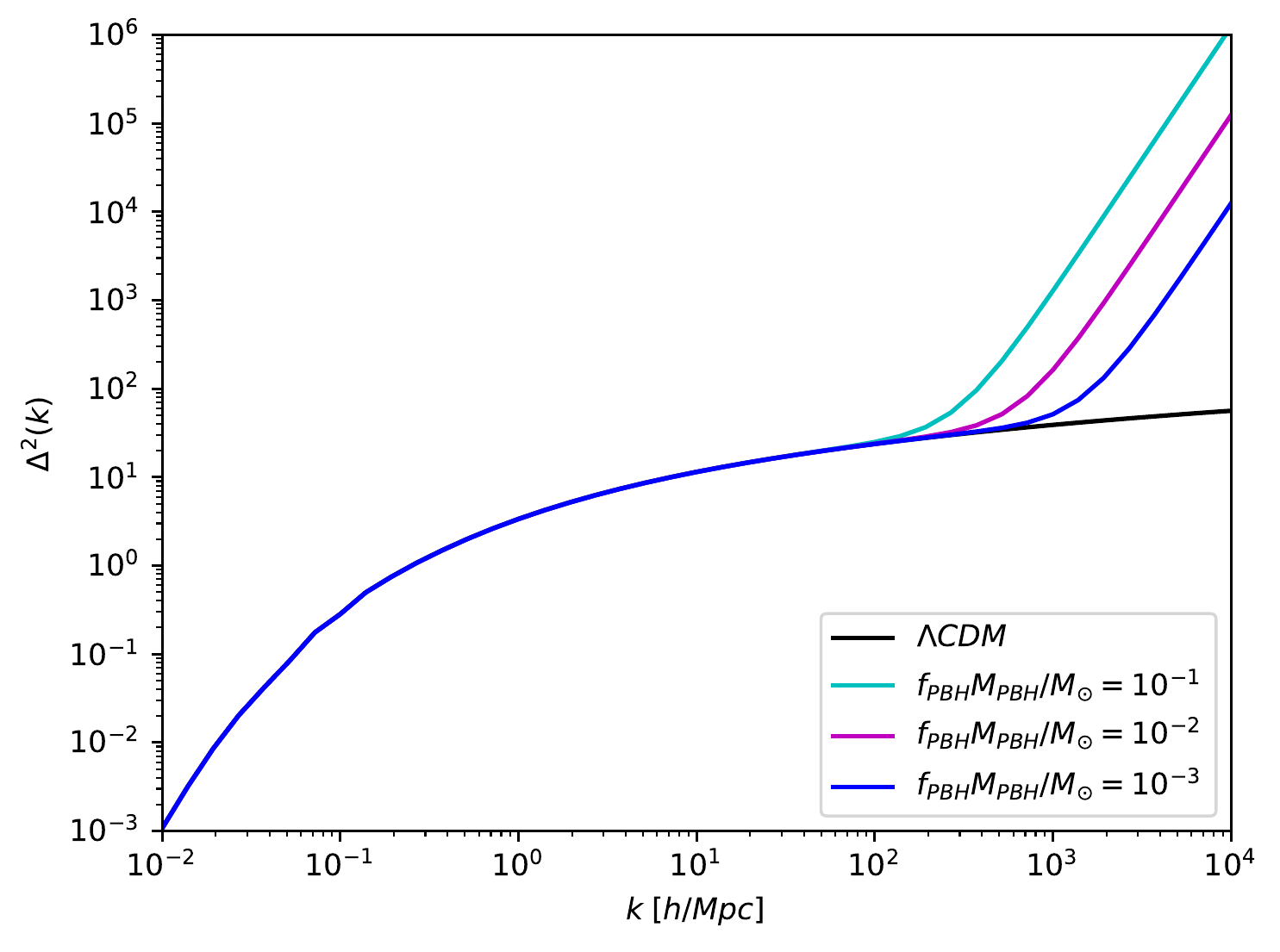}
\includegraphics[scale=0.43]{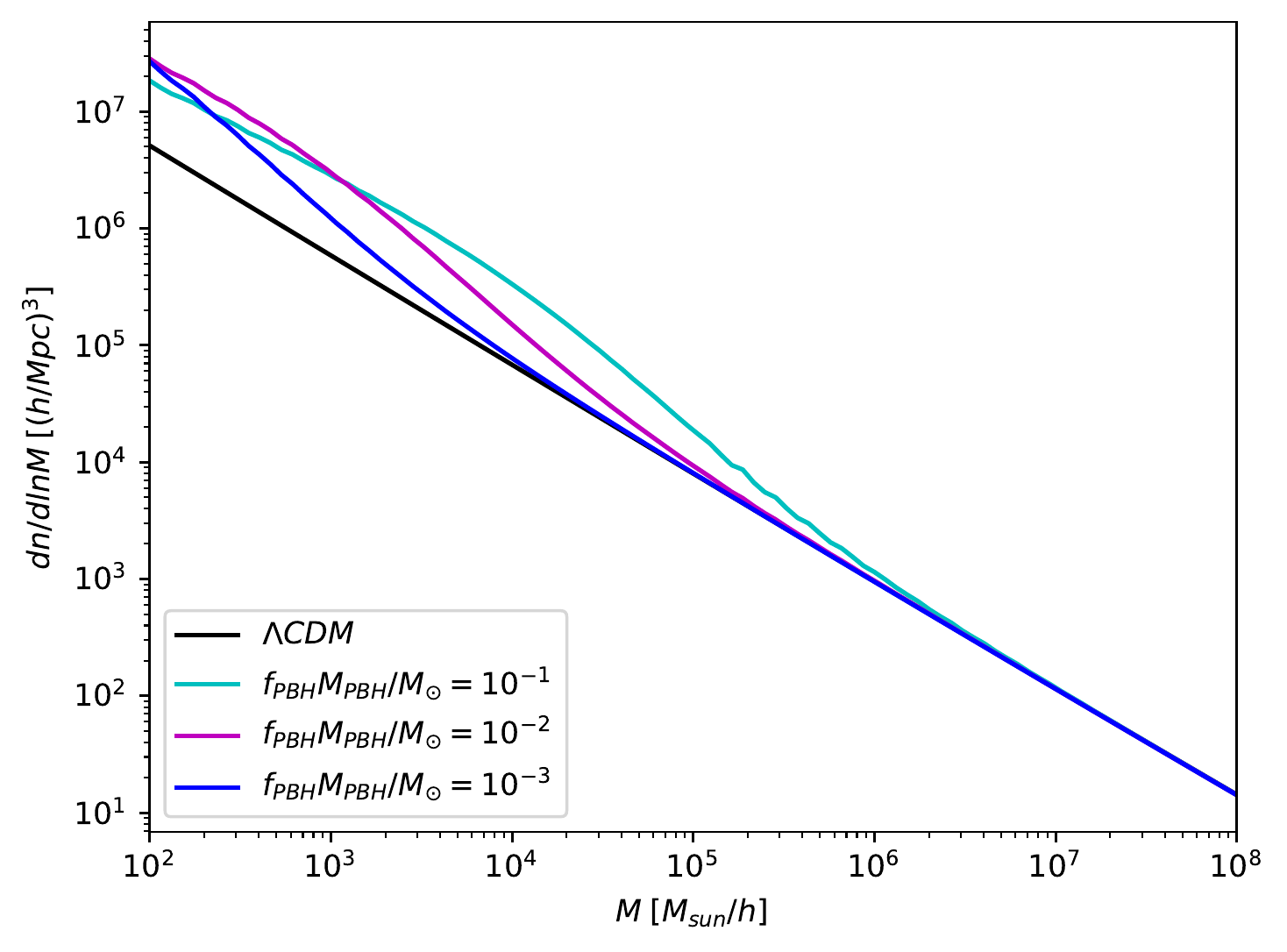}
\caption{Dimensionless power spectrum (left) and halo mass function (right) at $z=0$, for the $\Lambda$CDM case and for three PBH models with the shot-noise contribution, enhancing the power at small scales and increasing the number of low mass halos. Note that, since $P \propto k^0$ at high $k$, then $\Delta^2 \propto k^3$.}
\label{fig:shotnoise}
\end{center}
\end{figure}

The inclusion of this term modifies the total matter power spectrum, and thus, the halo mass function on small scales, in analogy with the WDM and IDM cases (see Chapter \ref{chap:DarkMatter}). However, contrary to these previous DM models, low-mass halos are now enhanced instead of suppressed. For this reason, this contribution becomes only relevant for low-mass halos not large enough to cool and collapse to form stars, which are commonly known as \textit{minihalos}. Fig. \ref{fig:shotnoise} shows the dimensionless power spectrum and the halo mass function at $z=0$ for three models accounting for this contribution. Note that the virial mass at $T_{\rm vir} = 10^4$ K, $M \sim 10^8 \, M_\odot$, the relevant scale for star-forming halos, is way above these modifications (see Eq. \eqref{eq:Tvir}).

In Part \ref{partII} of this thesis, the shot-noise impact on the 21 cm signal is discussed in detail \cite{Mena:2019nhm}. Since the standard 21 cm signal is mostly ruled by the most massive star-forming halos (as shall be discussed in Chapter \ref{chap:IGM}), a contribution from a high number density of minihalos may become relevant \cite{Gong:2017sie, Gong:2018sos}. However, we shall demonstrate that previous estimates neglected the effect of the heating of the IGM, which, when is consistently accounted for, suppresses the minihalo contribution, making it mostly negligible.

\section{Observational constraints on PBHs as DM}
\label{sec:constraintsPBH}

\begin{figure}[!tbph]
\begin{center}
\includegraphics[scale=0.7]{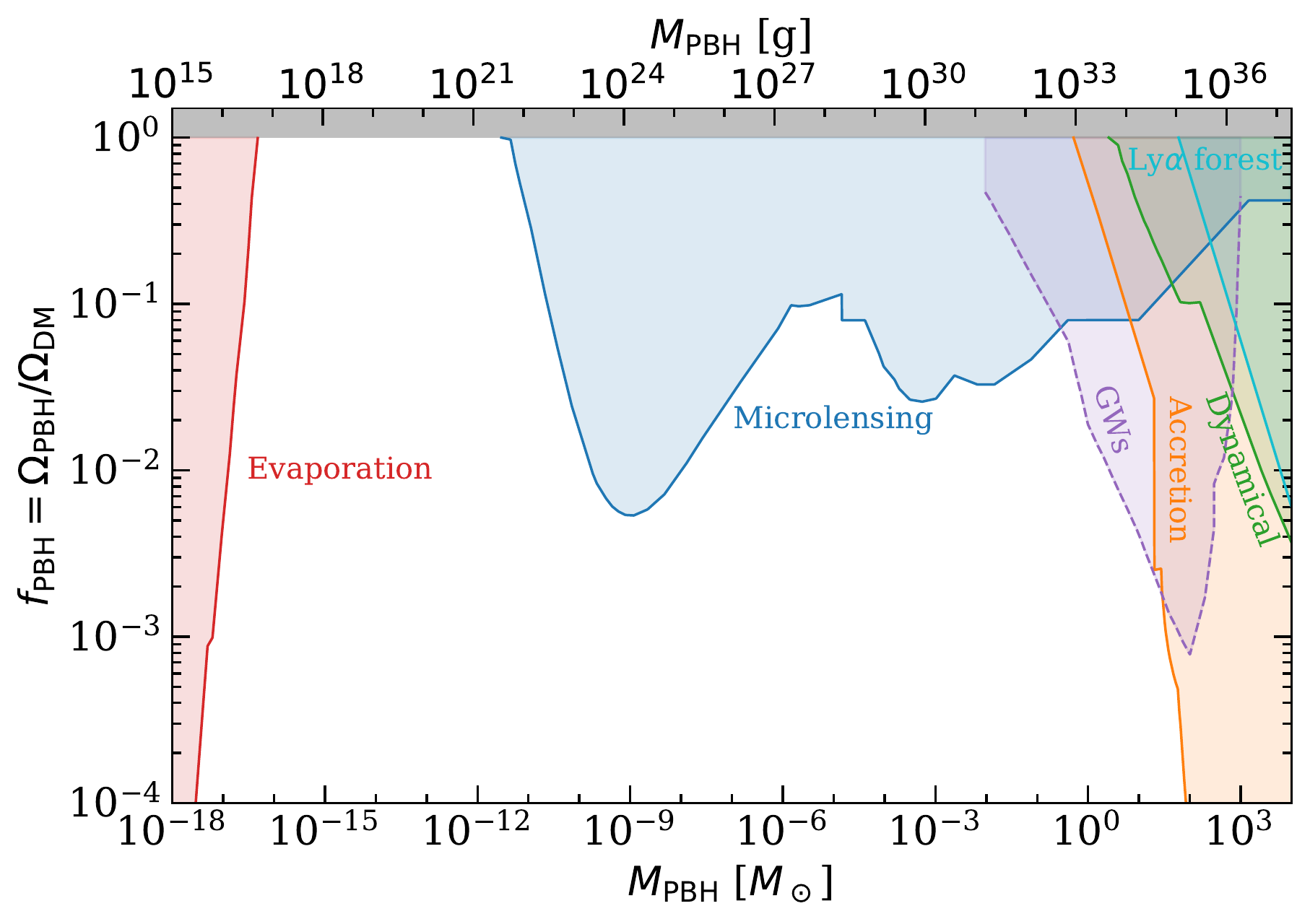}
\caption{Compilation of current constraints on the PBH fraction (with respect to DM) as a function of the PBH mass, assuming a monochromatic mass function. The different probes considered are: evaporation of BHs due to Hawking radiation (red), non-observation of microlensing events (blue), merger rates from gravitational waves measurements (purple), effect of accretion on the CMB (orange), dynamical constraints, such as disruption of stellar systems by the presence of PBHs (green), and the effect of the Poisson shot noise power spectrum in the Ly$\alpha$ forest (cyan), see the text for more details. Limits from GWs appear with dashed lines since they could be invalidated \cite{Boehm:2020jwd}. Figure created with the publicly available \texttt{Python} code \texttt{PBHbounds}\protect\footnotemark   \cite{bradley_j_kavanagh_2019_3538999}. }
\label{fig:pbhbounds}
\end{center}
\end{figure}

PBHs can impact cosmology and astrophysics in a wide range of ways, leaving different observational effects which allow constraining their properties. In this section, we review the most important bounds on the current fraction of PBHs as DM, $f_{\rm PBH} = \Omega_{\rm PBH}/\Omega_{\rm DM}$, for a wide range of masses $M_{\rm PBH}$. A collection of limits from the different probes is shown in Fig. \ref{fig:pbhbounds}. For a more comprehensive list of constraints, see, e.g.,  \cite{2010PhRvD..81j4019C, 2014arXiv1403.1198G, 2016PhRvD..94h3504C, Green:2020jor, 2020ARNPS..7050520C, 2020arXiv200212778C}.

\footnotetext{\href{https://github.com/bradkav/PBHbounds}{https://github.com/bradkav/PBHbounds}}

\begin{itemize}

\item \textbf{Evaporation}

As already seen, since BHs emit energy due to Hawking radiation, relatively light PBHs may be evaporating now. Those with a lifetime shorter than the age of the universe must have disintegrated nowadays, fact which forbids PBHs with $M_{\rm PBH}<M_* \simeq 4\times 10^{14}~{\rm g}$ to form part of the current DM \cite{1976PhRvD..14.3260P, 2016PhRvD..94d4029C}. Moreover,  PBHs with masses small enough, although still present, should emit a strong $\gamma$-ray and cosmic ray background which could be observed. Absence of its detection strongly constrains the range of masses $M_{\rm PBH} \lesssim 10^{17}$ g. Concretely, the maximum fraction allowed is $f_{\rm PBH} \lesssim 2 \times 10^{-8} (M_{\rm PBH}/M_*)^{3+\epsilon}$, with $\epsilon \sim 0.1 - 0.4$ \cite{2016PhRvD..94h3504C}. Comparable limits have been found from comparing the expected evaporation fluxes of positrons to the INTEGRAL observation of the Galactic Center 511 keV gamma-ray line \cite{DeRocco:2019fjq, Laha:2019ssq}. Diffuse supernova neutrino background searches at Super-Kamiokande are also able to set competitive constraints \cite{Dasgupta:2019cae}. Considering extra contributions from electron-positron annihilations and the background modeling may set stringent bounds from the hard X-ray band \cite{Iguaz:2021irx}.

\item \textbf{Microlensing}

If a compact object crosses the line of sight of a star, it may produce a so-called microlensing effect, which implies a transient and achromatic amplification of its flux. The range of masses of the objects which can produce it span from $5 \times 10^{-10}$ to $\sim 100 \, M_\odot$ \cite{1986ApJ...304....1P, Green:2020jor}. The non-detection of these events leads to stringent bounds on the maximum abundance of PBHs, as the ones found by the MACHO \cite{2001ApJ...550L.169A} and EROS \cite{2007A&A...469..387T} surveys in the Large and Small Magellanic Clouds, the Subaru Hyper Suprime Cam (HSC) in M31 (Andromeda) \cite{2019NatAs...3..524N} and the Optical Gravitational Lensing Experiment (OGLE) in the Galactic bulge \cite{2019PhRvD..99h3503N}. The above collaborations have placed upper bounds on the allowed fraction around $f_{\rm PBH} \lesssim 0.01-0.1$ in the aforementioned range of masses. Nonetheless, the existence of Earth-mass PBHs ($M_{\rm PBH} \sim 10^{-5} M_\odot$) with a fraction $f_{\rm PBH} \sim 0.03$ could explain the observation of 6 microlensing events found in the OGLE data \cite{2017Natur.548..183M}, being consistent with other constraints at these range of masses \cite{2019PhRvD..99h3503N}. Although this may constitute a hint of their existence, it cannot be regarded as a detection of PBHs, since these microlensing observations could also be explained by free-floating planets. There are, however, some caveats regarding the results of the MACHO collaboration \cite{2015A&A...575A.107H}, since the limits reported are model dependent and could be biased by the assumption of an over-massive halo. Moreover, the results of the MACHO and EROS projects have been found to be statistically incompatible. Therefore, these bounds are not completely reliable, and PBHs could not be definitely ruled out within these range of masses.

\item \textbf{Other lensing events}

Besides stellar microlensing, other variations of the lensing effect have been proposed to constrain such compact objects. For instance, non-smoothly distributed PBHs would produce a magnification effect of the observed light from supernovae type Ia, which constrains the PBH fraction to $f_{\rm PBH} \lesssim 0.4$ for $ \gtrsim 10^{-2} \, M_\odot$ \cite{2018PhRvL.121n1101Z}. However, it has been argued \cite{2017arXiv171206574G} that former work did not account properly for the size of supernovae, which significantly weakens the bounds in the region of first LIGO detections.\footnote{The preprint of Ref. \cite{2018PhRvL.121n1101Z} was titled \textit{No LIGO MACHO}, after which the former PhD advisor of one its authors replied with Ref. \cite{2017arXiv171206574G}, named \textit{LIGO Lo(g)normal MACHO}. Both titles represent a pun in Spanish language.} On the other hand, PBHs of masses $ \gtrsim 10-100 \, M_\odot$ may produce strong gravitational lensing of extragalactic Fast Radio Bursts (FRB). A compact object passing through could cause two different images of the same burst separated by a short time delay. The lack of observed events by future experiments like CHIME would lead to upper bounds around $f_{\rm PBH} \lesssim 0.01$ \cite{PhysRevLett.117.091301}. Femtolensing (i.e., lensing when the magnification depends on the wavelength of the observed light) of Gamma-Ray Bursts (GRBs) by asteroid-mass ($5 \times 10^{17} - 10^{20}$ g) compact objects has been also proposed to constrain their abundance \cite{2012PhRvD..86d3001B}, but posterior analyses showed that GRBs are too large to be modelled as point-sources, as had been previously  assumed, invalidating those bounds \cite{2018JCAP...12..005K}.

\item \textbf{Gravitational waves}

The observation of BH mergers by LIGO and Virgo collaborations can be employed to constrain the allowed number of PBHs. To do so, it is demanded that the predicted merger rates of PBH binaries cannot exceed the ones measured by gravitational waves. Employing data from the O1 LIGO run, tight upper bounds of $f_{\rm PBH} \lesssim 0.01$ have been found at masses between 1 and 300 $M_\odot$ \cite{2017PhRvD..96l3523A}. However, these constraints present some caveats. It was assumed that PBH binaries are not disturbed between the moment of their formation and their merger, something which has to be yet confirmed by numerical simulations. Moreover, to derive these limits, BHs have been treated as Schwarzschild BHs, while it should be more appropriate to employ cosmological BH solutions embedded in a FLRW metric, such as the Thakurta metric \cite{1981InJPh..55..304T}. This implies a time-dependent mass, and that PBH binaries created before galaxy formation would have merged at much earlier times, resulting in merger rates consistent with the LIGO data and completely avoiding these constraints \cite{Boehm:2020jwd}. Besides BH mergers, the non-observation of a stochastic gravitational wave background of mergers expected from a population of PBHs has also been used for constraining their abundance \cite{2020JCAP...08..039C}. Nonetheless, the same assumptions than for those mergers have been employed, and for the same reason, they could be invalidated \cite{Boehm:2020jwd}.

\clearpage

\item \textbf{Dynamical constraints}

Due to two-body interactions, kinetic energies of systems of different masses usually become balanced and match. If a stellar system presents in addition a MACHO population, its stars would gain kinetic energy and, due to the virial theorem, the system would expand. Therefore, the presence PBHs will dynamically heat the star clusters, making them larger and with higher velocity dispersions, leading to an eventual dissolution into its host galaxy. Populations with high mass to luminosity ratios are more sensitive to this effect, as happens with Ultra Faint Dwarf Galaxies (UFDW), which would be disrupted by the presence of PBHs. Making use of this fact, tight bounds have been obtained,  $f_{\rm PBH} \sim 10^{-3}$ at $M_{\rm PBH} \sim 10^4 \, M_\odot$, weakening at lower masses down to $f_{\rm PBH} \lesssim 1$ at $M_{\rm PBH} \sim 10 \, M_\odot$ \cite{2016ApJ...824L..31B}. In a similar way, wide binary stellar systems may be perturbed by compact objects, potentially being disrupted after multiple encounters. The separation distribution of wide binaries can lead to (somewhat weak) bounds on the PBH fraction, from $f_{\rm PBH} \lesssim 1$ for $M_{\rm PBH} \simeq 3 \, M_\odot$ up to $f_{\rm PBH} \lesssim 0.1$ at $M_{\rm PBH} \gtrsim 70 \, M_\odot$ \cite{2014ApJ...790..159M}.

\item \textbf{CMB constraints}

As already stated, PBHs can emit highly energetic radiation, either as Hawking radiation from their evaporation, or from the radiation emitted by accretion. Both processes may affect the CMB spectrum mostly in two different ways: producing spectral distortions and modifying temperature anisotropies. Spectral distortions are deviations from a thermal spectrum caused by the energy injection. On the other hand, the energetic radiation can enhance the ionization rate, delay Recombination and shift the peaks of the CMB anisotropy spectrum, as well as induce more diffusion damping. The polarization spectrum can also been modified, since the increase of the fraction of free electrons would increase the Thomson optical depth and enhance the Reionization bump at large angular scales (see Sec. \ref{sec:thomsonopdep} for more information).

As discussed in Sec. \ref{sec:acc}, although early CMB analyses \cite{2008ApJ...680..829R} found very stringent bounds on the allowed abundance of accreting PBHs, later works revisited these computations and found much milder constraints, ranging from $f_{\rm PBH} \lesssim 1$ for $M_{\rm PBH} \sim 10$ to $f_{\rm PBH} < 3 \times 10^{-9}$ at $M_{\rm PBH} \sim 10^4 \, M_\odot$ \cite{Ali-Haimoud:2016mbv, 2016arXiv161207264H}. The former bounds strongly depend on whether the accreted matter is ionized by collisional ionization (i.e., by collisions with free electrons) or photoionization (due to the absorption of ionizing UV photons), with the real scenario between both cases. On the other hand, while the former constraints rely on the assumption of spherical accretion, accreting disks have been argued to be more realistic for PBHs.  Limits in these scenarios are even tighter, finding $f_{\rm PBH} \lesssim (2M_\odot/M_{\rm PBH})^{1.6}$ \cite{Poulin:2017bwe}. Taking into account that PBHs would be immersed in DM halos with higher densities than the background, their accretion rates would be increased, also leading to more stringent constraints \cite{2020PhRvR...2b3204S}. CMB limits from accretion are currently the most stringent ones for masses $\gtrsim 10 \, M_\odot$. The caveat is their dependence upon some details of the accretion mechanisms, such as the effective velocity and the accretion rate, which may be not well understood yet. 

On the other hand, the energy injection coming from evaporation would produce anisotropies and spectral distortions in the CMB spectrum, which also would bound the maximum abundance, leading to similar constraints to those obtained from the extra-galactic $\gamma$-ray background commented above \cite{Poulin:2016anj, Clark:2016nst, Acharya:2019xla}. Besides energy injection from accretion of BH evaporation, spectral distortions can also be produced by other means, such as the diffusion of photons due to Silk damping at small scales. This fact allows translating constraints on spectral distortions from FIRAS to stringent upper bounds on the PBH abundace, for masses $M_{\rm PBH}>10^5M_\odot$ \cite{Nakama:2017xvq}.

\item \textbf{Ly$\alpha$ forest}

As discussed in Sec. \ref{sec:shot}, the discrete nature of PBHs would lead to a shot-noise contribution to the power spectrum, enhancing small scale fluctuations. Observations of the Ly$\alpha$ forest, which traces matter distribution at the smallest galactic scales, have been employed to extract limits of the maximum allowed fraction of PBHs \cite{2003ApJ...594L..71A}. As already commented, the shot-noise power spectrum depends on the joint product of $f_{\rm PBH}M_{\rm PBH}$, for which the upper bound $f_{\rm PBH}M_{\rm PBH} \leq 60 \, M_\odot$ has been obtained \cite{2019PhRvL.123g1102M}. The drawback of this method is on the priors of the Reionization modeling and, as any Ly$\alpha$ forest analysis, depends upon the details of the post-Reioniaztion IGM evolution.




\end{itemize}

Besides the aforementioned current constraints, other future probes have been suggested to constrain the abundance of PBHs. Among them, the 21 cm cosmological signal stands out as one of the most promising approaches.
As already mentioned, the radiation from the hyperfine structure of the hydrogen is highly sensitive to the thermal state of the IGM, and thus, energy injection from PBH accretion or evaporation may leave strong observable signatures. The first claimed measurement of a global absorption dip by the EDGES collaboration \cite{Bowman:2018yin} may lead to competitive bounds on the PBH abundance, either from accretion processes \cite{Hektor:2018qqw} or PBH evaporation \cite{Clark:2018ghm, Halder:2021rbq}. It must be noted, however, that the EDGES signal has not been confirmed yet by other experiments, and it has been argued that it could be explained by alternative mechanisms, as will be discussed in Sec. \ref{sec:edges}. In Part \ref{partII} of this thesis, it is shown that forecasts of the 21 cm power spectrum data with future experiments such as HERA and SKA could potentially improve the bounds up to $f_{\rm PBH}< 10^{-2}-10^{-6}$ for masses above $M_\odot$ \cite{Mena:2019nhm}. The 21 cm forest observed as absorption troughs in the spectra of radio sources at $z\sim 10-15$ could also provide similar limits on the abundance, due to the Poisson shot noise and to the accretion heating effect, as shown in Ref. \cite{Villanueva-Domingo:2021cgh}, not included in this thesis.

The extremely rich physics involved in the formation, evolution and distribution of PBHs implies a large number of observable effects which allow probing them. Although a myriad of constraints are present for a large range of masses, in recent years some of these limits, such as the ones coming from microlensing, femtolensing, CMB accretion or BH mergers, have been revisited and, after more detailed computations, significantly weakened or even removed. This allows opening windows in the parameter space where PBHs could still form a substantial part of the DM, if not all. On the other hand, future experiments with larger sensitivities may be able to reach yet unexplored regions in the parameter space and tighten up the current bounds. New probes, such as the 21 cm line pursued in radio interferometers like SKA, will present a promising powerful way to proof or refute the existence of solar mass BHs formed in the early universe.

\begin{comment}

\bibliographystyle{../../jhep}
\bibliography{../../biblio}

%% file: Chapters/Chapter_21cm/Chapter_21cm.tex
\end{comment}

\chapter{21 cm Cosmology}


\label{chap:21cm}

In this chapter we review the fundamentals on the 21 cm, or HI, redshifted cosmological signal. This signal arises from the spin-flip transitions in the lowest state of the Hydrogen atom. Despite its faintness due to the low emission rate, the ubiquity of Hydrogen atoms through the IGM makes it still observable with high sensitivity instrumentation. As we shall see, measuring and understanding either its global average or its fluctuations can provide highly precious information regarding the astrophysical processes which take part in the IGM, and moreover, about the DM nature, which is the focus of this thesis. In the following sections, we solve the radiative transfer equation in order to predict the 21 cm intensity, followed by an overview of the most relevant processes which determine the excitation levels of the Hydrogen. Finally, we summarize the current and forthcoming experiments devoted to detect this signal. For more details see also Refs.~\cite{Pritchard:2011xb, Furlanetto:2006jb, 2019cosm.book.....M} for comprehensive reviews about this topic, and, e.g., Ref. \cite{1986rpa..book.....R} for fundamentals on radiative transfer.

\section{Radiative transfer in the expanding universe}
\label{sec:radtransfer}

In order to predict the intensity of any radiation field, we must take into account its evolution through the IGM, accounting for emissions and absorptions in the medium. This is done by solving its Boltzmann kinetic equation, also known in this context as radiative transfer equation. The fundamental quantity is the \textit{specific intensity} (by energy) $I(t,\nu)$ at a time $t$ and frequency $\nu$, with dimensions of erg s$^{-1}$ cm$^{-2}$ Hz$^{-1}$ sr$^{-1}$.\footnote{The intensity can be related to the distribution function, or the photon occupation number, $f(t,\nu)$ as $I=(2h\nu^3/c^2)f$, where the factor of 2 accounts for the polarizations of the photon. With this definition, one can write the energy density of the radiation field as $u(t) =c^{-1}\int d\nu d\Omega I(t,\nu)=2\int (d^3p )/(2\pi \hbar)^3 f(t,\nu)$, with $\Omega$ the solid angle and $p=h\nu/c$ the photon momentum.} In the homogeneous expanding universe, the evolution of $I(t,\nu)$ is described by the radiative transfer equation:
\begin{equation}
\frac{\partial I(t,\nu)}{\partial t} +3H(t) I(t,\nu) - H(t) \nu \frac{\partial I(t,\nu)}{\partial \nu}  = - c\kappa(t,\nu) I(t,\nu) + \frac{c\epsilon(t,\nu)}{4\pi},
\label{eq:radcosmo}
\end{equation}
where $H(t)$ is the Hubble rate, $\kappa(t,\nu)$ is the opacity or absorption rate, and $\epsilon(t,\nu)$ is the emissivity. The left-hand side accounts for the adiabatic evolution of the spectrum due to the expansion of the universe, while the first and second right-hand side terms account for absorptions of photons in the medium and emission of new photons, respectively. Therefore, the variation of the intrinsic intensity along its path is a competition between the loss of energy due to the interaction with the medium and the energy gained from the existing sources. Making use of the method of characteristics, we can obtain a formal solution to Eq. \eqref{eq:radcosmo}:
\begin{equation}
\begin{split}
I(t,\nu) &= \left( \frac{a(t_*)}{a(t)} \right)^3 I\left(t_*,\frac{\nu a(t)}{a(t_*)} \right) e^{-\tau(t_*,t,\nu)}\\
& + \frac{c}{4\pi}\int_{t_*}^{t}dt' \left( \frac{a(t')}{a(t)} \right)^3 \epsilon\left(t',\frac{\nu a(t)}{a(t')}\right)e^{-\tau(t',t,\nu)},
\end{split}
\label{eq:radcosmosol}
\end{equation}
where $t_*$ is a reference early time, and the \textit{optical depth}, $\tau$, is defined by
\begin{equation}
\tau(t',t,\nu)  =  c\int_{t'}^{t}dt''\kappa\left(t'', \frac{\nu a(t)}{a(t'')}\right).
\label{eq:opticaldepth}
\end{equation}
The above dimensionless quantity denotes a measure of the total absorption along a path from time $t'$ to $t$. The total intensity can therefore be interpreted as the joint combination of the initial spectrum $I(t_*,\nu a(t)/a(t_*))$ at time $t_*$ (corresponding to a redshift $z_*$) damped by absorptions in the medium with the factor $e^{-\tau(t_*,t,\nu)}$, plus the integral over time of the emitted photons, whose absorptions across the medium are also taken into account through the factor $e^{-\tau(t',t,\nu)}$. It is easy to show that Eq. \eqref{eq:radcosmosol} is a solution of the radiative transfer equation, Eq. \eqref{eq:radcosmo}. Defining the operator $\mathcal{L}[...]=\frac{\partial }{\partial t} - H \nu \frac{\partial }{\partial \nu}$, notice that $\mathcal{L} [F(\xi)]=0$, where $\xi=\nu a(t)$ and $F(\xi)$ is any function of $\xi$. Therefore, note that $\mathcal{L}[\tau(t',t,\nu)]=c\kappa(t,\nu)$. Writing the above solution in terms of the comoving quantities $I_c=a^3I$ and $\epsilon_c=a^3\epsilon$ and acting $\mathcal{L}$ over Eq. \eqref{eq:radcosmosol}, we get
\begin{align}
\mathcal{L}[I_c(t,\nu)] &= -I_c\left(t_*,\frac{\nu a(t)}{a(t_*)} \right) e^{-\tau(t_*,t,\nu)}\mathcal{L}\left[\tau(t_*,t,\nu)\right] \\
& - \frac{c}{4\pi}\int_{t_*}^{t}dt' \epsilon_c\left(t',\frac{\nu a(t)}{a(t')}\right)e^{-\tau(t',t,\nu)}\mathcal{L}\left[\tau(t',t,\nu)\right] + \frac{c}{4\pi} \epsilon_c(t,\nu) \\
&= - c\kappa(t,\nu) I_c(t,\nu) + \frac{c}{4\pi}\epsilon_c(t,\nu),
\end{align}
which is nothing but Eq. \eqref{eq:radcosmo} in a comoving form.
To solve the integral of Eq. \eqref{eq:radcosmosol}, the specific forms of the opacity and absorption are required to be known, and those can be very different depending on the processes considered. This solution not only applies to the 21 cm line, but also to the radiation fields relevant in the IGM evolution, those whose spectra span the ranges of X-rays and UV. As we shall see in Chapter \ref{chap:IGM}, in these cases, photons are generated in astrophysical sources as stars, quasars and galaxies. However, in the case of the 21 cm line, emissions are produced by Hydrogen atoms throughout the neutral IGM. Henceforth, we focus on atomic transitions between a higher, excited state and a lower, ground state one, with number densities $n_1$ and $n_0$ respectively, separated by an energy $h\nu_0$. In the context of the hyperfine transition, its splitting in energy is $h\nu_0=5.9 \times 10^{-6}$ eV, which corresponds to a frequency of $\nu_0=1420$ MHz or a wavelength of 21 cm.

In order to have a measure for the relative occupation of the levels, it is customary to define the \textit{spin temperature} $T_S$ through the ratio between the number densities of the excited and de-excited states: 
\begin{equation}
\frac{n_1}{n_0}=\frac{g_1}{g_0} e^{-T_*/T_S}=3e^{-T_*/T_S},
\label{eq:Tspindef}
\end{equation}
where $T_*=h\nu_0/k_B=0.068$ K, while $g_1=3$ and $g_0=1$ denote the number of internal degrees of freedom of each state. These numbers arise from the triplet and singlet structures of the excited and de-excited levels, respectively. In all cosmological applications, $T_S \gg T_*$, implying that three out of four Hydrogen atoms are in the excited state approximately, and only one of four in the ground one. It is important to note that $T_S$ does not refer to an actual thermodynamic temperature, but to a measure of the relative ocupation of the hyperfine levels. Equation \eqref{eq:Tspindef} allows to write the population levels in terms of the spin temperature and the neutral Hydrogen number density, $n_{\rm HI}$:
\begin{equation}
n_1 = n_{\rm HI} \; \frac{3e^{-T_*/T_S}}{1+3e^{-T_*/T_S}}\, ; \; \; \; n_0 = n_{\rm HI}\; \frac{1}{1+3e^{-T_*/T_S}}\, .
\label{eq:n1n2}
\end{equation}
%
The cross section of an atomic line transition is usually sharply peaked around its central frequency (although it may present relevant tails, as we shall see with the Ly$\alpha$ line). Thus, it is customary to write the cross section of the transition $0 \rightarrow 1$ (being $0$ and $1$ the ground and excited states respectively) as $\sigma(\nu) = \sigma_{21}\phi(\nu)$, with $\phi(\nu)$ the normalized line profile, $\int d\nu \phi(\nu) = 1$, and $\sigma_{21} = 3c^2 A_{10}/(8\pi\nu_0^2)$, where $A_{10}=2.85 \times 10^{-15} s^{-1}$ is the Einstein coefficient for spontaneous emission.\footnote{Note that $\sigma_{21}$ has units of area over time, rather than cross section units.} From detailed balance, it can be related to the Einstein coefficient for stimulated emission, $B_{10}$, and absorption, $B_{01}$, as $A_{10}=2h\nu_0^3/c^2B_{10}$ and  $g_1B_{10}=B_{01}g_0$, while the normalization factor $\sigma_{21}$ can also be written as $\sigma_{21} = h \nu_0/(4 \pi)B_{01}$.\footnote{It is worth noting that other definition of the Einstein coefficients present in the literature employs the energy density, instead of the specific intensity used here, which implies an extra factor of $4\pi/c$ (see, e.g., Ref. \cite{2011piim.book.....D}).} Therefore, the opacity reads 
\begin{align}
\kappa(\nu)&=\frac{h \nu_0}{4 \pi} \left( B_{01} \, n_0 \, \phi(\nu) - B_{10} \, n_1 \, \psi(\nu) \right)\\
&=\frac{h \nu_0}{4 \pi}  B_{01} \, n_0 \, \phi(\nu)\left(1 -  \frac{g_0 n_1}{g_1 n_0} e^{h (\nu_0 - \nu)/T_S} \right),
\label{eq:opacity}
\end{align}
with $\psi(\nu)=\phi(\nu)e^{h (\nu_0 - \nu)/T_S}$ the stimulated emission profile. The first term stands for standard absorptions by Hydrogen atoms in the ground level, while we also take into account stimulated emission by the higher level in the second term, which acts as an effective negative absorption. On the other hand, the emissivity can be written as
\begin{equation}
\epsilon(\nu)= h \nu_0\left(\frac{\nu}{\nu_0}\right)^3 \psi(\nu) A_{10} n_1,
\label{eq:emissivity}
\end{equation}
Note that in Eqs. \eqref{eq:opacity} and \eqref{eq:emissivity}, the factors $(\nu/\nu_0)^3$ and $e^{h (\nu_0 - \nu)/T}$ must be present to ensure detailed balance, i.e., in equilibrium, $\kappa(t,\nu)I_{eq}(t,\nu) = \epsilon(\nu)/(4\pi)$ implies that the equilibrium intensity $I_{eq}(t,\nu)$ becomes a black body spectrum. However, in practice, the radiation field is evaluated close to the central frequency $\nu_0$, being thus $(\nu/\nu_0)^3\simeq 1$ and $\psi(\nu)\simeq \phi(\nu)$. For this reason, these factors are usually dismissed. For the following derivations, it is convenient to define the source function $S(\nu)$, as
\begin{equation}
S(\nu)=\frac{\epsilon(\nu)}{4\pi\kappa(\nu)}=\frac{2h \nu^3}{c^2}\frac{1}{e^{h\nu/T_S}-1} \simeq \frac{2\nu^2}{c^2}T_S ,
\end{equation}
where in the last equality, we have applied the Rayleigh-Jeans approximation, valid at low frequencies, where $h\nu/T_S \ll 1$. Note that the source function represents a black body spectrum with a temperature $T_S$. When equilibrium holds, the intensity becomes $I_{eq}(\nu) = S(\nu)$.

We shall explicitly evaluate now Eq. \eqref{eq:radcosmosol}. Note that the frequency arguments in Eq. \eqref{eq:radcosmosol} are redshifted as $\nu' = \nu (1+z')/(1+z)$. It is appropriate then to change coordinates to integrate over frequencies, with $|dt'|=dz'/((1+z')H(z'))=d\nu'/(\nu'H(z'))$. For the hyperfine transition, the shape of the line profile $\phi(\nu)$, which includes Doppler broadening due to the thermal motions of atoms, is expected to be quite sharp, so it is safe to approximate it as a Dirac delta function around the frequency $\nu_0$, $\phi(\nu) \simeq \delta(\nu-\nu_0)$. Exploiting this fact, we can  evaluate the emission integral of Eq. \eqref{eq:radcosmosol} as
\begin{align}
& \frac{c}{4\pi}\int_{t_*}^{t}dt' \epsilon_c(t',\nu a(t)/a(t'))e^{-\tau(t',t,\nu)} = c\int_{\nu}^{\nu_*}\frac{d\nu'}{\nu'H(z')} S_c(t',\nu') \kappa(t',\nu') e^{-\tau(t',t,\nu)} \\
&\simeq S_c(t_0,\nu_0) \int_0^{\tau(t_*,t,\nu)} d\tau e^{-\tau} = S_c(t_0,\nu_0) (1-e^{-\tau(t_*,t,\nu)}),
\label{eq11}
\end{align}
where $\nu_* = \nu (1+z_*)/(1+z)$, $t_0$ is the time corresponding to $1+z_0=(\nu_0/\nu)(1+z)$, and in the second equality we have approximated the line profile in the opacity as a delta function. On the other hand, given that $T_S \gg T_*$, the opacity from Eq. \eqref{eq:opacity} can be approximated as
\begin{equation}
\kappa(\nu) \simeq \frac{h \nu_0}{4 \pi}  B_{01} \, n_0 \, \phi(\nu) \frac{h\nu}{T_S} \simeq \frac{3 c^2 A_{10}}{32 \pi \nu_0^2} \, n_{\rm HI} \, \phi(\nu) \frac{h\nu}{T_S} ~,
\end{equation}
where in the second equality we have written $n_0 \simeq n_{\rm HI}/4$ from Eq. \eqref{eq:n1n2}, valid up to leading order in $T_*/T_S$. Therefore, integrating Eq. \eqref{eq:opticaldepth}, we obtain
\begin{equation}
\tau(z,\nu)  =  \frac{3c^3A_{10}}{32 \pi \nu_0^3  } \frac{x_{\rm HI}(z_0) n_{\rm H}(z_0)}{H(z_0)} \frac{T_*}{T_S(z_0)}\; ,
\label{eq:opdep21_0}
\end{equation}
where we have defined the \textit{fraction of neutral Hydrogen} $x_{\rm HI}$ with respect to the total Hydrogen number density $n_{\rm H}$ as $x_{\rm HI}=n_{\rm HI}/n_{\rm H}$. For the shake of simplicity, this derivation is being limited to the homogeneous case, but in reality, inhomogeneities and peculiar velocities of the gas must be taken into account. It leads to a slight modification of the argument of the line profile, induced by a Doppler kinematic effect. The approximation of $\phi(\nu)$ by a delta function $\delta(\nu-\nu_0)$ considered above must then be replaced by $\delta(\nu-\nu_0(1-\vec{\Omega}\cdot \vec{v}))$, with $\vec{\Omega}$ is the unity normalized vector along the line of sight and $\vec{v}$ the peculiar velocity of the gas. Integration of the optical depth modifies the result with an extra term depending on the derivative of the radial component of the velocity along the line of sight, $\partial_r v_r$ \cite{Furlanetto:2006jb},
\begin{align}
\tau(z,\nu)  &=  \frac{3c^3A_{10}}{32 \pi \nu_0^3  } \frac{x_{\rm HI}(z_0) n_{\rm H}(z_0)}{H(z_0)} \frac{T_*}{T_S(z_0)}\frac{1}{\left( 1+\frac{\partial_r v_r(z_0)}{H(z_0)} \right)} \\
& \simeq 9.2 \times 10^{-3} \; (1+\delta(z_0)) x_{\rm HI}(z_0) (1+z_0)^{3/2} \frac{\left( 1 \, {\rm K}/T_S\right)}{\left( 1+\frac{\partial_r v_r(z_0)}{H(z_0)} \right)} \; ,
\label{eq:opdep21}
\end{align}
where we have written the Hydrogen number density as the background value $\propto (1+z)^3$ times the factor accounting for the overdensity $\delta$, $n_{\rm H}= \bar{n}_{\rm H}(1+\delta)$. Since $T_* \ll T_S$ for the relevant regimes, the above optical depth is always much lower than unity, $\tau \ll 1$. In the following, we assume that the Hydrogen (and baryon) overdensity equals the DM one, which is an accurate approximation at the cosmological scales and redshifts of interest.

On the other hand, for the initial spectrum at $t_*$, the only relevant background radiation at high redshift at these frequencies is the CMB, since radio emission from astrophysical sources is expected to be several orders of magnitude lower. Therefore, we may take $I(t_*,\nu) = I_{CMB}(t_*,\nu) \simeq (2 \nu^2/c^2 ) T_\gamma$, with $T_\gamma = 2.73 \; {\rm K} \; (1+z)$ the CMB temperature. In the spirit of the Rayleigh-Jeans approximation, it is customary to define the \textit{brightness temperature} $T_b(t,\nu)$ as $I(t,\nu)=(2\nu^2/c^2)T_b(t,\nu)$. Thus, in terms of brightness temperature, Eq. \eqref{eq:radcosmosol} reads
\begin{equation}
T_b(z,\nu) = T_\gamma(z)e^{-\tau(z,\nu)} + T_S(z)(1-e^{-\tau(z,\nu)}).
\end{equation}
Since we are interested in the deviations from the CMB spectrum around the 21 cm line, the relevant quantity is the comoving \textit{differential brightness temperature} $\delta T_b$, as the comoving temperature relative to the CMB one evaluated at the 21 cm frequency $\nu_0$:
\begin{align}
\delta T_b(z) &= \frac{T_b(z,\nu_0)-T_\gamma(z)}{1+z} \simeq \frac{(T_\gamma(z)-T_S(z))\tau(z,\nu_0)}{1+z} \\
\begin{split}
& \simeq 27 \, \textrm{mK} \, x_\textrm{HI}(z) \, (1+\delta(z))\,  \left( 1 - \frac{T_\gamma(z)}{T_S(z)}\right) \left( \frac{1+z}{10}\right)^{1/2} \\
& \times \left(\frac{0.15}{\Omega_m h^2} \right)^{1/2} \left( \frac{\Omega_b h^2}{0.023}\right)\frac{1}{\left( 1+\frac{\partial_r v_r(z)}{H(z)} \right)}, 
\end{split}
\label{eq:dTb}
\end{align}
where in the second identity we have used $\tau \ll 1$ and Eq. \eqref{eq:opdep21}. This is the main equation for predicting the 21 cm cosmological signal, accounting for photons absorbed or emitted with frequency $\nu_0$ at redshift $z$, or equivalently, being observed now at a frequency $\nu_0/(1+z)$. Thus, the observation of a frequency spectrum from the cosmological 21 cm signal actually provides a record of the history of the IGM over a range of redshifts.

Several important facts can be noted by examining Eq. \eqref{eq:dTb}. Since the signal is proportional to the neutral fraction $x_\textrm{HI}$, it can be observed only if there is a significant fraction of neutral Hydrogen in the IGM. This makes it a sensitive probe for the ionization state of the medium, being a promising tool for understanding the evolution of the EoR, when the intergalactic gas becomes fully ionized. It may become especially useful to trace the end of this process, when the neutral fraction goes to zero and the overall signal vanishes, providing a picture of the evolution of $x_\textrm{HI}$. In addition, the signal gets enhanced in overdense regions as indicated by its dependence on $\delta$, becoming also a probe for the matter density fluctuations. It must be remarked that there will be a signal only if the spin temperature differs from the CMB temperature, being observed in \textit{absorption} if $T_S<T_{\gamma}$ ($\delta T_b$ negative), or in \textit{emission} if $T_S > T_{\gamma}$ ($\delta T_b$ positive). In the limit $T_S \gg T_{\gamma}$, the signal saturates, and becomes independent on $T_S$. Finally, it is convenient to note that the velocity gradient term, negligible in the homogeneous limit, accounts for redshift distortions along the line of sight \cite{Furlanetto:2006jb, Pritchard:2011xb}. In order to compute the brightness temperature for a given redshift, it is mandatory to know how $T_S$ evolves in redshift, which will be studied in the next section. 

Although not covered in this thesis, it is worth mentioning another situation of interest. Instead of employing the CMB as the radio background, one can also employ a loud radio point source, such as a quasar, as the backlight, changing $T_\gamma$ in the previous analysis by the brightness temperature of the radio radiation. In that case, $T_\gamma \gg T_S$, and the signal will be seen always in absorption. The presence of clouds of neutral Hydrogen along the line of sight would imprint a set of absorption lines in the redshifted quasar spectrum, given by the distance of the gas cloud. This is the so-called \textit{21 cm forest}, in analogy with the Ly$\alpha$ forest, which works under the same principle with the Ly$\alpha$ transition line, and stands as a promising tool to probe small scales. Since it makes use of a source with high emission in radio, it has the advantage that foregrounds are less important that when observing the signal from the IGM, which are one of its main experimental problems, as shall be discussed in Sec. \ref{sec:foregrounds}. Moreover, it allows tracing higher redshifts than the Ly$\alpha$ forest due to the low 21 cm optical depths. However, the main drawback of this approach is the high uncertainty on the existence of bright enough radio sources at high redshift. See, e.g., Refs. \cite{2002ApJ...579....1F, Furlanetto:2006dt, Pritchard:2011xb} for more details.

\section{Spin temperature}
\label{sec:tspin}

There are three main processes which can modify the population of the hyperfine levels and therefore determine the spin temperature, namely:
\begin{enumerate}
\item Absorption/emission of 21 cm photons from/to the CMB.
\item Collisions with electrons, protons and other Hydrogen atoms.
\item Resonant scattering of Ly$\alpha$ photons that cause a spin flip via an intermediate excited state (Wouthuysen-Field effect).
\end{enumerate}
Each of these processes may produce spin-flip transitions, exciting or de-exciting the Hydrogen atom. The kinetic equations for the population of each level must account for all these effects, and can be written as
\begin{equation}
\begin{split}
\dot{ n}_1 +3Hn_{1}&=-(\dot{ n}_0 +3Hn_{0})\\
&=(C_{01}+P_{01}+B_{01}\bar{I})n_0 - (C_{10} + P_{10} + A_{10} + B_{10}\bar{I})n_1~,
\end{split}
\end{equation}
where the term $3H$ accounts for the adiabatic expansion, $\bar{I}=(1/4\pi)\int d\Omega \int d\nu \phi(\nu) I(\nu)\simeq I(\nu_0)$ is the angle-averaged radiation field evaluated at the 21 cm line, and $C_{10}$ and $P_{10}$ are the de-excitation rates (per atom) from collisions and Ly$\alpha$ scattering, respectively, while $C_{01}$ and $P_{01}$ are their corresponding excitation rates. Given that the rates of these processes are much faster than the Hubble timescale, the spin temperature is given to a very good approximation by the equilibrium balance of these effects,
\begin{equation}
n_1(C_{10} + P_{10} + A_{10} + B_{10}\bar{I})=n_0(C_{01}+P_{01}+B_{01}\bar{I})~,
\label{eq:tspineq}
\end{equation}
Due to detailed balance in collisional coupling, we can relate the de-excitation and excitation coefficients to the \textit{kinetic temperature} $T_{K}$ of the gas. For that, consider only collisional coupling, where the evolution equation takes the form: $\dot{ n}_1 +3Hn_{1}=C_{01}n_0-C_{10}n_1$. In the equilibrium case at temperature $T_K$ (indicated by the superindex $eq$), the left hand side of this equation vanishes, leading to the relation between the excitation and de-excitation rates:
\begin{equation}
\frac{C_{01}}{C_{10}}=\left( \frac{n_1}{n_0} \right)^{eq}=\frac{g_1}{g_0}e^{-T_*/T_K} \simeq 3\left(1-\frac{T_*}{T_K}\right)~,
\end{equation}
where in the last equality we have used the fact that $T_* \ll T_K$ for all temperatures of interest. Analogously, we can \textit{define} the effective \textit{color temperature} of the UV radiation field, $T_c$, as
\begin{equation}
\frac{P_{01}}{P_{10}} \equiv \frac{g_1}{g_0}e^{-T_*/T_c} \simeq 3\left( 1-\frac{T_*}{T_c} \right),
\label{eq:Plyalpha}
\end{equation}
which can be related to the temperature of the Ly$\alpha$ radiation field, as indicated in Sec. \ref{sec:color}. From Eqs. \eqref{eq:Tspindef} and \eqref{eq:tspineq}, employing the relations above, together with the fact that $T_* \ll T_S$, and using the Rayleigh-Jeans approximation for the CMB radiation field, $\bar{I}\simeq 2k_B T_{\gamma} \nu^2_0/c^2$, it is straightforward to show that the spin temperature $T_S$ can be written as~\cite{Pritchard:2011xb}
\begin{equation}
T_S^{-1}=\frac{T^{-1}_{\gamma}+x_cT^{-1}_K+x_{\alpha}T^{-1}_c}{1+x_c+x_{\alpha}}~,
\label{eq:spintemp}
\end{equation}
where we have defined the collisional and Ly$\alpha$ \textit{coupling coefficients} $x_c$ and $x_{\alpha}$ as\footnote{The spin temperature can also be written as $T_S=(T_{\gamma}+y_cT_K+y_{\alpha}T_c)/(1+y_c+y_{\alpha})$, with the coupling coefficients given by $y_i = x_i T_\gamma/T_K$. This convention is also widely employed in the literature and will be used in some articles of Part \ref{partII}.}
\begin{equation}
x_c=\frac{C_{10} T_*}{A_{10} T_{\gamma}} ~\textrm{;} \quad x_{\alpha}=\frac{P_{10} T_*}{A_{10} T_{\gamma}}~.
\label{eq:coup_coef}
\end{equation}
From Eq. \eqref{eq:spintemp}, it can be seen that if collisions or Ly$\alpha$ are efficient enough, the spin temperature \textit{couples} to the kinetic or color temperature, respectively, being close to the CMB one otherwise.
In most cases of interest, since the Ly$\alpha$ optical depth is large, Ly$\alpha$ scattering is efficient enough to maintain the equilibrium with the gas, and the color temperature is close to the kinetic one, $T_c \simeq T_K$. We compute the color temperature properly from the radiation field in Sec. \ref{sec:color} and show the limit $T_c \simeq T_K$ explicitly. Then, generally, $T_S$ will take values between the kinetic and the CMB temperatures.

In the next subsections we discuss in some detail the physics involved in the above coupling mechanisms. Before that, it is convenient to emphasize that, although we have several quantities with units of temperature, some of them are not truly thermodynamic temperatures. For instance, $T_b$ is a measure of radio intensity, $T_S$ measures the relative occupation numbers of the two hyperfine levels, and $T_c$ describes the photon distribution in the vicinity of the Ly$\alpha$ transition, behaving as an effective brightness temperature for the Ly$\alpha$ radiation. Only the CMB blackbody temperature $T_{\gamma}$ and the gas kinetic one $T_K$ are genuine thermodynamic temperatures.

\subsection{Collisional coupling}
\label{sec:collcoup}

\begin{figure}
\centering
\includegraphics[scale=0.75]{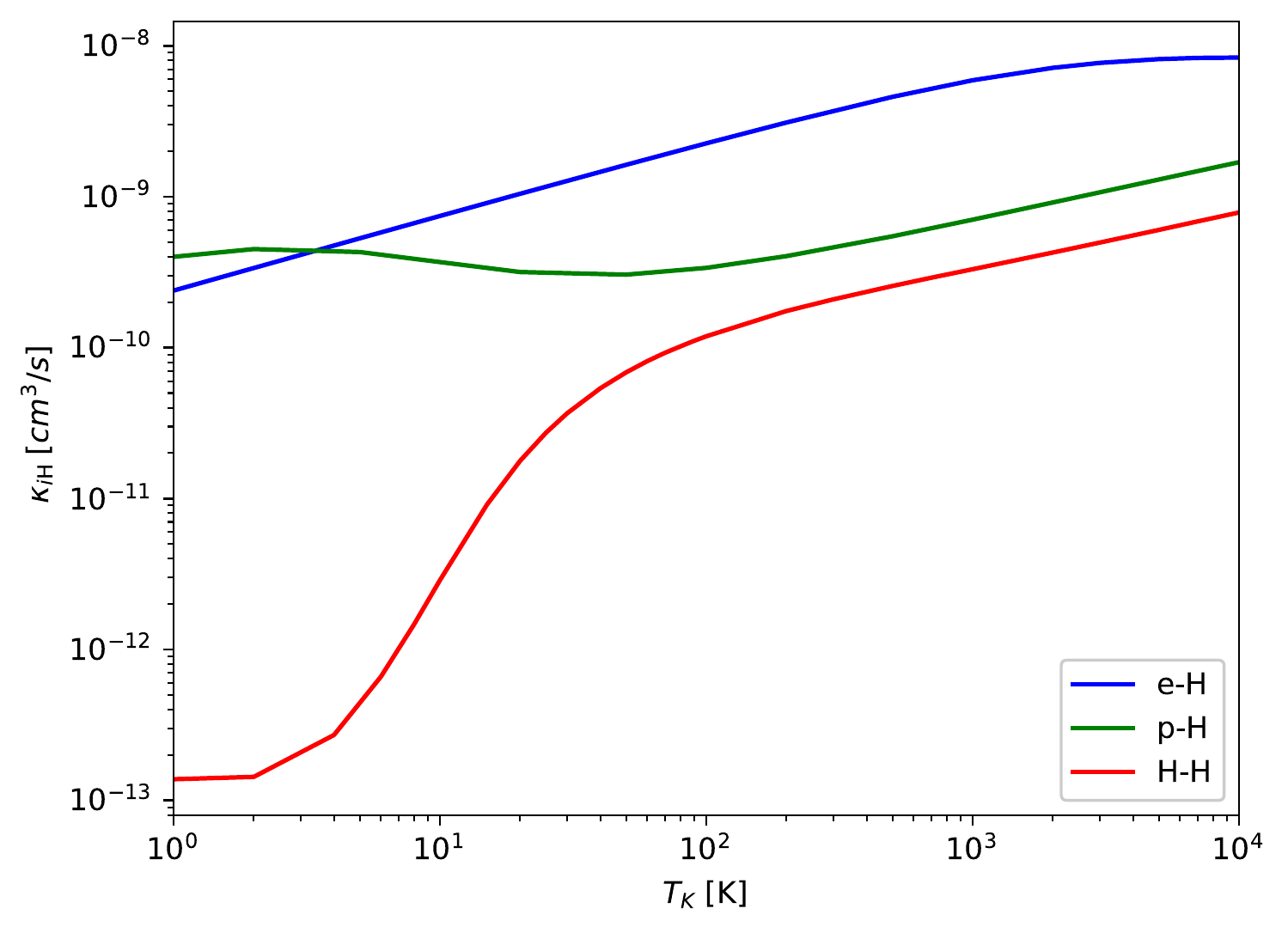}
\captionof{figure}{Scattering rates between the different species as a function of temperature: e-H (\cite{2007MNRAS.374..547F}, blue line), p-H (\cite{2007MNRAS.379..130F}, green line), and H-H (\cite{2005ApJ...622.1356Z}, red line) collisions.}
\label{fig:kappacoll}
\end{figure}

Collisions of neutral Hydrogen atoms with other IGM species, such as free protons, electrons and other H atoms, may induce spin-flips transitions in the hyperfine levels. These processes dominate during the Dark Ages, when the background gas density is still high, but become negligible at the epoch of star formation. The collisional de-excitation coefficient $C_{10}$ must include the three species mentioned above, and can be written as
\begin{equation}
C_{10}=\sum_i n_i \kappa_{i {\rm H}}=n_{\rm HI} \kappa_{\rm HH}+n_e \kappa_{e{\rm H}} + n_p \kappa_{p{\rm H}}~,
\label{eq:c10}
\end{equation}
where $n_i$ is the number density of each species $i$ and $\kappa_{i{\rm H}}$ the scattering rate between the neutral Hydrogen and each species $i$. These rates can be obtained integrating the cross section of each process, by a quantum mechanical calculation, as is done in \cite{Furlanetto:2006jb,1966P&SS...14..929S}. We make use of the tabulated values as a function of $T_K$ provided by Ref. \cite{2007MNRAS.374..547F} for the electron-Hydrogen rate $\kappa_{e{\rm H}}$, and by Refs. \cite{2005ApJ...622.1356Z, 1969ApJ...158..423A} for Hydrogen-Hydrogen collisions, $\kappa_{\rm HH}$ (see also the fits proposed in Refs. \cite{2001A&A...371..698L, 2006ApJ...637L...1K}). The proton-Hydrogen rate $\kappa_{pH}$ can be approximated for $T_K \gtrsim 100$ K as $\kappa_{pH} \simeq 2.0 \, \kappa_{HH}$~\cite{2007MNRAS.379..130F}, although for quantitative computations the tabulated data from Ref. \cite{2007MNRAS.379..130F} is used. However, since in Eq. \eqref{eq:c10} is weighted by the ionized fraction, it becomes subdominant at low ionization states, becoming relevant only at low temperatures. Figure \ref{fig:kappacoll} shows the scattering rates as a function of temperature from the tabulated data. We can safely neglect collisions with other species, such as Helium, since it is usually in a spin singlet state. This fact forbids spin exchange, unless it can be excited to the triplet state, something very unlikely in the thermal state of the IGM \cite{Furlanetto:2006jb,2007MNRAS.375.1241H}. It must be remarked that, although the collisional coupling process becomes irrelevant after the Dark Ages for the global 21 cm signal, it can be locally important at overdense regions, where collisions are enhanced by the term $1+\delta$ present in the number densities.

\subsection{Wouthuysen-Field effect}
\label{sec:wf}

\begin{figure}
\centering
\includegraphics[scale=0.3]{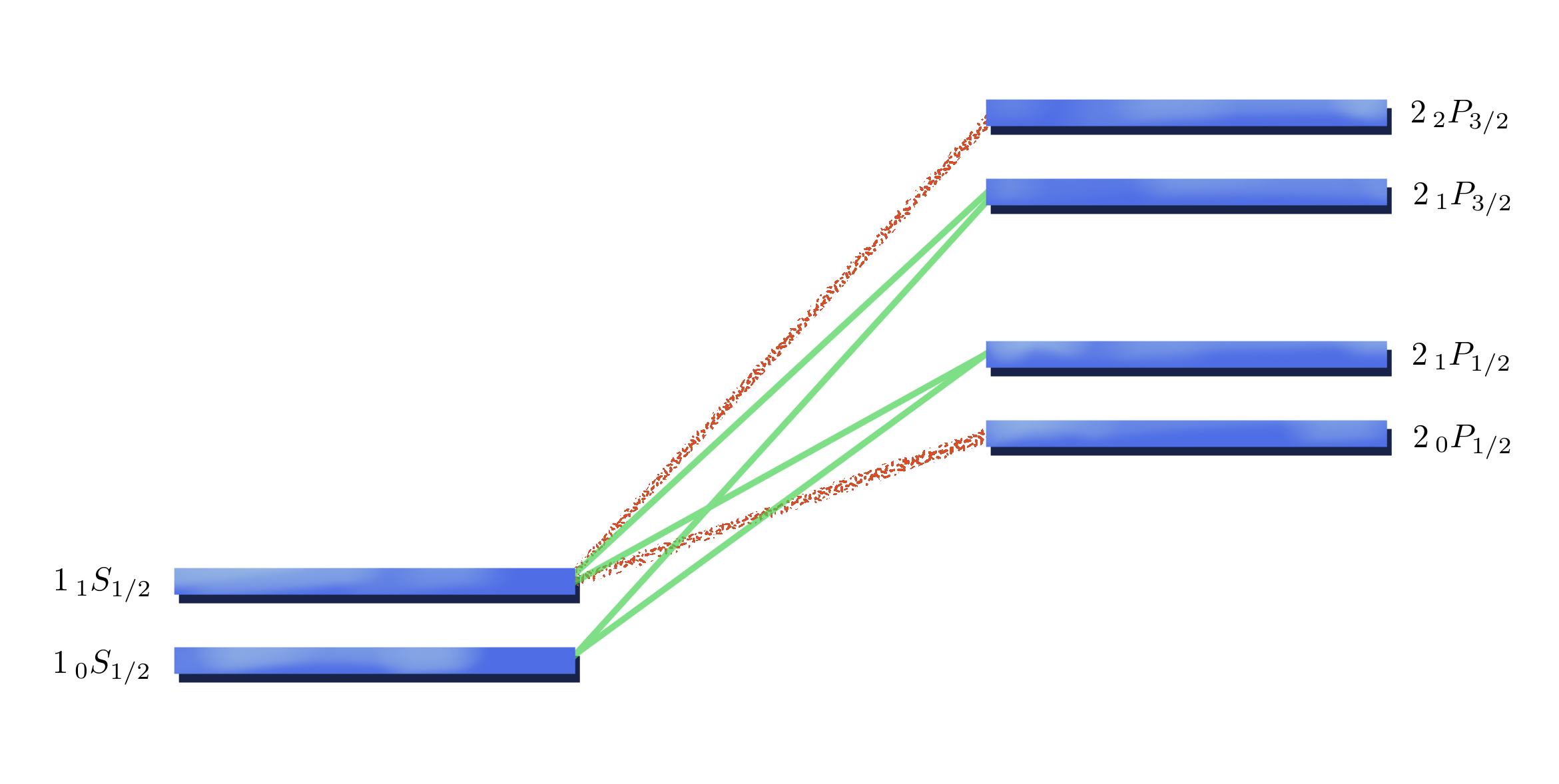}
\captionof{figure}{Hyperfine structure of the 1S and 2P levels of the Hydrogen atom and relevant transitions for the Wouthuysen-Field effect. Solid green lines represent transitions which cause spin flips, while dotted red lines are allowed transitions not contributing to spin flips.}
\label{fig:WFtransitions}
\end{figure}

Collisional coupling of the 21 cm line can only be efficient at large densities, and thus, in the bulk of the IGM becomes negligible at late times, after $z \sim 30$. However, once star formation begins, resonant scattering of Ly$\alpha$ photons provides a second channel to modify the spin temperature. This coupling mechanism was discovered independently by Siegfried A. Wouthuysen in 1952 \cite{1952AJ.....57R..31W} and by George B. Field in 1958 \cite{1958PIRE...46..240F}, and named after them as the Wouthuysen-Field (WF) effect. It enables transitions between hyperfine levels of the ground 1S state to the 2P level, which can cause an effective spin flip. With $F$ denoting the hyperfine spin state, the dipole selection rules impose restrictions over the possible spin changes $\Delta F$, allowing those transitions which fulfill $\Delta F= 0, \pm 1$, except $F = 0 \rightarrow 0$. These are showed in Fig.~\ref{fig:WFtransitions}, with the notation $n_FL_J$, being $n$, $L$ and $J$ the principal, orbital and total angular momentum quantum numbers. For example, suppose a Hydrogen atom initially in the singlet state $1_0 S_{1/2}$ which absorbs a Ly$\alpha$ photon, being excited into any of the central 2P hyperfine states. The de-excitation due to emission of another Ly$\alpha$ photon may relax the atom to its initial state, but also to the triplet ground state $1_1 S_{1/2}$, causing an effective spin flip. Note that not all the allowed transitions lead to a spin flip in the 1S state, i.e. the levels $2_0P_{1/2}$ and $2_2P_{3/2}$ do not contribute to the spin flips.

We can outline a simple example to illustrate mathematically the behavior of this mechanism. Consider an atom with the hyperfine states of the ground state, $n_1$, $n_0$, and also an effective excited state, $n_2$, which corresponds to the 2P levels. Neglecting stimulated emission\footnote{This is a fair approximation at Ly$\alpha$ frequencies for the relevant temperatures of the IGM post-Recombination, since $e^{h\nu_{Ly\alpha}/T_K} \gg 1$, being thus equivalent to replace the Bose-Einstein distribution function by the Maxwell-Boltzmann one.} and collisional coupling, the dynamical evolution of the occupation number is driven by the equation:
\begin{equation}
\dot{n}_2+3Hn_2=B_{02}I_{02}n_0+B_{12}I_{12}n_1-(A_{21}+A_{20})n_2~,
\end{equation}
where $B_{ij}$, $A_{ij}$ and $I_{ij}$ are the absorption and emission Einstein coefficients, and the radiation flux evaluated at the transition $i \rightarrow j$. Given the short lifetime of the 2P state, we can approximate the above equation by the equilibrium case, solving for $n_2$:
\begin{equation}
n_2 \simeq \frac{B_{02}I_{02}n_0+B_{12}I_{12}n_1}{A_{21}+A_{20}}~.
\end{equation}
Therefore, the change in $n_1$ due to transitions between the excited levels is given by:
\begin{equation}
\begin{split}
&(\dot{n}_1 +3Hn_1)\mid_{{\rm Ly} \alpha} =A_{21}n_2 -B_{12}I_{12}n_1 \\ &\simeq A_{21}\frac{B_{02}I_{02}}{A_{21}+A_{20}}n_0 - A_{20}\frac{ B_{12}I_{12}}{A_{21}+A_{20}}n_1 \equiv P_{01}n_0 - P_{10} n_1~,
\end{split}
\end{equation}
where in the last equality, we have related the flux and Einstein coefficients for the $0, 1 \leftrightarrow 2$ transitions with the excitation and de-excitation coefficients $P_{10}$ and $P_{01}$. In reality, the situation is slightly more complicated since there is not only one excited state but two hyperfine levels in the 2P state which contribute to spin flips, $2_1P_{3/2}$ and $2_1P_{1/2}$, but the reasoning is the same. Relating the Einstein coefficients for the different states, it is straightforward to obtain the final de-excitation rate as \cite{2000pras.conf...37M}
\begin{equation}
P_{10}=\frac{4 P_{\alpha}}{27}~,
\label{eq:wf4over27}
\end{equation}
where $P_{\alpha}$ is the Ly$\alpha$ scattering rate, defined as:
\begin{equation}
P_{\alpha}=4\pi\int d\nu \sigma_{\alpha}(\nu)J(\nu)~,
\label{eq:lyalscatrate}
\end{equation}
being $J$ the specific intensity by number, related to $I(\nu)$ as $I=h\nu J(\nu)$. The cross section of the Ly$\alpha$ transition, $\sigma_{\alpha}(\nu)$, can be written as $\sigma_{\alpha}(\nu)=\sigma_0 \varphi_{\alpha}(\nu)$, being $\sigma_0 = \frac{\pi e^2}{m_e c} f_{\alpha}$, with $f_{\alpha}=0.4162$ the oscillator strength of the Ly$\alpha$ transition, and $\varphi_{\alpha}(\nu)$ its line profile.\footnote{One can also define the Ly$\alpha$ cross section analogously to the 21 cm line with the spontaneous emission coefficient $A_{\alpha}=6.25 \times 10^8$ s$^{-1}$, as $\sigma_{0} = 3c^2 A_{\alpha}/(8\pi\nu_\alpha^2)$.} Since $\varphi_{\alpha}(\nu)$ is a peaked function around the Ly$\alpha$ transition, $P_{\alpha}$ can be estimated to be proportional to the radiation flux evaluated at the Ly$\alpha$ frequency, $J_\alpha$, as $P_\alpha \sim 4\pi \sigma_0 J_\alpha$. However, Doppler broadening and the tails of the distribution are actually important. Furthermore, resonant scattering around the Ly$\alpha$ line induces an absorption feature which must be properly calculated. Hence, the determination of the radiation spectrum becomes mandatory for a precise evaluation of $P_{\alpha}$, and is treated in the following section.

\section{Ly$\alpha$ line scattering}
\label{Lyalphatrans}

As stated in the previous section, the spin temperature dynamics, and therefore the 21 cm signal, strongly depends on the Ly$\alpha$ flux, its correct estimation being necessary. It is thus compulsory to solve the radiative transfer around that line, by employing a kinetic equation similar to Eq. \eqref{eq:radcosmo}. These transitions are, however, short lived, and absorptions of photons are quickly followed by emissions, being the full picture a scattering process which redistributes frequencies of the spectrum. Scattering around the Ly$\alpha$ line has been broadly studied in the literature for decades \cite{1994ApJ...427..603R,Chen:2003gc,Hirata:2005mz,Furlanetto:2006fs,Rybicki:2006th}. The kinetic equation for the comoving flux of photons $J$ (with units of s$^{-1}$ cm$^{-2}$ Hz$^{-1}$ sr$^{-1}$) can be written as
\begin{equation}
\frac{\partial J(\nu)}{\partial t} - H\nu  \frac{\partial J(\nu)}{\partial \nu} = c \, n_{\rm HI}\, \sigma_0 \, \left[ \int d\nu' \mathcal{R}(\nu',\nu)J(\nu') -\varphi(\nu) J(\nu) + \mathcal{C}\Delta\nu_D^{-1}\psi(\nu) \right],
\label{eq:scatLyaint}
\end{equation}
where the two first terms in the right-hand side correspond to the scattering part, while the last term accounts for line injection. Here, as in most of applications, isotropic scattering has been assumed, although angle-dependent cross section may be considered \cite{1994ApJ...427..603R}. Note also that stimulated emission has been neglected, as usual when dealing with the Ly$\alpha$ line, although it is easy to include its effect in the kinetic equation \cite{Rybicki:2006th}. The \textit{scattering kernel}, or redistribution function, $\mathcal{R}(\nu',\nu)$, drives the frequency variation of the scattered photons, and fulfills the normalization condition $\int d\nu' \mathcal{R}(\nu,\nu') = \varphi(\nu)$, which ensures number conservation in the scattering term. The specific shape of $\mathcal{R}(\nu,\nu')$ depends upon the line profile $\varphi(\nu)$ and the details of the redistribution mechanism \cite{1992ApJ...387..248H}. Regarding the last term, relative to injection of photons in the line, $\mathcal{C}$ is the rate at which they are produced, while $\psi(\nu)$ stands for the spectral profile, usually taken as a Dirac delta function (normalized to $\int d\nu \psi(\nu)=\Delta\nu_D$ for later convenience).

Rather than working with the frequency, it is useful to employ as variable the normalized frequency $x=(\nu-\nu_\alpha)/\Delta\nu_D$, defining the Doppler broadening frequency as $\Delta\nu_D= \nu_\alpha \sqrt{2k_B T_K/(m_H c^2)}$, which characterizes the width of the line due to thermal Doppler effect\footnote{If one computes the same quantity for the 21 cm line, given that it linearly scales with the frequency, one finds that it is a factor $\sim 6 \times 10^{-7}$ lower than the Ly$\alpha$ Doppler broadening. That is one of the reasons why one can safely approximate the 21 cm line profile as a Dirac delta function for most of applications in Sec. \ref{sec:radtransfer}, while the broadening of the Ly$\alpha$ line can be of great relevance, as shown in this section.}. Accordingly, we define the dimensionless line profile and scattering kernel as $\phi(x) = \Delta \nu_D \varphi(\nu)$ and $R(x',x) = (\Delta \nu_D)^2 \mathcal{R}(\nu',\nu)$ respectively, with the proper normalization conditions $\int dx \, \phi(x)=1$ and $\int dx'R(x',x)=\phi(x)$. Given that the frequency variation during these scattering processes is small, the redistribution function can be expanded following a Fokker-Planck procedure, which allows performing the integral of Eq. \eqref{eq:scatLyaint} as \cite{1994ApJ...427..603R,Chen:2003gc}
\begin{equation}
\int dx' R(x',x)J(x') \simeq \phi(x) J(x) + \frac{1}{2}\frac{\partial}{\partial x} \left[ \phi(x) \left( \frac{\partial J(x)}{\partial x} +2\eta(x) J(x)  \right) \right],
\label{eq:FPexpansion}
\end{equation}
where the second term in the expansion accounts for the thermal recoil, with $\eta(x) = h\nu_\alpha/\sqrt{2m_H c^2 k_B T_K} - (x+x_0)^{-1}$, being $x_0 = \nu_\alpha/\Delta\nu_D$. The last term $(x+x_0)^{-1}$ in $\eta$ ensures detailed balance, although for the forthcoming computations can be neglected. The Fokker-Planck expansion of Eq. \eqref{eq:FPexpansion} simplifies enormously the treatment of the scattering process, since, instead of an integro-differential equation, it allows writing a differential one, much easier to solve. As we are interested in the spectrum around the Ly$\alpha$ line, we can approximate the term $\nu\partial/\partial \nu J(x)\simeq (\nu_\alpha/\Delta\nu_D)\partial/\partial x J(x)$ in Eq. \eqref{eq:scatLyaint}. 
On the other hand, a relevant quantity in Ly$\alpha$ radiative transfer is its optical depth, usually known as the \textit{Gunn-Peterson optical depth}, which can be computed from Eq. \eqref{eq:opticaldepth} for the Ly$\alpha$ transition approximating the line profile as a Dirac delta function \cite{2013fgu..book.....L}:
\begin{equation}
\tau_{\rm GP} = \frac{c \, \sigma_0 \, n_{\rm HI}(z)}{H(z) \nu_\alpha} \simeq 5.0 \times 10^5 \, x_{\rm HI}(z)(1+\delta) \left(\frac{1+z}{7}\right)^{3/2}.
\label{eq:GPopticaldepth}
\end{equation}
This quantity is of great importance for the Ly$\alpha$ forest and the study of the ionization state of the IGM, as shall be discussed in Chapter \ref{chap:IGM}. Within this section, we neglect the overdensity factor $\delta$ since we focus on the homogeneous computation. Defining the \textit{Sobolev parameter} $\gamma_{S}$ as the inverse of the Gunn-Peterson optical depth, $\gamma_{S} = \tau_{\rm GP}^{-1} = H \nu_\alpha/(c \sigma_0 n_{\rm HI} )$, the final line scattering equation reads
\begin{equation}
\frac{\Delta\nu_D}{c n_{\rm HI} \sigma_0} \frac{\partial J(x)}{\partial t} - \gamma_{S}  \frac{\partial J(x)}{\partial x} =  \frac{1}{2}\frac{\partial}{\partial x} \left[ \phi(x) \left( \frac{\partial J(x)}{\partial x} +2\eta J(x)  \right) \right] + \mathcal{C}\psi(x).
\label{eq:fokkerplanck}
\end{equation}
Note that Eq. \eqref{eq:fokkerplanck} is consistent with detailed balance, i.e., the equilibrium spectrum $J(\nu)=2\nu^2/c^2 \exp[ -h\nu/(k_BT_K)]$ is a solution when the right hand side is equal to zero and there are no injections. Furthermore, when integrating over frequency, the scattering part (i.e., setting aside the injection term) fulfills number conservation by construction (the frequency integral of Eq. \eqref{eq:fokkerplanck} vanishes), given that the right-hand side term is a total derivative. Spin exchange due to hyperfine interactions can affect the above result, but it is easily included by making the substitution $\gamma_S \rightarrow \gamma_S(1+T_{se}/T_K)^{-1}$ and $\eta \rightarrow \eta(1+T_{se}/T_S)/(1+T_{se}/T_K)$, with $T_{se}=(2/9)T_K\nu_{21}/\Delta\nu_D^2 = 0.40$ K the spin-exchange characteristic temperature \cite{Chuzhoy:2005wv, Furlanetto:2006fs, Hirata:2005mz}.

\subsection{Modification of the spectrum}
\label{sec:lyspectrum}

\begin{figure}
\centering
\includegraphics[scale=0.75]{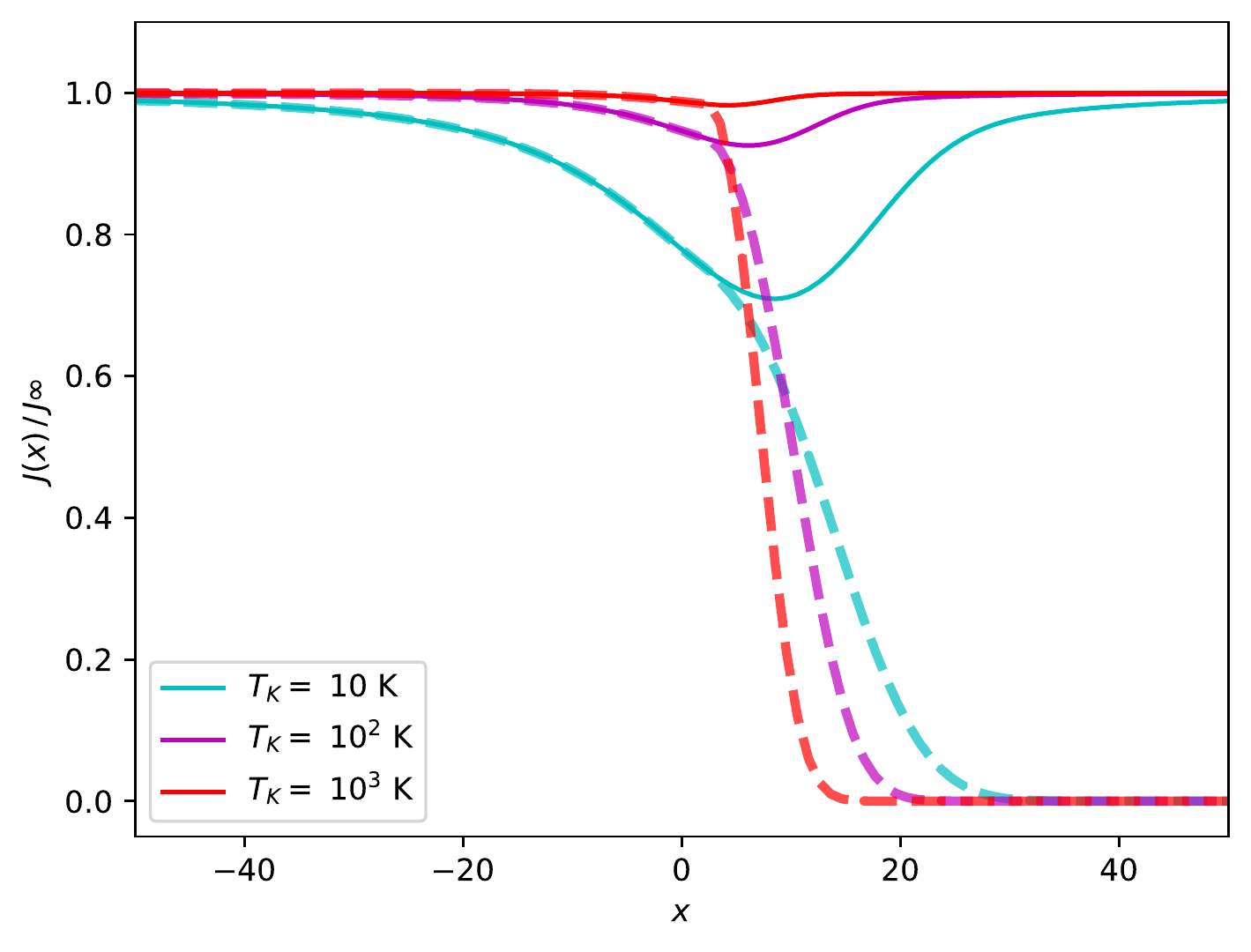}
\captionof{figure}{Spectrum for continuum (continuous lines) and injected (dashed lines) photons around the Ly$\alpha$ line, for three different temperatures at redshift $z=10$.}
\label{fig:Jcontinj}
\end{figure}

For the time scales of interest, we can just focus on stationary solutions, when the time derivative of Eq. \eqref{eq:fokkerplanck} can be neglected:
\begin{equation}
\frac{\partial }{\partial x}\left[ \phi(x) \frac{\partial J(x)}{\partial x} + 2 (  \gamma_{S} + \eta \phi(x) ) J(x) \right] =  -2\mathcal{C}\psi(x).
\end{equation}
The above equation is easily integrated, and since it is linear in $J$, our complete solution will be given by the sum of the continuum spectrum, $J_{c}(x)$ (i.e., the solution of the homogeneous equation, without injections), plus the injected photons near the line center, $J_{i}(x)$ (the non-homogeneous solution). Continuum photons correspond to those emitted between Ly$\alpha$ and Ly$\beta$ transitions, and then redshift to the Ly$\alpha$ line. On the other hand, injected photons are those between the Ly$\beta$ and the Lyman limit (13.6 eV), which can excite an atom reaching a Lyman series line, and thereafter being converted to Ly$\alpha$ photons when the atom decays \cite{Chen:2003gc}. It is worth noting that some radiative cascades can finish in the two-photon transition 2S $\rightarrow$ 1S, and do not contribute to Ly$\alpha$ injections \cite{Hirata:2005mz, Pritchard:2005an}. Imposing the proper boundary conditions, far away from the center of the line, the integration constants can be determined. In the continuum case, (\textit{i.e.}, $\mathcal{C}=0$), there are photons with energies between Ly$\beta$ and Ly$\alpha$ redshifting, and thus far enough from the resonance, we have $J_{c}(x) \rightarrow  J_{c,\infty}$ for $\nu \rightarrow \pm \infty$, getting
\begin{equation}
\phi(x) \frac{\partial J_{c}(x)}{\partial x} + 2 (  \gamma_{S} + \eta \phi(x) ) J_{c}(x) = 2 \gamma_{S} J_{c,\infty}.
\label{eq:Jcont}
\end{equation}
On the other hand, in the injected case, new photons are produced at the Ly$\alpha$ line, which can redshift towards lower energies, but higher energy photons cannot be produced (except from the scattering process, only relevant near the resonance). Thus, for $\nu \rightarrow \infty$, we expect $J_{i}(x) \rightarrow 0$, while for $\nu \rightarrow -\infty$ we write $J_{i}(x) \rightarrow  J_{i,\infty}$\footnote{Note the slight abuse of notation, denoting $J_{i,\infty}$ instead of $J_{i,-\infty}$, in order to unify notation with the continous case.}. Hence, taking $\psi(x)=\delta_D(x)$ as the injection profile\footnote{Employing other broader distributions has little impact on the resulting spectrum \cite{Chen:2003gc}.}, one finds $\mathcal{C}=\gamma_{S} J_{i,\infty}$ and
\begin{equation}
\phi(x) \frac{\partial J_{i}(x)}{\partial x} + 2 (  \gamma_{S} + \eta \phi(x) ) J_{i}(x) = 2 \gamma_{S} J_{i,\infty} (1- \Theta(x) ).
\label{eq:Jinj}
\end{equation}
Note that if the scattering terms are neglected (i.e., the terms with $\phi(x)$), the injected spectrum can be easily solved, being $J_i(x) = J_{i,\infty} (1- \Theta(x))$. The above equations can be integrated, obtaining the solution for continuum photons, and for injected photons with $x<0$, as
\begin{equation}
J(x) = J_\infty \left(1-2 \,\eta \, \int_0^{\infty}dy \exp{\left[ -2\, \eta\, y -2\, \gamma_S\, \int^{x}_{x-y} \frac{dx'}{\phi(x')}\right]} \right),
\label{eq:Jcontsol}
\end{equation}
while the solution for injected photons with $x>0$ is
\begin{equation}
J(x)=J(0) \exp{\left[ -2\, \eta \, x -2 \, \gamma_S \int_0^x \frac{dx'}{\phi(x')}\right] }.
\label{eq:Jinjsol}
\end{equation}
While for some results we have assumed $\phi(x)$ to be very sharp, the evaluation of the spectrum requires a more precise form of the line profile. Its proper shape implies a Lorentzian function, accounting for the resonance, convolved with a Maxwell-Boltzmann distribution, responsible for the thermal broadening, giving rise to the so-called \textit{Voigt distribution} (e.g., Ref. \cite{Chen:2003gc})
\begin{equation}
\phi(x) = \frac{a}{\pi^{3/2}}\int_{-\infty}^{\infty}dy \, \frac{e^{-y^2}}{(x-y)^2 + a^2},
\label{eq:voigt}
\end{equation}
where $a=A_\alpha/(8\pi\Delta\nu_D)$. Examples of the solutions of Eq. \eqref{eq:Jcontsol} and Eq. \eqref{eq:Jinjsol} employing the Voigt distribution for three temperatures at a fixed redshift $z=10$ are shown in Fig. \ref{fig:Jcontinj}. Note that the absorption feature due to scattering effects is more prominent for lower temperatures in the continuum case, while it broadens the suppression in the injected spectrum. At higher temperatures, the modification of the continuum spectrum gets milder, while the falloff in injected photons is sharper. In both cases, scattering effects become weaker at higher temperatures, approaching the no-scattering limit (constant and step function spectrum, respectively for continuum and injected photons).

\subsection{Ly$\alpha$ scattering rate}
\label{sec:lyscatteringrate}

\begin{figure}
\centering
\includegraphics[scale=0.7]{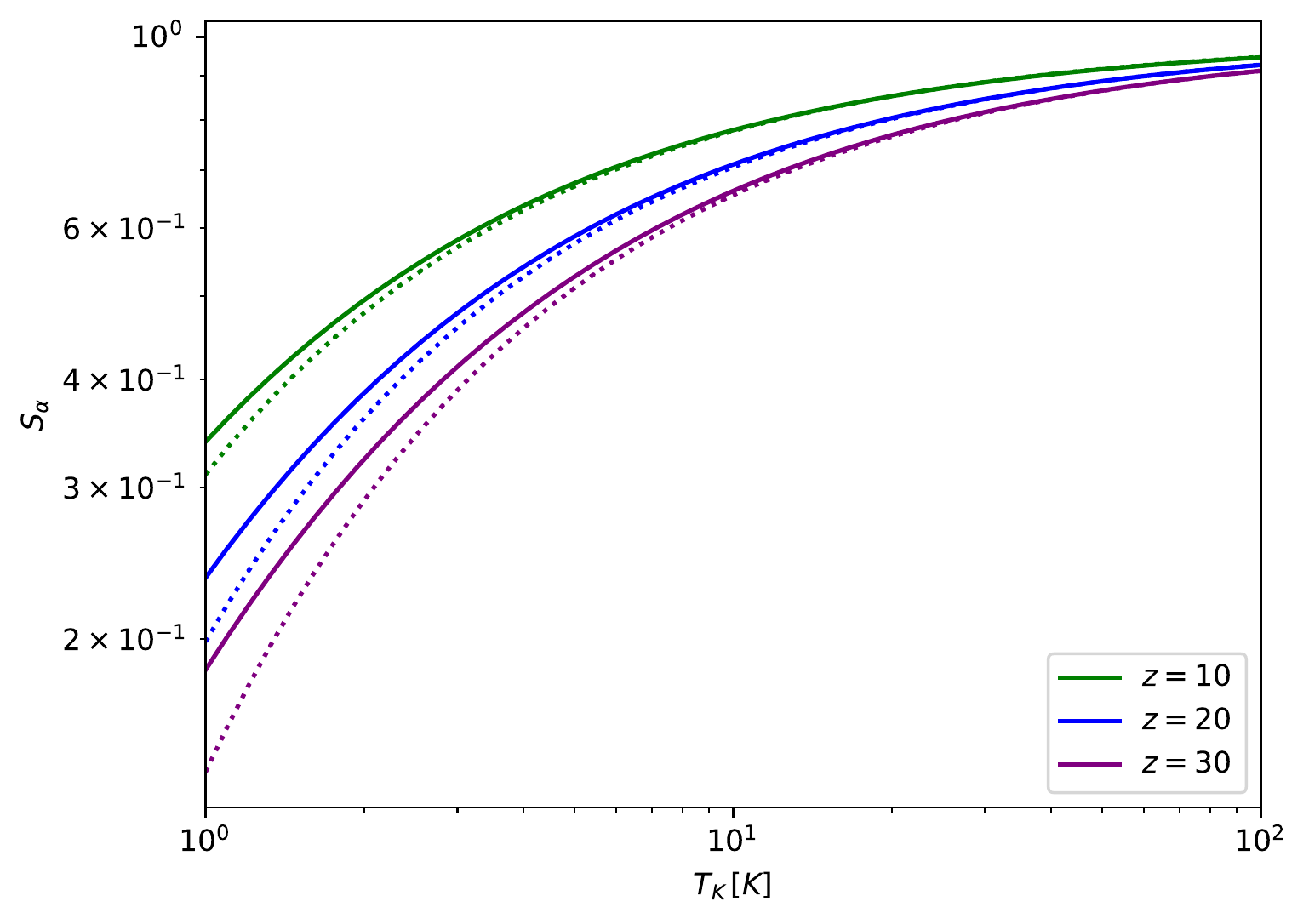}
\caption{$S_\alpha$ (for continuum photons) as a function of the kinetic temperature $T_K$, at several redshifts. Continuous lines stand for the exact case, while dotted lines correspond to the simple fit from Eq. \eqref{eq:fitSalphaCS} \cite{Chuzhoy:2005wv}.}
\label{fig:Salpha}
\end{figure}

The deviation of the Ly$\alpha$ intensity due to line scattering can be quantified defining the quantity
\begin{equation}
S_{\alpha,a} = \int dx \phi(x) \frac{J_{a,\nu}(x)}{J_{a,\infty}},
\end{equation}
for $a= c, i$. Note that $S_{\alpha,a}$ depends only on the shape of the spectrum and not on its normalization. Then, the Ly$\alpha$ scattering rate can be written from Eq. \eqref{eq:lyalscatrate} as
\begin{equation}
P_\alpha = 4\pi \sigma_0 \int dx \phi(x) J(x) = 4\pi \sigma_0  S_{\alpha}J_{\alpha, \infty},
\label{eq:Palpha_final}
\end{equation}
where it has been implicitly summed over the continuum and injected contributions. Hence, from Eqs. \eqref{eq:coup_coef}, \eqref{eq:wf4over27} and \eqref{eq:Palpha_final}, the Ly$\alpha$ coupling coefficient can be written as
\begin{equation}
x_\alpha = \frac{16\pi \sigma_0  T_*}{27 A_{10} T_\gamma} S_{\alpha}J_{\alpha, \infty},
\label{eq:x_alpha}
\end{equation}
which shall be used to compute the spin temperature. The flux far from the line, $J_{\alpha, \infty}$, is determined by the astrophysical model and cosmological evolution, and it will be computed in the next chapter. However, the factor $S_\alpha$ is only determined by the integrated spectrum, and therefore it does not depend on the specific astrophysical parameters, being a function of the temperature and redshift (or, more specifically, of temperature and the Sobolev parameter $\gamma_S$). This quantity accounts for scattering effects in the flux, and is usually of order $\sim 1$. The integrals of Eqs. \eqref{eq:Jcont} and \eqref{eq:Jinjsol} cannot be solved analytically, although there are numerical fits for $S_{\alpha,a}$ in the literature, such as the one from Ref. \cite{Hirata:2005mz}, employed in the code {\tt 21cmFAST} \cite{Mesinger:2010ne} (widely used in Part \ref{partII}). It is possible however to approximate the shape of the line profile to obtain analytic results, as done in the so-called \textit{wing approximation}, with a remarkable agreement with numerical computations \cite{Chuzhoy:2005wv,Furlanetto:2006fs}. Although that solution is given in terms of hypergeometric functions, an accurate fit following this approximation can be derived \cite{Chuzhoy:2005wv}, writing
\begin{equation}
S_\alpha \simeq \exp{[-1.79 \, \alpha]},
\label{eq:fitSalphaCS}
\end{equation}
with $\alpha = \eta \left( 3A_\alpha /(8\pi \Delta \nu_D \gamma_S)\right)^{1/3} =0.717 T_K^{-2/3}(\tau_{GP}/10^6)^{1/3}$ \footnote{Other approximate fits slightly more accurate can be found in Ref. \cite{Furlanetto:2006fs}.}. From this it can be seen that $S_\alpha$ is close to unity at large temperatures and low optical depths (which correspond to low redshifts in a mostly neutral medium, see Eq. \eqref{eq:GPopticaldepth}), being damped in the opposite case. This behavior can be verified in Fig. \ref{fig:Salpha}, which depicts the variation of $S_\alpha$ with the temperature at different redshifts, for the exact calculation (i.e., integrating Eq. \eqref{eq:Jcontsol}) and for the fit in Eq. \eqref{eq:fitSalphaCS}. In the shown cases, spin exchange effects have been neglected, which is a good enough approximation as long as $T_K, T_S \gg T_{se}=0.40$ K. Otherwise, $S_\alpha$ depends on $T_S$, via the modifications of $\eta$ and $\gamma_S$ described above.

\subsection{Color temperature}
\label{sec:color}

Although the Ly$\alpha$ spectrum is not an equilibrium spectrum, and therefore it does not have a definite temperature, one can define an effective \textit{color temperature} as \cite{Furlanetto:2006fs}
\begin{equation}
\frac{h}{k_B T_c(\nu)} \equiv -\frac{1}{f(\nu)}\frac{df(\nu)}{d\nu},
\label{eq:tempcol}
\end{equation}
where $f(\nu)=J/(2\nu^2/c^2)$ is the photon occupation number. Note that, with the above definition, if $T_c$ is a constant of the frequency, one recovers a thermal spectrum with that temperature (a Maxwell-Boltzmann distribution for high energies, since stimulated emission is neglected). It must be emphasized that the color temperature from Eq. \eqref{eq:tempcol} would not have to be necessarily the same as the one implicitly defined in Eq. \eqref{eq:Plyalpha}. However, it can be shown that both temperatures coincide if the Ly$\alpha$ line profile is narrow enough \cite{Breysse:2018slj}. One can employ the calculations from the previous section to predict the color temperature for Eq. \eqref{eq:tempcol} (and equivalently for Eq. \eqref{eq:Plyalpha} if the previous argument holds). Employing Eq. \eqref{eq:Jcont}, this temperature can be written as 
\begin{equation}
\frac{h}{k_B T_c(x)} =\frac{2\eta}{\Delta\nu_D} +\frac{2\gamma_S}{\Delta\nu_D\phi(x)}\left( 1 - \frac{J_{\infty}}{J(x)} \right).
\end{equation}
Including the effect of spin exchange (\textit{i.e.}, substituting $\gamma_S \rightarrow \gamma_S(1+T_{se}/T_K)^{-1}$ and $\eta \rightarrow \eta(1+T_{se}/T_S)/(1+T_{se}/T_K)$, as commented above), and evaluating the color temperature at the center of the line, one gets
\begin{equation}
T_c = T_K \left[ \left( \frac{1+T_{se}/T_S }{1+T_{se}/T_K} \right)  +  \frac{\gamma_S \left( 1+T_{se}/T_K \right)^{-1} }{\eta \phi(0) }\left( 1 - \frac{J_{\infty}}{J(0)} \right) \right]^{-1},
\end{equation}
where it can be shown that the second term becomes negligible at both low and high temperatures \cite{Furlanetto:2006fs}. The above one is an implicit equation: one needs $T_c$ to compute $T_S$ with Eq. \eqref{eq:spintemp}, but $T_c$ also depends on $T_S$. Therefore, it has to be solved by an iterative procedure, as done in {\tt 21cmFAST} \cite{Mesinger:2010ne}. From the previous equation, we can see that when the coupling to the kinetic temperature is efficient enough, $T_c \simeq T_K$ is a good approximation, as stated in Sec. \ref{sec:tspin}. In any case, since $T_K, T_S \gg T_{se}$ in most cases of interest, $T_K$ is a very good approximation for the color temperature even without a strong Ly$\alpha$ coupling.

\subsection{Ly$\alpha$ heating}
\label{sec:lyalphaheating}

Scattering around the Ly$\alpha$ line, besides modifying the spectral shape of the radiation field, can also produce an additional heating contribution to the gas temperature. Part of the energy of the Ly$\alpha$ photon can be left in the atom as thermal energy, although depending on the gas temperature, the atoms can also lose energy in this process, becoming a cooling mechanism. This contribution was firstly studied in Ref. \cite{Madau:1996cs}, under the assumption of neglecting the thermal motion of the atoms. This simplification overestimated the real heating rate, as was shown by the authors of Ref. \cite{Chen:2003gc}, who calculated it properly from the kinetic equation.

In order to derive the heating contribution from scatterings around the Ly$\alpha$ line, we have to multiply the kinetic equation Eq. \eqref{eq:fokkerplanck} by the energy $h\nu$ and integrate over frequency. In the case of injected photons, we have to take into account the energy carried by the photons injected in the line, $h\nu_\alpha \dot{n}_\alpha$ \cite{Oklopcic2013}, where $ \dot{n}_\alpha = 4\pi n_{\rm HI} \sigma_0 \mathcal{C}$ is the rate of change in the number of photons, obtained from the integration of Eq. \eqref{eq:fokkerplanck} over frequency. This contribution vanishes in the case of considering only continuum photons, since scattering processes conserve photon number. Thus, for either the continuum, $k=c$, or injected case, $k=i$, the heating rate of the gas by Ly$\alpha$ scattering reads
\begin{align}
\mathcal{Q} |_{\rm Ly\alpha,k} &= -\frac{4\pi}{c} \int d\nu \; h\nu \; \left( \frac{\partial J}{\partial t} - H\nu  \frac{\partial J}{\partial \nu} \right) + h\nu_\alpha \dot{n}_\alpha \\
& = -4\pi n_{\rm HI} \sigma_0 \int d\nu \; h\nu \;  \frac{1}{2}\frac{\partial}{\partial x} \left[ \phi \left( \frac{\partial J}{\partial x} +2\eta J  \right) \right] \\
& = 4\pi n_{\rm HI} \sigma_0 h\Delta\nu_D \int d\nu \; \frac{1}{2} \phi \left( \frac{\partial J}{\partial x} +2\eta J  \right),
\end{align}
where in the last equality we have integrated by parts. We have also taken into account the fact that the spectral profile vanishes far away from the resonance. Employing Eq. \eqref{eq:Jcont} for continuum photons, or Eq. \eqref{eq:Jinj} for injected ones, we can write the heating rate for each channel as  \cite{Chen:2003gc, Furlanetto:2006fs, Venumadhav:2018uwn}
\begin{equation}
\mathcal{Q} |_{\rm Ly\alpha,c} = \frac{4\pi H h\nu_\alpha \Delta\nu_D}{c} \; \mathcal{I}_k \; J^{(k)}_{\infty},
\end{equation}
where $k=c,i$, and we have defined the continuum and injected heating integrals, $\mathcal{I}_c$ and $\mathcal{I}_i$ respectively, as 
\begin{equation}
\mathcal{I}_c = \int_{-\infty}^{\infty} dx \; \frac{\left(J_{c,\infty} - J_{c}(x) \right)}{J_{c,\infty}}
\end{equation}
and
\begin{equation}
\mathcal{I}_i = \int_{-\infty}^{0} dx \; \frac{\left(J_{i,\infty} - J_i(x) \right)}{J_{i,\infty}} - \int^{\infty}_{0}dx \frac{ J_i(x)}{J_{i,\infty}}.
\end{equation}
The integrals $I_c$ and $I_i$ encode the details on the scattering effects and, as happened with the $S_\alpha$ factor, do not depend on the normalization, but only on the redshift and the temperature, or equivalently, on the temperature and the Sobolev parameter $\gamma_S$. Numerical fits for these quantities, based on the wing approximation, can be found in Ref. \cite{Furlanetto:2006fs}. The importance of this heating mechanism in the evolution of the IGM is however mild, as shall be discussed in the next chapter.

\section{21 cm power spectrum}
\label{sec:21ps}

The most obvious measurement of the 21 cm signal comes from averaging the incoming radiation from all sky directions, which is known as the \textit{global signal} $\overline{\delta T_b}$, corresponding to take the mean value of Eq. \eqref{eq:dTb}. While this spatially averaged signature could provide a superb source to understand the IGM evolution, it lacks some information about inhomogeneities. The signal is enhanced in overdense regions, and thus sensitive to the pattern of the cosmic web and distribution of galaxies and halos. Processes such as X-ray heating and reionization are expected to be highly inhomogeneous, since radiation fields would expand around galaxies, and their complex behavior cannot be completely encoded in a spatially averaged quantity. Moreover, the main experimental caveats to measure the global signal are the strong foregrounds from the galactic synchrotron emission and the effect of the ionosphere, which are more easily removed if different spatial scales are considered, as shall be discussed in Sec. \ref{sec:obs}. It is therefore mandatory to study not only the global signal but also its spatial fluctuations, via higher order statistics. The two-point correlation function is the most obvious choice. Given that radiointerferometers extract information about the Fourier modes of the signal, it is natural to work in Fourier space, rather than in real space, employing thus the power spectrum, as done in Sec. \ref{sec:lineargrowth} for the matter density field. The two-point statistics are of remarkable importance, since for gaussian fields, they present all the information of the distribution. 

Defining the normalized fluctuation of the 21 cm signal from Eq. \eqref{eq:dTb} as
\begin{equation}
\delta_{21} (z,\mathbf{r}) = \frac{\delta T_b (z,\mathbf{r})-\overline{\delta T_b}(z)}{\overline{\delta T_b}(z)},
\end{equation}
one can take its Fourier transform, as defined in Chapter \ref{chap:DarkMatter}, denoted as $\widetilde{\delta}_{21} (z,\mathbf{k})$. Therefore, equivalently to Eq. \eqref{eq:PS}, the 21cm differential brightness temperature power spectrum is defined as 
\begin{equation}
\langle  \widetilde{\delta}_{21} (z,\mathbf{k})  \widetilde{ \delta}_{21}^* (z,\mathbf{k}^\prime) \rangle \equiv (2\pi)^3 \delta^{\rm D} (\mathbf{k} - \mathbf{k}^\prime) P_{21}(z,k) ~.
\label{eq:21cmps}
\end{equation}
In the previous equation, the delta function accounts for invariance under translations. On the other hand, invariance under rotations imply that the power spectrum only depends on the modulus of the wavevector $\mathbf{k}$. These facts are the consequences of statistical homogeneity and isotropy, respectively. Instead of $P_{21}$, it is customary to work with the dimensionless power spectrum, defined as $\Delta^2_{21} (k,z) = (k^3/2 \pi^2) P_{21}(k,z)$. Since this quantity may present singularities when $\overline{\delta T_b}(z)$ vanishes, we rather employ the product $\overline{\delta T_b}^2(z) \Delta^2_{21} (k,z)$, which has units of temperature squared.

One can gain insight into the different contributions to the brightness temperature field by linearizing Eq. \eqref{eq:dTb}. Employing the spin temperature equation, Eq. \eqref{eq:spintemp}, the fluctuation of the 21 cm signal can be written as \cite{Furlanetto:2006jb,Pritchard:2011xb}
\begin{equation}
\delta_{21} = \beta \delta + \beta_x \delta_x + \beta_\alpha \delta_\alpha + \beta_T \delta_T - \delta_{\partial v},
\end{equation}
with each $\delta_i$ accounts for the fluctuations of the different fields: $\delta$ for the matter density field, $\delta_x$ for the neutral fraction, $\delta_\alpha$ for the Ly$\alpha$ flux, $\delta_T$ for the kinetic temperature of the gas and $\delta_{\partial v}$ for the line-of-sight gradient of the peculiar velocities. The functions $\beta_i$ weight the importance of each fluctuation in the 21 cm one, and a straightforward computation shows
\begin{equation}
\beta = 1 +\frac{x_c}{x_{\rm tot}(1+x_{\rm tot})},
\end{equation}
\begin{equation}
\beta_x = 1 + \frac{x_c^{\rm HH}-x_c^{\rm eH}}{x_{\rm tot}(1+x_{\rm tot})},
\end{equation}
\begin{equation}
\beta_\alpha = \frac{x_\alpha}{x_{\rm tot}(1+x_{\rm tot})},
\end{equation}
\begin{equation}
\beta_T = \frac{T_\gamma}{T_K-T_\gamma} + \frac{1}{x_{\rm tot}(1+x_{\rm tot})}\left( x_c^{\rm eH}\frac{ d \, \ln \kappa^{\rm eH}_{10}}{ d\,\ln T_K} + x_c^{\rm HH} \frac{ d \, \ln \kappa^{\rm HH}_{10}}{ d\,\ln T_K} \right),
\end{equation}
with $x_{\rm tot}=x_\alpha+x_c$, and $x_c^{\rm eH}$ and $x_c^{\rm HH}$ accounting only for the eH and HH contributions to the collisional coupling, respectively. Therefore, fluctuations in the brightness temperature can be regarded as a weighted sum over the fluctuations of the different fields on which it depends. To derive the previous equations, $T_c=T_K$ has been assumed for the sake of simplicity, which is a reasonable assumption, as stated in the previous section.

The unity terms of $\beta$ and $\beta_x$ come from the proportionality of the brightness temperature to the density and the ionization fraction, respectively, while the second term includes the contribution from collisional coupling, only relevant at high redshifts or overdense regions. The first term of the temperature coefficient $\beta_T$ accounts for how fast the spin temperature changes under fluctuations in $T_K$, while the second one comes from the temperature dependence on the collisional coupling coefficients. The first one presents a singularity when $T_K=T_\gamma$, but this is unphysical and disappears in the quantity $\overline{\delta T_b}^2(z) \Delta^2_{21} (k,z)$. The $\beta_\alpha$ is proportional to the Ly$\alpha$ coupling coefficient. Understanding the behavior of these coefficients can shed light on the evolution of the power spectrum, which shall be discussed in the next chapter.

In principle, it is possible to study the evolution of first order perturbations, by linearizing the evolution equations of the temperature and ionization fraction (which shall be explained in the next chapter), as done with the matter density field. This approach has been followed in the literature \cite{2005MNRAS.362.1047N, 2007MNRAS.376.1680P}, but since IGM processes become non-linear at some moment, its range of validity is limited. Although we shall compute the 21 cm power spectrum from the non-linear evolved fields in our numerical simulations, it is useful to overview these linear quantities to qualitatively understand the evolution of the power spectrum.

It can be argued that the different $\delta_i$ may be proportional to the overdensity $\delta$ in the linear limit. This is easily seen in the case of the temperature perturbation $\delta_T$, before the onset of heating, as long as the sound speed is spatially constant. Writing the pressure as $P=\rho k_BT_K/\mu$, with $\mu$ the molecular weight and $\rho$ the mass density, $c_s^2=dP/d\rho = k_BT_K/\mu + (\rho k_B/\mu) dT_K/d\rho$ and thus $\delta_T = (\rho/T_K) dT_K/d\rho  \delta= \left[ c_s^2/(k_BT_K/\mu) -1 \right] \delta$ \cite{2005MNRAS.362.1047N}. From the linearized ionization equation, before the galactic sources are switched on, the evolution is dictated by recombinations, whose rate is proportional to $\delta$, and thus $\delta_x \propto \delta$ (the details of ionization evolution shall be properly discussed in the next chapter). When X-rays are starting to be efficiently emitted, the previous arguments would not be valid. Nevertheless, X-ray photons are radiated from the galaxies placed at the overdensities, and thus would also be related to $\delta$. An analogous reasoning can be applied to the Ly$\alpha$ flux, $\delta_\alpha$. Finally, in linear theory, the redshift-space distortions term can be easily computed, getting $\delta_{\partial v} \simeq -{\rm cos}^2\theta f \delta$, where $\theta$ is the angle between the direction of the line of sight and the wavevector, and $f=d \, \ln D/d \, \ln a$ ($=1$ in the Einstein-de Sitter universe) \cite{2010gfe..book.....M}, with $D$ the growth factor defined in \ref{sec:lineargrowth}. Therefore,  the 21 cm power spectrum approximately traces the matter power spectrum, weighted by combinations of the $\beta_i$ functions. However, the actual behavior of the different fluctuations implies extra dependences with the scale for each $\delta_i$, since astrophysical processes are highly inhomogeneous \cite{2007MNRAS.376.1680P}. For this reason, the matter power spectrum cannot be trivially extracted from the 21 cm one.

Finally, it is noteworthy that, unlike the CMB, the 21 cm signal is expected to be non-gaussian. This is due to the growth of perturbations into the non-linear regime, where deviations from gaussianity are developed (as happens in standard perturbation theory at higher order, see, e.g., \cite{2002PhR...367....1B}). Therefore, the power spectrum could not provide the full statistical information about its fluctuations. Although not considered in this thesis, other summary statistics have been exploited in the literature, and could be extremely useful to understand the signal. For instance, the bispectrum (three-point correlation function in Fourier space) encodes information not only about the magnitude of the field but also about its complex phases (which is related to the filamentarity) \cite{Majumdar:2017tdm,Majumdar:2020kpt,Saxena:2020san}. On the other hand, Minkowski functionals supply information regarding the morphology and shapes of the structures, especially useful to gain insight into the growth of HII bubbles during the EoR \cite{Yoshiura:2016nux,Gleser:2006su,Kapahtia:2017qrg,Kakiichi:2017aqi}.

\section{Observational tests of the 21 cm cosmological signature}
\label{sec:obs}


The observation of the redshifted HI cosmological signal is one of the greatest challenges in current observational astronomy. Thus, the requisites to measure this signature need to be critically evaluated. Given the strong foregrounds and the faintness of the signal, achieving a great sensitivity is a crucial point, and large collecting areas and long integration times are needed. Unlike the CMB, 21 cm photons may provide snapshots for different redshifts, which allows attempting a 3D tomographic view of the IGM. In order to make a 3D map of the universe and to obtain information from different epochs, broadband antennas are required, spanning over a large range of frequencies. As already stated, photons observed at a frequency $\nu$ are redshifted from its absorption or emission by the hyperfine transition with a frequency $\nu_0=1420$ MHz at redshift $z$, such that $\nu=\nu_0/(1+z)$. Therefore, the range of frequencies observed provides information from a given epoch (although higher redshifts can also be reached employing Fourier modes along the line of sight in scale dependent measurements \cite{2020PASP..132f2001L}).

Depending on the task, either single stations or interferometric arrays of antennas could be employed to detect radio frequencies. One of the key differences between both approaches is the spatial resolution. In a single antenna, usually designated as \textit{dish} (although often they do not have that shape), the angular resolution $\theta$ is determined by the wavelength $\lambda$ of the observed radiation and the characteristic size of the antenna $L$, as \cite{2020PASP..132f2001L}
\begin{equation}
\theta \simeq \frac{\lambda}{L} \, .
\end{equation}
On the other hand, instead of employing one unique dish, it is possible to employ interferometric techniques to correlate many of them, acting as a huge effective antenna which enlarges the collecting area. These telescopes are known as \textit{interferometers}, and are characterized, among other parameters, by the number of antennas and the maximum baseline distance, which is roughly the maximum length covered by the array. The observation of a given Fourier mode can be performed by the correlation of a pair of antennas separated by a baseline distance $b$. In this case, the angular resolution reads
\begin{equation}
\theta \simeq \frac{\lambda}{b} \, .
\end{equation}
While single dish experiments usually have sizes of tens of meters (being the Green Bank Telescope with 100 m the largest of the world \cite{5109717}), interferometers may spread over tens of kilometers, which allow them to reach much larger resolutions. Selecting pairs of antennas with different baseline distances, distinct Fourier modes could be probed. For this reason, in order to observe the 21 cm fluctuations, interferometers are the proper choice. However, these installations may be extremely expensive. Experiments committed to measure the global average signal do not need such a high resolution, and could work with single dishes. In the case of the interferometers, the maximum baseline is of great relevance since it sets the maximum angular resolution which can be achieved at a given wavelength. And equivalently, the minimum baseline between two antennas determines the largest spatial scales which interferometers are sensitive to. More explicitly, the wavenumber $k$ is proportional to the baseline distance $b$ \cite{2019cosm.book.....M}, and thus, inversely proportional to the angular resolution $\theta$. In interferometers, the size of antennas has also a great importance since it determines, together with the number of elements, the total collecting area. The larger it is, the better the sensitivity gets, and the fainter the signals that can be detected are. Finally, from the equations above it can be noticed that the wavelength of observation also determines the observable scale. As a consequence, low frequency radiation, redshifted from older epochs such as the Dark Ages, cannot be as well resolved as photons with larger frequencies.

\subsection{Foregrounds}
\label{sec:foregrounds}

Unlike the CMB, whose temperature is relatively free from foregrounds at the observed frequencies, in the case of the 21 cm line they become the major issue to overcome. These contaminants can be 3 or 4 orders of magnitude larger than the cosmological brightness temperature, which makes understanding its behavior mandatory in order to extract cosmological information.

There are two main sources of foregrounds from cosmic origins \cite{2019cosm.book.....M}: \textit{i)} galactic foregrounds, given by diffuse synchrotron and free-free emission from the Milky Way (such as thermal bremsstrahlung of free electrons scattering off free ions without being captured), and \textit{ii)} extragalactic foregrounds, as radio emission from Active Galactic Nuclei (AGN) and star-forming galaxies. While the former are mostly important at large scales, being important for angular scales larger than a degree, the latter dominates at smaller scales. The synchrotron emission in the Milky Way is produced by relativistic cosmic-ray electrons accelerated in supernovae remnants and other sources, which typically present a power law energy spectrum. This implies that their synchrotron emission has also a power-law spectral shape \cite{1986rpa..book.....R}. Since the galactic plane contains most of stars, supernovae are more common there, where cosmic rays originate. This leads to a stronger synchrotron emission along this plane. On the other hand, AGN are powered by accretion of matter onto a central supermassive BH, which can present strong radio emission. Star forming galaxies can also present synchrotron and free-free emission in a similar way to the Milky Way. In each case, power laws can describe their spectra. Therefore, in both galactic and extragalactic foregrounds, given the characteristic spectra of AGN, synchrotron and free-free emission, both contributions are expected to be smoothly varying in frequency, which is a key point for their removal. The characteristic decaying power-law shape provides larger foregrounds at lower frequencies, which makes harder to detect signals from distant epochs, such as the Dark Ages, whose photons have redshifted more.

Employing scale-dependent statistics, such as the power spectrum, allows removing the foregrounds much more easily than in the global signal case. This is because, at some scales, gradients in the inhomogeneous signal could present a sharp evolution. This is, for instance, what happens at the edges of the ionized bubbles, where the signal varies abruptly along short distances. These sudden changes contrast with the astrophysical foregrounds, which vary much more smoothly, due to the synchrotron and free-free spectra, which can be easily separated from the cosmological signal.

Besides the cosmic foregrounds, the presence of the ionosphere can also impact the observed signal. It is specially important at the lowest radio frequencies, becoming an even harder task to observe signals from distant epochs, such as the Dark Ages. The most effective way to mitigate their effect would be to observe from outside the Earth, as proposed in several telescopes discussed in the following sections.

In interferometric searches, an alternative to foreground removal is \textit{foreground avoidance}, based on observing modes where foregrounds are suppressed. The Fourier space can be decomposed in wavenumbers parallel and perpendicular to the line of sight, $(k_\|, k_\bot)$. The frequency axis  corresponds to the line of sight direction. Thus, spectrally smooth foregrounds would be mostly compacted in the lowest $k_\|$ modes. This defines a region in the Fourier space, where foregrounds are restricted, and then can be avoided by looking away. However, due to chromatic instrumental effects which imprint unsmooth spectral features, the dominance of foregrounds extends towards large $k_\bot$ at higher $k_\|$, shaping a \textit{wedge} in Fourier space \cite{2010ApJ...724..526D}. Luckily, this region is relatively well delimited, allowing to evade the contaminated modes. Foreground avoidance presents an advantage since allows to elude possible biases from foregrounds fitting. The drawback of this method is that the number of accessible scales becomes significantly reduced. The subtleties of this procedure are discussed in, e.g., Refs. \cite{2020PASP..132f2001L, Chapman:2014sfa, PhysRevD.89.023002}.

\subsection{Global signal experiments}

Attempting to measure the global averaged signal is one of the primary goals of observational 21 cm cosmology. These measurements correspond to the monopole in a spherical harmonics decomposition, typical in CMB studies. Experiments designed to detect the global signature do not need to resolve spatial details on the sky, and are thus commonly carried out by single antennas that have wide beams, rather than large interferometric arrays, as happens with experiments focused on measuring fluctuations (although there are exceptions, such as LEDA, as discussed below). For the very same reason, these experiments can also employ coarser frequency channels with larger bandwidths, with no need for a spectral resolution as large as the experiments searching for fluctuations. The mitigation of cosmic and ionosphere foregrounds, which requires a precise calibration, becomes the greatest challenge. However, an epoch of rapid reionization, with an abrupt drop in the signal, may be different enough from the smoothly varying foregrounds, fact which has been used to rule out too fast and sharp reionization processes, placing constraints on the timing and duration of the EoR. The most relevant experiments pursuing the global signature are listed in what follows (see, e.g., Refs. \cite{2019arXiv190804296K,2013PhRvD..87d3002L} for more details).

\begin{itemize}

\item The Experiment to Detect the Global EoR Signature (EDGES) \cite{2008ApJ...676....1B} is located at Murchison Radio-astronomy Observatory in Western Australia, and composed by a single dipole antenna. Early measurements were able to rule out very rapid EoR \cite{2010Natur.468..796B}. The most remarkable result was announced in 2018, when its team reported the evidence of an absorption profile in the low-band spectrum (50-100 MHz), centered around 78 MHz (corresponding to $z \sim 17$), with an amplitude of $\sim 0.5$ K \cite{Bowman:2018yin}. This deep trough is roughly twice the maximum predicted within the context of the $\Lambda$CDM model. The implications and caveats of this measurement are discussed in Sec. \ref{sec:edges}. Regardless of this absorption feature, high-band measurements spanning 90-190 MHz \cite{Monsalve:2016xbk} have been employed to constrain the astrophysical parameter space from Cosmic Dawn to the EoR \cite{2017ApJ...847...64M, Monsalve:2018fno, Monsalve:2019baw}.

\item The Shaped Antenna to measure the background RAdio Spectrum (SARAS) \cite{Patra:2012yw} was sited at the Raman Research Institute in India, although it has been its upgraded version, SARAS 2 \cite{2018ExA....45..269S} at Timbaktu Collective in Southern India, the one which have offered more competitive results. Using its observations at frequencies 110-200 MHz, rapid periods of reionization have been constrained \cite{2017ApJ...845L..12S}. Furthermore, SARAS 2 has ruled out several astrophysical models which present a very cold IGM at the epoch of the Cosmic Dawn, which would lead to large amplitudes in the global signal \cite{2017ApJ...845L..12S, Singh:2017cnp}. Note that this may be in tension with the EDGES results.

\item The Large-Aperture Experiment to Detect the Dark Ages (LEDA) \cite{Greenhill:2012mn, 2018MNRAS.478.4193P} is placed on the Long Wavelength Array stations at Owens Valley Radio Observatory (OVRO-LWA), in California, USA. Unlike other global signal single-antenna experiments, it is formed by $\sim 250$ dipole antennas spread over an area of $\sim 200$ m of diameter, constituting an interferometric array capable of observing frequencies between 30 and 88 MHz, which may probe the Cosmic Dawn epoch (redshifts $16< z <34$). LEDA observations allowed to place limits on the amplitude and width of the 21 cm absorption profile at $z\sim 20$
\cite{2016MNRAS.461.2847B}.

\item The Dark Ages Polarimeter Pathfinder (DAPPER) \cite{Burns:2019zia, Burns:2021ndk} is an ambitious space based project to put into orbit a spacecraft in the dark side of the Moon, avoiding therefore the ionosphere foregrounds. Unlike the previous experiments, it is at the planning stage and has not been built yet. It is designed to operate at very low frequencies, between 10-110 MHz, being able in principle to reach the Dark Ages, at redshifts $z\sim 36-83$. It was preceded by the proposal of the Dark Ages Radio Explorer (DARE) \cite{Burns:2011wf,2012AdSpR..49..433B}.

\end{itemize}

\subsection{The EDGES affair}
\label{sec:edges}

\begin{figure}
\centering
\includegraphics[scale=0.6]{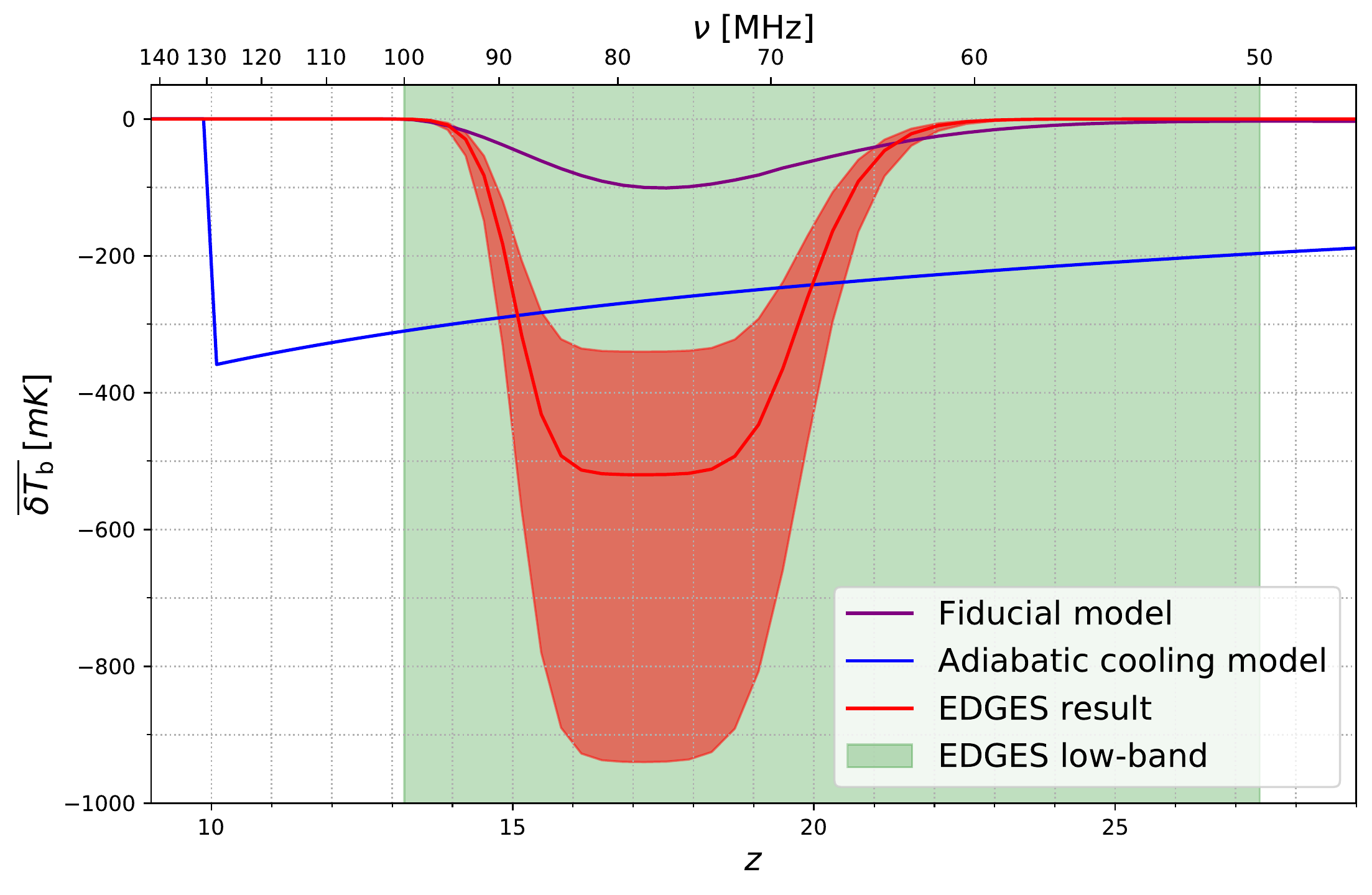}
\captionof{figure}{Global 21 cm brightness temperature for the best fit to explain the EDGES absorption feature shown in red, with its 99\% CL interval \cite{Bowman:2018yin}. EDGES low-band data lie in the range covered by the green region. For comparison, an extreme case with adiabatic cooling, $T_S$ completely coupled to $T_K$ and instantaneous reionization at $z\sim 10$ is also included in blue, as well as a fiducial model shown in purple computed from \texttt{21cmFAST} \cite{Mesinger:2010ne} such that it presents the absorption dip at roughly the same redshifts (see Chapter \ref{chap:IGM} for a detailed discussion on the evolution of the brightness temperature and the astrophysical processes involved).}
\label{fig:edges}
\end{figure}

The reported absorption profile by the EDGES collaboration in the low-band spectrum (50-100 MHz) has supposed a major result in the last years, given its unexpected shape. In order to explain such a deep absorption trough, there are two main possibilities, as can be seen from equation Eq. \eqref{eq:dTb}: either \textit{i)} the spin temperature (and thus the kinetic one) is very cold, or \textit{ii)} the CMB background $T_\gamma$ has extra contributions at radio frequencies.

Standard evolution histories within the $\Lambda$CDM scenario limit the coldest kinetic temperature attainable, since the only cooling process efficient at this regime is due to adiabatic expansion, $T_K \propto a^{-2}$. Figure \ref{fig:edges} shows a sample of such extremely optimistic scenario in blue, with the largest possible amplitude within the standard picture, compared to the EDGES result in red. Even with the spin temperature completely coupled to the adiabatic one at these epochs, the largest possible signals could not exceed $\sim -250$ mK at $z \sim 17$. The issue gets worse for more realistic scenarios, as the one shown in purple, where heating of the IGM and the non-perfect coupling of the spin temperature to the kinetic one have been taken into account properly (as shall be reviewed in Chapter \ref{chap:IGM}). The EDGES measurement may thus imply that extra cooling processes are present during the Cosmic Dawn and Dark Ages to drive the spin temperature to such a low value. Following this idea, many authors have proposed non-standard scenarios to achieve a cooler IGM. This is the case of DM-baryon interactions, which would transfer thermal energy from the gas to the DM particles \cite{Barkana:2018lgd, Munoz:2018pzp, Munoz:2018jwq, Fialkov:2018xre, Berlin:2018sjs, Mahdawi:2018euy}. An early Dark Energy period could provide an earlier decoupling of the gas temperature from the Compton interactions, giving more time to cool adiabatically \cite{Hill:2018lfx}.

The excess of radio radiation at the epoch of the Cosmic Dawn has also been suggested as a possible way to increase $T_\gamma$ and thus explain the EDGES signal \cite{2018ApJ...858L..17F}. However, it has been argued that the diffuse radio background from astrophysical origins, by either synchrotron or inverse-Compton scattering due to relativistic electrons, seems unlikely to explain the signal. This is because the radio emissivity required to enhance $T_\gamma$ should be enlarged by a factor from $10^3$ to $10^6$ respect to the local $z =0$ estimates \cite{2018MNRAS.481L...6S, Mirocha:2018cih}. However, the existence of soft photon emission at low radio frequencies from light DM, such as axion-like particles, may account for the required background \cite{Fraser:2018acy, Pospelov:2018kdh, 2018PhLB..783..301M}. Accretion onto intermediate-mass black holes has also been proposed to produce the required extra radio emission \cite{Ewall-Wice:2018bzf,Ewall-Wice:2019may}.

Besides trying to explain the signal, the EDGES result has also been used to constrain different DM models, such as DM annihilation \cite{DAmico:2018sxd, Liu:2018uzy}, DM decay \cite{Mitridate:2018iag}, WDM \cite{Safarzadeh:2018hhg,2018PhRvD..98f3021S}, PBHs \cite{Hektor:2018qqw, Clark:2018ghm, Halder:2021rbq} or FDM \cite{2018PhRvD..98b3011L, 2018PhRvD..98f3021S}. In Part \ref{partII} of this thesis, an example of this is shown, constraining IDM and WDM models making use of the timing of the EDGES signal \cite{Lopez-Honorez:2018ipk}. The consistency of the EDGES result with other experimental constraints from EoR and CMB have also been explored, even though that work is not included in this thesis \cite{Witte:2018itc}.

Thus, although some (exotic) solutions may be able to explain the EDGES result, it also presents some controversy within the scientific community for other reasons. It has not been confirmed yet by other experiments. Observations of SARAS 2 at similar range of frequencies have ruled out models which should show such a deep absorption trough, being in tension with the EDGES finding \cite{2017ApJ...845L..12S, Singh:2017cnp}. The shape of the signal extracted from the data is also highly non-trivial and not well understood. The absorption model to fit the data employed in \cite{Bowman:2018yin} was chosen to be a flattened gaussian, with no clear interpretation from physical grounds. The broad dip suggests a relatively extended epoch around $z \sim 17$ with constant temperature, which can not be explained within current IGM evolutionary models. In fact, when one attempts to accommodate the astrophysical modeling of the signal to the EDGES data, the width of the absorption dip can only be fitted in models with high heating, opposite to what is required to obtain large amplitudes, suggesting an additional inconsistency between the measurement and canonical astrophysical scenarios \cite{Witte:2018itc}.

In addition, several caveats regarding the treatment of foregrounds and calibration have been claimed, which could invalidate the measurement. Although the EDGES calibration has been carefully scrutinized by the team during 10 years, some concerns about possible systematics have been raised. A possible ground plane artifact may produce such a broad absorption signal, as was shown by Ref. \cite{2019ApJ...874..153B}. This would induce one or several resonances in the data, which could be confused with a cosmological signal. The authors found that three of these resonant features with no 21 cm signal provided a fit as good as the one obtained by EDGES. Remarkably, the best-fit resonant frequencies were in good agreement with those estimated for a ground plane artifact, considering the dimensions of the square patch and properties of the soil.  

Furthermore, Ref. \cite{2018Natur.564E..32H} found that different foreground and absorption models could fit the EDGES signal. It was shown that the best fit parameters for the ionospheric foregrounds employed by EDGES were not physical, since they imply optical depths and temperatures of the ionosphere with negative values. In addition, it was found that two gaussians or a sinusoidal feature with no absorption signal can provide a fit to the spectrum as good as the one presented by the EDGES collaboration. These non-cosmological features could be a consequence of some sort of overlooked systematics, such as the one proposed by \cite{2019ApJ...874..153B}.

The authors of Ref. \cite{2020MNRAS.492...22S} further analyzed these questions, by employing a Bayesian evidence-based comparison over a broad amount of models. They discussed different models for the signal, including physical motivated ones from simulations varying astrophysical parameters. Cosmic and ionospheric foregrounds, as well as possible calibration errors with sinusoidal shapes, were also considered. It was found that models including calibration systematics were decisively preferred. Moreover, models with amplitudes consistent with the standard cosmological scenario were present among the best fitting models. Furthermore, they did not find strong evidence to favor models exhibiting 21 cm signal over those without one. All these facts place reasonable doubts about whether it is possible to infer an actual 21 cm cosmological measurement from the EDGES data.

All the above references show the great impact that the EDGES result have left on the astrophysics and cosmology communities. However, the aforementioned caveats place important concerns regarding the validity of this measurement, which can only be confirmed by upcoming observations from future radiotelescopes.


\subsection{21 cm power spectrum experiments}

\begin{table}[t]
\begin{center}
\resizebox{\textwidth}{!}{

{\def\arraystretch{1.3}

\begin{tabular}{|c||c|c|c|c|c|}
\hline
Facility & Location & $\nu$ [MHz] & $z$ & $N_{\rm ant}$ & $b_{\rm max}$ [km]  \\
\hline
\hline
GMRT & India & 50-1420 & 0-27 & 30 & 30 \\
MWA & Australia & 70-90, 135-195 & 15-19, 6-10 & 128 & 5 \\
LOFAR & Netherlands & 30-80, 120-190 & 17-46, 6-11 & 50--60 & 50 \\
PAPER & South Africa & 110-180 & 7--12 & 64 & 0.210 \\
OVRO-LWA & USA & 27-88 & 15-52 & 288 & $<$ 1.5 \\
LEDA & USA & 45-88 & 15-30 & 256 & $<$ 10 \\
HERA* & South Africa & 50-250 & 5-27 & 350 & $<$ 1 \\
SKA-low* & Australia & 50-350 & 4-27 & $\sim 10^5$ & 65 \\
SKA-mid* & South Africa & 350-15.3$\times 10^3$ & 0-4 & 197 & 150\\
\hline
\end{tabular}
}
}
\captionof{table}{List of current and future interferometers attempting to measure the spatial fluctuations of the 21 cm signal, including its location, range of frequencies of operation $\nu$, accessible redshifts $z$, number of antennas $N_{\rm ant}$, and maximum baseline attainable $b_{\rm max}$. Experiments marked with an asterisk (*) are not completed yet. Note that LEDA, despite being an interferometer, is devoted to detect the global signal rather than 21 cm fluctuations.}
\label{tab:exps}
\end{center}
\end{table}

\begin{figure}
\centering
\includegraphics[scale=0.55]{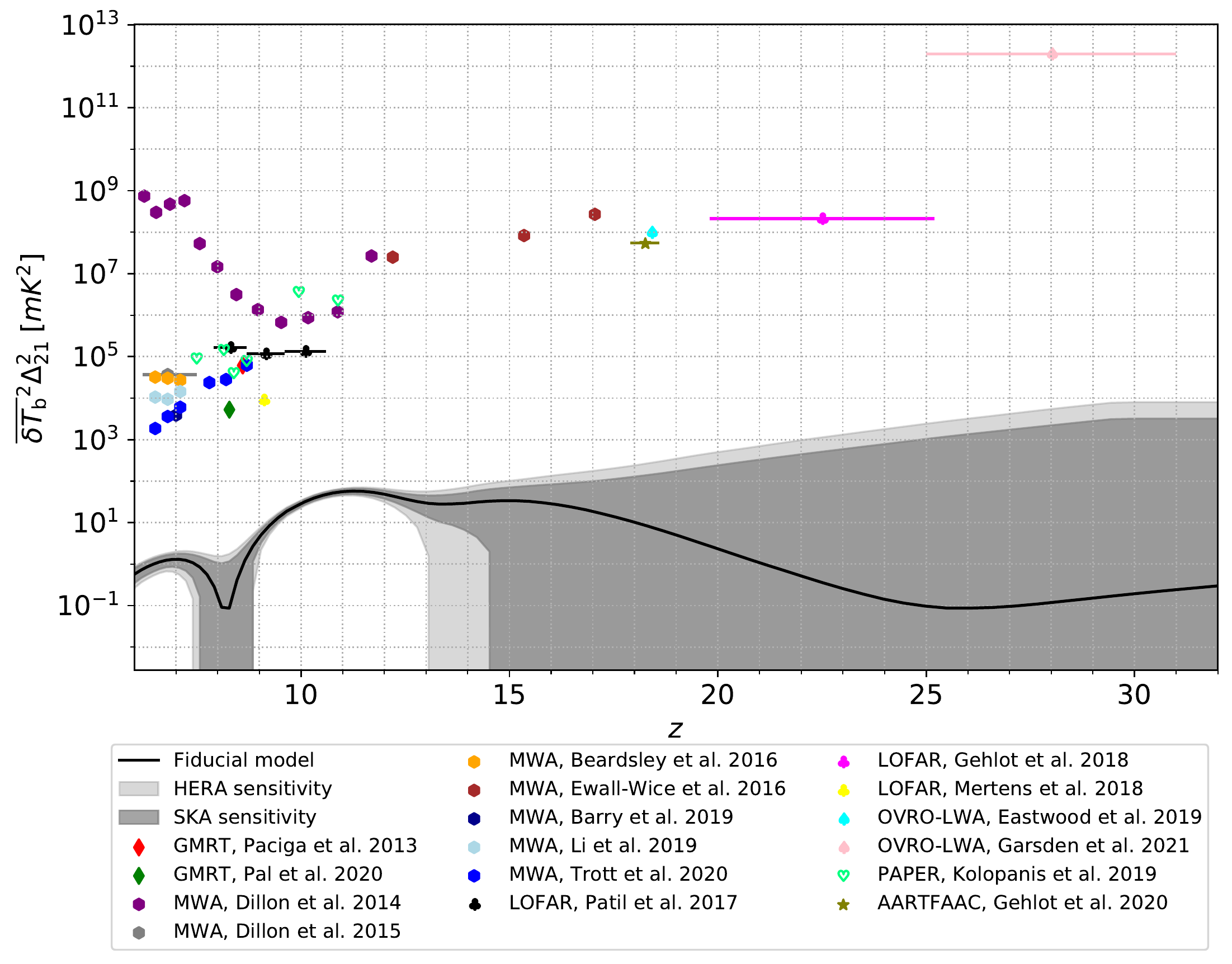}
\captionof{figure}{Summary of upper limits on the 21 cm power spectrum. The dots denote the bounds found by the experiments GMRT \cite{2013MNRAS.433..639P, Pal:2020urw}, MWA \cite{PhysRevD.89.023002, PhysRevD.91.123011, 2016ApJ...833..102B, 2016MNRAS.460.4320E, Barry:2019qxp, 2019ApJ...887..141L, Trott:2020szf}, LOFAR \cite{2017ApJ...838...65P, 2019MNRAS.488.4271G, Mertens:2020llj}, OVRO-LWA \cite{2019AJ....158...84E, Garsden:2021kdo}, PAPER \cite{2019ApJ...883..133K} and AARTFAAC (an extension to LOFAR) \cite{2020MNRAS.499.4158G}. Most of them lie between the wavenumbers $k=0.1$ and $0.5$ $h/$Mpc, except the ones from Refs. \cite{Pal:2020urw} and \cite{2019MNRAS.488.4271G} (note also that Ref. \cite{2017ApJ...838...65P} presents stronger limits at $k=0.053$ $h/$Mpc). The continuous black line corresponds to a fiducial theoretical model computed with \texttt{21cmFAST} \cite{Mesinger:2010ne}, at $k=0.2$ $h/$Mpc (details regarding its evolution and the astrophysical modeling are discussed in Chapter \ref{chap:IGM}). Gray bands depict the forecasted sensitivities for the HERA and SKA experiments, computed via the code \texttt{21cmSense}\protect\footnotemark \cite{2016ascl.soft09013P}. The script to generate this figure has been made publicly available\protect\footnotemark \cite{pablo_villanueva_domingo_2021_4579904}. }
\label{fig:21ps_constraints}
\end{figure}


Although detecting the global signature is of significative importance, the information which can be extracted from it is limited, and the experimental challenges and caveats are numerous. For this reason, most observational efforts are devoted to detect the spatial fluctuations of the 21 cm signal, which would enable a multi-scale view of the Hydrogen map, and an easier removal of the foregrounds. Interferometry based on large arrays of antennas conforms the proper tool to access small scales. Interferometers offer a natural way to directly estimate the power spectrum of an astronomical image. This is because the voltage induced in a pair of antennas can be correlated, providing the visibility, which is nothing but the Fourier transform of the perceived intensity of the radiation \cite{2012ApJ...753...81P, 2019cosm.book.....M}. Moreover, as already mentioned, foregrounds can be more easily treated than when dealing with global measurements.

The achieved sensitivity is subject to different aspects of the experimental design and systematics. It is also limited by the spatial scales and redshifts to be observed. Among the possible sources of noise, the most important one is thermal noise. Thermal fluctuations in the apparatus produce a gaussian white-noise signal (i.e., scale independent) characterized by its root-mean-square $T_{N,rms}$ \cite{2012ApJ...753...81P}. Thus, the thermal noise at a single baseline goes as $\overline{\delta T_b}^2(z) \Delta^2_{21} (k,z)\sim k^3 T_{N,rms}^2$. This implies that the sensitivity worsens at very small scales. Furthermore, due to the factor to convert from angles and bandwidth to cosmological distances, this noise also increases with redshift, roughly as $(1+z)^{5/2}$ \cite{2013AJ....145...65P}. Therefore, distant sources (i.e., remote epochs) become more difficult to observe by a limitation in the sensitivity, and not only by the effect of foregrounds. In order to calculate the sensitivity of a full interferometric array, one must sum up the contributions by each baseline over all the Fourier modes. Combining independent $k$ modes, the sensitivity can be improved, with a noise dependence of $\sim k^{5/2}$ rather than $k^3$. Moreover, if different baselines measure the same Fourier modes, having redundant observations, the total noise decreases with the number of antennas as $N_{\rm ant}^{-1}$. This dependence is easy to interpret, given that the power spectrum is a sort of variance, which typically decreases with $N_{\rm ant}^{-1}$ when independent measurements are taken into account. It is thus auspicious to have interferometers with a large number of dishes, placed at regular grids to probe the same Fourier modes, in order to improve the sensitivity, see Refs. \cite{2012ApJ...753...81P, 2013AJ....145...65P} for more details on these regards.

\footnotetext[11]{\url{https://github.com/jpober/21cmSense}}
\footnotetext[12]{\url{https://github.com/PabloVD/21cmBounds}}  

Thus far, none of the existing experiments has achieved a detection of the 21 cm fluctuations from the pre-EoR universe (although some cross-correlation with other probes have been detected from the post-EoR low redshift universe, as discussed below). However, current radiointerferometers have placed remarkable upper limits on the amplitude of the fluctuations, increasingly tighter over the years, which have constrained the possible thermal evolution histories. The current upper bounds are shown in Fig. \ref{fig:21ps_constraints}, together with a fiducial model of the 21 cm power spectrum. In the following, we comment on the interferometer experiments which have been able to place constraints on the 21 cm fluctuations amplitude from epochs between the Cosmic Dawn and the EoR, which are also listed in Tab. \ref{tab:exps}. For more information, see, e.g., Refs. \cite{2020PASP..132f2001L, 2019cosm.book.....M, 2019arXiv190804296K, Raste:2021auu}.

\begin{itemize}

\item The Giant Metrewave Radio Telescope (GMRT) \cite{10.2307/24094934} is an array located in Maharashtra, India, for low-frequency general purposes. It consists of 30 dishes of 45 m of diameter, spread over 25 km. It allows observing in 6 frequency bands, covering from 50 to 1420 MHz. Reionization and post-EoR epochs have been studied with this instrument, being pioneer in the treatment of foregrounds and their removal employing spatial correlations. Although not optimized for the cosmological 21 cm signal, it has been the first one to place upper limits at redshifts relevant for the EoR, at $z \sim 9$ \cite{2011MNRAS.413.1174P}, being subsequently revised (and significantly weakened) \cite{2013MNRAS.433..639P}. A stringent constraint has also been recently proposed at $z \simeq 8$ \cite{Pal:2020urw}.

\item The Murchison Widefield Array (MWA) \cite{Bowman_2013} is located in the Murchison Radio-astronomy Observatory, at the Western Australian desert. It consisted of 128 elements in its Phase I (2013-2016), being increased to 256 with longer baselines since 2016. Each of these elements is formed by a tile of $4\times4$ grid of crossed dipole antennas, ranging from 70 to 300 MHz. The MWA has published several upper limits covering a broad range of epochs, from the EoR, at redshifts $z \sim 6$ to $11$ \cite{PhysRevD.89.023002, PhysRevD.91.123011, 2016ApJ...833..102B, Barry:2019qxp, 2019ApJ...887..141L, Trott:2020szf}, to Cosmic Dawn \cite{2016MNRAS.460.4320E}, constraining redshifts $z\sim 12, 15$ and $17$.

\item The Donald C. Backer Precision Array for Probing the Epoch of Reionization (PAPER) \cite{Parsons:2009in, Parsons:2011ew} is an interferometer designed and optimized for observations of the EoR, operating between 100 and 200 MHz. It consisted of two stations, a preliminary one with 32 antennas sited at Green Bank, West Virginia, USA, for testing, and the main installations for scientific research with 64 antennas placed at the South African Karoo desert. Its team pioneered ways to improve the sensitivity, proposing methods such as maximizing redundant observations of the same Fourier modes to reduce the noise \cite{2012ApJ...753...81P}, and the so-called delay spectrum technique to avoid foregrounds \cite{2012ApJ...756..165P}. It provided several upper limits at the EoR \cite{Pober:2013ig, Pober:2014aca, Ali:2015uua}. However, posterior re-analysis of their data showed errors in their analysis related to the covariance matrices and to an underestimation of the level of signal loss, which invalidated these limits \cite{2018ApJ...868...26C, 2018ApJ...863..201A}. Nevertheless, they were later updated properly accounting for these errors, significantly weakening the bounds \cite{2019ApJ...883..133K} (see \cite{ 2019cosm.book.....M} for further details on this issue). The PAPER experiment is already deccomissioned, but its infrastructure in South Africa currently serves for the HERA interferometer.

\item The LOw Frequency Array (LOFAR) \cite{van_Haarlem_2013} is a multi-purpose interferometer focused on low frequencies sited at the Netherlands\footnote{The Netherlands has a long tradition related to the 21 cm signal. Among other achievements, it was Dutch astronomers, van de Hulst and Wouthuysen, who predicted the hyperfine transition and the Ly$\alpha$ coupling, respectively, becoming aware of the astronomical importance of the 21 cm signal in radioastronomy. Furthermore, Muller, Oort and Van de Hulst observed the first rotation curve and HI map of the Milky Way, showing for the first time its spiral arms. See Ref. \cite{2006JAHH....9....3V} for more information.}. It includes two types of antennas, covering high-band (110-250 MHz) and low-band (10-90 MHz) observations. While the low-band elements are standard dipole antennas, the high-band ones consist of a tile of 16 individual antennas. Upper limits on the 21 cm power spectrum have been placed using its observations, both at the EoR, covering redshifts $z \sim 8$ to $10$ \cite{2017ApJ...838...65P, Mertens:2020llj}, and at the Cosmic Dawn, reaching redshifts from $z \sim 19$ to $25$ \cite{2019MNRAS.488.4271G}. An add-on real time transient detector facility appended to LOFAR called Amsterdam-ASTRON Radio Transients Facility And Analysis Center (AARTFAAC) \cite{2016JAI.....541008P} has recently found a stringent upper limit at $z\simeq 18$ \cite{2020MNRAS.499.4158G}. An extension of LOFAR coined NenUFAR, located in Nançay, France, is ongoing, which may serve as a pathfinder for future experiments such as SKA \cite{7136773}.

\item The Owens Valley Radio Observatory Long Wavelength Array (OVRO-LWA) is sited in Owens Valley, California, USA. It is formed by 288 dipole antennas spanning from 27 to 85 MHz, arranged in an pseudo-random disposition in a compact core of 200 m of diameter, being thus well suited for transient searches such as gamma ray bursts or exoplanets. Besides that, it has also been employed for constraining the 21 cm fluctuations, finding an upper limit at $z \simeq 18$ \cite{2019AJ....158...84E}, and recently in the range $25 < z < 31$, being this one the bound located at the farthest redshift so far \cite{Garsden:2021kdo}. The LEDA facilities are also located at the OVRO-LWA.


\end{itemize}

All of the aforementioned upper limits on the amplitude of fluctuations are placed at a range of scales spanning $0.1\, h{\rm Mpc}^{-1} < k < 1\,h{\rm Mpc}^{-1}$ (although the LOFAR ones include also modes down to $k\sim 0.01$ $h{\rm Mpc}^{-1}$), and are about 2 orders of magnitude larger than the expected fiducial models for the 21 cm signal \cite{2020PASP..132f2001L}. Besides the interferometers mentioned above, some single dish experiments have been employed to constrain the 21 cm fluctuations at low redshift, from the post-reionization universe. Although these epochs lie beyond the scope of this thesis, it is worth mentioning some results due to their significance. The Green Bank Telescope (GBT) \cite{5109717}, located at Green Bank, West Virginia, USA, is the world's largest fully steerable radio telescope, with 100-meter diameter. Operating at frequencies between 100 and 115 GHz, it has been the first to detect 21 cm fluctuations in the post-EoR universe, via cross-correlations with other surveys: with the optical galaxy survey DEEP2 at $z \simeq 0.5-1.1$ \cite{2010Natur.466..463C}, and WiggleZ at $z\sim 0.8$ \cite{2013ApJ...763L..20M}. The Parkes Radio Telescope \cite{1996PASA...13..243S} in New South Wales, Australia, has 64-meter diameter, and detects frequencies from 1230 to 1530 MHz. Similarly to the GBT, the cross-correlation of its observations with those 
from the 2dF galaxy survey provided detections of the cosmological HI line at redshifts $0.057 < z < 0.098$ \cite{2018MNRAS.476.3382A}.

On the other hand, several radiointerferometers are expected to be built within this decade, which may have enough sensitivity to attain a positive detection of the signal in the next few years. Some of them are devoted to perform HI \textit{line intensity mapping}, i.e., measuring the integrated 21 cm emission from unresolved gas clouds, accessing clumpy neutral regions in the post-reionization era, up to $z \sim 2.5$. Examples of these are the Canadian CHIME \cite{2019clrp.2020...12L}, HIRAX in South Africa \cite{2016SPIE.9906E..5XN}, the Tianlai experiment in China \cite{2012IJMPS..12..256C} or the BINGO observatory in South America \cite{2019JPhCS1269a2002W}. Among the proposals for studying the Cosmic Dawn and EoR periods, here we highlight two interferometers which by their sensitivity and international effort, may become a great step forward towards the detection of the HI signature.

\begin{itemize}

\item The Hydrogen Epoch of Reionization Array (HERA) \cite{Beardsley:2014bea} is placed at the same facilities than PAPER in the Karoo desert, South Africa, specifically designed to observe the EoR. For that, it will observe in a range between 50 and 250 MHz, and is expected to have 350 dishes with 14 m diameter each, when completed. It has been deployed following the ideas developed for PAPER, with a highly redundant array to enhance the sensitivity. Although not finished yet, it is already running and collecting data. 

\item The Square Kilometer Array (SKA) \cite{Mellema:2012ht}, is the most ambitious radiointerferometer designed so far, with the world’s largest area ($\sim$ 1 km$^2$) when completed. It will consist of two different facilities operating at different frequencies. On the one hand, SKA-low will be comprised of $\sim 10^5$ antennas grouped into stations in the Murchison Radio-astronomy Observatory at the Western Australian desert, spanning frequencies from 50 to 350 MHz. The SKA-mid, on the other hand, will be formed by 197 dishes in the South African Karoo desert, observing frequencies from 350 MHz to 15.3 GHz. The pseudo-random arrangement of antennas will allow reaching baselines up to 150 and 65 km, respectively. It will employ the facilities previously occupied by MWA and HERA. Although far from being finished yet, pathfinder telescopes are already constructed and gathering data, such as MeerKAT in South Africa \cite{santos2017meerklass} and ASKAP in Australia \cite{McConnell_2020}.

\item The Farside Array for Radio Science Investigations of the Dark ages and Exoplanets (FARSIDE) is a proposal for building an interferometric array based on the Moon \cite{Burns:2021pkx}, in addition to the orbiting satellite DAPPER. It would be composed by 128 dipoles across a 10 km$^2$ area. Covering from $0.1$ to 40 MHz, it is aimed to detect the faint 21 cm power spectrum during the Dark Ages. Contrary to previous interferometers, it is still an idea under design and its construction has not started yet.

\end{itemize}

The above lists of current and future interferometers are not intended to be comprehensive, but to provide an outlook of experimental status, as well as about the increasing efforts of the scientific community for pursuing the cosmological 21 cm signal. A lot of work remains to be done, but these are exciting times for seeking the hyperfine line of the IGM.

\begin{comment}

\bibliographystyle{../../jhep}
\bibliography{../../biblio}

%% file: Chapters/Chapter_IGM/Chapter_IGM.tex
\end{comment}

\chapter{Evolution of the Intergalactic Medium}

\label{chap:IGM}


The physics of the IGM is highly rich and complex. Its study is fundamental in order to understand how and when galaxies formed, evolving from a mostly homogeneous fluid towards the cosmic web of galaxies and quasars immersed in an ionized medium we can see today. At $z\sim 1100$, the CMB decouples from the thermal plasma, marking the beginning of the so-called Dark Ages. After Recombination, most atoms become neutral, only a small fraction of free electrons remaining. The gas temperature keeps cooling, coupled to the photon temperature due to Compton scattering until $z\sim 150$, when these interactions become inefficient. Thereafter, the gas cools adiabatically. During the Dark Ages, overdensities around primordial seeds keep growing, leading eventually to the formation of galaxies. The newly born stars emit highly energetic radiation capable of altering the IGM evolution.

In order to understand the IGM evolution in detail, full numerical simulations are required, properly computing the hydrodynamics and radiative transfer, which are very expensive computationally wise. However, some semi-analytical approximations can be assumed, which simplify significantly the picture, and allow us to gain insight into the physical processes involved.
Simplifying the full picture, we can treat the IGM as a two-phase medium \cite{Pritchard:2011xb}. One of the phases is the mostly neutral IGM, characterized by the gas temperature $T_K({\bf x}, z)$, and the free electron fraction $x_e({\bf x}, z)$, which accounts for the free electrons remaining from Recombination. After the onset of star formation, energetic X-ray radiation can reach these neutral areas, heating and ionizing the gas, although still remaining mostly neutral. On the other hand, the immediately surrounding regions of galaxies and ionizing sources are expected to be almost completely ionized, forming HII bubbles. Once the star formation processes are efficient enough, these bubbles can grow and eventually merge among them, ionizing the full IGM during the so called Epoch of Reionization (EoR). We characterize the fraction of volume occupied by the HII bubbles with the ionized \textit{filling factor} $Q_i$, where the local ionization fraction is $\sim 1$. In these completely ionized regions, temperatures reach $\sim 10^4$ K, where atomic cooling peaks. Therefore, the global ionization fraction including the (mostly) neutral IGM plus the HII regions can be written as
\begin{equation*}
\bar{x}_i = Q_i + (1-Q_i)\bar{x}_e,
\end{equation*}
where $\bar{x}_e$ is the spatial average over $x_e({\bf x}, z)$ (in the neutral IGM). Hence, during the EoR, the mostly neutral IGM becomes completely ionized. The two-phase IGM is a simplification of the real IGM, but accurate enough and widely employed in semi-analytical computations, as in the code {\tt 21cmFAST} \cite{Mesinger:2010ne, Park:2018ljd}, which is widely employed throughout this thesis. In this chapter, we review the evolution of the ionized fraction and gas temperature at each phase, discussing the different sources of energy injection, as well as the current constraints on the reionization epoch. For further details, there are many reviews in the literature treating the physics of reionization and the IGM, see Refs. \cite{Barkana:2000fd, Meiksin:2007rz, 2009CSci...97..841C, 2016SAAS...43....1G, Ferrara:2014sda, Wise_2019}.

\section{Ionized IGM}
\label{sec:ionIGM}

In this section, we discuss the details of the reionization processes, starting by considering only the globally averaged ionized fraction, and commenting latter on the effects of inhomogeneities. Current observational constraints on the EoR are summarized at the end of the section.

\subsection{Global Reionization}
\label{sec:ionIGMglobal}


The evolution of the HII number density $n_{\rm HII}({\bf x},t)$ is given by its continuity equation (e.g., \cite{2009CSci...97..841C}):
%
\begin{equation}
\frac{d n_{\rm HII}}{d t} + 3H n_{\rm HII} =  \Gamma_{\rm HI} n_{\rm HI} - \alpha_A  n_{\rm HII} n_e,
\label{eq:nHII}
\end{equation}
where the left-hand side stands for the adiabatic evolution and the right-hand side accounts for the interplay between ionizations and recombinations, being $\Gamma_{\rm HI}$ the \textit{ionization rate} and $\alpha_A$ the so-called \textit{case-A recombination} coefficient. Peculiar velocities may also be taken into account, but we neglect them in this derivation for the sake of simplicity. The recombination coefficient considered depends upon the medium. In neutral IGM, ionizing photons are rapidly absorbed, and transitions to the ground state do not effectively contribute to the ionization. This is the so-called \textit{case-B recombination}, where recombinations to the ground state are excluded. Contrarily, the HII phase is a thin medium where these UV photons can freely travel through. For this reason, employing case-A recombination there is more adequate, such that recombinations to the lowest state are taken into account \cite{Furlanetto:2006jb}. In this case, the coefficient takes the value $\alpha_A = 4.2 \times 10^{-13}$cm$^3$ s$^{-1}$, evaluated at $T_K = 10^4$ K, a factor of $\sim 2$ larger than case-B.

We characterize the state of ionization of the IGM by means of the fraction of ionized volume, defining the filling factor as the volume-average of the HII fraction in the ionized medium: $Q_i = \langle x_{\rm HII} \rangle$, where $x_{\rm HII} = n_{\rm HII}/n_{\rm H}$. This quantity provides the fraction of volume of the IGM which is completely ionized by UV photons. To obtain an evolution equation for $Q_i$, we must spatially average the above equation. Since in this phase, both Hydrogen and Helium are mostly ionized, $n_e \simeq (1+\chi) n_{\rm HII}$, with $\chi = \mathfrak{f}_{\rm He}/\mathfrak{f}_{\rm H}$, with $\mathfrak{f}_{\rm H}=n_{\rm H}/n_b=0.92$ and $\mathfrak{f}_{\rm He}=n_{\rm He}/n_b=0.08$ the fraction of Hydrogen and Helium atoms per baryon, respectively.\footnote{Doubly-ionized Helium, HeIII, can be usually neglected for the redshifts of interest, due to its high ionization threshold.} Decomposing the HII number density as $n_{\rm HII} = x_{\rm HII}\, \bar{n}_{\rm H}\, (1+\delta)$, with $\delta$ the density contrast respect to the mean value, as defined in Sec. \ref{sec:lineargrowth}, the squared averaged HII density can be written as
\begin{equation}
\langle n_{\rm HII}^2 \rangle \simeq C \, Q_i \, \bar{n}_{\rm H}^2,
\label{eq:clumping}
\end{equation}
where we have approximated $ \langle x_{\rm HII}^2 \rangle \simeq \langle x_{\rm HII} \rangle = Q_i $, given that $x_{\rm HII}$ takes values very close to 1 in these mostly ionized regions. The \textit{clumping factor} $C = \langle n_{\rm H}^2 \rangle / \langle n_{\rm H} \rangle^2 = \langle (1+\delta)^2 \rangle$ accounts for the fact that the IGM is not completely homogeneous and has a density distribution. These inhomogeneities have an impact in the ionization history through this clumping factor, since recombinations are enhanced in overdense regions. There are several prescriptions to compute this coefficient, depending on the specific definition and method of simulation, and it depends upon the scale at which it is evaluated \cite{Finlator_2012, 2014ApJ...789..149S}. While the modeling of the clumping factor is highly challenging, most estimations predict values of order unity at the redshifts of interest \cite{Kaurov_2014, Kaurov_2015}.

We focus now on the ionizing term of Eq. \eqref{eq:nHII}. As long as the mean free path is much smaller than the size of the universe, or equivalently, the opacity $\kappa$ is larger than the expansion rate, $1/\kappa \ll c/H$, ionizing photons are absorbed soon after being released. This is the so-called \textit{on-the-spot} approximation, which can be safely used for the ionizing field at the EoR. Employing $J$ and $\varepsilon$ as the flux and comoving emissivity by number, respectively,\footnote{Note that in Sec. \ref{sec:radtransfer}, $I$ and $\epsilon$ were the flux and emissivity by energy, rather than by number, i.e., with an additional factor $h\nu$.} this simplification implies that Eq. \eqref{eq:radcosmo} can be replaced by 
\begin{equation}
0\simeq - c\kappa  J + \frac{c(1+z)^3\varepsilon}{4\pi}.
\end{equation}
Hence, the ionizing UV flux, $J_{\rm UV}(\nu)$ (in units of s$^{-1}$ keV$^{-1}$ cm$^{-2}$ sr$^{-1}$), reads
\begin{equation}
J_{\rm UV}(\nu) \simeq \frac{(1+z)^3\varepsilon(\nu)}{4 \pi \kappa(\nu)}
\label{eq:onthespot}
\end{equation}
with $ \kappa(\nu) = \sigma_{\rm HI}(\nu)n_{\rm HI}$ and $\sigma_{\rm HI}(\nu)$ the photoionization cross section for Hydrogen. Introducing the above flux in the source term of Eq. \eqref{eq:nHII}, one obtains: 
\begin{equation}
n_{\rm HI} \Gamma_{\rm HI}  = 4 \pi n_{\rm HI} \int^{\infty}_{\nu_{\rm HI}} d\nu \, \sigma_{\rm HI}(\nu) J_{\rm UV}(\nu) \simeq (1+z)^3 \int^{\infty}_{\nu_{\rm HI}} d\nu \, \varepsilon(\nu) = \dot{n}_{\gamma},
\label{eq:ionrate}
\end{equation}
with $\dot{n}_{\gamma}$ the proper rate density of emitted photons and $h\nu_{\rm HI}=13.6$ eV the threshold ionization energy for Hydrogen. Thus, spatially averaging Eq. \eqref{eq:nHII}, and making use of the results from Eqs. \eqref{eq:clumping} and \eqref{eq:ionrate}, we obtain the evolution equation for the filling factor,
\begin{equation}
\frac{d Q_i}{d t} = \frac{\dot{n}_{\gamma}}{\bar{n}_{\rm H}} - \frac{Q_i}{\bar{t}_{rec}},
\label{eq:dotQ}
\end{equation}
being $\bar{t}_{rec}^{-1} = \alpha_A (1+\chi) C \bar{n}_{\rm H}$ the mean recombination time. The source term $\dot{n}_{\gamma}$ is completely determined by the emissivity, which can be modeled as proportional to the comoving \textit{star formation rate} $\dot{\rho}_*$,
\begin{equation}
\varepsilon(\nu) = \mathcal{N}(\nu) f_{esc}\frac{\dot{\rho}_*}{\mu m_p} = \mathcal{N}(\nu) f_*f_{esc} \bar{n}_{b,0} \frac{df_{coll}(z,M_*)}{dt},
\label{eq:emissivityUV}
\end{equation}
with $f_*$ the fraction of baryon into stars, $f_{esc}$ the fraction of photons which escape from the galaxies to the IGM, $\mu$ the mean baryonic weight and $\mathcal{N}(\nu)$ the the number of emitted photons per baryon per frequency unit, which accounts for the energy spectrum and it is usually assumed to be a power-law function. In the second equality we have written the star formation rate $\dot{\rho}_*$ in terms of the derivative of the the fraction of mass collapsed in halos above a mass $M_*$, $f_{coll}$, defined in Eq. \eqref{eq:fcoll}. Here, since we assume that ionizing photons come from stars within galaxies, $M_{*}(z)$ should be taken as the minimum mass required to host star formation in a halo. Therefore, only halos with masses large enough to host star formation can contribute to the UV flux. Using the Press-Schechter prescription, Eq. \eqref{eq:PShmf}, the fraction of collapsed mass can be computed analytically, obtaining Eq. \eqref{eq:fcollPS}, while for the Sheth-Tormen function, Eq. \eqref{eq:SThmf}, it must be computed numerically. This threshold can be related to the \textit{minimum virial mass} $M_{\rm vir}^{\rm min}$ corresponding to a virial temperature $T_{\rm vir}^{\rm min}$, which can be defined from Eq. \eqref{eq:Tvir} as \cite{Barkana:2000fd}
 \begin{equation}
M_{\rm vir}^{\rm min} (z) = 10^8 \left(\frac{T_{\rm vir}^{\rm min}}{1.98 \times 10^4 \, {\rm K}} \frac{0.6}{\mu} \right)^{3/2} \left(\frac{1+z}{10}\right)^{-3/2} M_\odot/h ~,
\label{eq:m_vir}
\end{equation}
where $\mu$ is the mean molecular weight which is equal to $1.2$ ($0.6$) for a neutral (fully ionized) primordial gas. The minimum virial temperature is usually chosen at the atomic cooling threshold, $T_{\rm vir}^{\rm min} \simeq 10^4$ K, minimum value required to have an efficient atomic cooling, allowing the gas to fragment, condense and form stars. However, molecular cooling would be relevant in metal-free clouds, cradle of the so-called Population III stars, a hypothetical first generation of very massive stars lacking metals. The formation of H$_2$ would present the most important mechanism to lose thermal energy, leading to lower values of the temperature, down to $\sim 10^3$ K \cite{Barkana:2000fd, Wise_2019}. Given the uncertainty on the stellar population and cooling mechanism in halos, in the following we consider either $T_{\rm vir}^{\rm min}$ or $M_{\rm vir}^{\rm min}$ as a free parameter. The density rate of ionizing photons is then
\begin{equation}
\dot{n}_{\gamma}(z) = N_{\gamma/b} f_* f_{esc} \bar{n}_{b}(z) \frac{df_{coll}(z,M_{\rm vir}^{\rm min})}{dt},
\end{equation}
being $N_{\gamma/b} = \int_{\nu_{\rm HI}} d\nu \mathcal{N}(\nu)$ the number of ionizing photons per baryon. Note that this formalism is specially appropriate for high redshifts, since at later times, $z < 10$, mergers of halos can also contribute to star formation \cite{Pritchard:2011xb}. Eq. \eqref{eq:dotQ} can thus be rewritten as
\begin{equation}
\frac{d Q_i}{d t} = \xi \frac{d f_{coll}}{dt} - \frac{Q_i}{\bar{t}_{rec}},
\label{eq:dotQfinal}
\end{equation}
being $\xi$ the \emph{ionization efficiency}, defined as the product of the astrophysical parameters
\begin{equation}
\xi = N_{\gamma/b}f_* f_{esc}\frac{1}{\mathfrak{f}_{\rm H}}.
\end{equation}
This is the final equation that must be solved in order to predict the evolution of the ionized medium during the reionization era.\footnote{Eq. \eqref{eq:dotQfinal} can also apply to the post-reionization era by accounting for the residual neutral clouds in the ionized IGM as an extra term in the opacity, correctly reproducing the redshift evolution \cite{Madau:2017wny}.} A further simplification can be taken when recombinations are slow enough. Assuming that the product $N_{\gamma/b}f_* f_{esc}$ is redshift independent, we can formally solve Eq. \eqref{eq:dotQfinal} as \cite{Furlanetto:2006tf}
\begin{equation}
Q_i(z) = \frac{\xi \; f_{coll}(z,M_{\rm vir}^{\rm min})}{1+N_{rec}},
\label{eq:qifinal}
\end{equation}
where $N_{rec}=Q_i(z)^{-1}\int_z (dt/dz') dz'Q_i(z')\bar{t}_{rec}^{-1}(z')$ is the mean cumulative number of recombinations per Hydrogen atom. The filling factor is therefore proportional to the fraction of matter collapsed in star-forming halos, with an efficiency factor given by a product of astrophysical parameters. In principle, these parameters may be redshift and mass dependent, but due to the uncertainties in their values and for the sake of simplicity, we shall consider them as constant both in mass and in redshift. The number of ionizing photons per baryon $N_{\gamma/b}$ can be estimated from the spectrum and the initial mass function of stars, which depend upon the details of star formation (burst formation or continuous) and on their metallicity \cite{Bromm_2001, Zackrisson_2011}. Making use of the present-day initial mass function of low-metallicity stars, known as Population II stars, $N_{\gamma/b} \sim 4 \times 10^3$. Nonetheless, it can reach $\sim 10^5$ if metal-free and very massive stars of $> 10^2M_\odot$, the so-called Population III, are considered as the initial composition. Regarding the fraction of baryons into stars, previous works on the comparison of the star formation rate with the one derived from UV luminosity function measurements (see, e.g., Refs. \cite{Leite:2017ekt, Lidz:2018fqo} and references therein), as well as radiation-hydrodynamic simulations of high-redshift galaxies (see, e.g., Ref. \cite{Wise:2014vwa}), have found values of this quantity around $f_* \sim 0.01$. On the other hand, most photons produced in a galaxy would remain within, being consumed in ionizing its neutral matter. Nonetheless, a small fraction of photons, encoded by $f_{esc}$, could escape from the galactic boundaries into the IGM and contribute to the reionization mechanism. The estimation of this parameter is challenging since it depends on the distribution of column densities and galactic morphologies. Although it is neither well observed nor well modeled, it takes typical values of $f_{esc} \sim 0.1$ (see, e.g., Ref. \cite{2009CSci...97..841C} and references therein). Finally, in Eq. \eqref{eq:qifinal}, the mean number of recombinations $N_{rec}$ can also be taken as a free parameter, provided that it does not vary much with time. It is expected to take low values, $N_{rec}<1$ (e.g., Ref. \cite{Mesinger:2012ys}). Note that, at least as far as ionization is concerned, the relevant quantity is the product of these parameters, and not the specific value of each one. It also implies degeneracies between these parameters, which could only be broken if other observables besides the ionization fraction are considered. Due to the lack of observations of galaxies at high redshifts, these parameters are still poorly constrained and remain highly uncertain, and thus $\xi$ is usually taken as a free parameter. For the default values commented above with Population II stars, the fiducial value for the ionizing efficiency is $\xi \simeq 40$. Therefore, from Eqs. \eqref{eq:dotQ} or \eqref{eq:qifinal}, one can conclude that the parameters $\xi$ and $M_{\rm vir}^{\rm min}$ (or $T_{\rm vir}^{\rm min}$) rule the ionization history, and completely determine the filling factor in this simple picture of global ionization.

\begin{figure}
\centering
\includegraphics[scale=0.9]{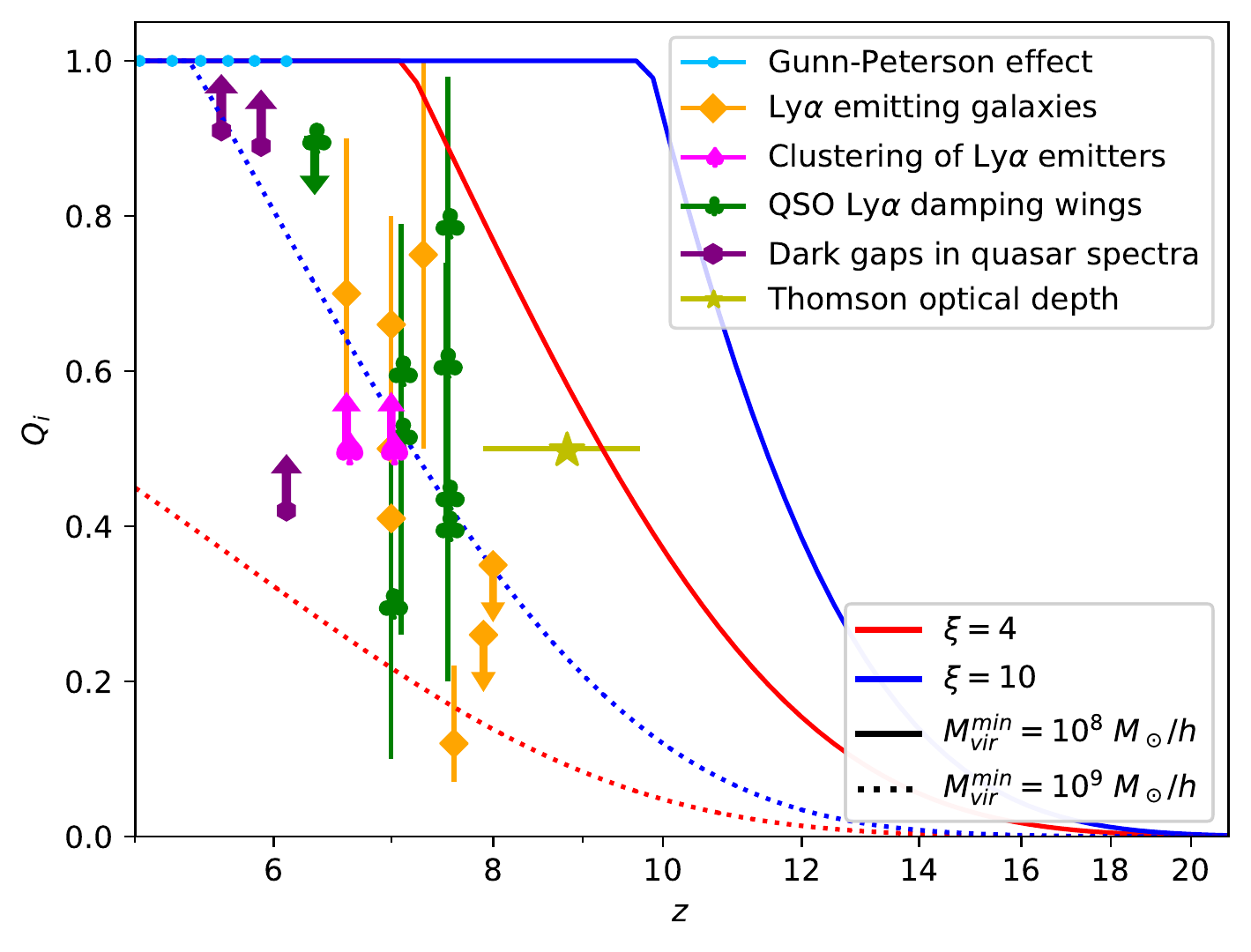}
\captionof{figure}{Evolution of the filling factor $Q_i$ as a function of redshift, for different values of $M_{\rm vir}^{\rm min}$ and $\xi$. The included observational data is listed in Tab. \ref{tab:tab_eor} and explained in Sec. \ref{sec:constraintsEoR}, arising from different probes: Gunn-Peterson effect, Ly$\alpha$ emission in galaxies, damping wings in Ly$\alpha$ spectra from QSOs, dark gaps in quasar spectra and estimates from the Thomson optical depth of the CMB. The upwards (downwards) arrows denote lower (upper) limits on $Q_i$.}
\label{fig:fillingfactor}
\end{figure}

To roughly estimate the conditions at which the IGM becomes completely ionized, one can see in Fig. \ref{fig:fcoll} that $f_{coll}(z,M_{\rm vir}^{\rm min}) \sim 0.1$ at $z\sim 6$ for $T_{\rm vir}^{\rm min}=10^4$ K. Thus, from Eq. \eqref{eq:qifinal}, to ensure that the IGM becomes completely ionized at that redshift, as expected from data, it is required that $\xi \gtrsim 10$, which roughly agrees with the above estimation with the fiducial values. To properly evaluate the evolution of the filling factor, one must numerically solve Eqs. \eqref{eq:dotQ} or \eqref{eq:qifinal}. Examples of ionization histories, computed with {\tt 21cmFAST} \cite{Mesinger:2010ne, Park:2018ljd}, can be found in Fig. \ref{fig:fillingfactor}, for several values of the relevant astrophysical parameters, $\xi$ and $M^{\rm min}_{\rm vir}$. The plot shows also the current observational constraints on the ionized fraction, which will be explained in Sec. \ref{sec:constraintsEoR}. The minimum virial masses of $M^{\rm min}_{\rm vir}=10^8$ and $10^9 \, M_\odot/h$ have been chosen to correspond to minimum virial temperatures of $T_{\rm min}^{\rm vir} = 10^4$ (atomic cooling threshold) and $10^5$ K respectively at $1+z \simeq 10$. On the other hand, the considered values of the ionizing factor $\xi =4$ and $10$ may be related to Population II ($N_{\gamma/b}=4 \times 10^3$) and Population III ($N_{\gamma/b}=10^5$) stars, respectively, in case $f_*\simeq 0.01$ and $f_{esc}\simeq 0.1$. Notice that both parameters are somewhat degenerate, since increasing (decreasing) the ionizing efficiency (the minimum virial mass) leads to an earlier reionization period. Efficiencies of $\xi \gtrsim 4$ with threshold masses of $\lesssim 10^9 \, M_\odot/h$ are needed in order to have the universe ionized at $z \sim 6$, as indicated by data (see Sec. \ref{sec:constraintsEoR}). Molecular cooling, with $T_{\rm min}^{\rm vir} \sim 10^3$ K, may lead to a very early reionization era inconsistent with observations, unless extremely low ionizing efficiencies were considered.

Many candidates have been considered as the sources of reionization. QSOs (Quasi-Stellar Objects, also known as quasars) have been largely considered, since they are among the brightest sources in the universe, being able to emit in the UV range. However, their number density drops at high redshifts, and seems to be not enough to contribute significantly to the global ionization of the IGM, accounting only for 1-5\% of the required ionizing background \cite{Grissom:2013mea}. Standard galaxies become then the preferred option, since we have observational confirmation of bright galaxies for $z \gtrsim 6$ up to $z\sim 11$ thanks to the Hubble Space Telescope (HST) Ultra Deep Field \cite{2013ApJ...763L...7E, Oesch_2016}. Future telescopes such as the JWST \cite{Gardner:2006ky} could be able to detect farther and fainter galaxies. Given the slope of the typical luminosity function, dwarf faint galaxies may be the most relevant ones for producing an ionizing background due to their high abundance \cite{Wise_2019}. Regarding the stellar components, early redshift stars responsible for reionization may belong to the Population II group, whose stars present low metallicity, about one tenth of the solar metallicity. However, the first stars of the universe could have had even lower metallicities, being mostly composed only by Hydrogen and Helium, the most abundant elements after Primordial Nucleosynthesis. The so-called Population III stars would constitute a metal-free first stellar generation, rather different to the other known stellar types \cite{Wise_2019}. The lack of heavy elements reduces the cooling capability of the initial collapsing cloud, leading to much more massive stars than the usually observed, with hotter central temperatures. As stated above, the number of emitted photons could be larger than in the Population II case. Moreover, the cooling of the gas happens by $H_2$ formation, leading to lower virial temperatures. This kind of stars has not been directly observed, yet some indirect evidences are already present \cite{Sobral:2015rka}.

\subsection{Inhomogeneous Reionization}

The discussion above stands for a global reionization scenario, and it is useful to predict the average ionized volume. However, numerical simulations agree on the fact that reionization is highly inhomogeneous \cite{Zahn_2011, Gnedin_2014}. In principle, only hydrodynamic simulations with detailed radiative transfer can account for the complex physics of the inhomogeneous reionization process. However, several semi-analytical procedures have been proposed in the literature in order to account for inhomogeneities with good agreement with numerical simulations (see, e.g., Ref. \cite{Kaurov_2016} for a comparison). In Ref. \cite{Furlanetto:2004nh}, the \emph{FZH model}\footnote{After its authors Furlanetto, Zaldarriaga and Hernquist.} was proposed in order to understand the growth of ionized regions, making use of the excursion set formalism. The basic idea is to promote Eq. \eqref{eq:qifinal} to be scale dependent, writing the filling factor $Q_i(z;\delta,M)$ of a region with a mass $M$ and an overdensity $\delta$ (smoothed within the scale given by $M$) as
\begin{equation}
Q_i(z;\delta,M) = \frac{\xi \; f_{coll,cond}(z,M_{\rm vir}^{\rm min};\delta,M)}{1+N_{rec}},
\end{equation}
where the global collapsed fraction has been replaced by the \emph{conditional} one, $f_{coll,cond}(z, M_{\rm vir}^{\rm min};\delta, M)$, the fraction of mass in collapsed objects above $M_{*}$ at a redshift $z$ within a region of mass scale $M$ and overdensity $\delta$. For the Press-Schechter function, this reads
\begin{equation}
f_{coll,cond}(z, M_{\rm vir}^{\rm min};\delta, M)= erfc \left(\frac{\delta_c(z)-\delta}{\sqrt{2(\sigma(M_{\rm vir}^{\rm min})^2-\sigma(M)^2)}}\right).
\end{equation}
This is equivalent to assume that the ionized mass is proportional to the mass of collapsed objects, with the proportionality constant $\xi/(1+N_{rec})$. Therefore, as done in the standard excursion-set formalism \cite{Bond:1990iw}, a point of space ${\bf x}$ with a density contrast $\delta({\bf x})$ smoothed over a scale $M$ is asigned to an ionized region of mass $M$ if $M$ is the largest scale at which the condition $Q_i(z;\delta({\bf x}),M) \geq 1$ is fulfilled. This is equivalent to the criterion $\xi f_{coll,cond}(z,M_{\rm vir}^{\rm min};\delta({\bf x}),M)/(1+N_{rec}) \geq 1$, where this last inequality imposes a condition over $\delta$ in the ionized bubbles, implying that only regions dense enough are capable of self-ionize, i.e the condition:
\begin{equation}
\delta \geq \delta_c(z) - erfc^{-1}\left((1+N_{rec})/\xi)\sqrt{2(\sigma(M_*)^2-\sigma(M)^2)}\right)
\end{equation}
must hold within the region in order to be ionized. Note that the expression above is valid only for the PS prescription. For these reasons, this is considerered as an \textit{inside-out} scenario of reionization, where, on average, high-density regions are firstly ionized. The threshold overdensity at a scale $M$ can be considered a moving barrier to construct a HII mass function in the excursion set formalism \cite{Furlanetto:2004nh}. This prescription is well suited for numerical implementation and it is computationallt very efficient, since it only needs to filter the density field, marking as ionized the cells above the threshold at a given filter scale. For this reason, it is the method used in many codes such as {\tt 21cmFAST}~\cite{Mesinger:2010ne}. It has been shown to be in good agreement with more computationally expensive radiative transfer codes at scales $\gtrsim 1$ Mpc \cite{Zahn:2010yw}.

There are other analytical models to treat inhomogenous reionization, such as the known as \emph{MHR model}\footnote{After Miralda-Escudé, Haehnelt and Rees.} proposed in \cite{2000ApJ...530....1M}. In this prescription, it is assumed that only regions with overdensities above a certain treshold remain neutral, as a consequence of an enhanced recombination rate. Contrarily to the FZH model, this is an \textit{outside-in} scenario, where the most dense regions ionize later. Although it may seem in contradiction with FZH, the application of MHR is appropriate in the later stages of reionization, when only small overdense spots of the IGM remain neutral. Moreover, it is evaluated in a single, small scale, instead of the multiscale approach of FZH, which is able to describe the morphology  of the reionization process on large scales. Both methods can be combined and provide an accurate description of the reionization from different perspectives \cite{Kaurov_2016}.

Finally, although our focus in this thesis is on semi-analytic methods to solve the ionization history, a proper determination of the IGM evolution requires more accurate though computationally expensive methods. Some of them involve N-body simulations to compute the density fields and halo population, combined with a detailed radiative transfer treatment, via, e.g., adaptive ray tracing algorithms \cite{Trac:2006vr, Dixon:2015fcm}. However, hydrodynamic radiative transfer simulations are the most reliable at the smallest scales, able to reach spatial resolutions of $\sim 100$ pc - $\sim 10$ kpc, although are limited to boxes of tens of Mpc. They account for different baryonic effects absent in other prescriptions, which may include stellar and supernovae feedback, star formation, fully coupled three-dimensional radiative transfer and detailed cooling and heating for the metal enriched gas \cite{Gnedin:2014uta, 2016ApJ...833...84X, Doussot:2017cee}. Radiative transfer can be treated by three main approaches, namely employing moments methods (e.g., \cite{2001NewA....6..437G}), Monte Carlo sampling (e.g., \cite{Ciardi:2000wv}) or ray tracing (e.g., \cite{2011MNRAS.414.3458W}). See Refs. \cite{2006MNRAS.371.1057I, 2009MNRAS.400.1283I} for detailed comparisons between codes and radiative transfer methods, and also, e.g., Refs. \cite{2011ASL.....4..228T, Mesinger:2018ndr} for overviews and more details on this topic.

\subsection{Constraints on the Epoch of Reionization}

\label{sec:constraintsEoR}

\begin{table}
\begin{center}
{\def\arraystretch{1.4}
\resizebox{\textwidth}{!}{
\begin{tabular}{|c||c|c|c| } 
 \hline
 \, Data \, & \, Redshift \, & \, $Q_{i}$ \, & \, Reference \, \\
 \hline
 \hline
 \multirow{6}{*}{Gunn-Peterson effect}
 & 5.03 & \, 0.9999451$_{-0.0000165}^{+0.0000142}$ \, & \multirow{6}{*}{Fan et al., 2006 \cite{Fan:2005es}} \\
 & 5.25 & 0.9999330$_{-0.0000244}^{+0.0000207}$ &  \\
 & 5.45 & 0.9999333$_{-0.0000301}^{+0.0000247}$  &  \\
 & 5.65 & 0.9999140$_{-0.0000460}^{+0.0000365}$  &  \\
 & 5.85 & 0.9998800$_{-0.0000490}^{+0.0000408}$  &  \\
 & 6.10 & 0.99957$\pm$0.00030 &  \\
 \hline
 \multirow{3}{*}{Dark gaps in quasar spectra}
 & 5.6 & $>$0.91 & \multirow{3}{*}{McGreer et al., 2015 \cite{McGreer:2014qwa}} \\
 & 5.9 &  $>$0.89 &  \\
 & 6.1 &  $>$0.42 &  \\
 \hline
 \multirow{8}{*}{\, Ly$\alpha$ Emission in Galaxies \, }
 & 7 & 0.66 $^{+0.12}_{-0.09}$ & \multirow{2}{*}{Schenker et al., 2014 \cite{Schenker:2014tda}} \\ 
 & 8 & $< 0.35$ & \\ 
 \cline{2-4}
  & 6.6 & $0.7\pm 0.2$ & Konno et al., 2018 \cite{2018PASJ...70S..16K} \\
   & 7 & $0.5^{+0.3}_{-0.1}$ & Inoue et al., 2018 \cite{2018PASJ...70...55I} \\
  & $\sim 7$ & $0.41^{+0.15}_{-0.11}$ & Mason et al., 2018 \cite{Mason:2017eqr} \\
 & 7.3 & $0.75\pm 0.25$ & Itoh et al., 2018 \cite{2018ApJ...867...46I} \\
  & $7.6 \pm 0.6$ & $0.12^{+0.10}_{-0.05}$ & Hoag et al., 2019 \cite{2019ApJ...878...12H} \\
  & $7.9 \pm 0.6$ & $<0.26$ & Mason et al., 2019 \cite{Mason:2019ixe} \\
  \hline
  \multirow{3}{*}{\, Clustering of Ly$\alpha$ Emitters \, }
  & 6.6 & $ \gtrsim 0.5$ & McQuinn et al., 2007 \cite{2007MNRAS.381...75M} \\
   & 6.6 & $ \gtrsim 0.5$ & Ouchi et al., 2010 \cite{2010ApJ...723..869O} \\
 & 7 & $ \gtrsim 0.5$ & Sobacchi and Mesinger, 2015 \cite{2015MNRAS.453.1843S} \\
  \hline 
  \multirow{8}{*}{\, QSO Ly$\alpha$ damping wings \, }
 & 6.24-6.42 & <0.9 (2$\sigma$) & Schroeder et al., 2013 \cite{Schroeder:2012uy} \\
 & 7.00 & $0.30^{+0.23}_{-0.20}$ & Wang et al., 2020 \cite{2020ApJ...896...23W} \\
  & 7.09 & $0.60^{+0.19}_{-0.21}$ & Greig et al., 2017 \cite{Greig:2016vpu} \\
  & 7.52 & $0.61^{+0.13}_{-0.22}$ & Yang et al., 2020 \cite{2020ApJ...897L..14Y} \\
   & 7.54 & $0.79^{+0.19}_{-0.17}$ & Greig et al., 2018 \cite{Greig:2018rts} \\
   & 7.54 & $0.44^{+0.18}_{-0.21}$ (95\% CL) & Bañados et al., 2018 \cite{Banados:2017unc} \\
   \cline{2-4}
    & 7.09 & $0.52 \pm 0.26$ & \multirow{2}{*}{Davies et al., 2018 \cite{Davies:2018yfp}} \\
 & 7.54 & $0.40^{+0.23}_{-0.20}$ & \\ 
  \hline
 \multirow{2}{*}{\, Thomson Optical Depth \,}
 & $8.8\pm 0.9$ & 0.5 & Adam et al., 2016 \cite{Adam:2016hgk} \\
 \cline{2-4}
 & \multicolumn{2}{c|}{$\tau_T = 0.0519^{+0.0030}_{-0.0079}$}  & Aghanim et al., 2020 \cite{Aghanim:2018eyx}\\ 
\hline
\end{tabular}
}
}
\caption{List of representative constraints on $Q_i$ at $1\sigma$ (unless stated otherwise). Data points are plotted in Fig. \ref{fig:fillingfactor}, see, e.g., Refs. \cite{Robertson:2013bq, Bouwens:2015vha} for comprehensive (though slightly outdated) compilations of data.}
\label{tab:tab_eor}
\end{center}
\end{table}

Currently, several estimations of the global ionized fraction of the IGM at different redshifts have been performed. Although the end of reionization is well determined, the precise evolution at lower redshifts becomes much more uncertain, being some of the estimations strongly dependent on the underlying assumed models. In this section we review some of the most important methods to bound the onset and timing of the EoR. Table \ref{tab:tab_eor} shows a summary of the most important constraints, which are also plotted in Fig. \ref{fig:fillingfactor}.

\subsubsection{Gunn-Peterson effect}

The first, and one of the most important probes of the EoR, is the Gunn-Peterson effect, named after the physicists who predicted it in 1965~\cite{1965ApJ...142.1633G}. Due to the fact that most of the Hydrogen atoms in the IGM are in their fundamental state, the relevant transitions are those which relate its ground state with the excited ones, as the Ly$\alpha$ transition. Many astrophysical objects like galaxies and quasars can emit high energy radiation which, due to the expansion of the universe, redshift to lower frequencies along its journey. In their passing across the IGM, these photons could eventually redshift to the wavelength corresponding to the Ly$\alpha$ transition. Thus, any neutral Hydrogen atom around could absorb it, preventing us from seeing it. Since the IGM is densely filled by Hydrogen atoms, these absorptions would be likely even if a considerable fraction of Hydrogen atoms were ionized. The efficiency of this absorption is given by the optical depth of the Ly$\alpha$ transition, the so-called Gunn-Peterson optical depth $\tau_{\rm GP}$, already computed in Eq. \eqref{eq:GPopticaldepth}, which can be approximated as $\tau_{\rm GP} \simeq 5.0 \times 10^5 \, x_{\rm HI} \left((1+z)/7\right)^{3/2}$. From that formula, one can clearly see that the medium is optically thick ($\tau_{\rm GP} > 1$) unless it is highly ionized, e.g. $x_{\rm HI} < 10^{-5}$ at $z \sim 6$. Hence, observing QSO spectra  embedded in a mostly neutral IGM, we would see an absorption trough (known as \textit{Gunn-Peterson trough}) just above the Ly$\alpha$ transition (since photons are redshifted from higher frequencies). However, we actually can observe light from relatively distant quasars without this expected suppression, up to $z \sim 6$, when a significant damping in the spectrum appears. This means that, for $z \lesssim 6$, essentially there is no neutral gas in the IGM, being most of the atoms in a highly ionized state. For higher redshifts, however, the medium appears to be still predominantly neutral. This transition between these two regimes is the so-called EoR. 

The first evidence of this GP trough at $z\sim 6$ was reported in Ref. \cite{2001AJ....122.2850B} from spectroscopic observations of quasars in the Keck Observatory. The Gunn-Peterson effect has been extensively used in order to constrain the end of this epoch, observing the light emitted by distant quasars. The detection of the Gunn-Peterson trough in the spectra allows us to determine the ionized fraction with great precision. This method was used in Ref. \cite{Fan:2005es} to estimate the neutral fraction $1-Q_i$ between $\sim 10^{-5}$ and $\sim 10^{-4}$ at six different redshifts $z = 5.03,$ $5.25,$ $5.45,$ $4.65,$ $5.85$ and $6.10$, by means of the observation of spectra of 19 bright quasars. These measurements are not, however, model-independent, since assumptions such as the temperature-density relation of the IGM or the PDF of overdensities were assumed.

On the other hand, there is a related method to constrain the ionized fraction less subject to the modeling of the IGM. In Ref. \cite{McGreer:2014qwa}, lower limits on $Q_{i}$ were placed from the distribution of dark gaps in quasar spectra at redshifts $5.6$, $5.9$ and $6.1$ \cite{McGreer:2014qwa} (indicated in Tab. \ref{tab:tab_eor} as \textit{dark gaps in quasar spectra}). These results were achieved by determining the fraction of binned spectra in Ly$\alpha$ and Ly$\beta$ forests which were dark, i.e., presented zero flux. Although less restrictive, these are robust limits obtained in a model-independent way, unlike the estimations from Ref. \cite{Fan:2005es}. All these data indicate that reionization has to be completed by $z \sim 6$. It is worth mentioning that observations of the Ly$\alpha$ forest work under the same idea than the GP effect. While the GP trough is a deep suppression of the spectrum due to a not completely neutral IGM, the Ly$\alpha$ forest arises from neutral clouds of gas embedded in the ionized IGM, which imprint narrow absorption dips in the QSO spectrum determined by their size. This allows to tightly constrain the residual neutral fraction in the post-EoR universe via estimations of the optical depth \cite{Songaila:2004gk, 2008ApJ...681..831F, Becker:2014oga}.

\subsubsection{Thomson Optical Depth}
\label{sec:thomsonopdep}

Once reionization has started, the increase in the number of free electrons makes Thomson scattering to be relevant again, as it happened before the Recombination era. This scattering isotropizes the radiation field, being the angular power spectrum of the CMB suppressed by a damping factor $e^{-2\tau_T}$, where $\tau_T$ is the Thomson optical depth, defined as
\begin{equation}
\tau_T=\int \frac{cdz}{(1+z)H(z)}\, \sigma_T \, n_e(z)~,
\label{eq:opdeptom}
\end{equation}
with $\sigma_T=0.6652 \times 10^{-24}$ cm$^2$ the Thomson cross section. This damping enforces the radiation field to become isotropic for large enough $\tau_T$, erasing anisotropies. As a consequence, the reionization epoch leaves imprints on the CMB spectrum, being $\tau_T$ the quantity which characterizes its impact. This is one of the six parameters of the baseline $\Lambda$CDM cosmological model, determined with great precision by the Planck mission \cite{Akrami:2018vks}. Although in the temperature $TT$ spectrum there is a degeneracy between the Thomson optical depth and the amplitude of the primordial fluctuations, it is broken in the polarization of the CMB spectrum, since Thomson scattering induces polarization at low multipoles \cite{Zaroubi_2012}. Parameterizing the evolution of $Q_i$ with $z$ as a $\tanh$ function, the latest Planck analysis offers a value for this quantity of~\cite{Aghanim:2018eyx}
\begin{equation}
\tau_T = 0.0519^{+0.0030}_{-0.0079}~,
\end{equation}
making use of the low-$l$ $EE$ modes likelihood. Previous Planck analyses \cite{Adam:2016hgk} evaluated also the beginning, the final and the duration of reionization based on a slightly larger, earlier estimation of the optical depth, $\tau_{T,2016}=0.058 \pm 0.012$. Within the $tanh$ parameterization, the reionization process occurs suddenly enough around a characteristic redshift of $z_{re}=8.8 \pm 0.9$ (where the ionization fraction reaches 50\%), with a total duration of $\Delta z< 4.6$ (at $95\%$ CL), with the additional constraint that the IGM must be fully ionized at $z<6$, due to the Gunn-Peterson effect. These results change only slightly if an asymmetric power-law parameterization is assumed, rather than a $tanh$ function \cite{Adam:2016hgk}. Besides the functional parameterization, other approaches have been used in the literature to exploit the $\tau_T$ constraint, such as the PCA (Principal Component Analysis) decomposition \cite{Hu:2003gh, Mortonson:2007hq,Heinrich:2016ojb, Villanueva-Domingo:2017ahx}, the PCHIP (Piecewise Cubic Hermite Interpolating Polynomial) \cite{Villanueva-Domingo:2017ahx} or CMB plus reionization modeling \cite{PhysRevLett.125.071301}. In Part \ref{partII} of this thesis, CMB data is exploited in order to reconstruct the ionization history by means of the aforementioned PCA and PCHIP analyses, studying the plausibility of an early reionization period \cite{Villanueva-Domingo:2017ahx}.

Measurements of the Thomson optical depth constitute a robust indication of reionization and allow us to parameterize some general properties of the EoR. However, in order to infer timing, duration or patchiness aspects of the EoR, specific parameterizations are needed, since this does not provide us with information about the evolution of the free electron fraction at each redshift, but instead about its integral along time.

\subsubsection{Other constraints}

While the Gunn-Peterson trough and the CMB measurements provide the most robust bounds on the EoR, other methods may also lead to constraints on the ionization history. The most relevant ones are summarized in the following.

\begin{itemize}
\item \textbf{Ly$\alpha$ emission in galaxies}

One of the most profitable approaches is based on the observation of Ly$\alpha$ emitting galaxies, usually presenting large star formation rates, which can constitute a successful approach for exploring how reionization has proceeded at higher redshifts ($z \gtrsim 7$) \cite{Dijkstra:2014xta}. The Subaru Hyper Suprime-Cam has provided measurements of thousands of luminosities of Ly$\alpha$ emitters through several wavelength bands, which allows placing constraints on the reionization history. Examples of this are the survey SILVERRUSH \cite{2018PASJ...70S..13O}, from which several estimates of the neutral fraction have been derived at redshifts $z=6.6$ \cite{2018PASJ...70S..16K} and $z =7.3$ \cite{2018PASJ...70...55I}, and the CHORUS survey \cite{2020PASJ...72..101I}, which has provided constraints at $z\simeq 7$ \cite{2018ApJ...867...46I}. Other constraints rely on Ly$\alpha$ emission from faint Lyman break galaxy candidates, those detected by a drop in the spectrum at frequencies above the Lyman limit, caused because ionizing photons are absorbed in their local environment.
The detection and non-detection of Ly$\alpha$ emission in such galaxies allows placing bounds around $z\sim 7$ \cite{Schenker:2014tda, Mason:2017eqr}, $z\sim 7.6$ \cite{2019ApJ...878...12H} and $z\sim 8$ \cite{Schenker:2014tda, Mason:2019ixe}. All the above results indicate that reionization is not yet complete at those epochs, $z \sim 7$. One of the caveats of employing Ly$\alpha$ emitters to infer the ionization fraction is that radiative transfer modeling is needed, these results being, therefore, quite sensitive to the assumptions considered and to the details of the numerical implementations.

\item \textbf{Clustering of Ly$\alpha$ emitters}

Constraints closely related to the above ones can be derived from the clustering properties of Ly$\alpha$ emitters, rather than their luminosity functions. An enhancement in the clustering of Ly$\alpha$ emitters could be regarded as a consequence of reionization, since radiation from clumped galaxies would suffer less absorption than from isolated ones. This effect can be hardly mimicked via other processes, thus indicating the progress of the EoR. It has allowed to place lower limits on the ionized fraction at $z=6.6$ \cite{2007MNRAS.381...75M, 2010ApJ...723..869O} and $z\simeq 7$ \cite{2015MNRAS.453.1843S}. The main difficulty of this method is that these estimates require a great number of Ly$\alpha$ emitters, in order to properly infer the clustering properties.

\item \textbf{Ly$\alpha$ damping wing}

Ly$\alpha$ photons absorbed in the tails (\emph{damping wings}) of the cross section instead of the resonance could also provide a powerful probe of the EoR. Ly$\alpha$ cross section can be modeled by a Voigt profile, Eq. \eqref{eq:voigt}, which near the resonance behaves similar to a Gaussian distribution, but far enough from the center of the line, decays much slower as a Lorentzian function. One can split the optical depth in two different contributions, one from the center of the line, and the other from the tails of the distribution, which is a smoother function of the frequency. The strength of the damping wing absorption strongly depends on the neutral fraction of the environment, fact which allows placing constraints on $Q_i$ \cite{1998ApJ...501...15M, 2004ApJ...613...23M}. Taking advantage of its spectral smoothness, it was possible to set upper limits on $Q_i$ between $z = 6.24$ and $6.42$ \cite{Schroeder:2012uy}. Stronger bounds can be obtained by studying the absorption on the red side of the Ly$\alpha$ line, where resonant absorption is negligible, and thus must be produced at the tails. The discovery of four of the farthest observed quasars has allowed placing constraints using that method at redshifts $z=7.00$ \cite{2020ApJ...896...23W}, $z = 7.09$ \cite{Greig:2016vpu,Davies:2018yfp}, $z=7.52$ \cite{2020ApJ...897L..14Y} and $z= 7.54$ \cite{Greig:2018rts, Banados:2017unc, Davies:2018yfp}. The derivation of all these constraints based on the Ly$\alpha$ damping wing also rely on radiative transfer and hydrodynamical simulations. For the same redshifts, different methodologies and model assumptions may provide differing constraints (see Tab. \ref{tab:tab_eor}).

\end{itemize}

On the other hand, there is recent evidence of overlapping ionized bubbles around three galaxies at $z\sim 7.7$ from spectroscopic observations \cite{Tilvi_2020}. Finally, as commented in the previous chapter, although the 21 cm signal have not been measured at these redshifts, some constraints have been obtained on the duration and completion redshift of the EoR. The EDGES high-band observations are capable of ruling out too abrupt and rapid reionization processes \cite{2017ApJ...847...64M}. Other compilations of ionization data can be found in Refs. \cite{Robertson:2013bq, Bouwens:2015vha}.

\section{Neutral IGM}

In this section, the relevant physics of the neutral IGM is outlined. Before ionized regions around each galaxy grow and merge, the surrounding medium is still in a neutral state, with a remnant ionized fraction from Recombination. The astrophysical processes taking part in galaxies and other sources can produce X-ray radiation capable of reaching areas far beyond the galaxies, altering its thermal and ionization state. Moreover, Ly$\alpha$ radiation is expected to fill the neutral medium, of great importance for the 21 cm signal. In the following, we comment on the heating, ionization and Ly$\alpha$ flux present in the mostly neutral IGM.

\subsection{Heating of the IGM}
\label{sec:heating}

From energy conservation, one can derive a differential equation which determines the kinetic temperature of the gas, $T_K$. The thermal evolution equation reads (e.g., Ref. \cite{1997MNRAS.292...27H})
\begin{equation}
\frac{d T_K}{dt} + 2 H T_K - \frac{2}{3} \frac{T_K}{(1+\delta)}\frac{d \delta}{dt}
 +\frac{T_K}{1+x_e}\frac{d x_e}{dt} = \frac{2  \mathcal{Q}}{3 \, n_b(1+x_e)}~,
\label{eq:eqT}
\end{equation}
where $\mathcal{Q}$ is the total heating rate per unit volume. The second and third terms of the left-hand side account for adiabatic expansion, the fourth one induces changes in the temperature given by the ionization variation, whereas the right-hand side term includes any possible interacting process which can cool down or heat up the gas. Note that, in the absence of heating/cooling $ \mathcal{Q}$ terms, the above equation can be anallitically solved, obtaining $T_K \propto a^{-2}(1+\delta)^{2/3}/(1+x_e)$. In the homogeneous, post-recombination limit, we have the well-known adiabatic cooling evolution $T_K \propto a^{-2}$ as a consequence of the expansion of the universe.

There are several possibles sources of heating and cooling in the IGM which contribute to $\mathcal{Q}$. In the following, we shall comment on the most relevant ones to take into account. While in this section, only the standard processes are considered, further heating terms may arise from exotic DM scenarios, such as accretion in PBHs. The heating effects in that case are widely discussed in Part \ref{partII} of this thesis \cite{Mena:2019nhm}.

\subsubsection{Compton cooling}
\label{sec:compton}

Compton scattering $e + \gamma \leftrightarrow e + \gamma$ is the responsible of the coupling between matter and radiation in the early universe. Even after the decoupling of the CMB, the gas is still affected by scatterings with photons, due to the large number of photons compared to baryons, enforcing the gas temperature $T_K$ to be equal to the CMB temperature $T_\gamma$. The corresponding cooling rate can be derived from the Boltzmann equation for the Compton scattering, after performing a Fokker-Planck expansion (similar to Eq. \eqref{eq:FPexpansion}) and obtaining the so-called Kompaneets equation \cite{1957JETP....4..730K}. Integrating it, the Compton cooling rate is obtained \cite{1993ppc..book.....P},
\begin{equation}
\mathcal{Q} |_{Compton} = \frac{4 \sigma_T n_e \rho_\gamma}{m_e c}   \left( T_\gamma - T_K \right),
\label{eq:comptoncool}
\end{equation}
where $\rho_\gamma = (\pi^2/15)T_\gamma^4$ is the CMB energy density. This result was first derived in Ref. \cite{1965PhFl....8.2112W}. At high redshifts, when Compton cooling is very efficient and there are no other relevant sources of heating, the gas temperature is coupled to the CMB temperature $T_K \rightarrow T_\gamma \propto (1+z)$. However, due to the expansion of the universe, there is a moment when the adiabatic cooling term becomes dominant, and the Compton interactions no longer can couple both temperatures. This thermal decoupling happens at a redshift $z_D$ when both rates (adiabatic and Compton) are similar
\begin{align}
& \left. 2HT_K \right|_{z_D} \sim \left. \frac{2\mathcal{Q} |_{Compton}}{3k_Bn_b(z_D)(1+x_e^\infty)} \right|_{z_D} \;\\
& \Rightarrow \;  1+z_D \simeq \left( \frac{45 m_e H_0 \sqrt{\Omega_m}}{4\pi^2 \sigma_T T_{\gamma,0} x_e^{\infty} } \right) \simeq 150,
\end{align}
where we have employed $x_e^{\infty} \simeq 3 \times 10^{-4}$ as the leftover free electron fraction from Recombination, which is roughly constant, as estimated in Sec. \ref{sec:xe} or computed via publicly available Recombination codes such as RECFAST \cite{1999ApJ...523L...1S}. Therefore, we expect an adiabatic evolution of the kinetic temperature $T_K \propto (1+z)^2$ for redshifts below $z_D \sim 150$, as long as no other heating sources are present.

\subsubsection{X-ray heating}
\label{sec:xrayheating}

After the onset of star formation, some astrophysical objects such as high mass X-ray binaries in starbust galaxies, supernova remnants or black holes in mini-quasars may emit a strong background radiation at UV and X-ray frequencies \cite{2001ApJ...553..499O, Pritchard:2011xb}. Given that the ionization cross section can be approximated as $\sigma(\nu) \propto \nu^{-3}$ (see, e.g., Ref. \cite{2010gfe..book.....M}), the mean free path of photons goes as $\lambda_{mfp} \sim \nu^3$. While UV photons have a lower mean free path and are absorbed in the neighborhood of astrophysical sources, ionizing their environment, the more energetic X-rays can travel much farther, being able to reach the mostly neutral IGM, distant from galaxies. Photons with $\sim$ keV energies can therefore be absorbed, transferring their energy either to ionization, heating or atomic transitions.

Solving the radiative transfer equation for X-rays, Eq. \eqref{eq:radcosmo}, it is possible to obtain obtain the specific X-ray intensity (by number) $J_{\rm X}$ (in units of s$^{-1}$ keV$^{-1}$ cm$^{-2}$ sr$^{-1}$),\footnote{Note that other convention in the literature denotes with $J_{\rm X}$ the flux by energy, rather than by number, having thus units of erg s$^{-1}$ keV$^{-1}$ cm$^{-2}$, and implying an extra factor of $h\nu$. For that quantity, we employ $I$.} which reads
\begin{equation}
J_{\rm X}(z,\nu) = \frac{(1+z)^2}{4\pi}\int_{z}^{\infty} dz' \frac{dt}{dz'} \, (1+z')\varepsilon_{\rm X}(\nu') \, e^{-\tau_{\rm X} (\nu, z, z')}  ~,
\label{eq:JX}
\end{equation}
where $\nu'=\nu(1+z')/(1+z)$ is the comoving frequency at emission, and $\tau_{\rm X}(\alpha,m_e)$ is the IGM optical depth from $z'$ to $z$. This factor accounts for the attenuation of the X-ray flux through the IGM. In analogy to Eq. \eqref{eq:emissivityUV}, the comoving specific X-ray emissivity $\epsilon_{\rm X}(z, \nu)$ is modeled as
\begin{equation}
\varepsilon_{\rm X}(\nu) = \frac{L_{\rm X}(\nu)}{h\nu \, {\rm SFR}} \dot{\rho}_* = \frac{L_{\rm X}(\nu)}{h\nu \,{\rm SFR}} f_* \bar{\rho}_{b,0} \frac{d f_{coll}(z,M_{\rm vir}^{\rm min})}{dt},
\label{eq:specific_emissivity}
\end{equation}
where $L_{\rm X}/{\rm SFR}$ is the specific differential X-ray luminosity normalized to the star formation rate escaping the host galaxies (in units of erg s$^{-1}$ Hz$^{-1}$ $M_\odot^{-1}$ yr), whose spectral shape can be assumed to be a power law, $L_{\rm X} \propto (\nu/\nu_0)^{-\gamma_{\rm X}}$, with $\nu_0$ a threshold frequency. The slope takes values around $\gamma_{\rm X} \simeq 1$, ranging between $0.8$ and $1.5$, depending on the specific X-ray source \cite{2001ApJ...553..499O, Furlanetto:2006tf, Sazonov:2017vtx}. High-redshift X-ray emitters are not directly observed yet, and thus we shall consider the luminosity normalization as a free parameter, given by the integrated luminosity (per SFR) over the energy band $(2 - 10~{\rm keV} )$, $L_{{\rm X} [2 - 10~{\rm keV}]}$/SFR.\footnote{There are other parameterizations to account for the normalization, such as those followed in the early versions of {\tt 21cmFAST} code~\cite{Mesinger:2010ne}. Instead of the luminosity, the heating efficiency $\xi_X$ was chosen, defined as the number of X-ray photons per solar mass in stars, which can be easily related to $L_{{\rm X} [2 - 10~{\rm keV}]}/{\rm SFR}$.} We shall take as a fiducial value the extrapolation from nearby starbusts galaxies, $L_{{\rm X} [2 - 10~{\rm keV}]}/{\rm SFR} = 3.4 \times 10^{40}$ erg s$^{-1}$  $M_\odot^{-1}$ yr \cite{2001ApJ...553..499O, Furlanetto:2006tf}. More recent estimates agree with that default value, showing luminosities from X-ray compact binaries between $10^{39}$ and $10^{40}$ erg s$^{-1}$  $M_\odot^{-1}$ yr between redshifts 0 and 4 \cite{Madau:2016jbv}. The threshold frequency $\nu_0$ is tipically assumed to lie between $h\nu_0=300$ and $500$ eV, but its specific value is less relevant given is degeneracy with $L_{{\rm X} [2 - 10~{\rm keV}]}/{\rm SFR}$ and the slope $\gamma_X$. Frequencies below $\nu_0$ are locally absorbed in the nearby region of the emitting source, and do not reach the IGM. Comparing to Eq. \eqref{eq:specific_emissivity}, one can see that the X-ray escape fraction is effectively set to 1, since most of the X-rays are able to get away galaxies due to the high mean free path.

The heating rate by X-ray sources is thus given by
\begin{equation}
\mathcal{Q} |_{X}(z) = 4\pi\, n_b(z)\int_{\nu_0}^{\infty} d\nu \, J_{\rm X}(z,\nu) \, \sum_{j}(h \nu-E_{j}^{\rm th})f_{\rm heat} \, f_{j} \, x_{j} \, \sigma_{j}(\nu) ~,
\label{eq:Xray_heating_rate}
\end{equation}
where $\sigma_j$ are the photoionization cross-sections of species $j$, and $f_{\rm heat}$ accounts for the fraction of energy from X-rays that is converted into heat in the IGM, using the fit 
from Ref. \cite{2010MNRAS.404.1869F}. Since evaluating the integrals of Eqs. \eqref{eq:JX} and \eqref{eq:Xray_heating_rate} together with the optical depth can be quite time-consuming, some numerical implementations (such as the one followed in the {\tt 21cmFAST} code~\cite{Mesinger:2007pd, Mesinger:2010ne, Park:2018ljd}) approximate the absorption factor $e^{-\tau_{\rm X}}$ to speed-up the computation. Given its exponential form, we can assume that all photons with optical depth large enough, $\tau_{\rm X} > 1$, are easily absorbed in the medium, whereas photons with $\tau_{\rm X} \le 1$ are able to travel without being absorbed. It is equivalent to approximate the absorption factor as a Heaviside step function, $e^{-\tau_{\rm X}} \simeq \theta\left[1 - \tau_{\rm X}(\nu, z, z')\right]$, avoiding to integrate over the optical depth,
\begin{equation}
\begin{split}
\mathcal{Q} |_{X}(z) &= (1+z)^2 n_b(z) \, \int_{z}^{\infty} dz' \frac{dt}{dz'}(1+z') \,\int_{Max[\nu_0,\nu_{\tau=1}]}^{\infty} d\nu  \, \sum_{j} (h \nu-E_{j}^{\rm th})f_{\rm heat} \\
& \times \, f_{j} \, x_{j} \, \sigma_{j}(\nu) \varepsilon_{\rm X}(\nu') ~,
\end{split}
\label{eq:Xray_heating_rate2}
\end{equation}
where $\nu_{\tau=1}$ is the frequency where the optical depth equals unity.

The function $f_{\rm heat}$ accounts for the fraction of X-ray energy deposited as heating. Other amounts of this energy are absorbed in form of ionization and excitation, having therefore the corresponding fractions $f_{\rm ion}$ and $f_{\rm exc}$. Although these functions depend on the X-ray energy and the ionization fraction $x_e$, they saturate at energies $ \gtrsim 100 $ eV, having a function of only $x_e$ \cite{1985ApJ...298..268S, 2010MNRAS.404.1869F}. A rough approximation gives \cite{Chen:2003gz} 
\begin{equation}
f_{\rm heat} \sim \frac{1+2x_e}{3}, \;\;\; f_{\rm ion} \sim f_{\rm exc} \sim \frac{1-x_e}{3}.
\label{ec:f_heationexc}
\end{equation}
Then, in mostly neutral media, the energy splits equally in the three channels, while it is mainly deposited as heating when the ionization is high.

\begin{figure}
\centering
\includegraphics[scale=0.7]{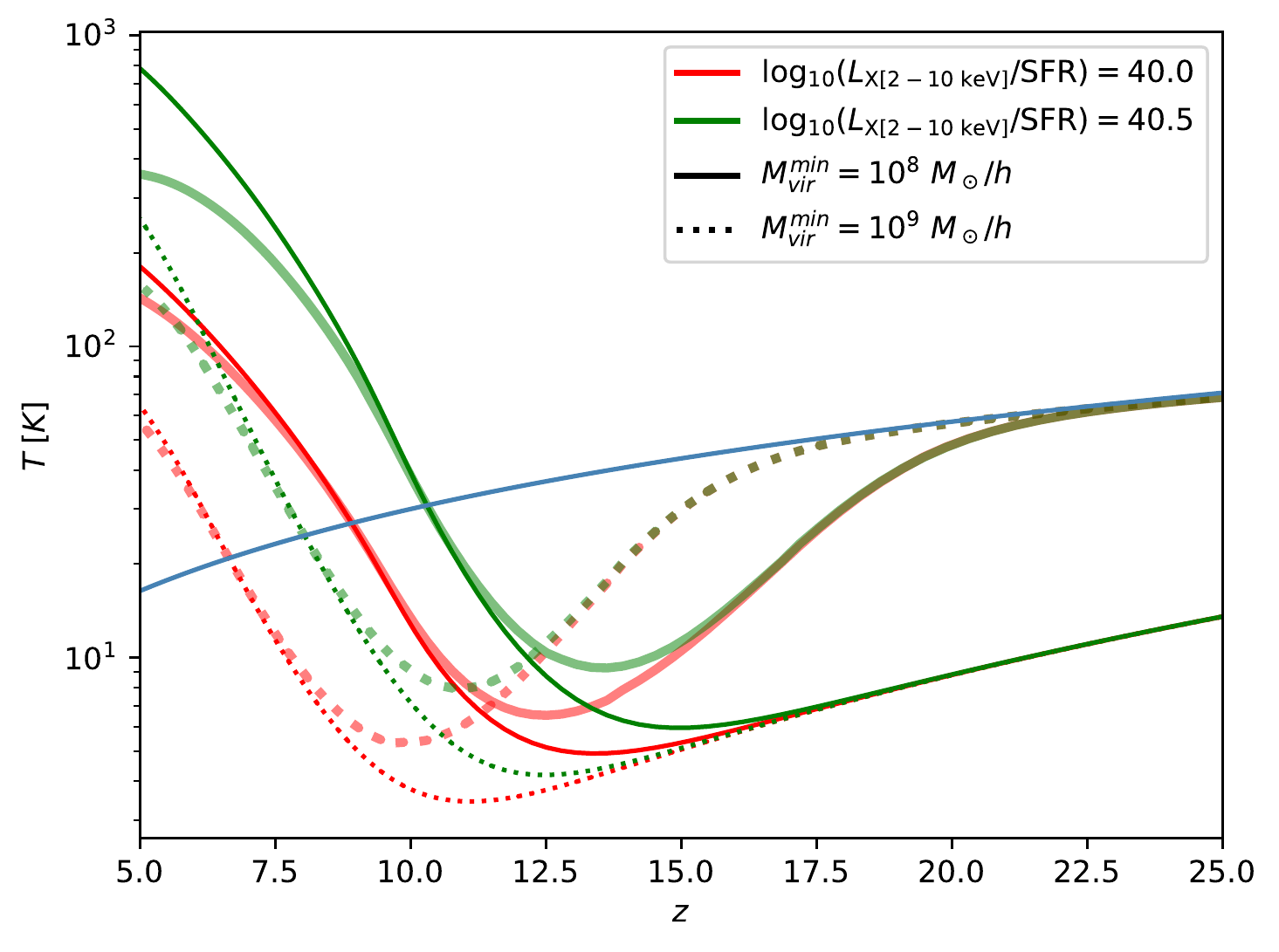}
\captionof{figure}{Evolution of the relevant temperatures as a function of redshift, for different values of the minimum virial mass and X-ray luminosity per star formation rate (in units of erg s$^{-1} M_\odot^{-1}$yr). Thin lines stand for the kinetic temperature of the gas $T_K$, thick lines for the spin temperature $T_S$, while the blue line corresponds to the CMB temperature $T_\gamma$.}
\label{fig:temperature}
\end{figure}

To gain insight about the impact of X-rays in the temperature evolution, we can estimate the total heating by taking the on-the-spot approximation in the flux (i.e., employing Eq. \eqref{eq:onthespot} in Eq. \eqref{eq:Xray_heating_rate}), getting a simpler heating rate (e.g., \cite{Furlanetto:2006tf}),
\begin{align}
\mathcal{Q} |_{X}(z) = \frac{L_{{\rm X} [2 - 10~{\rm keV}]}}{{\rm SFR}}\, f_* f_{\rm heat}\, \rho_b(z)\, \frac{d f_{coll}(z,M_{\rm vir}^{\rm min})}{dt}.
\label{eq:heatXray}
\end{align}
Thus, the increase in temperature per baryon in a Hubble time is
\begin{equation}
\begin{split}
\frac{2 \,\mathcal{Q} |_{X}}{3\, H \, n_b} & \simeq 10 \, {\rm K} \; \left( \frac{ -(1+z) d f_{coll}(z,M_*)/dz }{10^{-3}}\right) \\
& \times \left( \frac{ L_{{\rm X} [2 - 10~{\rm keV}]}/{\rm SFR} }{3.4 \times 10^{40} erg \cdot s^{-1} M_\odot^{-1} yr } \right) \left( \frac{f_*}{0.01} \right) \left( \frac{f_{\rm heat}}{0.2} \right),
\end{split}
\label{eq:heatXrayhubble}
\end{equation}
assuming $d f_{coll}(z,M_{\rm vir}^{\rm min})/dt/H(z)=-(1+z) d f_{coll}(z,M_{\rm vir}^{\rm min})/dz \simeq 10^{-3}$ as a fiducial value at $z \simeq 20$. Note that, given the growth of the collapse factor, the above formula increases by an order of magnitude at $z \simeq 15$, as can be noticed in Fig. \ref{fig:fcoll}. As a further estimation, ignoring adiabatic terms and changes in the ionization fraction in the temperature equation, Eq. \eqref{eq:eqT}, we can integrate the temperature taking Eq. \eqref{eq:heatXray} as the source in order to estimate the order of magnitude increase in temperature generated by X-rays, $\Delta T |_{X}(z)$,
\begin{align}
\Delta T |_{X}(z) &= \frac{2}{3}\frac{L_{{\rm X} [2 - 10~{\rm keV}]}}{{\rm SFR}}\, f_* f_{\rm heat}\,
 \mu m_p\, \Delta f_{coll}(z,M_*)  \\
 \begin{split}
 & \simeq 10^4 {\rm K} \; \Delta f_{coll}(z,M_{\rm vir}^{\rm min}) \left( \frac{ L_{{\rm X} [2 - 10~{\rm keV}]}/{\rm SFR} }{3.4 \times 10^{40} erg \cdot s^{-1} M_\odot^{-1} yr } \right)\\
 & \times \left( \frac{f_*}{0.01} \right) \left( \frac{f_{\rm heat}}{0.2} \right),
 \end{split}
\end{align}
where we have assumed the fiducial values $L_{{\rm X} [2 - 10~{\rm keV}]}/{\rm SFR} \simeq 3.4 \times 10^{40} erg s^{-1} M_\odot^{-1} yr $, $f_{\rm heat} \simeq 0.2$ and $f_* \simeq 0.01$ \cite{Furlanetto:2006tf}. Therefore, an increase in the collapse fraction of $\Delta f_{coll}(z,M_{\rm vir}^{\rm min})  \sim 1$ would lead to heat up the medium by $\sim 10^4$ K. Note however, that the on-the-spot scenario would only be accurate for low redshifts, when the mean free path is much shorter than the Hubble length, and for quantitative computations, Eq. \eqref{eq:Xray_heating_rate} (or Eq. \eqref{eq:Xray_heating_rate2}) must be used. Figure \ref{fig:temperature} shows the temperature prediction for different astrophysical models, computed with {\tt 21cmFAST}. As can be seen, the minimum virial temperature sets the timing of the onset of heating, being earlier for lower threshold masses, as lower mass halos are allowed to host star formation. On the other hand, the X-ray luminosity determines the strenght of heating. The spin temperature, treated in Sec. \ref{sec:tspin}, is also shown, being coupled to the kinetic one by means of scattering of Hydrogen atoms by Ly$\alpha$ radiation, whose flux is commented in Sec. \ref{sec:Lyalphaflux}.

\subsubsection{21 cm heating}

Recently, a new heating channel was proposed based on CMB photons interacting via the 21 cm line \cite{Venumadhav:2018uwn}. The absorption and emission of CMB photons by the hyperfine levels in Hydrogen atoms can also transfer some thermal energy in the process. Its corresponding heating rate can be written as \cite{Venumadhav:2018uwn}
\begin{equation}
\mathcal{Q} |_{21} = \frac{3}{4} \; n_H \; x_{\rm HI} \; x_{CMB} \; A_{10} \; T_* \; \left( \frac{T_\gamma}{T_S} -1 \right),
\label{eq:heat21}
\end{equation}
where $x_{CMB} = (1-e^{-\tau_{21}})/\tau_{21} \simeq 1$, where the 21 cm optical depth $\tau_{21}$ is small. Note that this heating implicitly depends on the Ly$\alpha$ flux, and therefore on the astrophysical parameters, through $T_S$. Given the factor $T_\gamma/T_S -1$, this heating rate turns on when $T_S$ deviates from $T_\gamma$, during the Ly$\alpha$ coupling period, vanishing however when the hyperfine transitions are driven by CMB photons. Since this epoch (partly) coincides with the onset of X-ray heating, it has a subdominant impact on the thermal history of the IGM, unless X-ray emitters are quite faint \cite{Villanueva_Domingo_2020}. Note that in the limit when the medium is very hot, and the CMB is cooler than the IGM, $T_\gamma/T_K< 1$, and cooling instead of heating takes place. The maximum possible cooling from this source appears in the limit when $T_\gamma/T_K$ can be neglected, and thus $T_\gamma/T_S-1 \simeq -1$.

One can easily estimate quantitatively this effect. Approximating the color temperature to the kinetic one (which is a fair assumption, as commented in Sec. \ref{sec:color}), we can write
\begin{equation}
\frac{T_\gamma}{T_S} -1 \simeq \frac{x_\alpha}{1+x_\alpha} \left(\frac{T_\gamma}{T_K} -1 \right).
\end{equation}
In Sec. \ref{sec:Lyalphafluxstars}, the Ly$\alpha$ coupling coefficient is estimated as $x_\alpha \sim 0.5$ at $z\sim 20$ in the fiducial model. The largest possible values for $T_\gamma/T_K$ are reached when heating sources are not efficient enough and the kinetic temperature still cools adiabatically, having $T_\gamma/T_K = (1+z_D)/(1+z)$. Then, at $z \sim 20$, $T_\gamma/T_K \sim 7$ and $T_\gamma/T_S -1 \sim 2$. In real scenarios, heating sources could be already important at these times to produce temperature deviations from the adiabatic evolution, lowering this value. Therefore, it must be taken as an upper limit. Hence, the heat per baryon and per Hubble time reads
\begin{align}
\frac{2 \,\mathcal{Q} |_{21}}{3\, H \, n_b} &\simeq \frac{1}{2} \; \mathfrak{f}_H  \; \frac{A_{10} \; T_*}{H}  \; \left( \frac{T_\gamma}{T_S} -1 \right) \\
&\simeq 3 \; {\rm K}  \; \left( \frac{21}{1+z} \right)^{3/2} \; \left( \frac{T_\gamma/T_S-1}{2} \right),
\label{eq:heat21cm_estim}
\end{align}
which is roughly an order of magnitude lower than the corresponding X-ray heat from Eq. \eqref{eq:heatXrayhubble}. For lower redshifts, this difference even increases: while the X-ray heating increases by an order of magnitude at $z\simeq 15$, Eq. \eqref{eq:heat21cm_estim} would reach as much as $\sim 20$ K at the same epoch, if the assumption of adiabatic evolution is still valid (which should not be true, since heating sources could already be important) and the Ly$\alpha$ coupling is completely efficient. These considerations make the 21 cm heating term negligible in most of astrophysical scenarios, and consequently, is usually not considered, as happens in {\tt 21cmFAST}. It must be remarked, however, that if X-ray luminosities are much lower than expected, this would be the dominant heating channel at high redshifts. Furthermore, it could also be relevant for even higher redshifts, deep in the Dark Ages, when collisional coupling becomes efficient, allowing the factor $T_\gamma/T_S-1$ to deviate from 0.

\subsubsection{Ly$\alpha$ heating}

Ly$\alpha$ scattering with atoms can transfer a fraction of energy from radiation to the gas, with a contribution from continuum and injected photons. The heating rate, already computed in Sec. \ref{sec:lyalphaheating}, reads
\begin{equation}
\mathcal{Q} |_{\rm Ly\alpha,k} = \frac{4 \pi H h\nu_\alpha \Delta\nu_D}{c} J_{\infty,k}I_k=n_bh\Delta\nu_D H \frac{J_{\infty,k}}{J_0}I_k~,
\label{eq:heatLya}
\end{equation}
where $k$ stands for $k=c$, \textit{continuum}, or $k=i$, \textit{injected} photons, and $J_0(z) = c \bar{n}_{b}(z)/(4\pi\nu_\alpha)$ the intensity corresponding to one photon per baryon \cite{Chen:2003gc}. The integrals $I_c$ and $I_i$ encode the details of the scattering effects and depend on the temperature and the Gunn-Peterson optical depth $\tau_{GP}$ (the optical depth of Ly$\alpha$ photons redshifting through the IGM). For these quantities, we use the fits provided by Ref. \cite{Furlanetto:2006fs}. 

It is straightforward to estimate the efficiency of Ly$\alpha$ photons to heat the medium. For the ranges of interest, epochs between $z\sim 30$ and $z\sim 10$, and for temperatures between $\sim 10$ K and $\sim 100$ K, the heating integrals $|I_k|$ are of the order of $\sim 10$ or lower \cite{Furlanetto:2006fs}. The rate is proportional to the energy $h\Delta \nu_D \simeq 0.5 \, {\rm K} \, \sqrt{T_K/(10^2 {\rm K})}$. Employing the estimate of $J_{\infty}/J_0 \sim 3 \times 10^{-2}$ at $z\sim 20$ from Sec. \ref{sec:Lyalphafluxstars}, we obtain a heating per baryon in a Hubble time of
\begin{align}
\frac{2 \,\mathcal{Q} |_{\rm Ly\alpha,k}}{3\, H \, n_b} &= \frac{2}{3} h\Delta\nu_D \frac{J_{\infty,k}}{J_0}I_k \\
&\simeq 0.1 \; {\rm K} \; \left(\frac{T_K}{10^2 {\rm K}} \right)^{1/2} \left(\frac{J_{\infty}/J_0}{3 \times 10^{-2}}\right)\left(\frac{I_k}{10} \right)~,
\label{eq:heatLya_estim}
\end{align}
Note that this is two orders of magnitude lower than the estimation from X-rays in Eq. \eqref{eq:heatXrayhubble}, and also lower than the 21 cm contribution, Eq. \eqref{eq:heat21cm_estim}. This difference even increases for lower redshifts, when X-rays are more efficient. Moreover, the term of injected photons cool the gas rather than heating it at these temperatures and redshifts, which counteracts in some amount the continuum term. Therefore, Ly$\alpha$ scatterings only mildly heat up the gas compared to absorption of X-rays, see Refs. \cite{Venumadhav:2018uwn, Villanueva_Domingo_2020} for further comparisons. As happens with the 21 cm heating rate, with realistic X-ray luminosities this heating channel becomes subdominant and in practice (as done in {\tt 21cmFAST}) is usually neglected.


\subsection{Ionization evolution}
\label{sec:xe}

The evolution of the local ionized fraction of the neutral IGM, $x_e({\bf x}, z) \equiv n_e/n_b$, is given by the interplay between ionizations of H and He atoms, and recombinations of positive ions. Part of the discussion is similar to that of Sec. \ref{sec:ionIGMglobal}, with few differences. Since the neutral IGM is a thick medium where most of ionizing photons are fastly absorbed, the recombinations to the ground state do not effectively account for the net ionized fraction, and then case B recombination must be employed, rather than case A \cite{Furlanetto:2006jb}. Moreover, the ionization of the neutral IGM is produced by X-ray photons with a mean free path long enough to escape beyond the galaxy environment, instead of ionizing UV photons locally absorbed. For that reason, the on-the-spot approximation is not as good estimate as in Sec. \ref{sec:ionIGMglobal}, and the proper ionization rate should be computed. Considering Eq. \eqref{eq:nHII} and the equivalent ones for HeI and HeIII, and taking into account that conservation of charge implies $n_e = n_{\rm HII} + n_{\rm HeII} +2n_{\rm HeIII}$, we can derive the evolution equation for $x_e$:
\begin{equation}
\frac{dx_e}{dt} = \Lambda_{\rm ion} - \alpha_{\rm B} \, C \, x_e^2 \, n_b \, \mathfrak{f}_{\rm H} ~,
\label{eq:xe}
\end{equation}
where $\alpha_{\rm B} \simeq 2.6 \times 10^{-13}(T_K/10^4 K)^{-0.7} cm^3 s^{-1}$ is the case-B recombination, and $C$ is the clumping factor (evaluated at the scale on the scale of the simulation cell). The total ionization rate $\Lambda_{\rm ion}$ is given by the joint contribution of the ionizing rates $\Gamma_j$ of each species $j$, with $n_b\Lambda_{\rm ion} = \sum_j \Gamma_j n_j$, and can be written as
\begin{equation}
\Lambda_{\rm ion} = \sum_{j} 4 \pi x_j \mathfrak{f}_j \int^{\infty}_{\nu_0} d\nu \, \sigma_j(\nu) \, F_j \, J_X(\nu),
\label{eq:lambda_ion}
\end{equation}
where $J_X$ is the X-ray flux, given by Eq. \eqref{eq:JX}, $F_j$ accounts for the contribution from primary and secondary ionizations, and the summation goes over the different degrees of ionization $j={\rm HI,HeI,HeII}$. In the neutral IGM, we can assume the same degree of ionization of the species HII and HeII, thus  $n_{\rm HII}/n_{\rm H} \simeq n_{\rm HeII}/n_{\rm He} \simeq x_e$, the fraction of HeIII being negligible. Therefore, in the equation above we can write $x_{\rm HI} = x_{\rm HeI} = 1-x_e$ and $x_{\rm HeII} = x_e$. The factors $F_j$ read \cite{Mesinger:2010ne}:
\begin{equation}
F_j= 1 + (h\nu -E^{th}_j)\left( \frac{f_{\rm ion,HI}}{E^{th}_{\rm HI}} + \frac{f_{\rm ion,HeI}}{E^{th}_{\rm HeI}} + \frac{f_{\rm ion,HeII}}{E^{th}_{\rm HeII}} \right),
\end{equation}
being $E^{th}_j$ the threshold energy for each species, and $f_{\rm ion, j}$ the fraction of the electron’s energy going into secondary ionizations of species $j$. We employ the fits derived in Ref. \cite{1985ApJ...298..268S} for the functions $f_{\rm ion, j}$, which can be roughly approximated by Eq. \eqref{ec:f_heationexc} for HI, being suppressed by an order of magnitude for Helium. The unity term in the equation above stands for primary ionizations, while the second term accounts for secondary ones. These secondary ionizations are produced by electrons which, after receiving energy from a X-ray photon, become very fast and non-thermal, colliding and distributing their energy in several channels until thermalizing.

It is possible to sketch the $x_e$ evolution in some approximate limits. Before the formation of X-ray sources and after Recombination, only a leftover fraction remained ionized. One can easily estimate the evolution of this remnant by neglecting ionizing sources in Eq. \eqref{eq:xe}. In this simple limit, the differential equation can be easily integrated, obtaining a slowly decaying function of redshift towards a constant value. For the recombination coefficient, it can be assumed that $T_K \simeq T_{\gamma,0} (1+z)$ (with $T_{\gamma,0}=2.73$ K) due to Compton cooling, valid for high redshifts (as shown in Sec. \ref{sec:compton}). Since Recombination takes place at $z_{\rm R}\sim 10^3$, when the universe was still completely ionized, $x_e(z_{\rm R}) \sim 1$, hence, the remnant ionized fraction for redshifts much lower than $z_{\rm R}$ is given by
\begin{equation}
x_e \simeq \frac{H_0 \sqrt{\Omega_m}}{\alpha_B(T_{\gamma,0})n_{\rm H,0}(1+z_{\rm R})^{0.8}} \simeq 3 \times 10^{-4},
\end{equation}
which remarkably coincides with the precise estimation from Recombination codes such as RECFAST \cite{1999ApJ...523L...1S}.\footnote{It has to be emphasized that at $z\sim 150$, the Compton coupling is not efficient anymore, and the kinetic temperature deviates from the CMB one, as commented above, but taking this into account would not alter the order-of-magnitude estimate.} On the other hand, this fraction grows again when starbust galaxies or other sources start injecting X-ray photons to the medium. Following the same procedure of Sec. \ref{sec:xrayheating}, we can treat the ionization rate using the on-the-spot approximation, to rouhgly estimate the impact of X-rays on $x_e$. Similarly to Eq. \eqref{eq:heatXray}, and neglecting secondary ionizations, we get
\begin{equation}
\Lambda_{\rm ion} \simeq \frac{L_{{\rm X} [2 - 10~{\rm keV}]}}{{\rm SFR}}\, f_* \frac{\mu m_p}{h \nu_0}\, \frac{d f_{coll}(z,M_*)}{dt}.
\end{equation}
Neglecting recombinations, we can solve Eq. \eqref{eq:xe}, obtaining
\begin{equation}
\begin{split}
\Delta x_e(z) &\sim 10^{-2} \Delta f_{coll}(z, M_*) \left( \frac{ L_{{\rm X} [2 - 10~{\rm keV}]}/{\rm SFR} }{3.4 \times 10^{40} erg \cdot s^{-1} M_\odot^{-1} yr } \right) \\
& \times \left( \frac{f_*}{0.01} \right) \left( \frac{500 eV}{h\nu_0} \right).
\end{split}
\end{equation} 
Note that for the fiducial values considered, the increase in $x_e$ would be somewhat mild. This fact, compared to the discussion in Sec. \ref{sec:ionIGM}, indicates that the progress of reionization processes must be mostly driven by the expansion and overlap of HII bubbles due to UV radiation, rather than the effect of X-rays in the mostly neutral regions, which has a moderate impact. For more quantitative computations, however, a full and precise determination of $x_e$ is required and must be computed properly.

\subsection{Ly$\alpha$ flux}
\label{sec:Lyalphaflux}

One of the most relevant quantities for the determination of the spin temperature is the Ly$\alpha$ flux, $J_{\alpha}$, responsible of the WF effect commented in Sec. \ref{sec:wf}. There are two main sources which are able to produce a relevant flux in the IGM at the redshifts of interest: the redshifted radiation from stellar emission, providing a flux $J_{\alpha, *}$, and X-ray excitation, $J_{\alpha, X}$. The total flux is then given by the sum $J_{\alpha} = J_{\alpha, *}+J_{\alpha, X}$. These mechanims are reviewed in the following.

\subsubsection{Stellar emission}
\label{sec:Lyalphafluxstars}

Once the first stars in the universe are formed, they release radiation background between the Lyman limit and the Ly$\alpha$ frequency. Due to the expansion of the universe, these photons can redshift to Ly-$n$ resonances, and eventually lead to a $2p \rightarrow 1s$ transition through a decaying cascade. Due to the high optical depth of the IGM, photons redshifting to Lyman resonances are always absorbed. When a photon reaches the Ly-$n$ resonance at redshift $z$, it can be emitted at $z_{max,n}>z$, which is the maximum emission redshift so that a photon reaches the Ly-n resonance at redshift z:
\begin{equation}
1+z_{max,n} = (1+z)\, \frac{\nu_{n+1}}{\nu_n}= (1+z)\,\frac{1-\frac{1}{(n+1)^{2}}}{1-\frac{1}{n^{2}}},
\end{equation}
where the Ly-$n$ frequency is $\nu_n=\nu_{LL}(1-n^{-2})$, with $\nu_{LL}=13.6$ eV$/h$ the Lyman limit frequency. As stated in Sec. \ref{sec:lyspectrum}, some radiative cascades do not contribute to the Ly$\alpha$ flux, since they terminate in the 2s $\rightarrow$ 1s transition. We thus define the probability to generate a Ly$\alpha$ photon after the absorption by the $n$ level as the recycled fraction of level $n$, $f_{rec}(n)$ \cite{Pritchard:2005an}. This quantity can be easily computed by means of an iterative algorithm from the selection rules and the decay rates. The total Ly$\alpha$ flux can be written in terms of the photon comoving emissivity $\varepsilon_\alpha(\nu,z)$ as \cite{Pritchard:2005an,Haardt_2012}
\begin{equation}
J_{\alpha,*}=\frac{c(1+z)^2}{4\pi} \sum_{n=2}^{n_{max}} f_{rec}(n) \int_z^{z_{max,n}} dz'\frac{\varepsilon_{\alpha}(\nu'_n,z')}{H(z')}~,
\label{eq:Jalphaflux}
\end{equation}
where the emission frequency is $\nu'_n= \nu_n(1+z')/(1+z)$.

Following the procedure of Sec. \ref{sec:ionIGM}, we can employ a simple model for the emissivity, proportional to the comoving star formation rate density $\dot{\rho}_*(z)$~\cite{Barkana:2004vb},
\begin{equation}
\varepsilon_{\alpha}(\nu,z) = \mathcal{N}(\nu) \frac{\dot{\rho}_*(z)}{\mu m_p} = 
 \mathcal{N}(\nu) f_* \bar{n}_{b,0}  \frac{df_{coll}(z,M_*)}{dt}~,
\end{equation}
where $\mathcal{N}(\nu)$ is the the number of emitted photons per frequency. The radiation spectral distribution in frequencies can be taken as a piecewise power law between each $n$ and $n+1$ Lyman levels,
\begin{equation}
\mathcal{N}(\nu)=N_n\frac{(\beta_n+1)\nu_{\alpha}^{\beta_n}}{(\nu_{n+1}^{\beta_n+1}-\nu_{n}^{\beta_n+1})} \left( \frac{\nu}{\nu_{\alpha}}\right)^{\beta_n}~, \;\;\;\; \nu_n \leq \nu \leq \nu_{n+1},
\label{eq:speclyalpha}
\end{equation}
with $\nu_{\alpha}=(3/4)\nu_{LL}$ the Ly$\alpha$ frequency, $N_n$ the number of photons emitted between the $n$ and $n+1$ resonances and $\beta_n$ the spectral index, whose numerical values can be derived from stellar synthesis models \cite{1999ApJS..123....3L,Barkana:2004vb}. The function above is normalized as $\int_{\nu_n}^{\nu_{n+1}} d\nu \mathcal{N}(\nu) = N_n$, while the normalization of the total spectrum gives the total number of photons emitted between the Ly$\alpha$ and the Lyman limit $\int_{\nu_{\alpha}}^{\nu_{LL}} d\nu \mathcal{N}(\nu) = \sum_n N_n = N_{\rm tot}$. For Population II (or low-metallicity) stars, $N_{\rm tot} \simeq 9690$, out of which 6520 photons are emitted between Ly$\alpha$ and Ly$\beta$ lines \cite{1999ApJS..123....3L,Barkana:2004vb}. For very massive and metal-free Population III stars, $N_{\rm tot} \simeq 4800$, with 2670 photons between Ly$\alpha$ and Ly$\beta$ \cite{Bromm_2001}. However, given the lack of observations of high-redshifts galaxies, the mass distribution, metallicity and therefore the spectrum of these sources is still highly uncertain. Hence, $N_{\rm tot}$ may be considered as a free parameter, which is usually proportional to the number of ionizing photons $N_{\gamma/b}$. In {\tt 21cmFAST}, the fiducial scenario assumes Population II stars, which have $N_{\gamma/b}=4 \times 10^3$, that, with the spectrum of Eq. \eqref{eq:speclyalpha}, corresponds to a value of $N_{tot} \simeq 9690$.

As a side note, an emissivity proportional to $\nu^{-1}$ (normalized as $\mathcal{N}(\nu) = N_{tot}/{\rm ln}(4/3)\,\nu^{-1}$) allows to solve analytically the integral of Eq. \eqref{eq:Jalphaflux}, obtaining the following Ly$\alpha$ flux \cite{Villanueva_Domingo_2020}:
\begin{equation}
J_{\alpha,*}(z)=\frac{c \bar{n}_{b}(z)f_* N_{tot}}{4\pi\nu_\alpha ln(4/3)} \sum_{n=2}^{n_{max}} f_{rec}(n) \frac{\nu_\alpha}{\nu_n} \left(f_{coll}(z,M_*)-f_{coll}(z_{max,n},M_*)\right)~.
\end{equation}
This flux differs by less than 10\% from the one computed with the piecewise power-law spectrum presented above, and given its analytical form, can provide insight about the flux evolution. Profiting from the formula above, it is possible to estimate the production of Ly$\alpha$ photons in the high redshift universe. Writing the intensity corresponding to one photon
per baryon $J_0(z) = c \bar{n}_{b}(z)/(4\pi\nu_\alpha)$ as a reference value \cite{Chen:2003gc}, it is is possible to give an order-of-magnitude approximation for the Ly$\alpha$ flux as
\begin{equation}
J_{\alpha,*}(z) \sim 3 \times 10^{-2} J_0(z) \left(\frac{N_{tot}}{9690}\right) \left(\frac{f_*}{0.01} \right) \left(\frac{f_{coll}(z,M_*)}{10^{-4}}\right),
\label{Jalphaestim}
\end{equation}
where the fiducial value taken for $N_{tot}$ corresponds to Pop II stars, as commented above, and $f_{coll}(z,M_{*}) \simeq 10^{-4}$ for a minimum virial temperature of $10^4$ K at $z\simeq 20$ (see Fig. \ref{fig:fcoll}). With this result, the Ly$\alpha$ coupling coefficient from Eq. \eqref{eq:x_alpha} reads
\begin{align}
x_\alpha &\simeq 14.5 \times S_\alpha \frac{J_{\alpha,*}(z)}{J_0(z)}\left( \frac{1+z}{21}\right)^2 \\
&\sim 0.5 \; S_\alpha \left( \frac{1+z}{21}\right)^2 \left(\frac{N_{tot}}{9690}\right) \left(\frac{f_*}{0.01} \right) \left(\frac{f_{coll}(z,M_*)}{10^{-4}}\right),
\end{align}
where, as stated in Sec. \ref{sec:lyscatteringrate}, $S_\alpha$  is an order 1 function accounting for the details of radiative transfer. Therefore, we see that, for these fiducial astrophysical parameters, the WF coefficient approaches $x_\alpha \sim 1$ at $z \simeq 20$, implying an efficient WF effect and thus a strong coupling between the spin temperature and the kinetic one from there on.

\subsubsection{X-ray excitation}

Similarly to what happens with secondary ionizations, absorption of energetic X-ray photons can lead to collisional excitation by fast electrons of Hydrogen atoms, producing a Ly$\alpha$ photon at the de-excitation \cite{2007MNRAS.376.1680P}. The corresponding flux of this mechanism can be written as
\begin{equation}
J_{\alpha, X}= \frac{c n_b(z)}{4 \pi h\nu_\alpha H \nu_\alpha} 4\pi\int_{\nu_0}^{\infty} d\nu \, J_{\rm X}(z,\nu) \, \sum_{j}(h \nu-E_{j}^{\rm th})f_{\rm Ly\alpha} \, f_{j} \, x_{j} \, \sigma_{j}(\nu),
\label{eq:JalphaXrays}
\end{equation}
where $f_{\rm Ly\alpha}$ is the fraction of electron energy which is deposited as Ly$\alpha$ excitations, whose value is close to the total excitations fraction, $f_{\rm exc}$, as $f_{\rm Ly\alpha} \simeq 0.8 f_{\rm exc}$ \cite{2010MNRAS.404.1869F}. In the high energy limit, where $f_{\rm exc}$ saturates to constant values of energy, the expression above can be approximated as
\begin{equation}
J_{\alpha, X} \simeq \frac{c}{4 \pi h\nu_\alpha H \nu_\alpha} \frac{f_{\rm Ly\alpha}}{f_{\rm heat}} \mathcal{Q} |_{X}(z) = J_0(z)\frac{f_{\rm Ly\alpha}}{f_{\rm heat}}\frac{\mathcal{Q} |_{X}(z)}{ H h \nu_\alpha n_b(z)} 
\label{eq:JalphaXraysaprox}
\end{equation}
with $f_{\rm Ly\alpha} \simeq f_{\rm heat}$ at high energies \cite{2010MNRAS.404.1869F}. As we have done for the heating and ionization in the previous subsections, we can apply the on-the-spot approximation in the X-ray flux of the equation above, to estimate the rough contribution of this term to the total Ly$\alpha$ background. Employing Eq. \eqref{eq:heatXray}, one can find
\begin{equation}
\begin{split}
J_{\alpha, X}(z) & \simeq 10^{-4} J_0(z) \; \left( \frac{ -(1+z) d f_{coll}(z,M_*)/dz }{10^{-3}}\right) \\
& \times \left( \frac{ L_{{\rm X} [2 - 10~{\rm keV}]}/{\rm SFR} }{3.4 \times 10^{40} erg \cdot s^{-1} M_\odot^{-1} yr } \right) \left( \frac{f_*}{0.01} \right) \left( \frac{f_{\rm Ly\alpha}}{0.2} \right),
\end{split}
\end{equation}
where we have evaluated $d f_{coll}(z,M_*)/dt/H(z)=-(1+z) d f_{coll}(z,M_*)/dz \simeq 10^{-3}$ as a fiducial value at $z \simeq 20$ for a minimum virial temperature of $10^4$ K. Note that this flux is about 2 orders of magnitude lower than the estimated one from Eq. \eqref{Jalphaestim}, and therefore, only appears as a subdominant contribution to the overall flux. However, quantitative computations must still consider it for a precise evaluation of the WF coupling epoch.

\section{Evolution of the 21 cm line}

\begin{figure}
\centering
\includegraphics[scale=0.8]{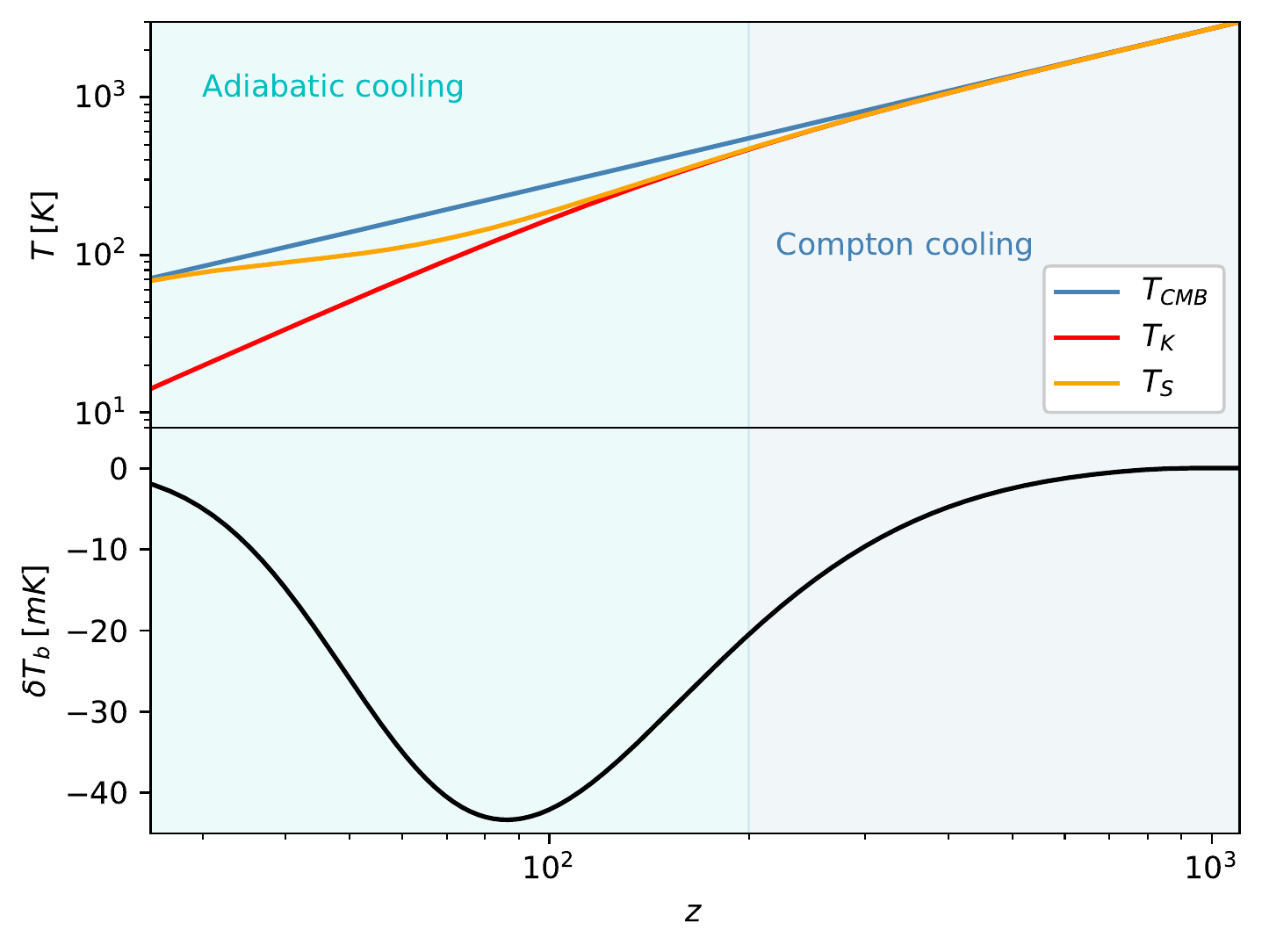}
\captionof{figure}{Evolution of the CMB, kinetic and spin temperatures (top panel) and global brightness temperature (bottom panel) as a function of redshift during the Dark Ages.}
\label{fig:darkages}
\end{figure}

\begin{figure}
\centering
\includegraphics[scale=0.99]{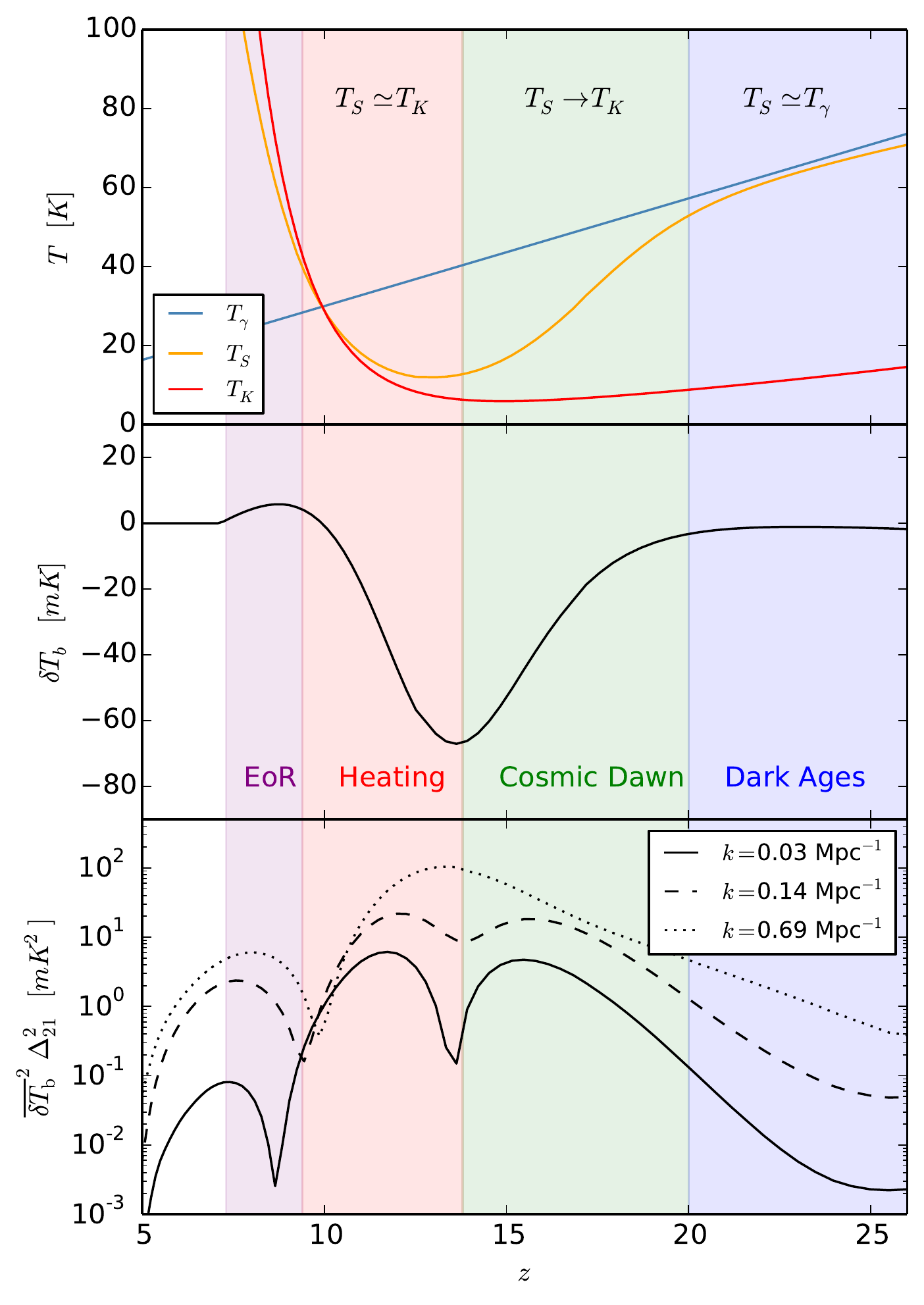}
\captionof{figure}{Evolution of the CMB, kinetic and spin temperatures (top panel), global brightness temperature (medium panel) and 21 cm power spectrum (bottom panel) as a function of redshift.}
\label{fig:plotTb}
\end{figure}

In this section an overview of the most important phases of the thermal history of the universe is presented. The different epochs, driven by the heating and ionization processes studied in the previous sections, may be probed by means of the 21 cm line, either from its global sky average or from the power spectrum described in Sec. \ref{sec:21ps}. It is worth writing here explicitly the global brightness temperature, $\overline{\delta T}_b$, taking the background quantities in Eq. \eqref{eq:dTb}\footnote{Strictly, the global signal must be obtained from averaging the brightness temperature fields from simulations, rather than by replacing in Eq. \eqref{eq:dTb} the background values, which may induce slightly different results. This average has been analytically treated, making use of the probability distribution for the density, in Ref. \cite{Villanueva_Domingo_2020}. In any case, the conclusions regarding the IGM history outlined in this section are unaltered.} and assuming the default cosmological parameters:
\begin{equation}
\overline{\delta T}_b \simeq 27 \, \textrm{mK} \, x_{\rm HI} \,  \left( 1 - \frac{T_\gamma}{T_S}\right) \left( \frac{1+z}{10}\right)^{1/2} 
\label{eq:dTbglobal}
\end{equation}
Figure \ref{fig:darkages} shows the evolution of the CMB, kinetic and spin temperatures (top panel) and global brightness temperature (bottom panel) during the Dark Ages. After that period, more complex astrophysical modeling is needed. Fig. \ref{fig:plotTb} presents the evolution of the CMB, kinetic and spin temperatures (top panel), global brightness temperature (medium panel) and 21 cm power spectrum (bottom panel) for a standard thermal history, computed from a {\tt 21cmFAST} simulation. The power spectrum is evaluated at three differentt scales, $k = 0.03, \, 0.14, \, 0.69 \, {\rm Mpc}^{-1}$. The specific astrophysical model assume Population II stars, with the parameters $M_{\rm vir}^{\rm min}=10^8 M_\odot$, $L_{{\rm X} [2 - 10~{\rm keV}]}/{\rm SFR} = 10^{40.5}$, $N_{\gamma/b}=4 \times 10^3$ and $f_*=0.01$. By inspecting the 21 cm signal, it is possible to identify the following epochs:

\begin{enumerate}

\item \textbf{Dark Ages and collisional coupling.} Firstly, after Recombination, happening at $z \sim 1100$, the gas temperature evolves still coupled to the CMB, due to Compton cooling (discussed in Sec. \ref{sec:compton}). When the kinetic temperature departs from the CMB one, evolving adiabatically, the still high density of the IGM induces collisional coupling discussed in Sec. \ref{sec:collcoup}. Hence, the spin temperature is expected to be coupled to the gas by that process, producing an absorption dip in the brightness temperature, Eq. \eqref{eq:dTbglobal}. This absorption regime persists as long as the gas is dense enough to produce sufficient collisions, whereupon CMB photons drive again the spin temperature and the signal reduces its amplitude. Such signal would be hard to detect due to its small amplitude and high redshift range (or equivalently, low frequencies). Figure \ref{fig:darkages} depicts this period, specifying when the gas temperature evolves by Compton cooling or adiabatically.

\item \textbf{End of the Dark Ages.} After the collisional coupling period but at high enough redshifts, $z > 20$, there are not enough stars formed yet, and thus the hyperfine transitions are mainly driven by absorption and emission of CMB photons. The spin temperature is therefore coupled again to the CMB temperature, which cools as $1+z$, and thus the brightness temperature remains close to zero. However, given the growth of fluctuations, the 21 cm power spectrum starts to increase. Meanwhile, since the gas is already thermally decoupled from the CMB, its kinetic temperature continues cooling adiabatically as $(1+z)^2$. The blue region in Fig. \ref{fig:plotTb} represents this era.

\item \textbf{Cosmic Dawn and WF effect.} Once the first stars are born, they start to emit UV radiation which redshift to Lyman resonances. These photons scatter with neutral Hydrogen atoms with a high cross section, causing spin-flip transitions and modifying the hyperfine level populations. This effect, the so-called WF effect already studied in Sec. \ref{sec:wf}, couples the spin temperature to the kinetic temperature of the gas, which in this epoch is still cooling adiabatically. The factor $T_{\gamma}/T_S$ of Eq. \eqref{eq:dTbglobal} leads to an absorption profile in the brightness temperature. The peak in the power spectrum indicates the summit of the Ly$\alpha$ fluctuations. In Fig. \ref{fig:plotTb}, this epoch is depicted by the green area.

\item \textbf{Heating of the IGM.} When there is enough X-ray radiation emitted from starbusts and other sources, the IGM begins to be heated, increasing $T_K$, and therefore the spin temperature too, which is expected to be mostly coupled to $T_K$ at this time due to the WF effect. This reduces the depth of the peak of the global signal, and eventually leads to a transition between the absorption regime to an emission regime when $T_S>T_{\gamma}$. The largest fluctuations in heating are indicated in the peak of the power spectrum. The red area corresponds to this phase in Fig. \ref{fig:plotTb}.

\item \textbf{Epoch of Reionization.} Finally, when the ionized bubbles around each light source grow enough to overlap, the IGM moves into a fully ionized phase. Since the global signal is proportional to the ionized fraction, it begins to decrease until it dissappears at the end of the EoR. The peak in the power spectrum roughly corresponds to this moment. The purple area of Fig. \ref{fig:plotTb} illustrates this period.
\end{enumerate}

It is possible to gain insight into the milestones in the evolution of the 21 cm power spectrum shown in Fig. \ref{fig:plotTb} by inspecting the behavior of the $\beta_i$ coefficients defined in Sec. \ref{sec:21ps}. As can be noticed from the examination of $\beta_\alpha$, this coefficient is small until the Ly$\alpha$ flux becomes strong enough. After that, when Ly$\alpha$ coupling dominates, $x_{\rm tot} \simeq x_\alpha$ and $\beta_\alpha$ decays again. This leaves a peak in $\beta_\alpha$, and also in the power spectrum around the onset of Ly$\alpha$ radiation (see, e.g., Fig. 9 of Ref. \cite{Furlanetto:2006jb}). This peak traces the epoch of the Cosmic Dawn, and is seen in Fig. \ref{fig:plotTb}. The $\beta_T$ coefficient presents a similar behavior. Neglecting the collisional term, with the proper normalization to avoid the singularity, it reads
\begin{equation}
\overline{\delta T_b}\beta_T \propto \left(1-\frac{T_\gamma}{T_S}\right)\frac{T_\gamma}{T_K-T_\gamma} =\frac{x_{\rm tot}}{1+x_{\rm tot}}\frac{T_\gamma}{T_K},
\end{equation}
The kinetic temperature is expected to cool adiabatically, until the X-ray radiation becomes efficient to heat the IGM, and $T_K$ grows. At that moment, $T_K$ gets its lower value, and thus $T_\gamma/T_K$ the maximum. On the other hand, the factor $x_{\rm tot}/(1+x_{\rm tot})\simeq x_{\alpha}/(1+x_{\alpha})$ grows until it saturates to $1$ when $x_\alpha \gg 1$. Both contributions place a maximum in this term of the power spectrum, which marks the heating epoch. This maximum does not have to occur exactly when $T_K$ is minimum, due to the $x_{\rm tot}/(1+x_{\rm tot})$ term. Finally, the existence of a third peak indicating the EoR can also be easily interpreted. One has to bear in mind that, as commented in Sec. \ref{sec:ionIGM}, the ionization process occurs due to the growth of completely ionized HII bubbles onto the neutral IGM, and is hence a highly inhomogeneous process. However, both the mostly neutral IGM and the universe after reionization present homogeneous ionization fractions ($\sim 10^{-4}$ and $\sim 1$, respectively), with small deviations from the mean ionization value. During the transition between the two periods, the non-homogeneous expansion of HII bubbles enhance the spatial fluctuations, which are reduced again when EoR is nearly complete. Fluctuations must present then a peak, which is also reflected on the 21 cm power spectrum. The qualitative behavior of the power spectrum at the different scales is similar, although at smaller scales, the heating and Ly$\alpha$ peaks spread and merge. 

As can be seen from the previous discussion, both the 21 cm global signal and its power spectrum could provide us with a superb tool to study the thermal history of the IGM: from the birth of the first stars (that can be related to the beginning of the absorption regime) to the end of reionization (which we could date when the signal vanishes), passing through the onset of the IGM heating (linked to the position of the absorption dip and the transition between absorption and emission regimes). Observing and understanding this evolution would provide us with highly relevant information about the astrophysical processes which take place during these times, and more importantly, about the processes of structure formation and the nature of the Dark Matter component, which is the focus of this thesis.

\begin{comment}

\bibliographystyle{../../jhep}
\bibliography{../../biblio}

%% file: Chapters/Others/Summary.tex
\lhead[{\bfseries \thepage}]{ \rightmark}
\rhead[Summary of the results \leftmark]{\bfseries \thepage}
\renewcommand{\headrulewidth}{0.4pt}

\chapter*{Summary of the results}
\addcontentsline{toc}{chapter}{Summary of the results} 


Part \ref{partI} of this thesis has been devoted to introduce the cosmological impact of several non-canonical DM models, as well as to overview the relevant physics of the IGM evolution and the 21 cm line. Part \ref{partII} comprises the original research work carried out during the development of this PhD thesis. In order to ease its reading and to provide a concise guideline including the key points of each article, we briefly summarize in what follows the research manuscripts that this thesis comprises.

\begin{enumerate}

\item \textbf{Warm dark matter and the ionization history of the Universe}

Reference \cite{Lopez-Honorez:2017csg} studies the impact of WDM scenarios on the thermal and ionization evolution of the IGM. Due to the free streaming effect of $\sim $ keV particles, the formation of structures is suppressed on small scales with respect to the standard CDM case, as described in detail in Sec. \ref{sec:WDM}. This implies a reduction in the number of star forming low-mass halos, capable of emitting UV and X-ray radiation responsible for ionizing and heating the IGM. The decrease of the number of such sources leads to a delay of the cosmic dawn, the heating epoch and the EoR. Making use of semi-analytical simulations, employing the formalism described in Chap. \ref{chap:IGM}, the evolution of the kinetic temperature and the ionization fraction is computed for several WDM scenarios, varying the WDM mass between $1$ and $4$ keV, a physically motivated range to account for the small-scale issues outlined in Sec. \ref{sec:smallscale}. Given that the thermal history of the IGM is very sensitive to the stellar and radiative processes involved, we also consider different astrophysical scenarios. Three astrophysical parameters are modified, namely the minimum virial temperature (related to the minimum mass to host star formation), the UV ionizing efficiency and the X-ray heating efficiency (directly related to the X-ray luminosity).

The results of the ionization fraction from simulations have been confronted with observational bounds on the state of ionization during the EoR. Two kinds of data are taken into account, a subset of the compilation discussed in Sec. \ref{sec:constraintsEoR}. On the one hand, constraints and lower limits on the ionization fraction at redshifts $ 8 \geq z \gtrsim 6$ from quasar spectra and Ly$\alpha$ emitting galaxies are considered. On the other hand, we also employ the Thomson optical depth inferred from the CMB anisotropies. It is shown that bounds employing only CMB data are less restrictive than those from the ionization state at some redshift, given that the Thomson optical depth is an integrated quantity, rather than a redshift dependent constraint. We find that some astrophysical parameters are strongly degenerate among them, and also with the WDM mass. More specifically, the delay of the EoR present in a WDM model can be approximately mimicked either by a high ionization efficiency (which boost an earlier reionzation) or by a lower minimum virial temperature (so that lower mass halos could form stars and contribute to emit ionizing radiation). For that reason, lower limits on the WDM mass obtained by this method are weaker than those from the Ly$\alpha$ forest, although still capable of bounding its value down to $\sim 1 $ keV.

\vspace{1cm}

\item \textbf{Warm Dark Matter and Cosmic Reionization} 

In order to dive deeper in the phenomenology of WDM, Ref. \cite{Villanueva-Domingo:2017lae} further investigates the impact of WDM models on the cosmological evolution and formation of structures. In this case, accurate hydrodynamical simulations are employed, rather than semi-analytic computations, providing thus more reliable results. We make use of the framework of Cosmic Reionization on Computers (CROC) \cite{Gnedin:2014uta}, which represents the state of the art in the numerical modeling of the EoR, properly modifying the initial power spectrum in order to include the characteristic suppression of WDM scenarios. A WDM mass of $m_{\rm WDM}=3$ keV is assumed, value which lies around the constraints from Ly$\alpha$ forest observations.

Different reionization observables are computed in order to explore the impact of the free streaming scale, such as the halo mass function and the galaxy UV luminosity functions, which show both a suppression (at low masses and high magnitudes, respectively) with respect to the CDM case. While current telescopes are not able to discriminate among CDM and WDM cases, the upcoming JWST telescope, capable of reaching the faint end of the galaxy luminosity function, could provide the key to constrain such WDM masses. The effect on the cumulative probability distribution function for the effective Ly$\alpha$ opacity $\tau_{\rm GP}$ in the post-EoR IGM is also discussed, finding that, compared to the CDM scenario at a given redshift, the WDM case presents more extended tails toward high $\tau_{\rm GP}$, which correspond to high density regions, since they are ionized later. Compared to Ly$\alpha$ forest data, it favors the CDM case, although it is not clear whether tunning the astrophysical parameters in WDM scenarios could improve its agreement with data. Finally, the 21 cm power spectrum is also computed, showing an interplay between a suppression due to the free streaming and an enhancement because of the delayed reionization in the WDM case. This leads to a non-trivial behavior, which makes it hardly detectable by the first generation of experiments.

\vspace{1cm}

\item \textbf{Was there an early reionization component in our universe?}

The ionization history of the universe during the EoR is far from being well determined. Although most of the bounds are consistent with a relatively fast reionization process ending at $z\sim 6$, an early contribution, for example from non-canonical DM scenarios injecting energy into the medium, may still be possible. In Ref. \cite{Villanueva-Domingo:2017ahx}, this question is examined under the light of Planck CMB data. The ionization history is modeled in order to be consistent with the the power spectrum of the EE polarization modes of the CMB, which is sensitive to reionization at low multipoles. Several methods are employed in order to emulate the evolution of the ionized fraction. Firstly, functional parameterizations are considered, such as the frequently used redshift-symmetric and the redshift-asymmetric parameterizations. On the other hand, an approach based on the Principal Component Analysis (PCA) is applied, extracting the eigenmodes from the Fisher matrix of the $C_l$ components of the EE spectrum. Those with the smallest variances are expected to carry the most valuable cosmological information. A subset of these eigenfunctions is employed as a basis to recover the ionization history around a given fiducial model, either assuming a constant value or one of the aforementioned parameterizations. A third method based on Piecewise Cubic Hermite Interpolating Polynomial (PCHIP) is employed, considering as free parameters the ionized fraction at several different redshifts, which are smoothly interpolated via the PCHIP method.

We compare the different approaches to recover the ionization history with a Monte Carlo Markov Chain (MCMC) analysis, showing that some of them prefer an early reionization component, including some oscillatory pattern, such as using the PCA combination summed to a fiducial model. However, when the PCA combination is included as an argument of the parameterization, exploiting its functional form, the resulting history is smoother and does not show an early onset. The application of the frequentist Akaike information and the Bayesian information criteria confirms that none of them in particular is preferred over the other possible formulations, and thus an early reionization is not preferred by CMB data, contrary to previous claims \cite{Heinrich:2016ojb}.

\vspace{1cm}

\item \textbf{A fresh look into the interacting dark matter scenario}

In Ref. \cite{Escudero:2018thh}, phenomenological DM models involving elastic scattering with photons, coined as Interacting Dark Matter (IDM), are discussed in the context of the thermal evolution of the IGM and the number of satellite galaxies of the MW. In this case, as shown in Sec. \ref{sec:IDM}, the coupling between DM particles and photons produces a collisional damping effect, inducing a suppression and an oscillatory pattern at the smallest scales, in a similar way as the power spectrum is modified by the baryon-photon coupling. This effect can reduce the number of small scale structures and low mass halos, similarly to WDM. The magnitude of this damping is determined by the elastic scattering cross section over the DM particle mass, $\sigma_{\gamma \rm DM}/m_{\rm DM}$. 

Semi-analytic simulations of the thermal and ionization state of the IGM have been carried out in order to investigate the signatures of IDM scenarios. Suppressing the growth of structures due to collisional damping retards the formation of first stars and galaxies, delaying all astrophysical processes such as X-ray heating and reionization. Analogously to the findings in Ref. \cite{Lopez-Honorez:2017csg} regarding the WDM mass, the IDM cross section is strongly degenerated with several astrophysical parameters, mainly the ionization efficiency and the minimum virial temperature. On the other hand, the number of expected subhalos is computed, which in an IDM scenario is reduced with respect to the CDM scenario. This fact allows to constrain the elastic cross section when compared to the observed MW satellite galaxies. The satellite galaxies constraint, however, strongly depends upon the assumed MW total mass, which is not completely well determined yet, and within its expected mass range, the bounds on the cross section can vary up to one order of magnitude. By combining this upper limit with the bounds from the IGM ionization history, some astrophysical degeneracies can be broken.

\vspace{1cm}

\item \textbf{Dark Matter microphysics and 21 cm observations}

The EDGES collaboration claimed the detection of the global 21 cm signal, presenting an absorption dip with an amplitude of roughly $500$ mK, which doubles the expectations in the standard $\Lambda$CDM. As discussed in Sec. \ref{sec:edges}, astrophysical modeling seems unlikely to account for such large amplitude, possibly requiring non-standard scenarios with extra cooling mechanism to explain this absorption. Nevertheless, the timing of the EDGES absorption signal, perfectly consistent with standard $\Lambda$CDM predictions, can be employed in order to constrain other non-canonical scenarios which may impact the IGM history. In Ref. \cite{Lopez-Honorez:2018ipk}, two aspects of the redshift location of the EDGES signal are exploited in order to constrain IDM and WDM models, which are capable of delaying the onset of star formation during the cosmic dawn.

On the one hand, the minimum of the global 21 cm brightness temperature computed from simulations is demanded to appear not later than the EDGES absorption dip, located at $z \simeq 17$. On the other hand, by examining the beginning of the absorption regime in the EDGES data, one can infer when the Ly$\alpha$ coupling must be strong enough to deviate the spin temperature from the CMB one via the WF effect (as reviewed in Sec. \ref{sec:tspin}). Following that idea, in order to be consistent with data, one can assume that the Ly$\alpha$ coupling coefficient must become order one at $z \simeq 20$. Therefore, it is possible to rule out IDM and WDM models which present a too delayed absorption epoch and cannot fulfill the above conditions. Both aforementioned criteria allow extracting competitive constraints on both the WDM mass and the IDM cross section. These bounds strongly depend upon the specific astrophysical scenario, weakening the limits in scenarios with low virial temperatures. Nonetheless, assuming the default scenario of Pop II stars with a virial temperature of $10^4$ K as the source of the Ly$\alpha$ radiation, the obtained bounds from the EDGES timing largely improve the ones from other astrophysical probes.

\vspace{1cm}

\item \textbf{Constraining the primordial black hole abundance with 21-cm cosmology}

Reference \cite{Mena:2019nhm} studies in detail the cosmological effects of the existence of PBHs as part of the DM in the 21 cm cosmological signal. As overviewed in Sec. \ref{sec:signatures}, there are two main consequences of solar mass PBHs for the IGM evolution. On the one hand, the presence of PBHs in the early universe would imply an additional contribution to the matter power spectrum given by a shot noise spectrum independent on the scale, due to their discrete nature, which evolves as an isocurvature mode. This would magnify the formation of structures at small scales, where this term becomes dominant with respect to the standard adiabatic fluctuations. Hence, this enhancement mostly affects minihalos, low mass halos not massive enough to initiate star formation. The expected 21 cm sky-average brightness temperature coming from these minihalos is computed, which differs from the standard IGM global signal. It is shown that, even with the shot noise enhancement, it would not be enough to produce a detectable signal, contrary to previous claims in the literature \cite{Gong:2017sie, Gong:2018sos}.

On the other hand, like standard astrophysical BHs, PBHs would accrete their surrounding matter. During the infall onto the PBH, particles may be greatly accelerated, releasing UV and X-ray radiation to the environment. These photons would inject energy into the medium when absorbed, further heating and ionizing it. We compute the 21 cm signal arising from the local environment of PBHs, demonstrating that it would be well below observable sensitivities. However, energetic radiation can escape from the nearby surroundings of PBHs and be deposited into the IGM, leaving a strong signature on the 21 cm global signal and power spectrum from the IGM. Due to the further ionization and the hotter medium, the absorption amplitude of the brightness temperature and its fluctuations would be suppressed, leading to an emission regime in the most extreme cases. We run a MCMC analysis with semi-analytical simulations of the IGM 21 cm signal varying the fraction of PBHs as DM, $f_{\rm PBH}$, and their mass, $M_{\rm PBH}$, together with four astrophysical parameters. The expected sensitivities of the upcoming interferometers HERA and SKA are employed in order to explore the region of the PBH parameter space which may be probed by such experiments. We show that current bounds on the PBH abundance in the range $1-10^3 M_\odot$, coming from microlensing and the impact of accretion of the CMB, could be potentially improved up to two orders of magnitude by future 21 cm power spectrum observations.

\vspace{1cm}

\item \textbf{Removing Astrophysics in 21 cm maps with Neural Networks} 
 
The 21 cm line stands as a promising tool to map the distribution of matter in the universe, given that it allows tracing the location of neutral Hydrogen, which tracks the DM clumps. This may allow eventually achieving tomographic 3D density maps of the IGM. However, the link between the HI and DM density distribution is not unequivocal, since the brightness temperature strongly relies on the thermal and ionization state of the medium. Furthermore, once reionization is advanced, the signal from HII regions is mostly suppressed. In Ref. \cite{2021ApJ...907...44V}, we attempt to recover the underlying matter density field from a 21 cm map, making use of machine learning methods. Specifically, a Convolutional Neural Network (CNN) based on the so-called U-Net architecture \cite{pablo_villanueva_domingo_2021_4569964} is trained with 1000 numerical simulations of the brightness temperature field and its corresponding density maps, aimed to learn how to recover the latter from the former. These simulations include different initial conditions for each realization, varying also three relevant astrophysical parameters, in order to become robust to distinct astrophysical scenarios.

It is shown that, once it is trained, the CNN is able to successfully extract the density field from a given 21 cm map. To test its accuracy, several statistics of the matter density are computed, such as the power spectrum, the cross-correlation coefficient or the probability distribution function, matching the corresponding ones from the true maps within a few percent down to scales $k \sim 2$ Mpc$^{-1}$. The network can be applied to maps on different redshift, although its performance worsens at low redshifts, $z \sim 10$, when reionization is already ongoing and the link between the HI and the DM distribution becomes more diffuse. Furthermore, part of the trained layers of the network are employed to predict the underlying astrophysical parameters, showing that the CNN actually learns astrophysical information, substracting it from the 21 cm signal in order to predict the density field.

\end{enumerate}

%% file: Chapters/Papers/Papers.tex
\renewcommand{\headrulewidth}{0.4pt}
\lhead[{\bfseries \thepage}]{Part II: Scientific Research}
\rhead[{Part II: Scientific Research}]{\bfseries \thepage}

\cleardoublepage
\phantomsection\addcontentsline{toc}{chapter}{Warm dark matter and the ionization history of the Universe}
\includepdf[pages=1,scale=0.8,pagecommand={\pagestyle{fancy}}]{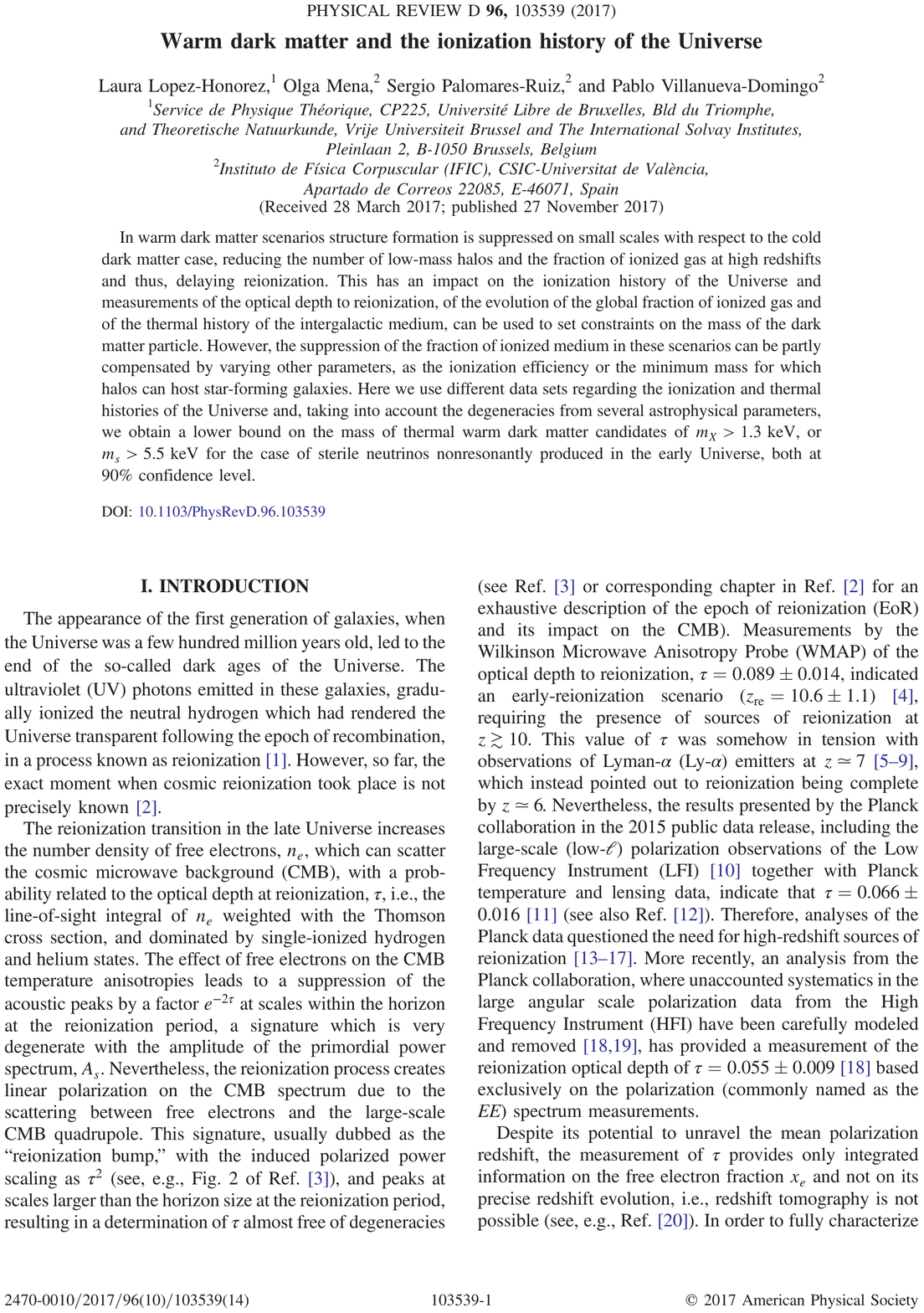}
\label{paper:WDM1}

\cleardoublepage
\phantomsection\addcontentsline{toc}{chapter}{Warm Dark Matter and Cosmic Reionization}
\includepdf[pages=1,scale=0.8,pagecommand={\pagestyle{fancy}}]{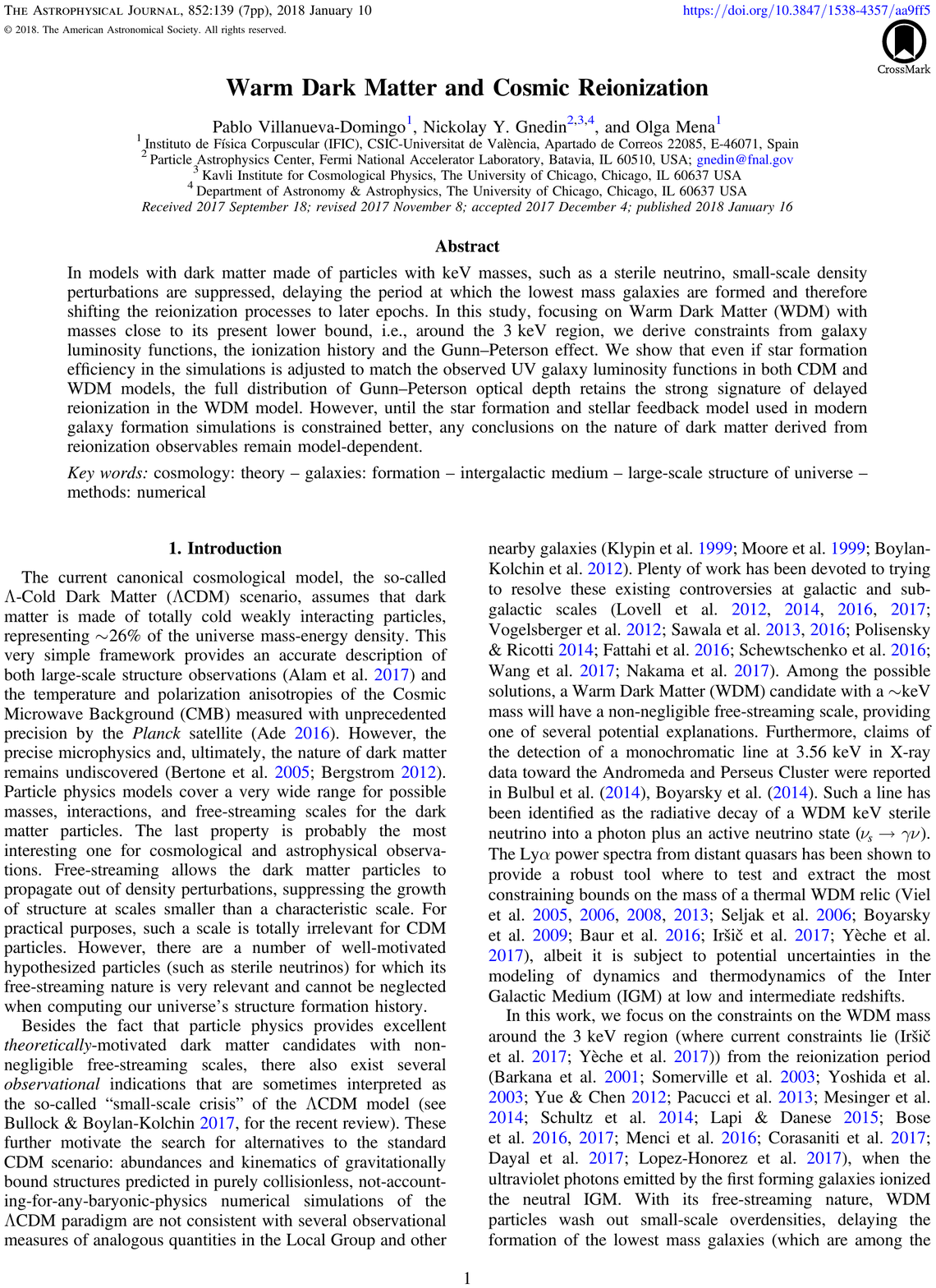}
\label{paper:WDM2}

\cleardoublepage
\phantomsection\addcontentsline{toc}{chapter}{Was there an early reionization component in our universe?}
\includepdf[pages=1,scale=0.8,pagecommand={\pagestyle{fancy}}]{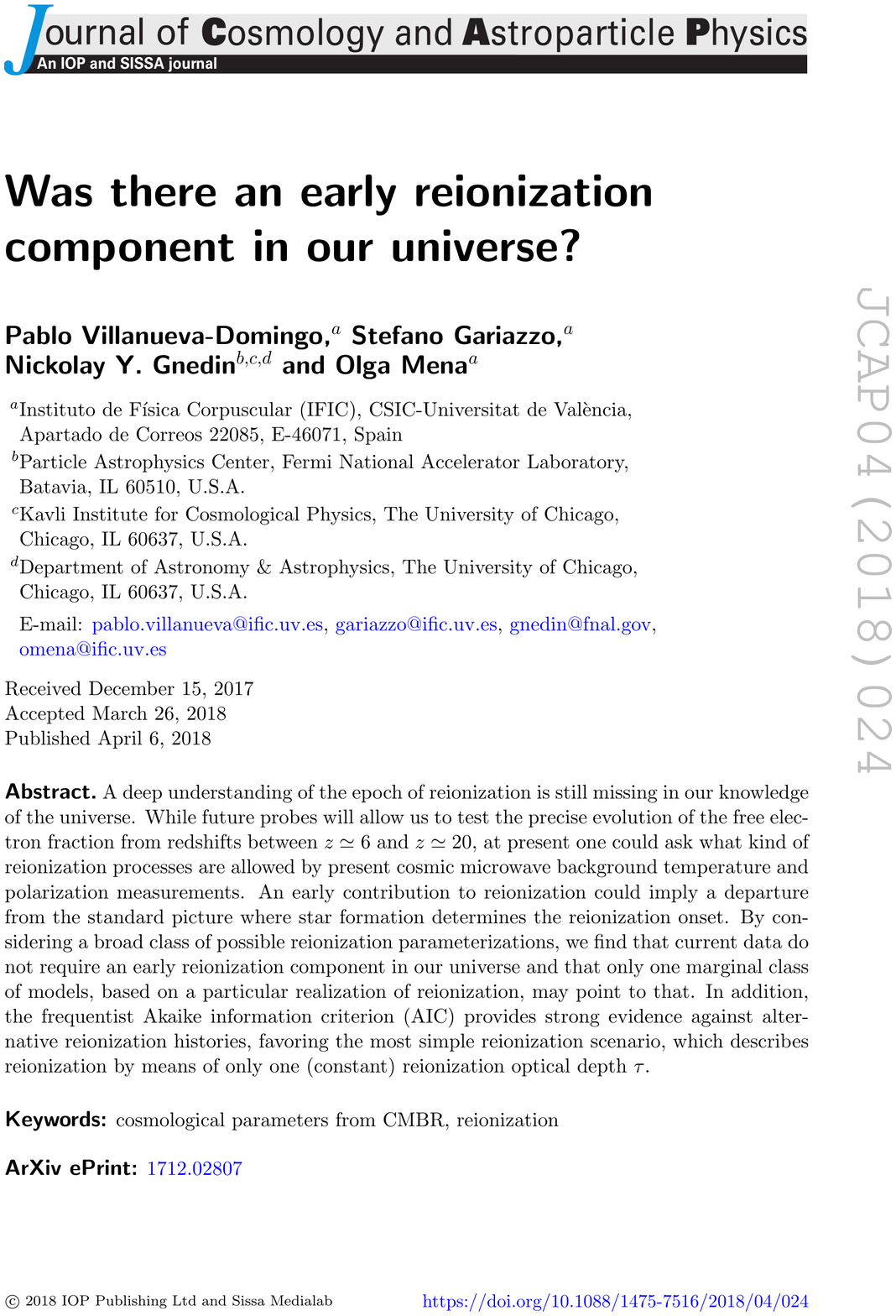}
\label{paper:earlyeor}

\cleardoublepage
\phantomsection\addcontentsline{toc}{chapter}{A fresh look into the interacting dark matter scenario}
\includepdf[pages=1,scale=0.8,pagecommand={\pagestyle{fancy}}]{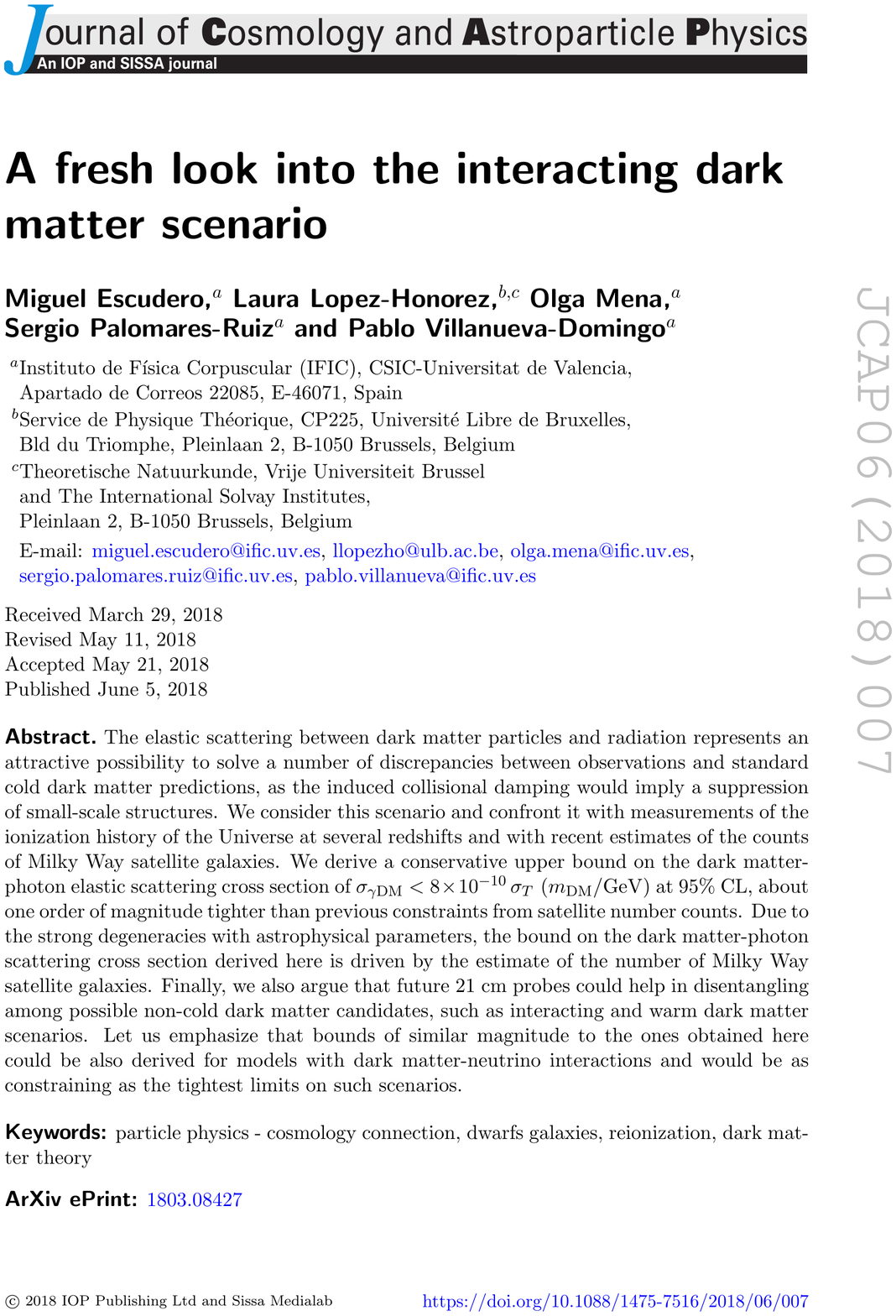}
\label{paper:IDM1}

\cleardoublepage
\phantomsection\addcontentsline{toc}{chapter}{Dark matter microphysics and 21 cm observations}
\includepdf[pages=1,scale=0.8,pagecommand={\pagestyle{fancy}}]{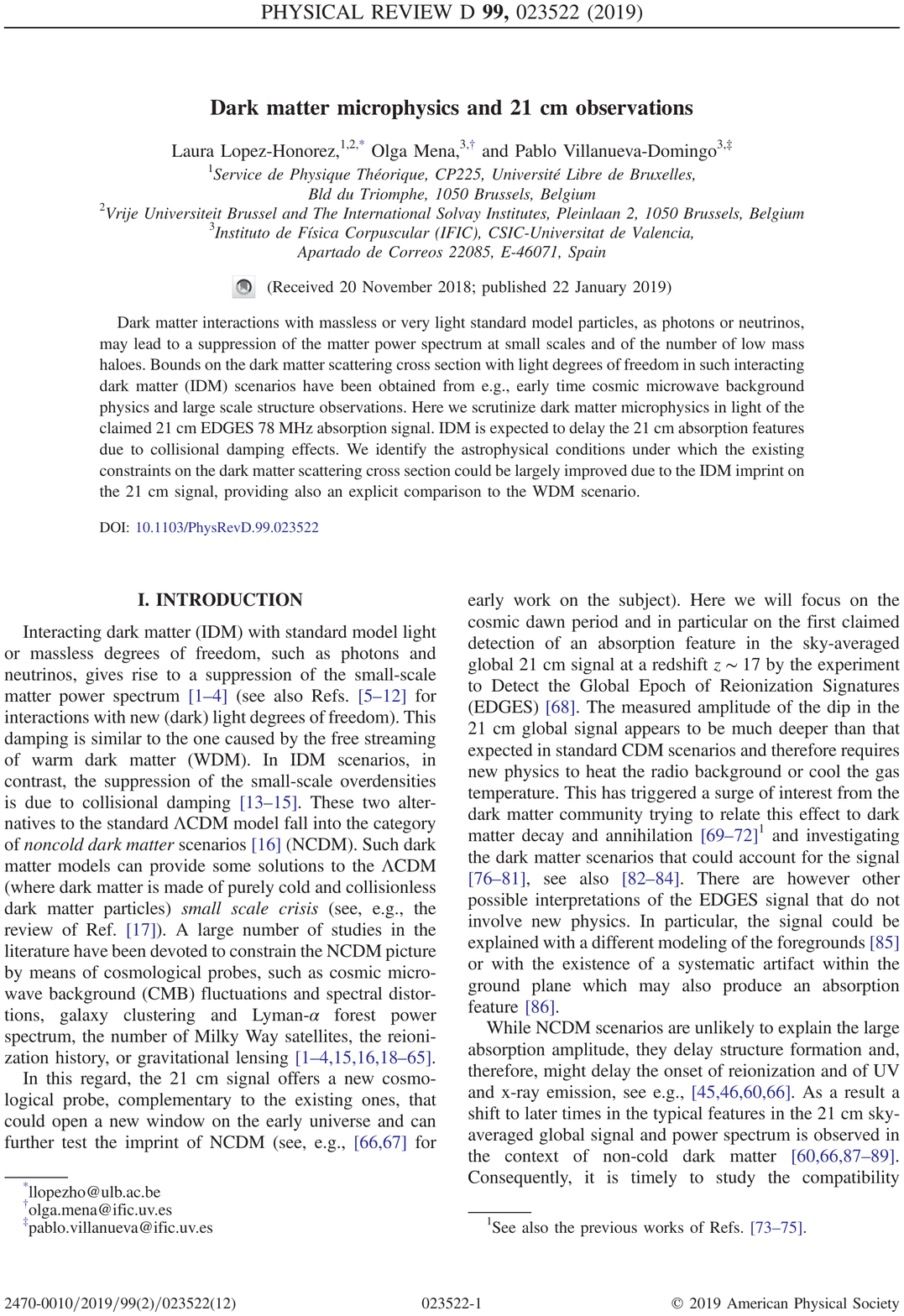}
\label{paper:IDM2}

\cleardoublepage
\phantomsection\addcontentsline{toc}{chapter}{Constraining the primordial black hole abundance with 21-cm cosmology}
\includepdf[pages=1,scale=0.8,pagecommand={\pagestyle{fancy}}]{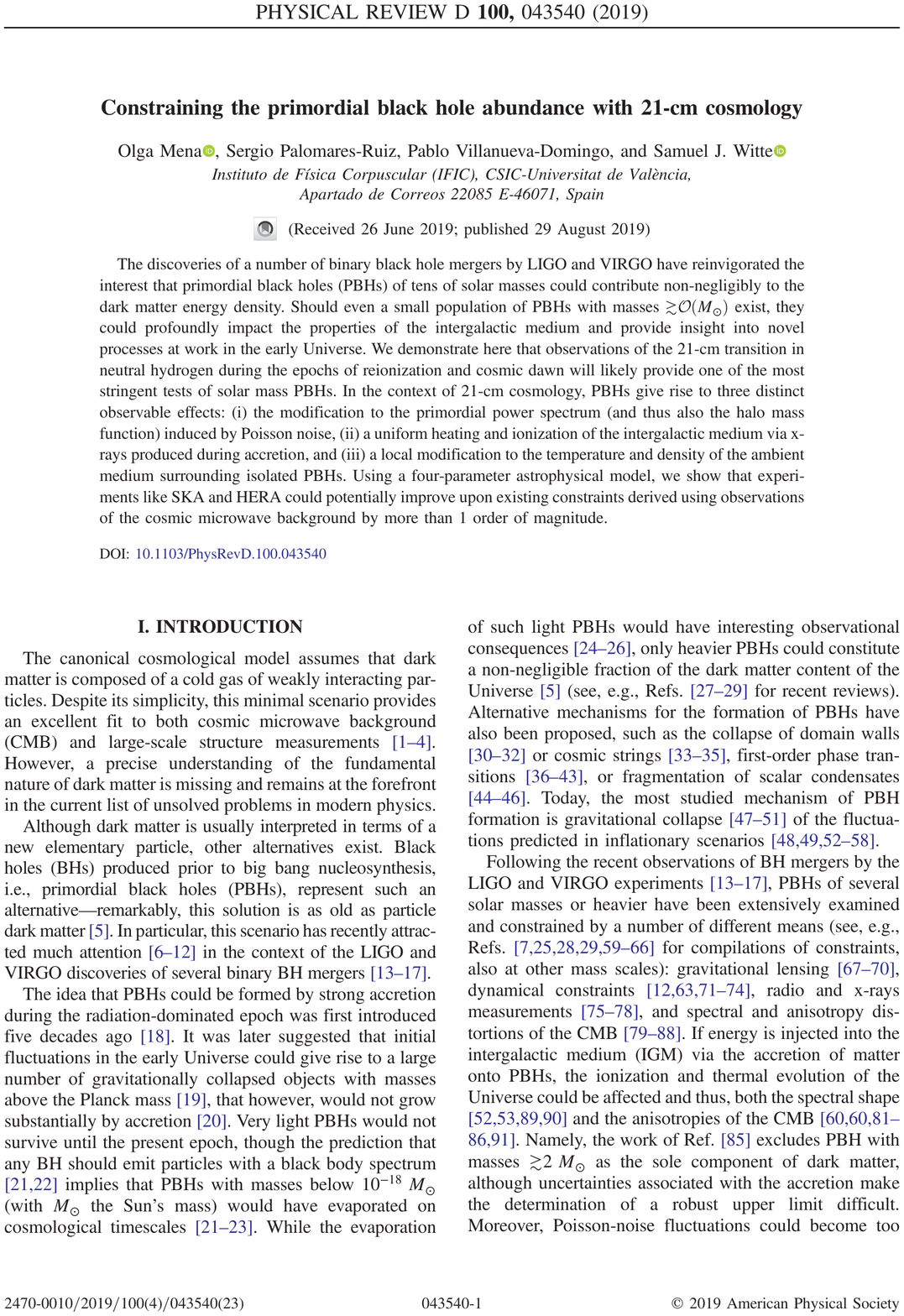}
\label{paper:PBH}

\cleardoublepage
\phantomsection\addcontentsline{toc}{chapter}{Removing Astrophysics in 21cm Maps with Neural Networks}
\includepdf[pages=1,scale=0.8,pagecommand={\pagestyle{fancy}}]{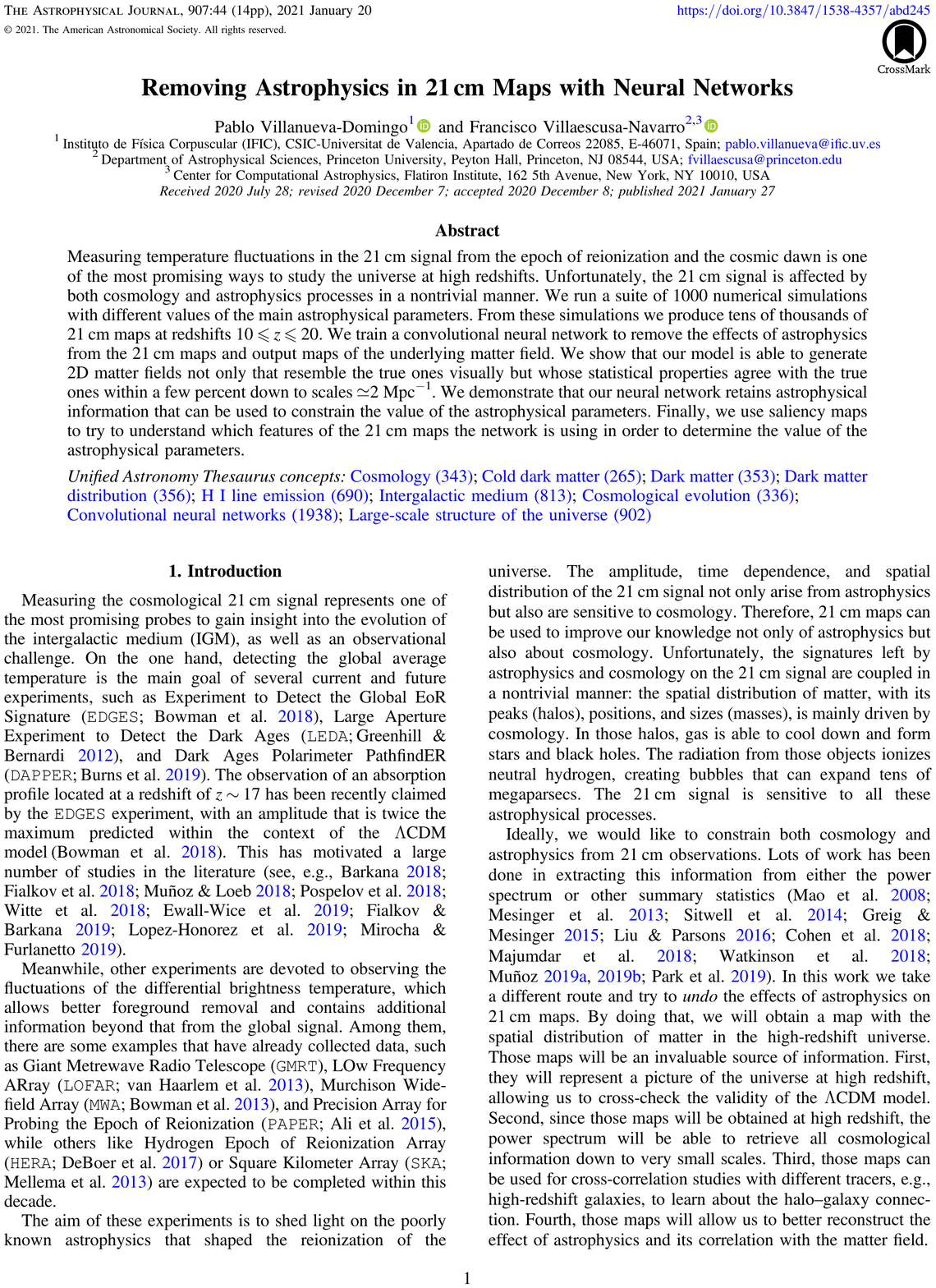}
\label{paper:CNN}

%% file: Chapters/Resumen_Tesis/Resumen_Tesis.tex
\lhead[{\bfseries \thepage}]{ \rightmark}
\rhead[Resumen de la tesis \leftmark]{\bfseries \thepage}
\part{Resumen de la tesis}
\label{partIII}

 \newcounter{alphasect}

 \renewcommand\thesection{%
 \ifnum\value{alphasect}=1%
A
 \else
\ifnum\value{alphasect}=2%
B
\else
\ifnum\value{alphasect}=3%
C
\else
\ifnum\value{alphasect}=4%
D
\else
 \arabic{section}
 \fi\fi\fi\fi}%

 \newenvironment{asection}{%
 \setcounter{alphasect}{1}
 }{%
 \setcounter{alphasect}{0}
 }%

 \newenvironment{bsection}{%
 \setcounter{alphasect}{2}
 }{%
 \setcounter{alphasect}{0}
 }%

\setcounter{section}{0}

\chapter*{Resumen de la tesis}
\addcontentsline{toc}{chapter}{Resumen de la tesis} 
\label{chap:resumen}

\section{Introducción}

Durante las últimas décadas, el enorme desarrollo de la cosmología
nos ha permitido alcanzar una profunda comprensión de nuestro universo, culminando en el establecimiento del paradigma estándar cosmológico $\Lambda$CDM. Este modelo asume diversos componentes para explicar la evolución del universo y las estructuras que observamos. Por un lado, la conocida como energía oscura, que conforma cerca de un 70\% de la energía actual. Esta presenta un comportamiento en la evolución cósmica equivalente a la constante cosmológica $\Lambda$, lo que causa la expansión acelerada del universo detectada a partir de supernovas lejanas. Por otra parte, diversas observaciones astrofísicas a lo largo del siglo XX, especialmente de las curvas de rotación galácticas y de cumúlos galácticos, hicieron evidente la presencia de un cierto tipo de materia no interactiva ni radiativa, conocida como materia oscura (\textit{Dark Matter}, DM), que comprende cerca de un 25\% de la energía cosmológica. Además, la mayor parte de observaciones sugieren que esta materia es relativamente fría, sin dispersión de velocidades térmica apreciable, siendo pues conocida como materia oscura fría (\textit{Cold Dark Matter}, CDM). Finalmente, la materia bariónica del Modelo Estándar de física de partículas tan solo contribuye en un 5\% a la fracción energética actual. Este paradigma cosmológico es capaz de explicar con precisión tanto observaciones de galaxias y supernovas lejanas como las anisotropías del Fondo Cósmico de Microondas (\textit{Cosmic Microwave Background}, CMB) y la nucleosíntesis primordial. Pese a ello, la naturaleza de la materia oscura y de la energía oscura permanecen esencialmente desconocidas.

Las observaciones de las anisotropías del CMB, el remanente fósil de fotones del universo primigenio, nos ofrecen información muy precisa sobre los diversos componentes del universo temprano, así como acerca del crecimiento de las perturbaciones primordiales. Por otra parte, los catálogos galácticos han permitido crear mapas cósmicos del universo cercano con gran detalle, los cuales permiten conocer cómo se distribuye la materia e inferir los procesos que han llevado a la formación de las estructuras cosmológicas que vemos actualmente. Sin embargo, entre el desacoplamiento del CMB y las galaxias de nuestro vecindario, existe un gran intervalo temporal esencialmente incierto y desconocido. Durante la época de dominación de materia, las regiones más densas del universo comienzan a colapsar formando las primeras estructuras cosmológicas. Este proceso se acrecienta durante las Edades Oscuras, periodo tras la Recombinación conocido así debido a la ausencia de fuentes de luz (estelares y primordiales) que permitan observarlo directamente. Con el paso del tiempo, llega un momento en que las inhomogeneidades son tan grandes que producen halos suficientemente masivos como para que la materia de su interior se enfríe y colapse formando estrellas. La época del nacimiento de las primeras estrellas se conoce como el Amanecer Cósmico, e implica la emisión por parte de las primeras galaxias de gran cantidad de radiación energética, en forma de rayos UV y X, que puede ser absorbida por los átomos del Medio Intergaláctico (\textit{Intergalactic Medium}, IGM). Esto se espera que caliente el gas difuso presente entre las galaxias y lo ionice. Cuando la formación estelar es lo suficientemente alta, el medio circundante alrededor de las galaxias se ioniza completamente, formando burbujas de hidrógeno ionizado que se expanden durante la Época de Reionización (\textit{Epoch of Reionization}, EoR), hasta que todo el medio intergaláctico alcance un estado ionizado. Si bien esta cronología cósmica es generalmente aceptada, todavía son desconocidos los detalles, duración y extensión de cada periodo. Aunque la observación del espectro de cuásares y galaxias, así como la polarización del CMB, pueden proporcionar cierta información relativa a la extensión y la consumación de la EoR, aún carecemos de una evaluación precisa de la historia de ionización del medio intergaláctico.

La señal cosmológica de 21 cm se constituye como una prometedora herramienta para entender la evolución del universo. Esta radiación emerge de la transición hiperfina del hidrógeno neutro, y dada la gran abundancia de este elemento en el universo, podría permitir realizar un cartografiado  de la distribución de materia en función del tiempo. Dada su sensibilidad respecto al estado térmico del medio, su detección aportaría valiosa información sobre la época de formación de primeras galaxias durante el Amanecer Cósmico. Además, dado que solo está presente en átomos neutros, proporcionaría un medio para poder conocer con precisión la historia de ionización durante la EoR. Por otra parte, la señal de 21 cm podría servir como una herramienta excepcional para ahondar en la naturaleza de la materia oscura. Aunque el modelo de materia oscura fría es el aceptado generalmente, existen otros candidatos bien motivados teóricamente que podrían proporcionar la materia oscura presente en el universo. Estos modelos de materia oscura alternativos podrían producir un impacto en la evolución del medio intergaláctico y tener por tanto efectos observables. Es por ello que la precisa determinación de las distintas épocas de la cronología cósmica resulta de gran importancia para acotar los distintos escenarios posibles de materia oscura. En esta tesis, se examina el potencial de la señal de 21 cm, junto con otras medidas del estado de ionización del medio intergaláctico, con el fin de acotar la composición, naturaleza y evolución de la materia oscura.

\section{Candidatos alternativos de materia oscura}

El caso paradigmático de materia oscura fría considera partículas masivas débilmente interactivas (\textit{Weakly Interacting Massive Particles}, WIMP), con masas usualmente del orden del GeV o superior, que estuvieron en equilibrio con el plasma cósmico en el universo primigenio. Estas interacciones dejarían de ser eficientes para mantener el acoplamiento con las partículas del Modelo Estándar en cierto momento, dejando un remanente que no interactúa con el resto de materia bariónica y radiación. Si bien esta hipótesis está bien motivada y es capaz de aportar la abundancia de materia oscura observada, las partículas constituyentes no han sido observadas en experimentos dedicados a su detección. Por otra parte, modelos de materia oscura fría presentan diversas discordancias con observaciones cosmológicas a escalas pequeñas, que pueden resumirse en el hecho de que simulaciones de materia oscura fría predicen más subestructuras de las halladas en las observaciones. Pese a que estos problemas podrían ser solucionados recurriendo a ciertos efectos astrofísicos, como retroalimentación estelar o de supernovas, otra posibilidad consiste en considerar modelos alternativos de materia oscura capaces de solucionar dichas discrepancias.

Una de las propuestas más atractivas se basa en la \textit{materia oscura templada} (\textit{Warm Dark Matter}, WDM), cuyas partículas tendrían una masa del orden del keV. Ejemplos de este tipo de materia oscura podrían ser los gravitinos de teorías supersimétricas, o bien neutrinos estériles, presentes en muchas extensiones del Modelo Estándar de partículas. Debido a su masa, su dispersión de velocidades y su temperatura serían notablemente mayores a las del caso frío. Esto sería relevante en el crecimiento de las fluctuaciones de materia, actuando como una presión efectiva, y contrarrestando el colapso gravitatorio en escalas pequeñas. Como consecuencia, estructuras con tamaños por debajo de una cierta escala serían fundamentalmente suprimidas. Esta escala actuaría como una escala de Jeans efectiva, y estaría determinada principalmente por la masa de la materia oscura templada y su abundancia. En estos modelos, la cantidad de halos de masas bajas sería reducida, explicando así algunas de las discrepancias presentes con la materia oscura fría. Este hecho tendría a su vez un impacto en la evolución del medio intergaláctico, ya que produciría menos galaxias de las esperadas, retrasando ciertos periodos de la cronología, como el amanecer cósmico o la reionización. Por ello, es de esperar que el análisis de observables del estado de ionización y de la señal de 21 cm sean sensibles a dichos efectos, y puedan potencialmente constreñir modelos de materia oscura templada.

Otra candidato alternativo podría tratarse de \textit{materia oscura interactiva} (\textit{Interacting Dark Matter}, IDM) con radiación. Pese a que hay cotas muy restrictivas respecto a las interacciones de la materia oscura con fotones, no es posible excluirlas por completo. Colisiones elásticas de partículas de materia oscura interactiva con fotones en el universo temprano producirían un acoplamiento de los primeros a los segundos, reduciendo las fluctuaciones de la materia oscura interactiva a escalas pequeñas, e imprimiéndoles en cierta manera el espectro oscilatorio típico de las perturbaciones adiabáticas de la radiación. En consecuencia, esto resultaría en una supresión del número de estructuras cosmológicas en pequeñas escalas, de una manera similar al caso de materia oscura templada, lo que ha sido propuesto como solución a algunas de las discordancias del escenario de materia oscura fría con los datos. De la misma manera, el número de halos que albergase formación estelar se reduciría, y por tanto, observaciones de la absorción y emisión en 21 cm y límites sobre la historia de ionización serían capaces de acotar el espacio de parámetros de la materia oscura interactiva compatible con los datos.

Finalmente, cabe la posibilidad de que los constituyentes de la materia oscura no sean partículas microscópicas, sino objetos astrofísicos macroscópicos masivos, conocidos como MACHOs. La mayoría de estos candidatos, como podrían ser planetas gigantes y enanas marrones, son descartados por no ser capaces de originar las fluctuaciones primordiales patentes en el CMB y la cantidad de bariones en la nucleosíntesis primordial. No obstante, el colapso de \textit{agujeros negros primordiales} (\textit{Primordial Black Holes}, PBHs) durante el universo temprano sería consistente con dichas restricciones, por su naturaleza previa a la formación estelar, y podrían formar parte de al menos una fracción de la materia oscura. Esta hipótesis recobró especial atención tras las observaciones por parte de las colaboraciones LIGO y VIRGO de diversos eventos de emisión de ondas gravitatorias originadas en la fusión de agujeros negros de varias masas solares \cite{PhysRevLett.116.061102}. Contrariamente a los agujeros negros astrofísicos estándares, limitados a masas iguales o superiores a $\sim 3 M_\odot$, los agujeros negros primordiales podrían ser formados en principio con cualquier masa. Estos objetos podrían además actuar como germen de los agujeros negros supermasivos observados en los centros de la mayoría de las galaxias, cuyas masas son difícilmente explicables mediante mecanismos de acrecimiento estándares, a no ser que partiesen de una masa inicial anterior a la época de formación estelar. Uno de los aspectos más interesantes de los agujeros negros primordiales como candidatos de la materia oscura es la gran cantidad de efectos observables que su existencia implicaría, permitiendo acotar su abundancia hasta niveles potencialmente muy bajos. Ejemplos de ello son efectos de lente gravitatoria, la emisión de partículas debido a su evaporación, la inyección de energía en el medio circundante por el acrecimiento de materia, o el incremento en las fluctuaciones primordiales debido a su carácter discreto. Estos dos últimos efectos podrían afectar potencialmente a la evolución del medio intergaláctico, y por tanto, a la línea de 21 cm.

\section{Metodología}

La mayor parte de la tesis se focaliza en explorar diversos modelos de materia oscura alternativos al estándar de materia oscura fría mediante simulaciones de la evolución del universo, con el fin de contrastarlos con datos observacionales para acotar sus propiedades. Para ello, se han llevado a cabo simulaciones del crecimiento de las fluctuaciones del campo de materia y la formación de halos, así como de la evolución térmica, y de ionización del medio intergaláctico. En la mayor parte de los artículos incluidos, se ha hecho uso del programa de acceso libre {\tt 21cmFAST} \cite{Mesinger:2010ne}, que emplea una serie de aproximaciones semi-analíticas para calcular la evolución del medio intergaláctico. Esencialmente, este código resuelve la evolución del campo de densidad empleando teoría de perturbaciones lagrangiana de segundo orden. A partir del mapa de densidad, el programa calcula las regiones ionizadas alrededor de las zonas con mayor densidad, siguiendo la prescripción de Furlanetto-Zaldarriaga-Hernquist (FZH) \cite{Furlanetto:2004nh}, la cual se basa en la aplicación del formalismo de conjunto de excursiones (\textit{excursion set formalism}) \cite{Bond:1990iw}. Estas simulaciones concuerdan con cálculos más rigurosos empleando simulaciones hidrodinámicas o de N cuerpos para escalas relativamente grandes, a partir de $\sim 1 $ Mpc. Sin embargo, para el análisis presentado en la Ref. \cite{Villanueva-Domingo:2017lae}, se han empleado  simulaciones hidrodinámicas más precisas, basadas en el programa de Reionización Cósmica en Computadoras (\textit{Cosmic Reionization on Computers}, CROC) \cite{Gnedin:2014uta}, que emplean un tratamiento preciso de la evolución de la densidad, del transporte radiativo, y por tanto, un cálculo más robusto del campo de ionización. Ambos métodos asumen diversos modelos astrofísicos cuyos parámetros no están bien determinados por las observaciones, y por tanto, los resultados finales son sensibles a la elección específica de dichos parámetros.

Para el análisis de resultados y el contraste con datos observacionales, se emplea tanto un enfoque probabilista frecuentista, basado en el máximo de la función de verosimilitud (o, equivalentemente, en el mínimo del $\chi^2$), como métodos bayesianos, muestreando la función de probabilidad posterior mediante  Cadenas de Markov de Monte Carlo (\textit{Monte Carlo Markov Chains}, MCMC). Finalmente, en la Ref. \cite{2021ApJ...907...44V}, se hacen uso de métodos de aprendizaje profundo (\textit{deep learning}), empleando redes neuronales convolucionales.

\section{Resultados de la tesis}

Esta sección comprende un resumen de los trabajos de investigación originales que conforman el cuerpo de la tesis doctoral. A continuación, se presentan la motivación y los resultados más relevantes de cada artículo.

\subsection{Materia oscura cálida e historia de ionización del Universo}
\label{sec:wdm_spa}

En la Ref. \cite{Lopez-Honorez:2017csg} se estudia el impacto de modelos de materia oscura templada en la evolución térmica y de ionización del medio intergaláctico. Debido al efecto de dispersión de velocidades de partículas con masas $ \sim $ keV, la formación de estructuras es suprimida en escalas pequeñas con respecto al caso estándar de materia oscura fría. Esto implica una reducción en el número de halos de baja masa que pueden albergar formación estelar, capaces de emitir radiación UV y rayos X responsables de ionizar y calentar el medio intergaláctico. La disminución del número de tales fuentes conduce a un retraso del amanecer cósmico, de la época de calentamiento y de la EoR. Haciendo uso de simulaciones semi-analíticas, la evolución de la temperatura cinética y de la fracción de ionización se calcula en varios escenarios de materia oscura templada, variando la masa de sus partículas entre $ 1 $ y $ 4 $ keV, un rango motivado físicamente para tratar de resolver los problemas a pequeña escala presentes con la materia oscura fría. Dado que la historia térmica del medio intergaláctico es muy sensible a los procesos estelares y radiativos involucrados, también consideramos diferentes escenarios astrofísicos. Se modifican tres parámetros astrofísicos: la temperatura virial mínima (relacionada con la masa mínima del halo para producir formación estelar), la eficiencia de ionización de rayos UV y la eficiencia de calentamiento de los rayos X (relacionada con la luminosidad en rayos X de fuentes astrofísicas, como binarias de rayos X).

Los resultados obtenidos en lo que respecta a la fracción de ionización de las simulaciones se han confrontado con límites observacionales sobre el estado de ionización durante la EoR. Entre los tipos de datos tenidos en cuenta, se hallan las restricciones y los límites inferiores de la fracción de ionización en el rango de desplazamientos al rojo (\textit{redshifts}) $ 8 \geq z \gtrsim 6 $ de espectros de cuásares y las galaxias emisoras de radiación Ly$ \alpha $. Por otro lado, también empleamos la profundidad óptica de Thomson inferida de las anisotropías del CMB, obtenida mediante la integral de la fracción de ionización sobre el redshift, en lugar de una restricción dependiente del corrimiento al rojo. Se muestra que los límites que emplean solo datos de CMB son menos restrictivos que los obtenidos usando el estado de ionización a un tiempo dado, dado que la profundidad óptica de Thomson es una cantidad integrada, y no una función del tiempo.

Encontramos que algunos parámetros astrofísicos están fuertemente degenerados entre ellos, y a su vez, con la masa de la materia oscura templada. Más específicamente, el retraso de la EoR presente en un modelo de materia oscura templada puede ser contrarrestado aproximadamente por una alta eficiencia de ionización (que impulsa una reionización más temprana) o por una temperatura virial mínima más baja (lo que permite que halos de menor masa formen estrellas y contribuyan a emitir más radiación ionizante). Por esa razón, los límites inferiores de la masa de la materia oscura templada obtenidos por este método son más débiles que aquellos inferidos a partir del bosque de Ly$ \alpha $, aunque son capaces de limitar masas de la materia oscura templada menores de $ \sim 1 $ keV.

\subsection{Materia oscura cálida y reionización cósmica}

Con el fin de profundizar en la fenomenología de la materia oscura templada, en la Ref. \cite{Villanueva-Domingo:2017lae} se continúa investigando el impacto de estos modelos en la evolución cosmológica y la formación de estructuras. En este caso, se emplean precisas simulaciones hidrodinámicas, en lugar de cálculos semi-analíticos, proporcionando así resultados más fiables. Hacemos uso de simulaciones basadas en el programa \textit{Cosmic Reionization on Computers} (CROC) \cite{Gnedin:2014uta}, que representan la vanguardia en el modelado numérico de la EoR. En las condiciones iniciales de cada simulación, el espectro de potencias es modificado adecuadamente para incluir la supresión característica de escenarios de materia oscura templada. Se asume una masa de las partículas de $ m_{\rm WDM} = 3 $ keV, valor que se encuentra alrededor de las restricciones obtenidas mediante el bosque de Ly$ \alpha $.

Se calculan diferentes observables de reionización para explorar el impacto de la escala de supresión de estructuras, como la función de masa de los halos y las funciones de luminosidad UV de la galaxia, que muestran una supresión (en masas bajas y magnitudes altas, respectivamente) con respecto al caso estándar. Los telescopios actuales no parecen ser capaces de discriminar entre estos casos de materia oscura fría y templada. Sin embargo, el futuro  telescopio \textit{James Webb Space Telescope} (JWST) sería capaz de alcanzar el extremo de la función de luminosidad de galaxias de magnitudes débiles, siendo apto para restringir tales masas de la materia oscura templada. También se discute el efecto en la función de distribución de probabilidad cumulativa para la profundidad óptica Ly$ \alpha $ efectiva $ \tau_{\rm GP} $ en el medio intergaláctico posterior a la reionización. En comparación con el escenario de materia oscura fría en un tiempo dado, el caso de materia oscura templada presenta distribuciones con colas más extendidas hacia valores de $ \tau_{\rm GP} $ altos, que corresponden a regiones de alta densidad, ya que se ionizan posteriormente. Comparando con los datos observacionales del bosque de Ly$ \alpha $, el caso de materia oscura fría resulta preferido respecto al caso de materia oscura templada, aunque no está claro si el ajuste de los parámetros astrofísicos en el escenario de materia oscura templada podría mejorar su concordancia con los datos. Por último, también se calcula el espectro de potencias de la línea de 21 cm, que en el caso de materia oscura templada muestra un balance entre una supresión debida a la dispersión térmica de velocidades, y un realzamiento dado por la reionización retardada. Esto conduce a un comportamiento no trivial, que lo hace difícilmente detectable por la primera generación de experimentos de 21 cm.

\subsection{¿Hubo un componente de reionización temprano en nuestro universo?}

La historia de ionización del universo durante la EoR está lejos de encontrarse bien determinada. Aunque la mayoría de los límites son consistentes con un proceso de reionización relativamente rápido que termina a redshifts $ z \sim 6 $, aún puede ser posible una contribución temprana, por ejemplo, de escenarios de materia oscura no canónicos que inyecten energía en el medio, como materia oscura desintegrándose o agujeros negros primordiales. En la Ref. \cite{Villanueva-Domingo:2017ahx}, esta cuestión se examina a la luz de los datos de Planck del CMB. La historia de ionización se modela para que sea coherente con el espectro de polarización EE del CMB, que es sensible a la reionización en multipolos bajos. Se emplean varios métodos para emular la evolución de la fracción de ionización $x_e$. En primer lugar, se consideran dos parametrizaciones funcionales frecuentes en la literatura, como la parametrización simétrica y la asimétrica en $z$. Por otro lado, se aplica una reconstrucción de $x_e$ basada en el Análisis de Componentes Principales (\textit{Principal Component Analysis}, PCA), extrayendo los modos propios de la matriz de Fisher de los componentes $ C_l $ del espectro EE. Se espera que aquellos con las varianzas más pequeñas contengan la información cosmológica más valiosa. Un subconjunto de estas funciones propias se emplea como base para escribir como su combinación el historial de ionización alrededor de un modelo por defecto dado, ya sea asumiendo un valor constante o una de las parametrizaciones mencionadas anteriormente. Finalmente, se emplea un tercer método basado en el polinomio de interpolación de Hermite cúbico por partes (\textit{Piecewise Cubic Hermite Interpolating Polynomial}, PCHIP), considerando como parámetros libres la fracción ionizada en varios $z$ diferentes, que se interpolan suavemente a través del PCHIP.

Comparamos los diferentes enfoques para recuperar la evolución de la ionización con un análisis empleando Cadenas de Markov de Monte Carlo (MCMC), que muestra que algunos de ellos prefieren un componente de reionización temprano, incluido algún patrón oscilatorio, como el uso de la combinación de PCA sumada a un modelo por defecto. Sin embargo, cuando se incluye la combinación PCA como argumento de la parametrización, explotando su forma funcional, la evolución de $x_e$ resultante es más suave y no muestra un inicio temprano. La aplicación de los criterios frecuentista de información de Akaike y de información bayesiano confirman que ninguna de las formulaciones en particular es preferida por los datos sobre el resto, y por lo tanto, que los datos de CMB no indican una reionización temprana, contrariamente a afirmaciones anteriores en la literatura \cite{Heinrich:2016ojb}.

\subsection{Una nueva mirada al escenario de la materia oscura interactiva}

En la Ref. \cite{Escudero:2018thh}, los modelos fenomenológicos de materia oscura que involucran la dispersión elástica con fotones, conocidos como materia oscura interactiva, se discuten en el contexto de la evolución térmica del medio intergaláctico y del número de galaxias satélites de la Vía Láctea. En el caso de la materia oscura interactiva, el acoplamiento entre partículas de materia oscura y fotones produce un efecto de amortiguación debido a las colisiones, induciendo una supresión y un patrón oscilatorio en las escalas más pequeñas en el espectro de potencias, de manera similar a los inducidos por el acoplamiento barión-fotón en el caso estándar. Este efecto puede reducir el número de estructuras a pequeña escala y halos de baja masa, de manera similar a lo que ocurre con la materia oscura templada. La magnitud de este amortiguamiento está determinada por la sección eficaz de dispersión elástica sobre la masa, $ \sigma_{\gamma \rm DM} / m_{\rm DM} $.

Se han realizado simulaciones semi-analíticas del estado térmico y de ionización del medio intergaláctico para investigar el impacto de escenarios con materia oscura interactiva. La supresión del crecimiento de estructuras debido a la amortiguación de colisiones retrasa la formación de las primeras estrellas y galaxias, retardando todos los procesos astrofísicos como el calentamiento por rayos X y la reionización. Análogamente a lo visto en la sección \ref{sec:wdm_spa} (Ref. \cite{Lopez-Honorez:2017csg}) con respecto a la masa de la materia oscura templada, la sección eficaz de la materia oscura interactiva está fuertemente degenerada con varios parámetros astrofísicos, principalmente con la eficiencia de ionización y la temperatura virial mínima. Por otro lado, se calcula el número de subhalos esperados, que en un escenario de materia oscura interactiva se reduce con respecto al escenario de materia oscura fría. Este hecho permite acotar la sección eficaz elástica al comparar las predicciones teóricas con las galaxias satélite de la Vía Láctea observadas. Sin embargo, la restricción a partir de las galaxias satélite depende en gran medida de la masa total asumida de la Vía Láctea, que no está completamente determinada con certeza. Dentro de su rango de masa esperado, los límites de la sección eficaz pueden variar hasta un orden de magnitud. Combinando este límite superior con los límites de la historia de ionización del medio intergaláctico, se pueden romper algunas degeneraciones astrofísicas.

\subsection{Microfísica de materia oscura y observaciones de 21 cm}

La colaboración de EDGES presentó la supuesta primera detección de la señal global de 21 cm, presentando un espectro de absorción con una amplitud de aproximadamente $ 500 $ mK, que duplica las expectativas en el caso estándar $ \Lambda $CDM. Esto resulta ser difícilmente explicable solo a partir de distintos modelos astrofísicos, posiblemente requiriendo escenarios no estándar con un mecanismo de enfriamiento adicional para explicar esta absorción, como materia oscura interaccionando con bariones. Sin embargo, la localización en el tiempo de la señal de absorción medida por EDGES, perfectamente consistente con las predicciones estándar en escenarios $ \Lambda $CDM, se puede emplear para restringir otros escenarios no canónicos de materia oscura que pueden afectar la evolución del medio intergaláctico. En la Ref. \cite{Lopez-Honorez:2018ipk}, se explotan dos aspectos de la ubicación temporal de la señal de EDGES para restringir modelos de materia oscura interactiva y templada, que son capaces de retrasar el inicio de la formación estelar durante el amanecer cósmico.

Por un lado, se exige que el mínimo de la temperatura global de brillo de 21 cm calculada a partir de simulaciones aparezca no más tarde del pico de absorción de EDGES, situado en $ z \simeq 17 $. Por otro lado, al examinar el comienzo del régimen de absorción en los datos de EDGES, se puede inferir cuándo el acoplamiento Ly$ \alpha $ debe ser lo suficientemente fuerte como para acoplar la temperatura de espín con la temperatura cinética del medio intergaláctico. Siguiendo estas ideas, para ser coherente con los datos, se puede suponer que el coeficiente de acoplamiento Ly$ \alpha $ debe alcanzar el orden uno a $ z \simeq 20 $. Por tanto, es posible descartar los modelos materia oscura interactiva y templada que presentan una época de absorción demasiado retrasada y no pueden cumplir las condiciones anteriores. Ambos criterios mencionados permiten extraer restricciones competitivas en la masa de la materia oscura templada y en la sección eficaz de la interactiva. Estos límites son fuertemente dependientes del escenario astrofísico específico, debilitando los límites en escenarios con bajas temperaturas viriales. No obstante, asumiendo el escenario predeterminado de estrellas de población II (de baja metalicidad) con una temperatura virial de $ 10^4 $ K como fuente de radiación Ly$ \alpha $, los límites obtenidos mediante este método pueden mejorar en gran medida aquellos obtenidos a partir de otros análisis.

\subsection{Restringiendo la abundancia de agujeros negros primordiales con cosmología de 21 cm}

En la Ref. \cite{Mena:2019nhm} se estudia en detalle los efectos cosmológicos de la existencia de agujeros negros primordiales como parte de la materia oscura en la señal cosmológica de 21 cm. Hay dos consecuencias principales de agujeros negros de origen primitivo de masa solar o superior para la evolución del medio intergaláctico. Por un lado, la presencia de agujeros negros primordiales en el universo temprano implicaría una contribución adicional al espectro de potencias de la materia dado por un espectro de ruido blanco de Poisson, independiente de la escala, debido a su naturaleza discreta, evolucionando como un modo de isocurvatura. Esto magnificaría la formación de estructuras a pequeña escala, donde este término se vuelve dominante con respecto a las fluctuaciones adiabáticas estándares. Por lo tanto, este incremento afectaría principalmente a los minihalos, aquellos halos de baja masa que no son lo suficientemente masivos para iniciar la formación de estrellas. La temperatura de brillo promedio de 21 cm esperada proveniente de estos minihalos es calculada, que difiere de la señal global estándar del medio intergaláctico. En el artículo se demuestra que, incluso con el incremento en el espectro de potencias a pequeñas escalas debido a la contribución de Poisson, no sería suficiente para producir una señal detectable, contrariamente a lo que se ha afirmado anteriormente en la literatura \cite{Gong:2017sie, Gong:2018sos}.

Por otro lado, al igual que los agujeros negros astrofísicos estándares, los agujeros negros primordiales acrecerían materia circundante debido a su atracción gravitatoria. Durante la caída hacia el PBH, las partículas podrían acelerarse enormemente, liberando radiación UV y de rayos X al ambiente circundante. Estos fotones inyectarían energía en el medio cuando se absorbieran, calentándolo e ionizándolo aún más. En el artículo, calculamos la señal de 21 cm que surge del entorno local de los PBH, mostrando que estaría muy por debajo de las sensibilidades observables. Sin embargo, la radiación energética puede escapar de los alrededores cercanos de los agujeros negros primordiales y depositarse en el medio intergaláctico, afectando notablemente la señal global de 21 cm y el espectro de potencias. Debido a la ionización adicional y al medio más caliente, se suprimiría la amplitud de absorción de la temperatura de brillo y sus fluctuaciones, implicando una transición a un régimen de emisión en los casos más extremos. Realizamos un análisis MCMC con simulaciones semi-analíticas de la señal del medio intergaláctico de 21 cm, variando la fracción de agujeros negros primordiales como materia oscura, $ f_{\rm PBH} $, y su masa, $ M_{\rm PBH} $, junto con cuatro parámetros astrofísicos. Las sensibilidades esperadas de los próximos interferómetros HERA y SKA se emplean para explorar la región del espacio de parámetros de agujeros negros primordiales que puede ser testada por tales experimentos. Finalmente, mostramos que los límites actuales en la abundancia de agujeros negros primordiales en el rango de $ 1-10 ^ 3 M_\odot $, provenientes del efecto de lente gravitatoria y el impacto de la acrecimiento en el CMB, podrían mejorarse potencialmente hasta en dos órdenes de magnitud con observaciones del espectro de potencias de 21 cm.

\subsection{Eliminando la astrofísica en mapas de 21 cm con redes neuronales}
 
La línea de 21 cm se erige como una prometedora herramienta para mapear la distribución de materia en el universo, dado que permite localizar la ubicación del hidrógeno neutro en los cúmulos de materia oscura. Esto podría permitir lograr mapas tomográficos de densidad en 3D del medio intergaláctico. Sin embargo, la relación entre la distribución de hidrógeno neutro y densidad de materia oscura no es unívoca, ya que la temperatura de brillo depende en gran medida del estado térmico y de ionización del medio. Además, una vez que la reionización es muy avanzada, la señal de las regiones de hidrógeno ionizado se suprime en su mayor parte. En la Ref. \cite{2021ApJ...907...44V}, se trata de recuperar el campo de densidad de materia subyacente a partir de un mapa de 21 cm, utilizando métodos de aprendizaje automático. En concreto, se entrena una red neural convolucional  basada en la arquitectura U-Net \cite{pablo_villanueva_domingo_2021_4569964} con 1000 simulaciones numéricas del campo de temperatura de brillo y sus correspondientes mapas de densidad, con el objetivo de aprender a recuperar este último a partir del mapa de 21 cm. Estas simulaciones incluyen diferentes condiciones iniciales para cada realización, variando también tres parámetros astrofísicos relevantes, con el fin de volverse robustos a distintos escenarios astrofísicos.

Se demuestra que, una vez entrenada, la red neuronal es capaz de extraer con éxito el campo de densidad de un mapa de 21 cm dado. Para probar su precisión, se calculan varios estadísticos del campo de densidad de materia, como el espectro de potencias, el coeficiente de correlación cruzada o la función de distribución de probabilidad, coincidiendo las predicciones con las correspondientes de los mapas reales con alta precisión hasta escalas tan pequeñas como $ k \sim 2 $ Mpc $^{- 1} $. La red se puede aplicar a mapas en diferentes momentos de la evolución cósmica, aunque su rendimiento empeora con instantes más recientes, $ z \sim 10 $, cuando la reionización ya está en curso y el vínculo entre la distribución de hidrógeno neutro y materia oscura se vuelve más difuso. Además, parte de las capas entrenadas de la red se emplean para predecir los parámetros astrofísicos subyacentes, lo que muestra que la red neuronal realmente aprende información astrofísica, con el fin de sustraerla de la señal de 21 cm para predecir el campo de densidad.

\section{Conclusiones}

Los trabajos de investigación incluidos en la presente tesis doctoral pretenden ejemplificar diversos enfoques para estudiar la naturaleza y propiedades de la materia oscura a partir del estudio de la evolución del medio intergaláctico. Diferentes modelos alternativos de materia oscura son considerados, como materia oscura templada, interactiva y agujeros negros primordiales, los cuales presentan diferentes efectos en la formación de estructuras y en la historia térmica. Los casos de materia oscura templada e interactiva presentan una supresión en las fluctuaciones a pequeñas escalas, dada por la dispersión de velocidades térmica y por el amortiguamiento colisional, respectivamente, retrasando la formación de estructuras cosmológicas y con ello todos los procesos astrofísicos. Por otra parte, agujeros negros primordiales de masa solar o mayor pueden acrecer materia, radiando e inyectando posteriormente energía en el medio, calentándolo e ionizándolo. Todos estos efectos dejarían un gran impacto en la evolución del medio intergaláctico, afectando a la época de formación de primeras estrellas, y a la EoR. Como se ha mostrado, tests observacionales del progreso de la reionización, como las anisotropías del CMB y los espectros de absorción de quásares lejanos, podrían por tanto ser capaces de conferir cotas sobre los diferentes modelos de materia oscura. Las restricciones sobre el espacio de parámetros de la materia oscura obtenidas de este modo son, sin embargo, altamente sensibles al modelado astrofísico del medio intergaláctico, y una precisa determinación de los distintos procesos y fuentes astrofísicas es necesaria para obtener límites robustos. Datos de la historia de ionización como el CMB también pueden aplicarse a casos donde un modelo específico no es asumido, estudiando por ejemplo procesos de reionización temprana, que podrían ser inducidos por partículas de materia oscura que inyectasen energía en el medio.

Especial atención merece la señal cosmológica de 21 cm, dada su gran sensibilidad al estado térmico y de ionización del medio, siendo un observable de gran interés. Empleando la localización temporal de la señal global de absorción detectada por EDGES, es posible acotar diversos modelos de materia oscura como templada e interactiva, mejorando algunas de las restricciones actuales mediante otros métodos. Sin embargo, aún más prometedor resulta el espectro de potencias de 21 cm, el cual sería capaz de arrojar información sobre las inhomogeneidades del medio intergaláctico, además de ser más fácilmente separable de los contaminantes astrofísicos, que pueden exceder en varios órdenes de magnitud la señal cosmológica. Como ejemplo de su potencial, se han calculado las posibles futuras cotas sobre la abundancia de agujeros negros primordiales como materia oscura asumiendo la sensibilidad de detección de la siguiente generación de radiointerferómetros, como SKA, mostrando que dichas restricciones podrían mejorar en varios órdenes de magnitud las presentes actualmente en el mismo rango de masas mediante otros métodos.

Además, se presentan aplicaciones de aprendizaje automático en cosmología, con el fin de obtener el campo de densidad de materia subyacente y los parámetros astrofísicos a partir de un mapa de 21 cm dado. Esto muestra, por un lado, el gran potencial de la señal hiperfina del hidrógeno como herramienta para estudiar el crecimiento de las inhomogeneidades y la formación y distribución de estructuras. Y por otra parte, la gran capacidad y aplicaciones que pueden tener los métodos basados en redes neuronales para extraer información cosmológica, la cual puede ser inalcanzable de otro modo.